\documentclass[11pt,a4paper]{article}

\usepackage[T1]{fontenc}
\usepackage[utf8]{inputenc}
\usepackage{lmodern}
\usepackage{amsmath,amssymb}
\usepackage{graphicx}
\usepackage{booktabs}
\usepackage{array}
\usepackage{siunitx}
\usepackage[super,square,sort&compress]{natbib}
\usepackage{microtype}
\usepackage{float}
\usepackage{caption}
\usepackage{xcolor}
\usepackage{ragged2e}
\usepackage{multirow}
\usepackage{threeparttable}
\usepackage{geometry}
\usepackage[colorlinks=true,allcolors=blue!55!black]{hyperref}
\graphicspath{{figures/}}
\hypersetup{
  pdftitle={MERCE: Method-Space Correlation Energy Extrapolation toward High-Rank Coupled Cluster Accuracy},
  pdfauthor={Mateusz Witkowski, Szymon Smiga, Pavlo O. Dral, Ireneusz Grabowski},
  pdfkeywords={electron correlation, coupled cluster theory, method-space extrapolation, effective rank, full configuration interaction, benchmark quantum chemistry}
}

\newcommand{\MERCE}{MERCE}
\newcommand{\Ecorr}{E^{\mathrm{corr}}}

\newcommand{\microEh}{\mu E_{\mathrm{h}}}
\newcommand{\kcalmol}{\mathrm{kcal\ mol^{-1}}}

\newcommand{\threesubsection}[1]{\subsection{#1}}

\title{MERCE: Method-Space Correlation Energy Extrapolation toward High-Rank Coupled Cluster Accuracy}
\author{Mateusz Witkowski\textsuperscript{1,2} \and
Szymon \'Smiga\textsuperscript{1,2} \and
Pavlo O. Dral\textsuperscript{1,2,3,4} \and
Ireneusz Grabowski\textsuperscript{1,2,*}}
\date{}

\begin{document}
\maketitle
\begin{center}
\small
\textsuperscript{1}Institute of Physics, Faculty of Physics, Astronomy, and
Informatics, Nicolaus Copernicus University in Toru\'n,
ul. Grudzi\k{a}dzka 5, 87-100 Toru\'n, Poland\\
\textsuperscript{2}Institute of Advanced Studies, Nicolaus Copernicus
University in Toru\'n, ul. Wile\'nska 4, 87-100 Toru\'n, Poland\\
\textsuperscript{3}State Key Laboratory of Physical Chemistry of Solid
Surfaces, Department of Chemistry, College of Chemistry and Chemical
Engineering, and Fujian Provincial Key Laboratory of Theoretical and
Computational Chemistry, Xiamen University, Xiamen, Fujian 361005, China\\
\textsuperscript{4}Aitomistic, Shenzhen 518000, China\\[0.35em]
\textsuperscript{*}Corresponding author: \href{mailto:ig@fizyka.umk.pl}{ig@fizyka.umk.pl}
\end{center}

\begin{center}
\small
\textbf{ORCID:}
M. Witkowski, \href{https://orcid.org/0000-0002-1944-2293}{0000-0002-1944-2293};
S. \'Smiga, \href{https://orcid.org/0000-0002-5941-5409}{0000-0002-5941-5409};
P. O. Dral, \href{https://orcid.org/0000-0002-2975-9876}{0000-0002-2975-9876};
I. Grabowski, \href{https://orcid.org/0000-0002-5661-5659}{0000-0002-5661-5659}.
\end{center}

\begin{center}
\small\textbf{Keywords:} electron correlation; coupled cluster theory;
method-space extrapolation; effective rank; full configuration interaction;
benchmark quantum chemistry
\end{center}

\begin{abstract}
Energy extrapolation is a well-established strategy for approaching
quantum-chemical limits without performing prohibitively expensive
calculations. Complete basis set extrapolation is widely used, whereas
extrapolation across electronic structure methods remains much less developed.
Method-Space Effective-Rank Correlation Energy Extrapolation (MERCE) is
introduced as an adaptive effective-rank framework for estimating high-rank
finite-basis correlation energies from second-order M{\o}ller--Plesset
perturbation theory (MP2), coupled cluster with single and double excitations
(CCSD), and CCSD with perturbative triples [CCSD(T)]. The method assigns
adaptive effective ranks to MP2 and CCSD(T) from the local
MP2--CCSD--CCSD(T) correlation energy pattern and fits a compact three-point
extrapolation form for each system. A hierarchical model-selection protocol
balances cross-dataset transferability, maximum dataset-level mean absolute
error (MAE) ratios, robustness on developmental sets referenced to full
configuration interaction (FCI), size-consistency defects, and performance for
A24 noncovalent interaction energies. Across the chemically diverse
benchmarks used during model development, the selected MERCE model
substantially reduces CCSD(T) errors and frequently improves upon the tested
higher-rank coupled cluster baselines. Transferability is further assessed
using calculations performed only after the functional form and parameters
were frozen. Formal nonadditivity is quantified using artificial
noninteracting pairs and the actual A24 and selected S66 component
calculations.
\end{abstract}

\section{Introduction}

Accurate electronic energies are needed to predict reaction energies, noncovalent interactions, ionization potentials, and bond-breaking processes. In a fixed one-particle basis, full configuration interaction (FCI) gives the exact solution of the electronic Schr\"odinger equation, but its factorial growth confines direct use to very small systems.\cite{KnowlesHandy1984} Coupled cluster (CC) theory supplies the most successful practical hierarchy for single-reference calculations: CCSD includes connected singles and doubles, CCSDT adds connected triples, CCSDTQ adds connected quadruples, and the full cluster operator recovers FCI in the same orbital space.\cite{PurvisBartlett1982,BartlettMusial2007,Kallay2004MRCC} In routine practice, however, calculations rarely go beyond CCSD(T), in which
connected triples are treated perturbatively at far lower cost than in iterative
CCSDT; the affordable methods in everyday use are second-order M{\o}ller--Plesset
perturbation theory (MP2), CCSD, and CCSD(T).\cite{Moller1934,Raghavachari1989}

This gap between routine and high-rank methods is not merely formal. Recent composite kinetics calculations provide a concrete illustration:
high-rank post-CCSD(T) contributions altered predicted reaction barriers and
rate constants substantially, while composite coupled cluster protocols were
used to approach the finite-basis FCI limit.\cite{Zhang2026Iodide} Post-CCSD(T) correlation can be important in high-precision thermochemistry, noncovalent interactions, stretched bonds, diradicals, ionization potentials, and other regimes in which perturbative triples do not represent a smooth remainder.\cite{Lyakh2012,Karton2017,Semidalas2025PostCCSDT} Yet direct access to these higher-rank contributions is
generally restricted to small molecules, reduced basis sets, or highly
specialized calculations. The central question of this work is therefore
whether the correlation energy pattern already encoded in a low-cost
MP2--CCSD--CCSD(T) sequence can be converted into a quantitative estimate of
the missing high-rank contribution.

This question is related to earlier efforts to infer higher-level correlation
energies from lower-level information. Goodson showed that short coupled cluster
sequences can be extrapolated toward the FCI limit, with continued-fraction
forms giving useful estimates from sequences such as self-consistent-field (SCF)--CCSD--CCSD(T).\cite{Goodson2002,Goodson2012}  Related empirical approaches
include the scaling-all-correlation (SAC) method and its
multi-coefficient extension (MCSAC), which rescale a correlation energy or combine successive correlation
increments using fitted coefficients.\cite{GordonTruhlar1986SAC,Fast1999MCSAC}  In the field of noncovalent interactions, highly accurate benchmark energies
are also often obtained by composite strategies rather than by a single direct
calculation at the target level.\cite{Santra2022S66x8,Semidalas2025PostCCSDT}
For example, the S66 reference energies were constructed using a
CCSD(T)-based complete basis set (CBS) scheme that combines basis set
extrapolation with lower-cost corrections.\cite{Rezac2011S66,Rezac2011S66Extensions,
Kesharwani2018S66F12,Santra2022S66x8,Semidalas2025PostCCSDT} Recent S66-inclusive tests further showed that small-basis evaluation of the
CCSD(T)--MP2 correction can be accurate with a diffuse-augmented compact
valence basis, whereas the corresponding unaugmented basis is appreciably less
reliable.\cite{Sun2026SmallBasisCorrection} These approaches demonstrate that carefully selected lower-level information can be
converted into useful high-level estimates.

The present work follows this general philosophy but differs in scope. We
introduce Method-Space Effective-Rank Correlation Energy Extrapolation
(\MERCE{}), an adaptive approach that constructs a system-dependent coordinate
in method space from the local MP2--CCSD--CCSD(T) correlation energy pattern.
Unlike basis-space extrapolation,\cite{Dunning1989,Helgaker1997} \MERCE{} does
not extrapolate with respect to the basis set cardinal number or apply a fixed
linear scaling to one or more correlation contributions. The comparison with
CBS extrapolation is conceptual: the effective-rank
coordinate is an empirical continuation variable and is not formally
equivalent to the CBS cardinal number. Throughout this work, ``method space''
denotes a fixed-basis correlation theory coordinate associated with the
MP2--CCSD--CCSD(T) input sequence and the higher-rank coupled cluster hierarchy;
it does not refer to joint variation of basis set and electronic-structure
method.

\begin{figure}[H]
\centering
\includegraphics[width=0.92\linewidth]{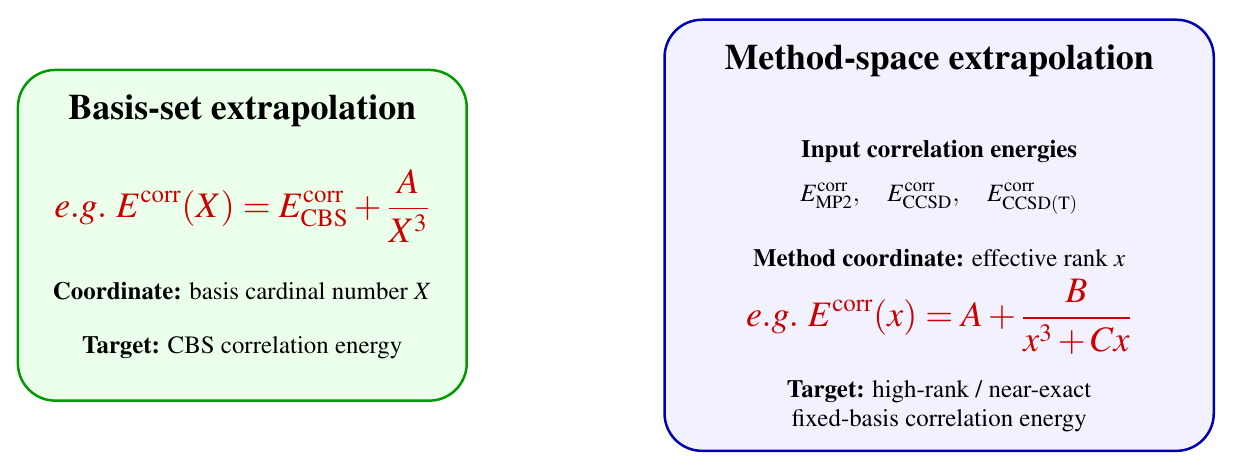}
\caption{Basis-space and method-space extrapolation approaches. The
CBS route depends on the cardinal number $X$. The \MERCE{} method
extrapolates along an adaptive effective-rank coordinate $x$ constructed from MP2, CCSD, and CCSD(T) correlation energies. The extrapolated value is evaluated at a chosen target rank: most commonly \(x=4\), representing a nominal quadruple-excitation-level estimate, or \(x=5\) for the five-correlated-electron FCI developmental systems.}
\label{fig:concept}
\end{figure}

The construction is inspired by CBS extrapolation, in which the
basis cardinal number $X$ provides a physically motivated coordinate related to
the maximum angular momentum represented in a correlation-consistent basis
set.\cite{Dunning1989,Helgaker1997} Here, an analogous coordinate is introduced
in method space, as illustrated in Figure~\ref{fig:concept}. For iterative
coupled cluster truncations, the maximum excitation rank supplies natural
integer positions: $x=2$ for CCSD, $x=3$ for CCSDT, and $x=4$ for CCSDTQ. MP2
and CCSD(T) cannot be assigned these integer ranks directly because their
treatments of double and triple excitations are approximate. Our proposed procedure 
therefore
assigns them system-dependent effective positions derived from the MP2--CCSD
and CCSD--CCSD(T) correlation energy increments.

These integer positions motivate the coordinate but do not establish that the
fitted \MERCE{} coordinate is a literal excitation-rank axis. In the
applications considered here, $x=4$ is used as a nominal CCSDTQ-like target.
The value $x=5$ coincides with the maximum excitation rank required for
finite-basis FCI only for systems containing exactly five correlated
electrons. More generally, $x$ should be regarded as an empirical continuation coordinate.

Figure~\ref{fig:concept} illustrates the central concept. In \MERCE{}, the
MP2--CCSD--CCSD(T) sequence is treated as a short trajectory through method
space. In this representation, MP2 and
CCSD(T) are assigned adaptive, system-dependent effective ranks, CCSD is kept
as an anchor point ($x=2$), and a compact three-point extrapolation form is fitted to estimate the correlation energy at a chosen higher effective rank. 
\MERCE{} is developed as an algebraically gated empirical post-CCSD(T) estimator. A representative model is selected using a hierarchy of criteria that includes cross-dataset transferability, between-dataset performance, robustness on the FCI-based developmental sets, size-consistency behavior, and property-level performance. Importantly, the numerical search does not identify a sharply isolated optimum. Instead, it reveals a broad performance plateau spanning multiple functional forms and parameter combinations that yield very similar errors. The final MERCE model should therefore be regarded as a robust, structurally simple representative of this plateau rather than a uniquely optimal parameterization. After the functional form and all parameters are frozen, its transferability
is tested on additional coordinate scans whose geometries and computed
energies played no role in model development.

\section{Results and Discussion}

\subsection{Benchmark Sets and Model-Selection Strategy}

Before discussing the final extrapolation model, we define the roles of the
benchmark datasets used throughout this work. The model-development protocol
combines several finite-basis reference sources that probe complementary
aspects of the extrapolation problem, including behavior near finite-basis FCI
for small systems, transferability to chemically different systems,
performance along bond-stretching coordinates, property-level energy
differences, and size consistency. Because most of these datasets contributed
to model screening or selection, they are treated as developmental or
cross-dataset model-selection benchmarks rather than as fully external
validation sets. Additional tests performed only after the functional form and global
parameters had been frozen are discussed separately.

Throughout this work, DLPNO denotes the domain-based local pair natural
orbital framework. DLPNO-CCSDT and DLPNO-CCSDTQ refer, respectively, to its
iterative-triples and full iterative-quadruples variants.
\cite{Jiang2026DLPNOCCSDTQ}

\begin{table*}[h!]
\centering
\caption{Roles of the datasets used in MERCE development and assessment. The table distinguishes the FCI-based developmental sets, cross-dataset model-selection benchmarks, property-level tests, size-consistency diagnostics, additional geometry scans, and fixed-basis applications. }
\label{tab:dataset_roles}
\footnotesize
\setlength{\tabcolsep}{4.5pt}
\renewcommand{\arraystretch}{1.02}
\resizebox{0.98\textwidth}{!}{%
\begin{tabular}{p{2.2cm} p{3.0cm} p{3.7cm} p{3.3cm} p{3.8cm}}
\toprule
\textbf{Role} &
\textbf{Dataset or analysis} &
\textbf{Reference or target} &
\textbf{Data source} &
\textbf{Use in this work} \\
\midrule

FCI-based development &
FCI-full-core and FCI-frozen-core &
Finite-basis FCI; \MERCE{} evaluated at $x=5.0$ for five-correlated-electron systems &
This work &
Constrains behavior near the exact finite-basis limit and enters the developmental-set robustness screen. \\

\midrule
\multirow[t]{4}{2.2cm}{Cross-dataset model selection} &
W4 subset &
CCSDTQ; \MERCE{} evaluated at $x=4.0$ &
Literature data from Karton et al.\cite{Karton2017} &
Tests whether the model transfers from small FCI developmental systems to chemically different thermochemical systems. \\

&
Rings &
CCSDTQ; \MERCE{} evaluated at $x=4.0$ &
Literature data from Damour et al.\cite{Damour2021RingsFCI} &
Tests transferability to conjugated five- and six-membered ring systems. \\

&
Correlation Energy Per Bond (CEPB) subset &
DLPNO-CCSDTQ; \MERCE{} evaluated at $x=4.0$ &
Canonical MP2, CCSD, and CCSD(T) data from previous CEPB work;\cite{CEPB} DLPNO-CCSDT and DLPNO-CCSDTQ data computed in
this work &
Tests transferability to larger systems outside the FCI developmental domain. \\

&
One-bond stretching scans &
DLPNO-CCSDTQ; \MERCE{} evaluated at $x=4.0$ &
This work &
Tests transferability along changing electronic structure as individual bonds are stretched. \\

\midrule
Property-level model selection &
23 reference-bearing A24 noncovalent interactions &
CCSDT(Q) correlation contributions to interaction energies;
\MERCE{} evaluated at $x=4.0$ &
Literature data from Lambie et al.\cite{Lambie2025} &
Tests small energy differences and enters the late-stage A24 property-level
screen and final tie-break. \\

\midrule
Derived property-level assessment &
Vertical ionization potentials &
Finite-basis FCI energy differences; \MERCE{} evaluated at $x=5.0$ or $x=6.0$
according to the number of correlated electrons &
This work &
Tests cancellation of neutral and cationic correlation energy errors; the
property differences did not enter the selection of the functional form or global
parameters. \\

\midrule
Size-consistency diagnostic &
Artificial noninteracting pairs &
Additive high-level reference energies &
Constructed in this work from the benchmark datasets above &
Quantifies the nonadditivity introduced by the nonlinear effective-rank mapping. \\

\midrule
Additional geometry scans &
LiF, CH$_3$F, and CH$_3$OH one-bond stretching; NH$_3$ planarization scan &
DLPNO-CCSDTQ; \MERCE{} evaluated at $x=4.0$ &
This work &
Assesses transfer after the functional form and global parameters were frozen;
none of these scan data entered model selection. \\

\midrule
Fixed-basis application test &
Selected S66/S66x8 complexes &
No direct S66/CBS error target; fixed-basis post-CCSD(T) corrections &
Geometries from S66/S66x8;\cite{Rezac2011S66,Santra2022S66x8} correlation energies computed in this work &
Illustrates practical finite-basis \MERCE{} corrections for realistic noncovalent complexes. \\

\bottomrule
\end{tabular}
}
\end{table*}

Table~\ref{tab:dataset_roles} summarizes the role of each dataset in the
development, model selection, size-consistency assessment
 and the application parts of the study.
Two FCI-based developmental sets are used to evaluate performance near the exact
finite-basis limit for small molecules. The first one, denoted FCI-full-core,
contains systems with only five fully correlated electrons: the B atom, the BeH
molecule, the molecular ions BH\textsuperscript{+},
BeH\textsubscript{2}\textsuperscript{+}, and
Li\textsubscript{2}\textsuperscript{+}, and the C\textsuperscript{+} cation.
The second one, denoted FCI-frozen-core, contains larger species treated with
the frozen-core approximation, again leaving five correlated electrons:
AlH\textsubscript{2} and BH\textsubscript{2}, the molecular ions
CH\textsubscript{2}\textsuperscript{+},
SiH\textsubscript{2}\textsuperscript{+}, NH\textsuperscript{+}, and
PH\textsuperscript{+}, the N atom, the SiH and CH radicals, and the
F\textsuperscript{2+} and O\textsuperscript{+} cations. For both developmental
sets, correlation energies were computed in a series of Dunning-type basis sets, ranging from cc-pVDZ or aug-cc-pVDZ to cc-pV5Z or aug-cc-pV5Z whenever
computationally feasible\cite{Dunning1989}.

Cross-dataset transferability during model selection is assessed using several
chemically distinct benchmark sets. The first is a ``W4.3'' subset of the W4-17 thermochemical benchmark \cite{Karton2017},
denoted hereafter as W4. The original subset contains 65 molecules
computed with the frozen-core VTZ(f,p) basis and reference CCSDTQ correlation
energies. Seven molecules (BN, C$_2$, CO$_2$, N$_2$, N$_2$O, NO$_2$, and
O$_3$) were excluded because their MP2--CCSD--CCSD(T) correlation energy
sequences violate the required monotonic ordering. H$_2$ was omitted because,
as a two-electron system, it cannot probe higher-than-double coupled cluster
excitations, while CN was removed because its CCSDT calculation did not
converge. The resulting W4 subset, therefore, contains 56 molecules. The second is a conjugated-ring benchmark, denoted Rings, containing five-membered
rings (cyclopentadiene, furan, imidazole, pyrrole, and thiophene) and
six-membered rings (benzene, pyrazine, pyridazine, pyridine, pyrimidine,
$s$-tetrazine, and $s$-triazine), computed in frozen-core cc-pVDZ with
reference CCSDTQ correlation energies.\cite{Damour2021RingsFCI} The third, denoted CEPB, is an 18-entry subset constructed from canonical MP2,
CCSD, and CCSD(T) correlation energies in the source dataset underlying our
previous CEPB work.\cite{CEPB} For the same system--basis-set entries,
DLPNO-CCSDT and DLPNO-CCSDTQ correlation energies were computed in this work,
with the latter used as an approximate high-rank reference.
\cite{Jiang2026DLPNOCCSDTQ} The subset contains nine entries in aug-cc-pVDZ,
eight in aug-cc-pVTZ, and one in aug-cc-pVQZ.

The original STRETCH model-selection benchmarks comprise one-bond scans for
hydrogen fluoride (HF), H$_2$O with one stretched O--H bond, CH$_4$ with one
stretched C--H bond, and C$_2$H$_4$ with a stretched C=C bond. These scans were computed in
the cc-pVTZ basis without the frozen-core approximation, so that all electrons
are correlated consistently along the stretching coordinate. They test whether the extrapolation remains stable as the electronic structure changes along a
molecular coordinate and are referenced to high-rank or approximate high-rank
coupled cluster data.

Property-level testing is performed on A24 noncovalent correlation contributions to interaction energies \cite{Lambie2025}
and vertical ionization potentials. These tests are deliberately different from
absolute correlation energy benchmarks because the target quantities are small
energy differences obtained by combining several total or correlation energies.
The ionization-potential set includes BH/BH\textsuperscript{+},
C/C\textsuperscript{+}, BeH\textsubscript{2}/BeH\textsubscript{2}\textsuperscript{+},
Li\textsubscript{2}/Li\textsubscript{2}\textsuperscript{+},
O/O\textsuperscript{+}, B\textsubscript{2}/B\textsubscript{2}\textsuperscript{+},
CH\textsubscript{2}/CH\textsubscript{2}\textsuperscript{+},
PH/PH\textsuperscript{+}, and
SiH\textsubscript{2}/SiH\textsubscript{2}\textsuperscript{+}. Finally, size
consistency is examined through artificial noninteracting pairs constructed
from the benchmark systems.

Together, these datasets separate near-FCI developmental behavior from
cross-dataset transferability, property-level error cancellation, and formal
nonadditivity. Their distinct roles, rather than any single aggregate metric,
determine the hierarchical selection described below; complete compositions,
references, and raw numerical tables are provided in the Supporting
Information.

\subsection{Method-Space Effective-Rank
Correlation Energy Extrapolation (MERCE)}
One of the most important aspects of the MERCE approach is the introduction and definition of effective positions in method-space (effective ranks).
Rather than assigning
fixed ranks in the extrapolation procedure to MP2 and CCSD(T), \MERCE{} first checks how the MP2--CCSD--CCSD(T) correlation energy
sequence behaves for the particular system and basis set under consideration, by calculating the relative sizes of the MP2--CCSD and CCSD--CCSD(T)
correlation energy increments. This approach is motivated by preliminary results indicating that extrapolations based on fixed ranks for MP2 and CCSD(T) cannot accurately predict higher-order correlation energies.

\begin{figure}[H]
\centering
\includegraphics[width=1.00\linewidth]
{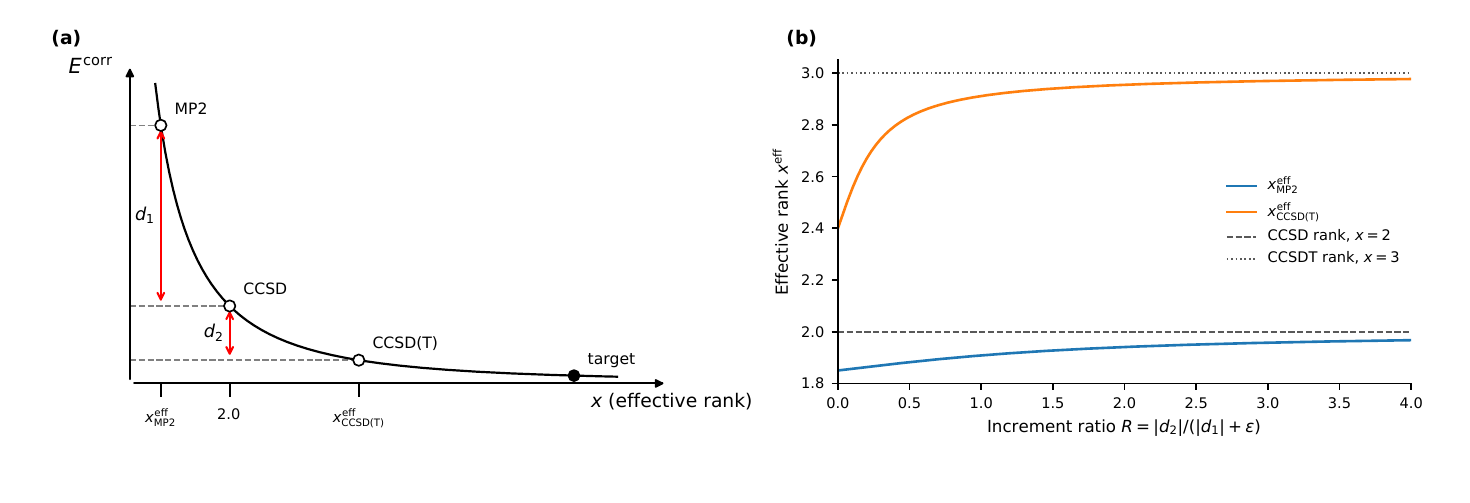}
\caption{Visualization of the \MERCE{} effective-rank construction.
(a) Schematic method-space correlation energy trajectory. The MP2, CCSD, and CCSD(T) input points define a monotonic decrease
along the effective-rank coordinate. The increments $d_1$ and $d_2$ characterize
the local shape of the trajectory and determine the adaptive placement of MP2 $(x_{\mathrm{MP2}}^{\mathrm{eff}}$)
and CCSD(T) ($x_{\mathrm{CCSD(T)}}^{\mathrm{eff}})$, while CCSD remains fixed at $x=2$. The filled marker illustrates
the evaluation of the fitted extrapolation curve at a chosen higher target rank.
(b) Final adaptive effective-rank mappings used in the selected \MERCE{} model.
The MP2 effective rank,
$x_{\mathrm{MP2}}^{\mathrm{eff}}$, moves slowly toward the CCSD anchor as $R$
increases, whereas the CCSD(T) effective rank,
$x_{\mathrm{CCSD(T)}}^{\mathrm{eff}}$, responds more rapidly and approaches the
upper bound $x=3$. The plotted mappings correspond to
$\alpha=1.85$, $\gamma=2.40$, $k_1=0.700$, and $k_3=4.250$.}
\label{fig:illustrative_merce_trajectory}
\end{figure}

For a given system and basis set, the three input correlation energies (MP2, CCSD, CCSD(T)) define
adjacent increments
\begin{equation}
 d_1 = \Ecorr_{\mathrm{CCSD}}-\Ecorr_{\mathrm{MP2}},\qquad
 d_2 = \Ecorr_{\mathrm{CCSD(T)}}-\Ecorr_{\mathrm{CCSD}} .
\label{eq:increments}
\end{equation}
The first increment measures how far the MP2 correlation energy lies from the iterative
singles-and-doubles CC, whereas the second one is the perturbative
triples correction to the CCSD on the correlation energy scale (see Figure~\ref{fig:illustrative_merce_trajectory} for the visualization). Their relative magnitude
therefore contains useful information about the local shape of the
method-space trajectory. In the present approach, this information is
compressed into the dimensionless ratio
\begin{equation}
 R = \frac{|d_2|}{|d_1|+\varepsilon},
\label{eq:ratio}
\end{equation}
where $\varepsilon=10^{-16}\,E_{\mathrm h}$ prevents numerical division by
zero. All correlation energy increments are expressed in Hartree. Before $R$ is
formed, inputs with $|d_1|<10^{-14}\,E_{\mathrm h}$ are rejected as
numerically degenerate. This safeguard prevents division by an effectively
vanishing MP2--CCSD increment and avoids assigning effective ranks from a
numerically unstable ratio.

The applicability gate is defined by the strict ordering
\begin{equation}
\Ecorr_{\mathrm{MP2}}
>
\Ecorr_{\mathrm{CCSD}}
>
\Ecorr_{\mathrm{CCSD(T)}}.
\label{eq:input_ordering}
\end{equation}
An extrapolation is rejected if this ordering is not satisfied or if
$|d_1|<10^{-14}\,E_{\mathrm h}$. Rejection returns no \MERCE{} estimate; the
input sequence is not reordered and no extrapolated value is forced through
the point. The fitted branch must additionally reproduce the three input
energies within the numerical tolerance and pass the finite-value and
rational-pole checks described below and in the Supporting Information.

The ratio $R$ should not be interpreted as a universal electronic-structure
diagnostic, but as a simple descriptor of the local MP2--CCSD--CCSD(T)
correlation energy trajectory that controls the adaptive placement of the input
methods. Small values of $R$ correspond to cases where the perturbative triples correction is small compared with the
MP2--CCSD displacement, so CCSD(T) remains close to the CCSD anchor ($x=2$) in
method space. Conversely, $R$ grows when the triples correction is large or
when the MP2--CCSD difference becomes unusually small; in the limiting case
$d_1\to 0$ with finite $d_2$, the ratio tends to infinity. Thus, a large value of $R$ does not have a unique physical meaning, but
indicates that the effective-rank positions of MP2 and CCSD(T) should be
adjusted for the given system rather than fixed a priori.

This behavior also motivates the functional form of the effective-rank mapping.
Here, $x$ is the effective-rank coordinate used for the
subsequent method-space extrapolation. When $d_1$ becomes small, MP2 and CCSD provide nearly redundant
correlation energies and should therefore occupy nearby positions along this
coordinate. In this large-$R$ regime, the effective MP2 rank ($x_{\mathrm{MP2}}^{\mathrm{eff}}$) is driven smoothly
toward the $x=2$, CCSD anchor. At the same time, the numerical scans showed that transferability improves when the CCSD(T) point is allowed to move away from the CCSD anchor in cases where
the perturbative triples increment remains large relative to the MP2--CCSD
displacement. However, the
effective position assigned to CCSD(T)
($x_{\mathrm{CCSD(T)}}^{\mathrm{eff}}$) is constrained not to exceed $x=3$,
the integer rank associated with full iterative triples in CCSDT. This is a
bounded modeling choice rather than a claim that CCSD(T) obeys a universal
energetic ordering relative to CCSDT. We therefore use a
bounded, monotonic arctangent mapping, which provides a smooth response at small
and intermediate $R$, while preventing uncontrolled growth of the effective
ranks as $R\rightarrow\infty$. This upper bound should not be interpreted as
assuming that CCSD(T) becomes equivalent to full CCSDT. Rather, it only limits
the effective position of the available CCSD(T) input point; the high-rank target is introduced separately by evaluating the fitted function at the
desired value of $x$.

Finally, the adaptive effective ranks are assigned as
\begin{align}
 x_{\mathrm{MP2}}^{\mathrm{eff}}
 &= \alpha + (2-\alpha)\frac{2}{\pi}\arctan(k_1R),
\label{eq:mapping_mp2}\\
 x_{\mathrm{CCSD(T)}}^{\mathrm{eff}}
 &= \gamma + (3-\gamma)\frac{2}{\pi}\arctan(k_3R).
\label{eq:mapping_ccsdpt}
\end{align}
Here, $\alpha$ and $\gamma$ are the lower-bound effective ranks assigned to
MP2 and CCSD(T), respectively, while $k_1$ and $k_3$ control how rapidly the
MP2 and CCSD(T) effective ranks respond to the increment ratio $R$. CCSD is
kept fixed as the integer-rank anchor at $x=2$. The arctangent form is bounded and monotonic, which preserves the imposed
requirements on the effective-rank behavior.
As $R$ increases, $x_{\mathrm{MP2}}^{\mathrm{eff}}$ approaches the CCSD
anchor from below, whereas $x_{\mathrm{CCSD(T)}}^{\mathrm{eff}}$ approaches
the upper bound $x=3$. Thus, the input points are displaced
smoothly in response to the local increment pattern, without allowing the
effective ranks to diverge. For each system, a three-parameter function is then fitted to the three points
$(x_{\mathrm{MP2}}^{\mathrm{eff}},\Ecorr_{\mathrm{MP2}})$,
$(2,\Ecorr_{\mathrm{CCSD}})$, and
$(x_{\mathrm{CCSD(T)}}^{\mathrm{eff}},\Ecorr_{\mathrm{CCSD(T)}})$. This construction and the role of the increments $d_1$ and $d_2$ are illustrated
schematically in Figure~\ref{fig:illustrative_merce_trajectory}.

After defining the effective-rank coordinate, the remaining choice is the extrapolation function used to continue the three-point trajectory toward higher
rank correlation energy. We examined a diverse library of three-parameter closed forms, summarized in the Supporting Information, together with broad ranges of the effective-rank mapping parameters. The search revealed a relatively flat region of model space in which several functional forms and numerous combinations of \(\alpha\), \(\gamma\), \(k_1\), and \(k_3\) produced closely comparable performance. Consequently, the available benchmark data do not support identifying a unique optimal functional form or parameter set. We therefore selected a representative model using a hierarchical procedure that considered cross-dataset transferability, maximum dataset-level MAE ratios, robustness on the FCI-based developmental sets, size-consistency defects, property-level performance, and structural simplicity.

This selection procedure, described in Section~\ref{sec:model_selection} and documented in detail in the Supporting Information, identified the inverse-cubic-linear form of the extrapolation function
\begin{equation}
 \Ecorr(x)=A+\frac{B}{x^3+Cx}.
\label{eq:final_model}
\end{equation}
For each system in a given basis set, the coefficients $A$, $B$, and $C$ are fitted from the
corresponding MP2, CCSD, and CCSD(T) correlation energies. 
The analytic form and the effective-rank mapping parameters in
Eqs.~\ref{eq:mapping_mp2} and \ref{eq:mapping_ccsdpt} are fixed globally in the
final \MERCE{} procedure.
The selected values were obtained from
a discrete scan over 40 candidate functional forms, and the parameter ranges
$1.20 \leq \alpha \leq 1.90$, $2.30 \leq \gamma \leq 2.90$,
$0.4 \leq k_1 \leq 0.9$, and $2.75 \leq k_3 \leq 4.50$. The fixed values \((\alpha,\gamma,k_1,k_3)=(1.85, 2.40, 0.700,4.250)\)  define the representative model used for all subsequent calculations. They correspond to a robust region of the explored parameter space and should not be interpreted as uniquely determined physical constants. 

Unless stated otherwise, ``the \MERCE{} model'' 
denotes this complete final representative protocol, including both the effective-rank mapping and
the inverse-cubic-linear extrapolation form. The electronic-structure thresholds, high-rank proof-of-principle results,
and complete candidate-function library are reported in
Tables~S1--S3, respectively. The multistart fitting procedure,
branch-selection algorithm, and model-selection hierarchy are documented in
the Supporting Information.

\subsection{MERCE model selection procedure}
\label{sec:model_selection}

With the benchmark roles defined in Section~2.1, we now describe how the final
representative MERCE model Eq.~\ref{eq:final_model} was selected. The search was guided by the observation
that genuine coupled cluster excitation-rank sequences, such as
CCSD--CCSDT--CCSDTQ, show regular convergence toward the finite-basis FCI limit.
This behavior motivates the use of smooth three-point analytic forms for
continuing a correlation energy trajectory to higher effective rank. In the
practical MERCE model, however, the fitted trajectory is constructed only from
the routinely accessible MP2, CCSD, and CCSD(T) correlation energies.

The final extrapolation function, Eq.~\ref{eq:final_model}, was chosen from a
broad library of three-point analytic forms designed to reproduce the
MP2--CCSD--CCSD(T) input sequence and to approach a finite high-rank limit as
$x$ increases. The tested library included several families of analytic forms.
Representative examples include exponential forms, $A+B\exp(-Cx)$;
inverse-power forms, $A+B/x^C$; and inverse-polynomial or rational forms,
$A+B/(x^m+Cx^n+d)$. Other smooth saturating forms were also tested, as detailed
in the Supporting Information. Candidate forms were rejected when they produced
nonphysical behavior, including real poles between the fitted effective-rank
interval and the target rank. The complete candidate-function library and the
algorithmic details of the nonlinear branch selection are provided in the
Supporting Information.

\begin{figure}[H]
\centering
\includegraphics[width=0.82\linewidth]{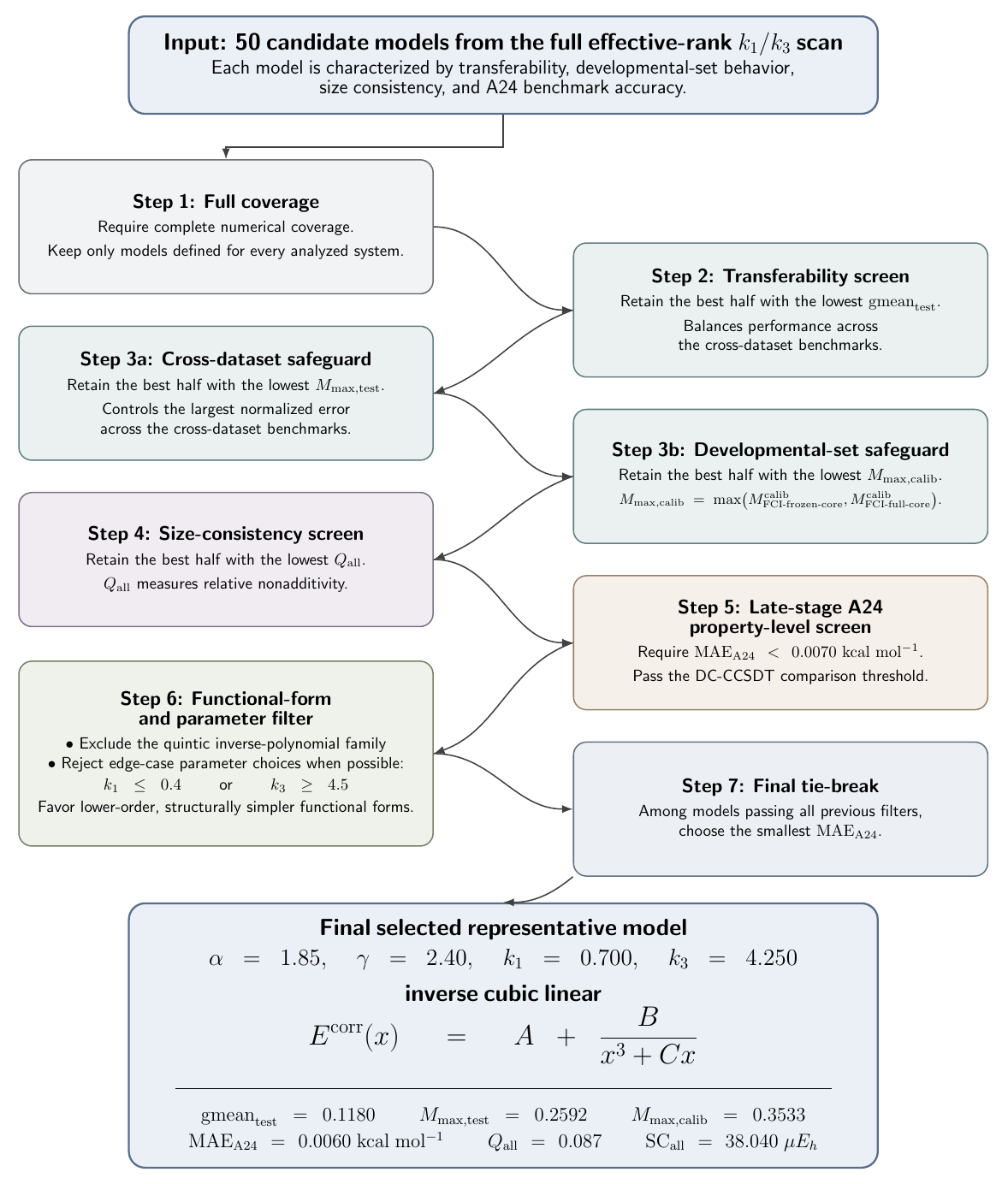}
\caption{Hierarchical selection protocol used to identify the final
representative \MERCE{} model. Candidate models must provide complete coverage,
show transferable and robust behavior, pass the developmental-set and
size-consistency safeguards, satisfy the late-stage A24 property-level screen,
and favor lower-order, simpler extrapolation forms. Step~5 is a pass/fail A24
screen, whereas Step~7 uses A24 only as a final tie-break among candidates that
have already passed the transferability, developmental-set, size-consistency,
and structural filters.}
\label{fig:selection}
\end{figure}

Figure~\ref{fig:selection} summarizes the selection protocol. Candidate models
are first required to give complete numerical coverage over the eligible entries. They are then filtered according to transferability across the cross-dataset model-selection benchmarks, protection against poor performance for an entire benchmark family, robustness on both FCI-based developmental sets, controlled size-consistency defects,
late-stage property-level screening on A24, and structural simplicity. The
selected model is therefore not the absolute minimum of any single ranking
column. It is a physically motivated compromise among accuracy, between-dataset robustness, controlled size-consistency defects, and interpretability.

The selection quantities used in Figure~\ref{fig:selection} are defined as
follows.

\textit{Step~1} enforces full numerical coverage: a candidate model is retained
only if it gives valid finite results for all systems in all analyzed benchmark
sets. This removes models that fail for some systems because of unstable fits,
unphysical branches, or singular behavior.

\textit{Step~2} defines the cross-dataset transferability score. For each cross-dataset model-selection benchmark $d$, the mean absolute error
(MAE) of a candidate model is normalized to the corresponding CCSD(T) error,
\begin{equation}
M_d =
\frac{\mathrm{MAE}_d(\mathrm{model})}
{\mathrm{MAE}_d(\mathrm{CCSD(T)})}.
\label{eq:md}
\end{equation}
Thus, $M_d<1$ means that the candidate improves on CCSD(T) for dataset $d$.
Step~2 measures transferability by the geometric mean of these normalized
errors over the cross-dataset model-selection benchmarks,
\begin{equation}
\mathrm{gmean}_{\mathrm{test}} =
\left(
\prod_{d\in \mathcal{D}_{\mathrm{test}}} M_d
\right)^{1/|\mathcal{D}_{\mathrm{test}}|},
\label{eq:gmean_test}
\end{equation}
where $\mathcal{D}_{\mathrm{test}}$ contains Rings, W4, CEPB, and the four
original one-bond stretching scans. We refer to the latter subgroup as
STRETCH. The geometric mean is used because the quantities $M_d$ are positive,
dimensionless error ratios and therefore combine naturally on a multiplicative
scale. Equivalently, $\ln(\mathrm{gmean}_{\mathrm{test}})$ is the arithmetic
mean of $\ln M_d$, so relative improvements and deteriorations are compared in
logarithmic rather than absolute-error space. This choice nevertheless defines
a particular weighting of the benchmark datasets and does not remove the
dependence among geometries belonging to the same scan. The subscript
``test'' is retained as an internal computational label for consistency with the
numerical analysis. Because the datasets in $\mathcal{D}_{\mathrm{test}}$
contributed to model selection, they are cross-dataset model selection
benchmarks rather than independent external-validation sets. The score favors
models that perform consistently across chemically different datasets rather
than models that are excellent for only one subset.

\textit{Step~3a} limits poor performance for any complete benchmark family by
considering the largest normalized error over the same cross-dataset
model-selection benchmarks,
\begin{equation}
M_{\max,\mathrm{test}} =
 \max_{d\in \mathcal{D}_{\mathrm{test}}} M_d .
\end{equation}
This criterion prevents a model with good average transferability from being
selected if it performs poorly on one benchmark family.

\textit{Step~3b} applies the analogous safeguard to the two FCI-based
developmental sets. To preserve direct correspondence with the released
analysis files, the historical subscript ``calib'' is retained in the metric
names. For each developmental set $c$, we define
\begin{equation}
 M_c^{\mathrm{calib}}=
 \frac{\mathrm{MAE}_c(\mathrm{model})}
      {\mathrm{MAE}_c(\mathrm{CCSD(T)})},
\end{equation}
and then use
\begin{equation}
M_{\max,\mathrm{calib}} =
 \max\left(
 M_{\mathrm{FCI\text{-}frozen\text{-}core}}^{\mathrm{calib}},
 M_{\mathrm{FCI\text{-}full\text{-}core}}^{\mathrm{calib}}
 \right).
\end{equation}
This step ensures that the model remains robust on both FCI-based
regimes and does not improve transferability at the expense of poor behavior
near the exact finite-basis limit.

\textit{Step~4} screens the size-consistency behavior. For each artificial
noninteracting pair $p=(A,B)$, the size-consistency defect is
\begin{equation}
 \Delta_{\mathrm{SC}}(A,B)
 =
 E_{\mathrm{MERCE}}(A+B)
 -
 E_{\mathrm{MERCE}}(A)
 -
 E_{\mathrm{MERCE}}(B),
\label{eq:delta_sc}
\end{equation}
where $E_{\mathrm{MERCE}}(A+B)$ is obtained by applying the extrapolation once
to the summed MP2, CCSD, and CCSD(T) fragment energies. A more detailed description of the construction of noninteracting pairs is provided in Section~\ref{sec:size_consistency}. The target rank and comparison energy are inherited from the corresponding
benchmark set: $x=5.0$ for the five-correlated-electron FCI-based
developmental sets and $x=4.0$ for the canonical-CCSDTQ- or
DLPNO-CCSDTQ-based cross-dataset model-selection benchmarks. The mean absolute
size-consistency defect over all artificial pairs is
\begin{equation}
 \mathrm{SC}_{\mathrm{all}}
 =
 \frac{1}{N_{\mathrm{pair}}}
 \sum_{(A,B)\in \mathcal{P}_{\mathrm{all}}}
 \left|\Delta_{\mathrm{SC}}(A,B)\right|,
 \label{eq:sc_all}
\end{equation}
where $\mathcal{P}_{\mathrm{all}}$ is the set of all artificial noninteracting
pairs and $N_{\mathrm{pair}}=|\mathcal{P}_{\mathrm{all}}|$. The corresponding
pair-energy error is
\begin{equation}
 \Delta_{\mathrm{pair}}(A,B)
 =
 E_{\mathrm{MERCE}}(A+B)
 -
 \left[E_{\mathrm{ref}}(A)+E_{\mathrm{ref}}(B)\right],
 \label{eq:delta_pair}
\end{equation}
and its mean absolute value is
\begin{equation}
 \mathrm{pair\text{-}MAE}_{\mathrm{all}}
 =
 \frac{1}{N_{\mathrm{pair}}}
 \sum_{(A,B)\in \mathcal{P}_{\mathrm{all}}}
 \left|\Delta_{\mathrm{pair}}(A,B)\right|.
 \label{eq:pair_mae_all}
\end{equation}
The relative size-consistency measure used in the selection protocol is then
\begin{equation}
 Q_{\mathrm{all}}=
 \frac{\mathrm{SC}_{\mathrm{all}}}{\mathrm{pair\text{-}MAE}_{\mathrm{all}}}.
\label{eq:qall}
\end{equation}
Small values of $Q_{\mathrm{all}}$ indicate that the nonadditivity introduced
by the nonlinear MERCE mapping is small compared with the typical pair-energy
error.

\textit{Step~5} uses A24 as a late-stage property-level screen. Candidate models are
required to reach an A24 mean absolute error below
$0.0070\,\mathrm{kcal\,mol^{-1}}$, corresponding to the DC-CCSDT comparison
threshold.  This threshold was chosen because it corresponds to the DC-CCSDT MAE reported for the same A24 reference data, providing a stringent property-level comparison
below standard CCSDT and CCSD(T).

\textit{Step~6} applies a structural simplicity and parameter filter. In particular, the quintic inverse-polynomial family was excluded because powers as high as
$x^5$, together with additional lower-order or constant denominator terms,
introduced unnecessary complexity relative to the three-point input sequence.
Edge-case mapping parameters, such as $k_1\leq 0.4$ or $k_3\geq 4.5$ where
applicable, were also avoided. These choices were made after the accuracy,
transferability, and robustness screens; numerical finiteness and pole checks
were applied independently.

Finally, \textit{Step~7} selects, among the surviving
candidates, the model with the smallest A24 mean absolute error. The selected model is therefore
not the minimum of a single ranking column, but a physically motivated
compromise among transferability,  between-dataset robustness, and developmental-set
behavior, controlled size-consistency defects, A24 property-level performance,
and functional-form simplicity.

For the final \MERCE{} model,
$\mathrm{gmean}_{\mathrm{test}}=0.1180$ and
$M_{\max,\mathrm{test}}=0.2592$. Thus, relative to CCSD(T), the
selected model reduces the geometric mean of the normalized errors across the
cross-dataset model-selection benchmarks to approximately 12\%, while its
least favorable dataset retains approximately 26\% of the corresponding
CCSD(T) error. The FCI-based developmental safeguard gives
$M_{\max,\mathrm{calib}}=0.3533$. For the A24 property-level
assessment, the model yields an MAE of $0.0060\,\kcalmol$. The overall
size-consistency ratio is $Q_{\mathrm{all}}=0.087$, with
$\mathrm{SC}_{\mathrm{all}}=38.040\,\microEh$. These complementary quantities
document the balance among average transferability,  between-dataset behavior,
performance near the finite-basis FCI limit, property-level accuracy, and size
consistency that motivated the selection of the representative protocol. The complete fixed-rank and effective-rank rankings, mapping comparisons,
full $(k_1,k_3)$ scan results, and artificial-pair size-consistency analyses
are reported in Tables~S4--S18. The representative remains competitive when the benchmark weighting is
changed. Among the 960 candidates with complete dataset-level summaries in the
original scan, it ranks 6th using the primary seven-dataset geometric mean,
4th when the four STRETCH scans are collapsed into one equally weighted
family, and 7th when datasets are weighted by their numbers of accepted
entries. Replacing $Q_{\mathrm{all}}$ by the absolute size-consistency MAE at
Step~4 leaves the same four survivors. These sensitivity checks support the
plateau interpretation; they do not convert the reused benchmarks into
independent validation data. Full rankings and pointwise tail statistics are
given in the Supporting Information and in the deposited repository.

 As an additional fixed-library sensitivity analysis, we reranked the 957
candidates with complete pointwise coverage across all nine benchmark groups
from the original library of 960 scan candidates, omitting one benchmark
family at a time. The remaining three candidates were excluded from this
analysis because each lacked one pointwise prediction. The final \MERCE{}
model ranks 11th, 7th, and 10th when Rings, W4, and CEPB are omitted,
respectively, and 101st when the four original STRETCH scans are omitted
together. Nevertheless, every candidate among the top 25 in each reduced
ranking improves upon CCSD(T) for the omitted family. The final model gives
corresponding omitted-family MAE ratios,
$\mathrm{MAE}_{\mathrm{MERCE}}/\mathrm{MAE}_{\mathrm{CCSD(T)}}$, of 0.0539,
0.2593, 0.1921, and 0.1044 for Rings, W4, CEPB, and STRETCH, respectively.
These results support the broad-plateau interpretation but remain sensitivity
analyses of a fixed candidate library rather than nested or fully independent
validation exercises.

The final representative \MERCE{} model is therefore defined by the
inverse-cubic-linear extrapolation form in Eq.~\ref{eq:final_model}, the
effective-rank mappings in Eqs.~\ref{eq:mapping_mp2} and
\ref{eq:mapping_ccsdpt}, and the globally fixed parameters
$\alpha=1.85$, $\gamma=2.40$, $k_1=0.700$, and $k_3=4.250$.

After the effective-rank construction, model-selection workflow, and
representative protocol had been established, we performed a substantially
expanded post-selection robustness challenge. This search examined 180
three-point extrapolation shapes, six bounded effective-rank mappings, and
1024 Sobol-sampled combinations of the four global mapping parameters $(\alpha,\gamma,k_1,k_3)$ for
each shape--mapping pair,\cite{Sobol1967} corresponding to more than 1.1 million attempted
candidate models. The expanded search identified many distinct
functional-form and parameter regions with closely comparable performance.

To test whether the finite Sobol sampling had missed nearby high-performing
parameter values, the eligible candidates were ranked by
$\mathrm{gmean}_{\mathrm{test}}$. The leading candidate from each distinct
combination of functional shape, mapping family, and mapping exponent was
retained until 25 starting regions had been collected. Within each region, the
four global mapping parameters were then locally optimized with a bounded,
derivative-free procedure using the original parameter bounds. The refinement
objective was based on $\mathrm{gmean}_{\mathrm{test}}$ and included penalties
for exceeding the predefined $M_{\max,\mathrm{calib}}$ and
$M_{\max,\mathrm{test}}$ safeguards. This bounded local refinement
produced only modest improvements relative to the corresponding Sobol points.
The exact seed-selection rule, refinement settings, and stopping criteria are
given in the Supporting Information.
Some alternatives improved individual developmental-set or size-consistency
metrics, but these gains were accompanied by deterioration in other criteria.
Five predefined representatives from the expanded search were subsequently
evaluated on A24; none improved the selected model's A24 MAE. Because the full
expanded Pareto set was not reranked through every stage of the original
selection hierarchy, this result is a robustness challenge rather than proof
that no alternative could dominate under every possible decision rule. We
therefore retained the previously established representative rather than
redefining the decision rule after inspecting the expanded results. The expanded challenge supports interpreting the inverse-cubic-linear
model as a simple and balanced representative of a broad high-performance plateau
region rather than as a uniquely optimal parameterization.

To determine whether the favorable performance was restricted to
\MERCE{}-type three-point forms, we also carried out a separate post-selection
comparison with Goodson's parameter-free continued-fraction and Pad\'e
transformations and with zero-intercept SAC-, scaled-(T)-, and MCSAC-inspired
linear models refitted to the present finite-basis data.
\cite{Goodson2002,GordonTruhlar1986SAC,Fast1999MCSAC} On the common development
domain, \MERCE{} retains the lowest seven-dataset geometric-mean MAE ratio,
0.1180, compared with 0.1407 for the closest selection-aligned linear model
and 0.2952 for the strongest parameter-free Goodson transformation.
\MERCE{} also gives the lowest A24 MAE among the models examined. Some
alternative estimators perform better for individual properties or coordinate
scans, however, so this comparison supports the balanced cross-dataset
performance of \MERCE{} rather than universal superiority. Complete
definitions, fitting protocols, and numerical results are provided in the
Supporting Information.

Given MP2, CCSD, and CCSD(T) correlation energies computed in the same basis
set and with the same core convention, the \MERCE{} extrapolation procedure
estimates correlation energies at a chosen target rank, most commonly $x=4.0$
for a CCSDTQ-like finite-basis target.  The target rank is chosen according to the reference being approximated. For
benchmarks with CCSDTQ or DLPNO-CCSDTQ references, the natural target is
$x=4.0$, corresponding to a quadruple-excitation-level finite-basis estimate.
For the five-correlated-electron FCI developmental systems, the finite-basis FCI
limit is evaluated at $x=5.0$, because the exact excitation rank in these
systems cannot exceed the number of correlated electrons.   These values align the
coordinate numerically with the nominal excitation rank of the comparison
method; they do not establish that the adaptive coordinate is a literal
excitation-rank axis or that its $x\rightarrow\infty$ limit is FCI. In the S66 application (Section~\ref{sec:s66_application}),
larger values such as $x=10$ and $x=100$ are not assigned to physical
coupled cluster truncations; they are used only as internal checks of the
large-$x$ saturation behavior of the fitted trajectory.

\subsection{Performance Across Developmental and Cross-Dataset Model-Selection Benchmarks}

The selected \MERCE{} model should be assessed from two complementary
perspectives. First, it should remain well-behaved on the FCI-based
developmental sets used during model selection. Because these systems contain
five correlated electrons, the corresponding finite-basis FCI target is
evaluated at $x=5.0$ in the \MERCE{} effective-rank coordinate. Second, the model should exhibit transferable behavior across chemically
distinct cross-dataset model-selection benchmarks. These datasets probe
transfer beyond the small FCI developmental domain, but they are not fully
external because they contributed to candidate screening or ranking.

\begin{figure}[H]
\centering
\includegraphics[width=0.88\linewidth]{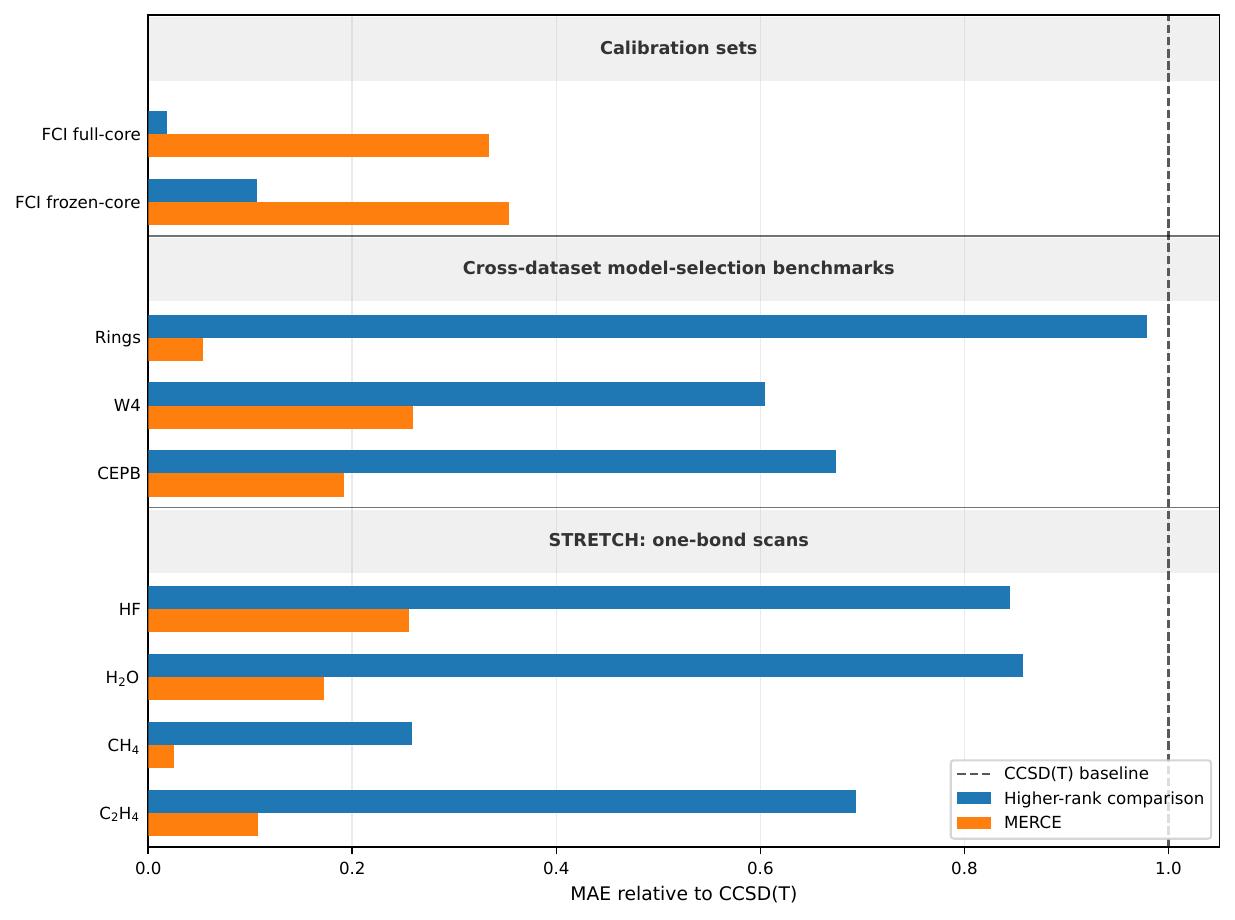}
\caption{Normalized errors on the FCI-based developmental sets and
cross-dataset model-selection benchmarks. Mean absolute errors are divided by
the corresponding CCSD(T) MAE according to Eq.~\ref{eq:md}, so the dashed
vertical line at 1.0 marks the CCSD(T) baseline. The orange bars show the final
\MERCE{} model, and the blue bars show the higher-rank comparison method:
CCSDT for the FCI-full-core, FCI-frozen-core, Rings, and W4 sets, and
DLPNO-CCSDT for CEPB and the STRETCH one-bond scans. The gray bands are visual group labels separating the FCI-based developmental
sets, the W4/Rings/CEPB molecular benchmarks, and the STRETCH coordinate scans.
STRETCH is shown separately for readability but belongs to the cross-dataset
model-selection category.}
\label{fig:transferability}
\end{figure}

Table~\ref{tab:calibration} summarizes performance on the FCI-based
developmental sets, Table~\ref{tab:external} collects the cross-dataset
model-selection results, and Figure~\ref{fig:transferability} shows the
corresponding mean absolute error (MAE) ratios relative to CCSD(T).
Root-mean-square (RMS) errors are reported as a complementary measure of the
error distribution. Additional tests performed after the functional form and
global parameters had been frozen are presented separately in
Section~\ref{sec:additional_coordinates}.

\begin{table}[H]
\centering
\caption{Performance of the final \MERCE{} model on the two FCI-based
developmental sets. The columns report mean absolute errors (MAEs) and
root-mean-square (RMS) errors in $\microEh$; $N$ denotes the number of
system basis set entries.}
\label{tab:calibration}
\begin{tabular}{l c rr rr rr rr}
\toprule
Dataset & $N$ &
\multicolumn{2}{c}{\MERCE{} 
} &
\multicolumn{2}{c}{CCSD(T)} &
\multicolumn{2}{c}{CCSDT} &
\multicolumn{2}{c}{CCSDTQ} \\
\cmidrule(lr){3-4}\cmidrule(lr){5-6}\cmidrule(lr){7-8}\cmidrule(lr){9-10}
& &
MAE & RMS &
MAE & RMS &
MAE & RMS &
MAE & RMS \\
\midrule
FCI-frozen-core & 66 & 191.89 & 222.17 & 543.08 & 583.97 & 58.00 & 68.74 & 0.77 & 1.06 \\
FCI-full-core   & 41 &  84.00 & 103.63 & 251.60 & 285.27 &  4.59 &  6.67 & 0.03 & 0.04 \\
\bottomrule
\end{tabular}
\end{table}

\begin{table}[H]
\centering
\caption{Performance (MAE [$\microEh$]) of the cross-dataset model-selection benchmarks.
Canonical CCSDTQ is used as the reference for Rings and W4, whereas
DLPNO-CCSDTQ is the reference
for CEPB and the
stretching scans.
$N$ denotes the number of retained entries: system basis set entries for the
molecular datasets and geometries for the bond-stretching scans. All \MERCE{}
estimates in this table were evaluated at $x=4.0$.
}
\label{tab:external}
\begin{tabular}{lrrrr}
\toprule
Dataset & $N$ & \MERCE{} 
& CCSD(T) & Higher-rank method \\
\midrule
Rings        & 12 & 148.96 & 2762.88 & 2704.63\textsuperscript{a)}\\
W4 subset    & 56 & 328.87 & 1268.55 &  766.31\textsuperscript{a)}\\
CEPB         & 18 & 237.81 & 1237.75 &  834.94\textsuperscript{b)}\\
HF scan      & 15 & 124.62 &  487.38 &  411.50\textsuperscript{b)}\\
H$_2$O scan  & 13 & 121.28 &  703.74 &  603.09\textsuperscript{b)}\\
CH$_4$ scan  & 12 &  18.16 &  721.05 &  186.12\textsuperscript{b)}\\
C$_2$H$_4$ scan & 10 & 169.13 & 1579.39 & 1095.60\textsuperscript{b)}\\
\bottomrule
\textsuperscript{a)} CCSDT \\
\textsuperscript{b)} DLPNO-CCSDT
\end{tabular}
\end{table}

On both FCI-based developmental sets, the \MERCE{} model 
remains
substantially more accurate than CCSD(T), reducing the MAE from
$543.08\,\microEh$ to $191.89\,\microEh$ for FCI-frozen-core and from
$251.60\,\microEh$ to $84.00\,\microEh$ for FCI-full-core. As expected,
explicit higher-rank coupled cluster methods remain more accurate on these small
finite-basis FCI reference systems, especially CCSDTQ, whose MAE relative to the FCI reference is
$0.77\,\microEh$ for FCI-frozen-core and $0.03\,\microEh$ for FCI-full-core. The important point is therefore not that \MERCE{} surpasses explicit
high-rank methods on the developmental sets, but that it achieves a clear and
consistent improvement over CCSD(T) while using only MP2, CCSD, and CCSD(T) as
inputs.

The cross-dataset model-selection benchmarks in
Table~\ref{tab:external} comprise the W4, Rings, CEPB, and original STRETCH
groups. The first three are molecular benchmark families, whereas STRETCH
contains one-bond coordinate scans that probe changes in electronic structure
along molecular distortions. Although STRETCH is displayed as a separate group
in Figure~\ref{fig:transferability} for clarity, all four groups contributed
to the same model-selection hierarchy. Because these datasets contributed to
candidate screening or ranking, they demonstrate cross-dataset robustness
rather than fully independent external validation. Within these cross-dataset model-selection benchmarks, the \MERCE{} model
reduces the CCSD(T) MAE for every included dataset. The effect is
particularly strong for aromatic and heteroaromatic rings, where the MAE
decreases from $2762.88\,\microEh$ for CCSD(T) and $2704.63\,\microEh$ for CCSDT
to $148.96\,\microEh$. On the W4 subset, the MAE is reduced by approximately a
factor of four relative to CCSD(T) and by more than a factor of two relative to
CCSDT. Similar improvements are observed for the CEPB subset and for the
one-bond stretching scans. Complete raw-rank and effective-rank scans,
reduced-scan comparisons, and detailed dataset-level tables are provided in the
Supporting Information.

\subsection{Energy Differences: Property-Level Tests}

Energy differences provide a complementary test because they probe whether
errors from multiple calculations cancel consistently. Depending on the
correlation of the component errors, cancellation can improve or degrade the
resulting property. We therefore examine A24 noncovalent correlation contributions to interaction energies 
and vertical ionization potentials separately from the absolute
correlation energy benchmarks.

A24 probes whether candidate models preserve small noncovalent
interaction energy differences. The assessment compares the Boys--Bernardi counterpoise-corrected (CP-corrected) correlation contributions to the
interaction energies, $E_{\mathrm{corr,int}}$. The CCSDT(Q), CCSDT,
CCSD(T), and DC-CCSDT comparison values were taken from
Lambie et al.,\cite{Lambie2025} whereas the corresponding \MERCE{}
quantities were evaluated in this work. A24 has also recently been used to assess transcorrelated coupled cluster
interaction energies, further illustrating the sensitivity of this benchmark
to balanced treatment of dimer and monomer correlation.
\cite{Hauskrecht2026A24}
The assessment contains the 23 entries for which
Ref.~\cite{Lambie2025} reports the complete four-method comparison ladder;
case~17 is omitted because the required reference values are unavailable. If these correlation contributions
are combined with a common CP-corrected Hartree--Fock (HF) interaction contribution,
differences relative to CCSDT(Q), and therefore the resulting MAEs and RMS
errors, are identical to those obtained from the corresponding total
interaction energies. Because A24 entered both the late-stage acceptance
screen and the final tie-break, the following values describe the selection set
performance and should not be interpreted as an independent validation
result. The A24 errors are reported in $\mathrm{kcal\,mol^{-1}}$, and the
corresponding values in $\microEh$ are given in parentheses to facilitate
direct comparison with the correlation energy benchmarks. The conversion used
is $1\,\mathrm{kcal\,mol^{-1}}\approx1593.6\,\microEh$.

Using CCSDT(Q) as the reference, the selected \MERCE{} model was evaluated at
$x=4.0$, treated as a nominal CCSDTQ-like target coordinate. For the 23-entry
subset it gives an MAE of $0.0060\,\kcalmol$ ($9.6\,\microEh$), compared with
$0.0133\,\kcalmol$ ($21.2\,\microEh$) for CCSD(T),
$0.0171\,\kcalmol$ ($27.3\,\microEh$) for CCSDT, and
$0.0070\,\kcalmol$ ($11.2\,\microEh$) for DC-CCSDT.
\cite{Lambie2025} The corresponding \MERCE{} RMS and maximum absolute errors
are $0.0134$ and $0.0588\,\kcalmol$, whereas the DC-CCSDT values are
$0.0127$ and $0.0375\,\kcalmol$; \MERCE{} has the smaller absolute error for
11 of the 23 entries. Thus, \MERCE{} lowers the selected-subset MAE, but it
does not improve the RMS or maximum error relative to DC-CCSDT. The largest
\MERCE{} deviation occurs for the formaldehyde dimer. Because these data
contributed to selection, the result documents property-level behavior of the
selected compromise rather than independent external validation.
System-resolved interaction energies, signed deviations, and the formal
nonadditivity audit are provided in Tables~S19--S21.

\begin{figure}[H]
\centering
\includegraphics[width=0.98\linewidth]{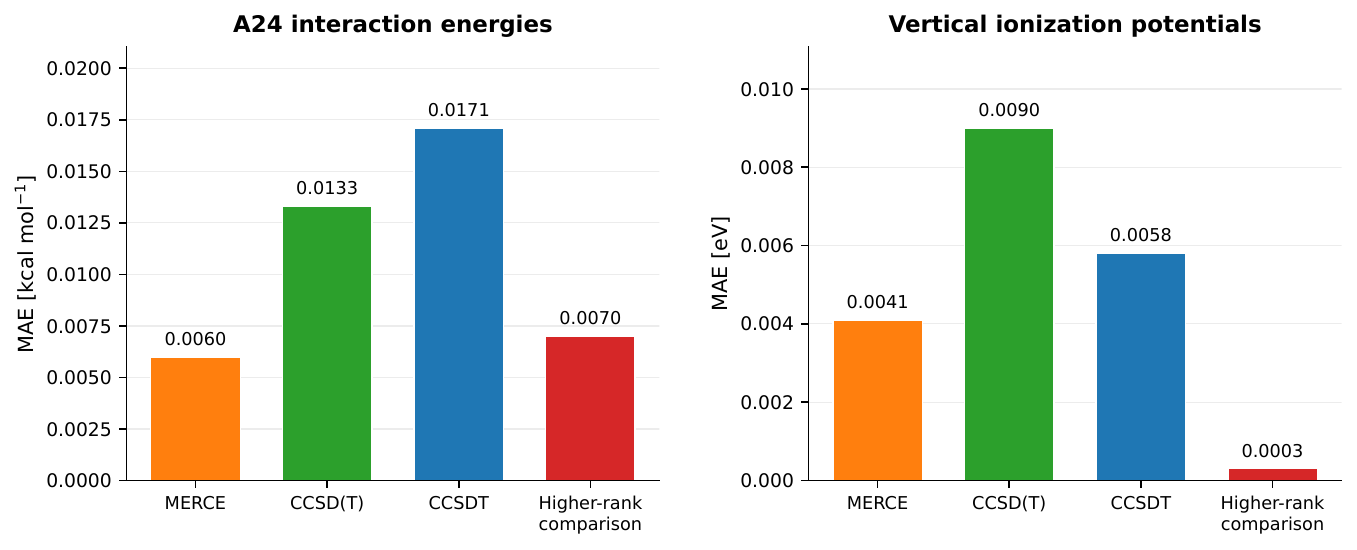}
\caption{Property-level performance on energy differences. A24 results are reported as MAEs of the CP-corrected correlation
contributions to the interaction energies, in $\kcalmol$, relative to
CCSDT(Q), whereas vertical ionization-potential errors are reported as MAEs
in eV relative to finite-basis FCI. A24 contributed to late-stage model selection and is therefore a
selection-set assessment rather than an independent validation test. The
higher-rank comparison is DC-CCSDT for A24 and CCSDTQ for the vertical
ionization potentials. Both quantities have dimensions of energy; the different units follow the
customary reporting of noncovalent interaction energies in $\kcalmol$ and
ionization potentials in eV. Corresponding values in $\microEh$ are provided
in the text to facilitate direct comparison.}
\label{fig:properties}
\end{figure}

Vertical ionization potentials provide a complementary property-level test
because neutral and cationic correlation energies must be combined in a
balanced manner. The test systems are constructed from the FCI-full-core and
FCI-frozen-core developmental sets, so the neutral and cationic species may
contain different numbers of correlated electrons. Accordingly, each \MERCE{} correlation energy component is evaluated at the
target coordinate corresponding
to its finite-basis FCI limit: $x=5.0$ for species with five correlated
electrons and $x=6.0$ for species with six correlated electrons. Against the resulting FCI references, the \MERCE{} model gives an MAE of
$0.0041\,\mathrm{eV}$ ($151\,\microEh$) and an RMS error of
$0.0088\,\mathrm{eV}$ ($323\,\microEh$). This is smaller than the
corresponding HF, MP2, CCSD, CCSD(T), and CCSDT errors and approaches the
accuracy of explicit high-rank coupled cluster corrections for these small
systems. System-by-system ionization potentials, signed errors, target-coordinate sensitivity, and the exploratory B$_2$-exclusion analysis are reported in
Tables~S22--S24. Figure~\ref{fig:properties} summarizes both
property-level tests.

As a target-coordinate sensitivity check, applying a common shift
$\Delta x\in[-0.5,0.5]$ to the neutral and cation coordinates changes the
ten-record \MERCE{} MAE only from 0.00407 to $0.00415\,\mathrm{eV}$ and the RMS
error from 0.00873 to $0.00890\,\mathrm{eV}$; the complete scan is reported in
the Supporting Information. The complete hierarchical-selection audit and associated robustness analyses
are reported in Tables~S25--S32. The expanded post-selection search and its
A24 follow-up are documented in Tables~S33--S36 and Figures~S1 and S2,
whereas the comparison with alternative sequence transformations and refitted
linear baselines is given in Tables~S37 and S38.

\subsection{One-Bond Stretching Deviation Profiles}

\begin{figure}[H]
\centering
\includegraphics[width=0.94\linewidth]{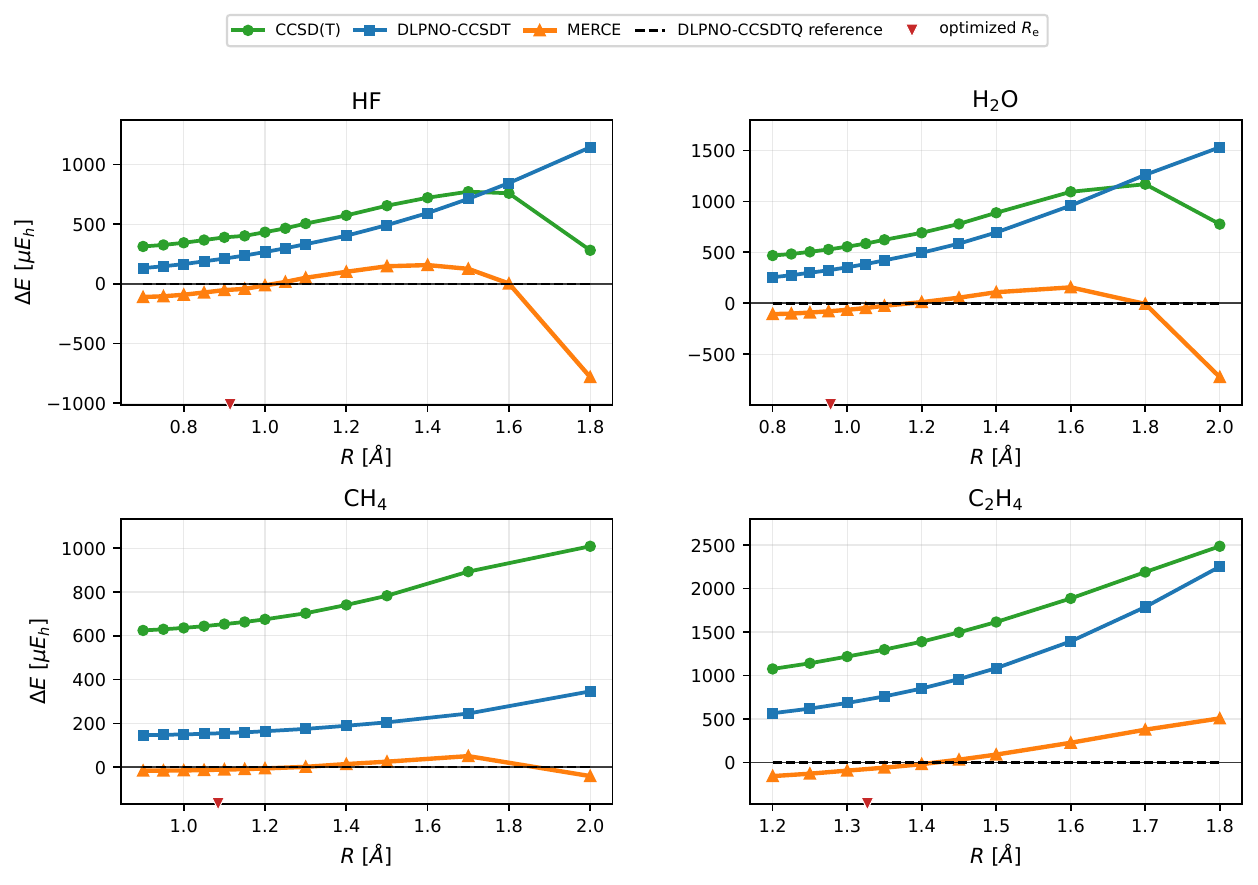}
\caption{One-bond stretching deviation profiles for HF, H$_2$O, CH$_4$, and
C$_2$H$_4$ relative to DLPNO-CCSDTQ. The curves therefore represent method
errors rather than total-energy profiles. Red markers on the $x$ axes indicate
the optimized parent-geometry bond lengths,
$R_{\mathrm{e}}=0.91423$~\AA{} for HF,
$0.95552$~\AA{} for O--H in H$_2$O, $1.08460$~\AA{} for C--H in CH$_4$,
and $1.32708$~\AA{} for C=C in C$_2$H$_4$; these exact coordinates were not
included in the sampled grids. The figure focuses on the high-rank comparison
by showing CCSD(T), DLPNO-CCSDT, and the final \MERCE{} model; MP2 and CCSD are
omitted for clarity. The \MERCE{} model remains close to DLPNO-CCSDTQ over the
near-equilibrium and moderately stretched regions, whereas the most elongated
HF and H$_2$O geometries expose deterioration of the present
single-reference-based mapping.}
\label{fig:stretching}
\end{figure}

One-bond stretching provides a complementary stress test because the electron
correlation character changes continuously along a molecular coordinate. In
contrast to equilibrium benchmark sets, these scans probe whether a single
global effective-rank mapping remains stable as the electronic structure moves
away from the near-equilibrium regime. We considered four representative
scans: HF bond elongation, stretching of one O--H bond in
H\textsubscript{2}O, stretching of one C--H bond in CH\textsubscript{4}, and
C=C bond stretching in C\textsubscript{2}H\textsubscript{4}. All four scans
were performed in the cc-pVTZ basis without the frozen-core approximation, so
that all electrons are correlated consistently along the stretching coordinate. The parent geometries for H$_2$O, CH$_4$, and C$_2$H$_4$ were taken from the
CEPB training set and were previously optimized at the all-electron CCSD/cc-pVTZ
level.\cite{CEPB} The corresponding HF geometry was taken from the NIST CCCBDB
at the same level of theory.\cite{NISTCCCBDB} Each scan was generated by
varying only the indicated bond length while retaining the remaining internal
coordinates of the parent geometry. Because rounded bond lengths were used to
construct the scan grids, none of the grids contains a calculation exactly at
the optimized parent-geometry value.
For the original STRETCH benchmarks, the corresponding $T_1$ and $D_2$
diagnostic profiles are shown in Figure~S3. The $T_1$ diagnostic is derived
from the single-excitation amplitudes,\cite{LeeTaylor1989} whereas the $D_2$
diagnostic is constructed from the double-excitation amplitudes and provides a
complementary measure of single-reference character.
\cite{JanssenNielsen1998,NielsenJanssen1999} These diagnostics are included as descriptive indicators of the
electronic-structure changes along the scanned coordinates. They were not used, though, as numerical thresholds for selecting or truncating the reported geometry ranges.
Figure~\ref{fig:stretching} shows method deviations relative to
the DLPNO-CCSDTQ reference. The plotted curves are therefore not potential-energy curves, but
errors with respect to the highest-level reference available for each geometry.
To focus on the high-rank regime while retaining sufficient resolution to
distinguish the most accurate methods, the figure compares CCSD(T),
DLPNO-CCSDT, and the \MERCE{} model. MP2 and CCSD are not shown because their
substantially larger deviations would dominate the vertical scale and obscure
the differences among the higher-rank approaches.

Across most scanned geometries, \MERCE{} remains much closer to the
DLPNO-CCSDTQ reference than CCSD(T) and is usually also closer than
DLPNO-CCSDT. This behavior is particularly visible for CH\textsubscript{4} and
C\textsubscript{2}H\textsubscript{4}, where the CCSD(T) and DLPNO-CCSDT
deviations increase monotonically with bond elongation, while the \MERCE{} curve stays close to zero over nearly the whole scan.

The most elongated geometries are especially informative because they expose
the boundary of applicability of the present model. For HF, the \MERCE{}
deviation changes sign at the final reported geometry. For H$_2$O, it becomes slightly negative at $R=1.80\,\text{\AA}$ and then
markedly more negative at $R=2.00\,\text{\AA}$, showing that the global
MP2--CCSD--CCSD(T) mapping begins to overcorrect in the strongly stretched
regime. At the same time, the $T_1$ and
$D_2$ profiles reported in the Supporting Information indicate increasing
departure from regular single-reference character as the bonds are elongated.
This correspondence supports, but does not by itself prove, the interpretation
that the deterioration of \MERCE{} is associated with the increasing
multireference character of the stretched-bond wave function.

This behavior is consistent with the method's construction. \MERCE{} uses
MP2, CCSD, and CCSD(T), all of which are most reliable when a
single-determinant reference remains qualitatively appropriate. During strong
bond elongation, static correlation becomes increasingly important, and one or
more of the input methods may cease to define a regular trajectory toward the
high-rank correlation limit. The resulting deterioration is therefore a
limitation inherited partly from the input sequence, rather than merely a
failure of the extrapolation function.

The stretching scans consequently demonstrate both the utility and the limits
of the present model. \MERCE{} substantially improves the description of
near-equilibrium and moderately stretched geometries, but it should not be
expected to remain reliable in the strongly stretched, increasingly
multireference regime. Such cases require either an explicit reliability flag,
a diagnostic-aware mapping, or an alternative trajectory constructed from
methods suitable for multireference electronic structure.

\subsection{Additional Tests on Molecular Coordinates}
\label{sec:additional_coordinates}

Most benchmark sets discussed above contributed, directly or indirectly, to
model screening or selection. We therefore performed additional
molecular-coordinate scans only after the \MERCE{} functional form,
effective-rank mapping, global parameters, target rank, and numerical domain
checks had been frozen. None of the scan geometries or energies described in
this subsection was used to modify the reported model. LiF, CH$_3$F, and
CH$_3$OH were absent from the model-selection datasets. NH$_3$ molecular
entries were present in the W4 and CEPB benchmarks, but none of the
umbrella-planarization geometries or their energies was used in fitting,
screening, or ranking.

\paragraph{Rigid bond-stretching scans.}

Three additional stretching coordinates were defined before any results from
these scans were available:
Li--F dissociation in LiF, C--F stretching in CH$_3$F, and C--O stretching
in CH$_3$OH. The reference Cartesian geometries were taken from calculated
geometry records in the NIST Computational Chemistry Comparison and Benchmark
Database (CCCBDB).\cite{NISTCCCBDB} In the CCCBDB notation, the optimization levels were CCSD(T)=FULL/aug-cc-pVQZ for LiF and CH$_3$F and
CCSD(T)=FULL/aug-cc-pVTZ for CH$_3$OH, where FULL denotes all-electron
correlation. The corresponding optimized bond lengths are
$R_{\mathrm{e}}=1.5562$~\AA{} for Li--F in LiF, $1.3815$~\AA{} for C--F
in CH$_3$F, and $1.4205$~\AA{} for C--O in CH$_3$OH. Each scan contained 17 predefined geometries at
\[
\begin{aligned}
R/R_{\mathrm e}={}&
0.80,\ 0.85,\ 0.90,\ 0.95,\ 1.00,\ 1.05,\ 1.10,\ 1.15,\\
&1.20,\ 1.30,\ 1.40,\ 1.50,\ 1.60,\ 1.75,\ 2.00,\ 2.25,\
\text{and}\ 2.50.
\end{aligned}
\]
The interval $0.80\leq R/R_{\mathrm e}\leq1.60$, comprising 13
geometries per system, was fixed before the calculations as the primary
assessment domain. The four longer-distance geometries were reserved as a
separate strong-stretch stress test and were not used to redefine the primary
range after inspecting the results.

The scans were rigid. For LiF, one atom was displaced along the molecular
axis. For CH$_3$F, the carbon atom and three hydrogen atoms were fixed while
fluorine was displaced along the C--F axis. For CH$_3$OH, the complete OH
fragment was translated relative to the fixed CH$_3$ fragment, preserving the
internal geometries of both fragments. All calculations in these three rigid bond-stretching scans employed a
restricted Hartree--Fock (RHF) reference, the cc-pVTZ basis, and all-electron
correlation. MP2, CCSD, and
CCSD(T) supplied the input trajectory, the frozen \MERCE{} model was evaluated
at $x=4$, and DLPNO-CCSDTQ served as the approximate high-rank comparison. DLPNO-CCSDTQ is used here as a computationally accessible approximate
high-rank comparator rather than as an exact canonical CCSDTQ reference.
Accordingly, the reported quantities measure deviations from DLPNO-CCSDTQ.

\begin{figure}[H]
\centering
\includegraphics[width=0.94\linewidth]
{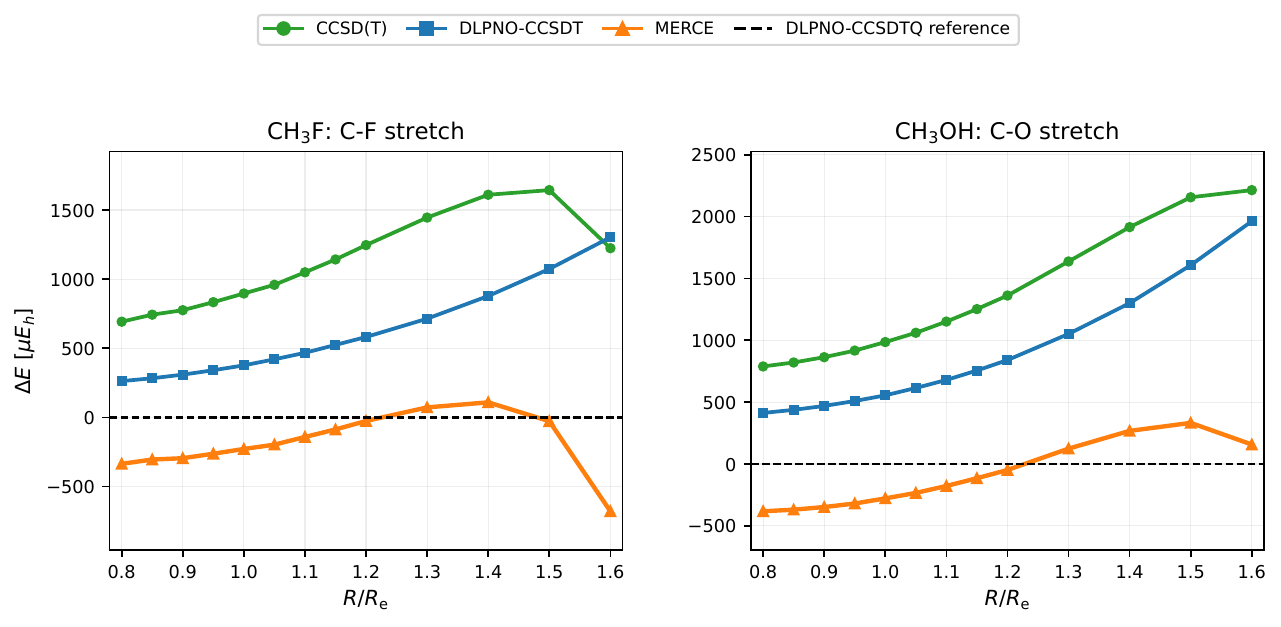}
\caption{Additional rigid one-bond stretching tests for CH$_3$F (left) and
CH$_3$OH (right) over the predefined primary domain
$0.80\leq R/R_{\mathrm e}\leq1.60$. Energy deviations are reported relative
to DLPNO-CCSDTQ. For CH$_3$F, the C and H atoms were fixed while F was displaced
along the C--F axis. For CH$_3$OH, the complete OH fragment was translated
relative to the fixed CH$_3$ fragment, preserving both fragment geometries.
The corresponding $T_1$ and $D_2$ diagnostic profiles are shown in
Figure~S4; the complete rigid-stretch audit, including the strong-stretch
region and the LiF rejection record, is given in Table~S39.}
\label{fig:additional_stretches}
\end{figure}

Figure~\ref{fig:additional_stretches} shows the predefined primary domains
for CH$_3$F and CH$_3$OH. All 13 DLPNO-CCSDTQ comparison calculations in each
primary domain were completed. For CH$_3$F, the \MERCE{} deviations have an
MAE of $212.4\,\microEh$, an RMS value of $271.3\,\microEh$, and a maximum
absolute value of $678.4\,\microEh$. The corresponding values are
$1097.1$, $1140.0$, and $1643.8\,\microEh$ for CCSD(T), and
$579.8$, $658.9$, and $1302.7\,\microEh$ for DLPNO-CCSDT. Thus, along this
previously untouched coordinate, the final MERCE model reduces the MAE of the
deviation from DLPNO-CCSDTQ by a factor of approximately 5.2 relative to
CCSD(T).

The independently specified CH$_3$OH scan gives a consistent result. The
\MERCE{} MAE, RMS deviation, and maximum absolute deviation are
$242.7$, $264.3$, and $382.6\,\microEh$, respectively, compared with
$1316.8$, $1403.6$, and $2213.7\,\microEh$ for CCSD(T), and
$860.9$, $980.2$, and $1962.3\,\microEh$ for DLPNO-CCSDT. The reduction in MAE
relative to CCSD(T) is a factor of approximately 5.4. Neither CH$_3$F nor CH$_3$OH appeared in the datasets used to select
the functional form or global parameters, and none of the geometries or
energies from these scans entered the selection procedure.  The results therefore support transfer of the frozen model
to moderate distortions of two additional polyatomic molecules, as measured
against the available DLPNO-CCSDTQ reference, without implying validity in
the dissociation limit or exact agreement with canonical CCSDTQ.

LiF gives a qualitatively different outcome. Except at
$R/R_{\mathrm e}=0.80$, its MP2--CCSD--CCSD(T) correlation energy sequence
fails the predefined monotonic-ordering requirement. The \MERCE{} calculation
is therefore rejected at 12 of the 13 primary geometries and at all four
strong-stretch geometries; no extrapolated values are forced through these
points. Because only one primary geometry is accepted, an aggregate LiF MAE
would not be meaningful and is not reported. This domain-rejection
result demonstrates both that the numerical safeguard operates independently
of subsequent reference errors and that the present MP2--CCSD--CCSD(T)-based
scheme is not universally applicable, even around the equilibrium region of
an ionic diatomic molecule. The complete LiF results and rejection record are
reported in the Supporting Information.

\paragraph{Rigid NH$_3$ umbrella planarization.}

A molecular deformation qualitatively different from bond stretching was
examined through a rigid, symmetry-preserving NH$_3$ umbrella-planarization
scan. The reference pyramidal geometry was defined using the equilibrium
structural parameters reported by Marquardt et al. for their global analytical
potential-energy surface of ground-state NH$_3$:
$r_{\mathrm{NH}}^{\mathrm e}=1.0128$~\AA{} and
$\angle\mathrm{HNH}^{\mathrm e}=107.03^\circ$.\cite{Marquardt2013NH3}

The dimensionless scan coordinate is
\begin{equation}
\lambda=\frac{h}{h_{\mathrm e}},
\label{eq:nh3_lambda}
\end{equation}
where $h$ is the perpendicular distance of the nitrogen atom from the plane
containing the three hydrogen atoms. For the reference pyramidal structure,
\begin{equation}
h_{\mathrm e}
=
r_{\mathrm{NH}}^{\mathrm e}
\sqrt{
\frac{
\cos\!\left(\angle\mathrm{HNH}^{\mathrm e}\right)+1/2
}{3/2}
}
=
0.37635446~\text{\AA}.
\label{eq:nh3_height}
\end{equation}
All three N--H distances were held fixed at
$r_{\mathrm{NH}}^{\mathrm e}$. 
The frozen scan comprised
\[
\begin{aligned}
\lambda={}&
1.00,\ 0.95,\ 0.90,\ 0.85,\ 0.80,\ 0.75,\ 0.70,\ 0.65,\ 0.60,\\
&0.55,\ 0.50,\ 0.45,\ 0.40,\ 0.30,\ 0.20,\ 0.10,\
\text{and}\ 0.00.
\end{aligned}
\]
Here, $\lambda=1$ denotes the reference trigonal-pyramidal
($C_{3v}$) structure and $\lambda=0$ the rigid planar ($D_{3h}$) structure.
The intermediate geometries were generated in this work and were not taken
from Ref.~\cite{Marquardt2013NH3}. Because the N--H distances remain fixed at
their equilibrium value, the planar endpoint is not the optimized inversion
transition structure reported by Marquardt et al. Consequently, the associated
energy difference is a rigid planarization energy rather than the physical
NH$_3$ inversion barrier.

The same all-electron RHF/cc-pVTZ electronic-structure protocol and frozen
$x=4$ \MERCE{} model were used as in the additional one-bond stretching tests.

\begin{figure}[H]
\centering
\includegraphics[width=0.96\linewidth]
{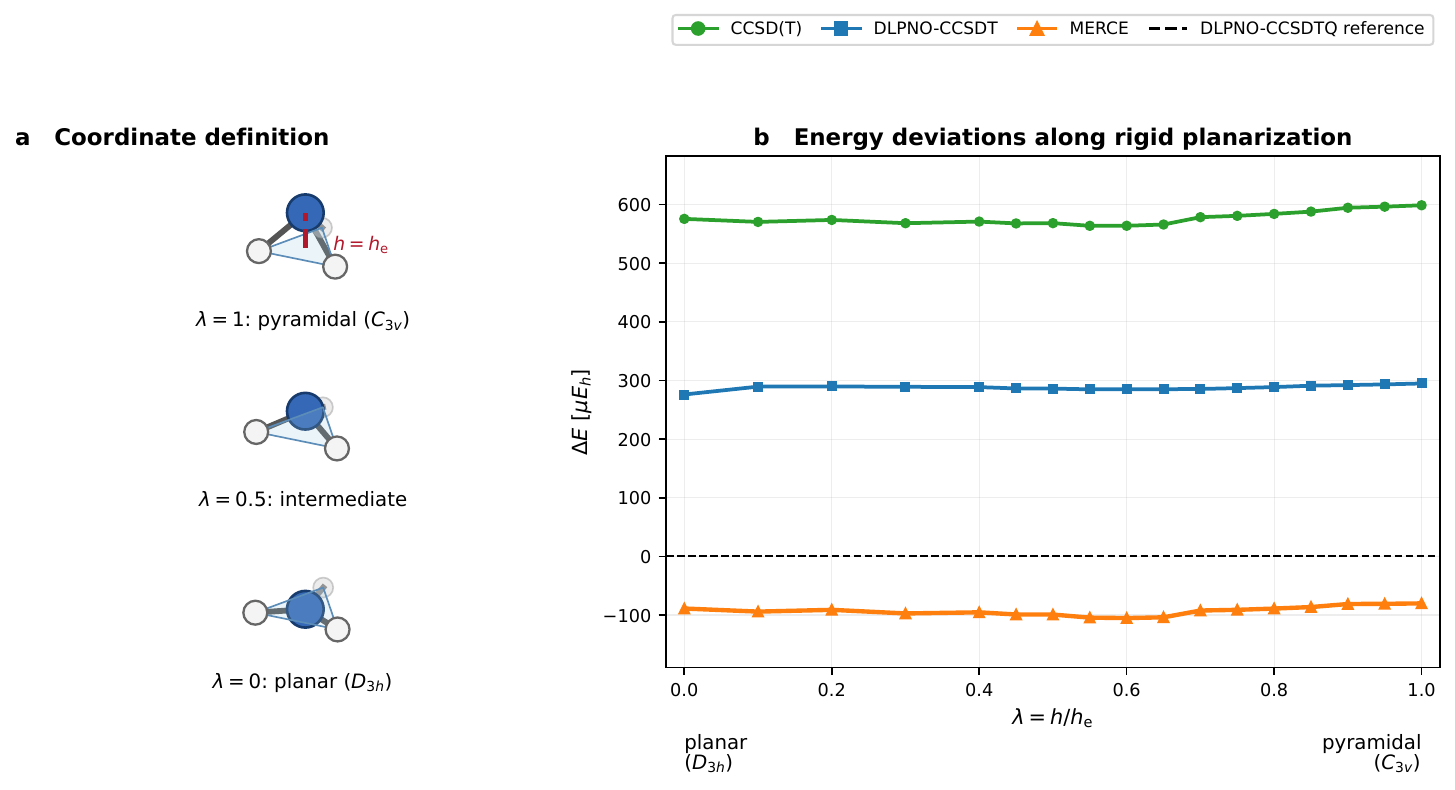}
\caption{Additional rigid NH$_3$ umbrella-planarization scan. (a) Definition
of $\lambda=h/h_{\mathrm e}$, where $h$ is the perpendicular displacement of
N from the plane containing the three H atoms and
$h_{\mathrm e}=0.37635446$~\AA{} is its value in the reference pyramidal
geometry. The N--H distances remain fixed at
$r_{\mathrm{NH}}^{\mathrm e}=1.0128$~\AA{} throughout the scan;
$\lambda=1$ denotes the pyramidal ($C_{3v}$) structure and $\lambda=0$ the
rigid planar ($D_{3h}$) structure. The reference structural parameters were
taken from Ref.~\cite{Marquardt2013NH3}. (b) Energy deviations relative to
DLPNO-CCSDTQ. All 17 predefined geometries, including the rigid planar endpoint at
$\lambda=0$, have converged DLPNO-CCSDTQ comparison values. The corresponding $T_1$ and $D_2$ diagnostic profiles are shown in
Figure~S5, and the complete numerical audit is given in Table~S40.}
\label{fig:additional_nh3}
\end{figure}

Figure~\ref{fig:additional_nh3} defines the coordinate graphically and shows
deviations from DLPNO-CCSDTQ along the rigid planarization scan. Across all 17
predefined geometries, the \MERCE{} deviations have an MAE of
$93.0\,\microEh$, an RMS value of $93.4\,\microEh$, and a maximum absolute
value of $105.3\,\microEh$. The corresponding values are $577.1$, $577.2$, and
$598.9\,\microEh$ for CCSD(T), and $287.8$, $287.9$, and
$294.9\,\microEh$ for DLPNO-CCSDT. The frozen \MERCE{} model therefore remains
in close and nearly uniform agreement with the DLPNO-CCSDTQ reference along a
deformation coordinate qualitatively different from bond stretching,
including the rigid planar endpoint.

\subsection{Size Consistency}
\label{sec:size_consistency}

MP2, CCSD, and CCSD(T) correlation energies are size-consistent for
noninteracting fragments when the same basis set, correlated-electron space, reference convention, and
computational protocol are used for the fragments and the formal supersystem.\cite{Pople1976SizeConsistency,BartlettMusial2007,Raghavachari1989}
The size-consistency issue considered here is therefore not caused by the
MP2, CCSD, and CCSD(T) input energies themselves, but by applying a nonlinear,
system-dependent extrapolation functional to a composite system. This type of
nonadditivity is a general risk for nonlinear extrapolation procedures with
system-dependent descriptors. Size consistency is therefore evaluated explicitly
using the artificial noninteracting pairs introduced in the model-selection
protocol.

To construct these tests, fragments are paired only within the same benchmark
set and the same one-particle basis; for example, cc-pVTZ data are paired only
with other cc-pVTZ data and are not mixed with cc-pVQZ or aug-cc-pVTZ
calculations. For a pair of fragments $A$ and $B$, the formal supersystem
$A+B$ is not recomputed as a real dimer. Instead, its MP2, CCSD, CCSD(T), and
reference correlation energies are constructed by adding the corresponding
fragment energies. The \MERCE{} estimate is then evaluated in two ways: first
by applying the extrapolation once to the combined $A+B$ input sequence, and
second by applying the same extrapolation separately to $A$ and $B$ and summing
the two results.

The difference between these two procedures is the size-consistency defect
$\Delta_{\mathrm{SC}}$ defined in Eq.~\ref{eq:delta_sc}. 
Similar corrections have been introduced in the contexts of approximate CC methods\cite{sscS} and adiabatic connection methods\cite{sscV}.

The corresponding
pair-energy error with respect to the additive high-level reference is
$\Delta_{\mathrm{pair}}$, defined in Eq.~\ref{eq:delta_pair}. Their mean
absolute values, $\mathrm{SC}_{\mathrm{all}}$ and
$\mathrm{pair\text{-}MAE}_{\mathrm{all}}$, together with the relative
size-consistency measure $Q_{\mathrm{all}}$, were defined in
Eqs.~\ref{eq:sc_all}--\ref{eq:qall} and used in the selection protocol. For the
additional stress-test analysis in this section, we also examine the combined
quantity
\begin{equation}
 \Delta_{\mathrm{comb}}=
 |\Delta_{\mathrm{pair}}|+|\Delta_{\mathrm{SC}}|,
\label{eq:delta_comb}
\end{equation}
which penalizes both the ordinary pair-energy error and the extrapolation-induced
nonadditivity.

The largest size-consistency defects occur for heterogeneous artificial pairs,
most prominently in the W4-derived subset. Here, ``heterogeneous'' does not
refer primarily to chemical composition, but to the effective-rank response of
the two fragments. The most problematic cases are mixed high-$R$+low-$R$
combinations, where one fragment samples the saturated part of the arctangent
mapping while the other remains in the low-response region. If the fragments
are extrapolated separately, each fragment is mapped according to its own local
increment ratio, $R_A$ or $R_B$. In the formal supersystem calculation, however,
the mapping is controlled by the ratio formed from the summed increments,
\[
R_{A+B} =
\frac{|d_2(A)+d_2(B)|}
{|d_1(A)+d_1(B)|+\varepsilon_{\mathrm{num}}},
\]
which generally does not reproduce the two fragment-specific responses.
Nonadditivity is therefore expected to be most visible when the fragments
occupy different nonlinear regions of the effective-rank mapping.

This interpretation is supported by the effective-rank mismatch analysis. For
the final representative \MERCE{} model, the absolute size-consistency defect
in the W4 artificial-pair set correlates strongly with the fragment mismatch
measure $|\ln R_A-\ln R_B|$, with a Pearson correlation coefficient of $0.812$.
By contrast, the ordinary pair-energy error has no analogous positive
correlation with this mismatch measure; the corresponding coefficient is
$-0.178$. Within the same W4-derived artificial-pair set, category-resolved
statistics for the final \MERCE{} model show that mixed
high-$R$+low-$R$ pairs
have a mean absolute size-consistency defect of $109.8\,\microEh$, compared
with $36.8\,\microEh$ for high-$R$+high-$R$ pairs,
$18.9\,\microEh$ for low-$R$+low-$R$ pairs, and only
$0.9\,\microEh$ for mid-$R$+mid-$R$ pairs. Thus, the dominant source of
nonadditivity is not the absolute size of the pair-energy error, but the
mismatch between the local effective-rank responses sampled by the two
fragments. Because many artificial pairs share one or both fragment calculations, the
individual pair records are not statistically independent. The reported
Pearson coefficients and category-resolved means should therefore be
interpreted as descriptive measures of the observed mismatch pattern rather
than as independent-pair significance tests.

Importantly, the largest size-consistency defects are not necessarily the
largest pair-energy errors. Even after applying the conservative combined-error
penalty defined in Eq.~\ref{eq:delta_comb}, extrapolated pair energies generally remain much more accurate than
CCSD(T) and usually more accurate than CCSDT. Nonadditivity is therefore a real
limitation of the present nonlinear protocol, but it is also a measurable one:
$\Delta_{\mathrm{SC}}$ can be reported alongside the pair-energy prediction as
an explicit nonadditivity diagnostic. It is not a calibrated error bound. The Supporting Information reports both the
final-model category analysis used for the numerical values above and an
aggregate analysis over ten high-performing models from the full $(k_1,k_3)$
scan, showing that the same $R$-mismatch pattern is not unique to the selected
inverse-cubic-linear model.

We also evaluated the formal nonadditivity directly for the component
calculations entering the A24 and selected-S66 interaction energy analyses.
Extending Eq.~\ref{eq:delta_sc} to an interaction energy convention
$Y\in\{\mathrm{CP},\mathrm{no\text{-}CP}\}$, we define
\begin{equation}
\Delta_{\mathrm{SC}}^{Y}(A,B;x)
=
E_{\mathrm{MERCE}}^{Y}(A+B;x)
-
E_{\mathrm{MERCE}}^{Y}(A;x)
-
E_{\mathrm{MERCE}}^{Y}(B;x).
\label{eq:property_nonadditivity}
\end{equation}
Here, $E_{\mathrm{MERCE}}^{Y}(A+B;x)$ is obtained by applying the \MERCE{}
extrapolation once, at target coordinate $x$, to the componentwise sums of the
MP2, CCSD, and CCSD(T) correlation energies of fragments $A$ and $B$ under
convention $Y$. The remaining two terms are obtained by extrapolating the two
fragment input sequences separately. In the A24 and selected-S66 analyses,
$x=4$. For the 23 A24 entries, the CP
mean absolute and maximum offsets are 2.67 and $32.00\,\microEh$,
respectively. Removing this formal offset changes the \MERCE{} A24 MAE from
0.00603 to $0.00644\,\kcalmol$, the RMS error from 0.01338 to
$0.01341\,\kcalmol$, and the number of lower-error cases relative to
DC-CCSDT from 11 to 10 of 23. For the 31 selected-S66 complex--basis entries,
the corresponding CP mean absolute and maximum offsets are 10.55 and
$107.83\,\microEh$. Thus, nonadditivity is small on average for A24 but is not
negligible for every property-level entry and should be retained as a reported
diagnostic.

\subsection{Domain of Validity}

The present \MERCE{} model has two distinct domain requirements. The first is
algorithmic: the MP2--CCSD--CCSD(T) sequence must satisfy the stated ordering,
increment, target-coordinate, residual, and pole checks. Passing this
algebraic applicability gate means only that the released estimator is
defined; it does not certify accuracy or diagnose single-reference character.
The second requirement is electronic-structural: MP2, CCSD, and CCSD(T) must
remain appropriate inputs for the system and geometry. Strong static
correlation can violate this requirement even when the algebraic gate passes.

This distinction is observed directly in the far-stretched CH$_3$F and
CH$_3$OH tails: the input ordering still passes the algebraic gate, but the
\MERCE{} deviations become larger than the corresponding CCSD(T) deviations.
Conversely, the LiF ordering failure activates the intended abstention. The
present gate should therefore be interpreted as a well-defined numerical
applicability test, not as a universal strong-correlation diagnostic.

The same caution applies to
systems with near-degeneracy, severe multireference character, or strongly
stretched bonds where the wave function loses single-determinant character. In
these regimes, a diagnostic-aware extension should either modify the mapping,
switch to a different sequence of input methods, or return an uncertainty flag
instead of a single extrapolated value.
A recently introduced density-difference diagnostic provides a complementary
mean-field-cost indicator of multireference character and can guide the choice
of electronic reference.\cite{Zamani2026KSDensity} Such diagnostics may help
identify systems near the present domain boundary, but they are not part of
the current \MERCE{} input gate.

Explicitly separating these two requirements is essential. Algebraic rejection
is reported as abstention, whereas the accumulated strong-stretch results
define an empirical limitation rather than a calibrated reliability threshold.

\subsection{Application to S66 Noncovalent Complexes: Fixed-Basis Post-CCSD(T) Corrections}
\label{sec:s66_application}

After the reference-based assessments and size-consistency analysis, the
\MERCE{} model was applied to selected chemically recognizable complexes from
the S66 set.\cite{Rezac2011S66} The purpose of this subsection is not to define
another accuracy benchmark against the standard S66 reference energies. Those
reference values are complete basis set interaction energies assembled through a
composite CCSD(T)-based protocol, whereas \MERCE{} predicts a finite-basis
post-CCSD(T) correction to a specified MP2--CCSD--CCSD(T) trajectory. The present results should therefore be interpreted as fixed-basis
post-CCSD(T) corrections to interaction energies evaluated using the
Boys--Bernardi counterpoise procedure to reduce basis set superposition
error (BSSE),\cite{BoysBernardi1970} rather than as errors relative to
the canonical S66/CBS reference values.

This distinction is important because post-CCSD(T) effects in S66 have recently
been examined explicitly by Semidalas et al.\cite{Semidalas2025PostCCSDT}
However, those calculations were carried out in a different computational setup,
including different basis set and interaction energy conventions. They provide
important qualitative context for the scale of post-(T) contributions, but they
are not directly equivalent to the present counterpoise-corrected fixed-basis
MERCE predictions.  Here, we use the equilibrium S66x8 geometries at the
$r/r_{\mathrm{e}}=1.00$ scaling factor and evaluate
counterpoise-corrected interaction energies for all seven selected complexes
in the cc-pVDZ, cc-pVTZ, aug-cc-pVDZ, and aug-cc-pVTZ basis sets.
For cases 21, 25, and 57, cc-pVQZ results are also included.

The selected subset covers chemically distinct noncovalent motifs, including
amide hydrogen bonding, aromatic and heteroaromatic $\pi$--stacking, mixed
$\pi$/hydrogen-bond interactions, and CH$\cdots$N contacts. The subset is not
intended to be statistically representative of the full S66 database. Its role
is to illustrate the type and magnitude of finite-basis post-CCSD(T)
corrections that \MERCE{} can provide for realistic noncovalent complexes when
direct fixed-basis CCSDTQ or CCSDT(Q) calculations are not practical.

In this analysis, $x=4$ is used as the nominal CCSDTQ-like target rank.
The larger values $x=10$ and $x=100$ do not correspond to physical
coupled cluster truncation levels. Instead, they are used solely to probe the
large-$x$ behavior of the fitted \MERCE{} trajectory and must not
be interpreted as independently validated post-CCSDTQ corrections. Accordingly, the
$x=4\rightarrow x=10$ difference measures the residual change predicted by
the fitted curve beyond the nominal quadruple-excitation-level target; it
should not be interpreted as an independently validated physical
post-CCSDTQ correction. A visible $x=4\rightarrow x=10$ change indicates that
the fitted trajectory has not yet reached its asymptotic regime at $x=4$.
Conversely, a small $x=10\rightarrow x=100$ difference indicates numerical
saturation of that trajectory. These quantities are therefore internal
trajectory-stability diagnostics and are not substitutes for explicit
higher-rank coupled cluster calculations.

\begin{table}[H]
\centering
\small
\setlength{\tabcolsep}{3.5pt}
\caption{Basis set and target rank dependence of CP-corrected \MERCE{}
post-CCSD(T) corrections for selected S66 noncovalent complexes. Energies are
given in kcal mol$^{-1}$. $E_{\mathrm{CCSD(T)}}^{\mathrm{int}}$ is the CP-corrected total
CCSD(T) interaction energy. The quantity
$\Delta_x=E_{\mathrm{MERCE}}^{\mathrm{int}}(x)
-E_{\mathrm{CCSD(T)}}^{\mathrm{int}}$ is the corresponding finite-basis
post-CCSD(T) correction.  Because \MERCE{} modifies only the correlation
contribution, the common HF term cancels, and
$\Delta_x=E_{\mathrm{MERCE}}^{\mathrm{corr,int}}(x)
-E_{\mathrm{CCSD(T)}}^{\mathrm{corr,int}}$. DZ, TZ, ADZ, and ATZ denote cc-pVDZ,
cc-pVTZ, aug-cc-pVDZ, and aug-cc-pVTZ, respectively. The $x=4$ column is the
nominal CCSDTQ-like correction, whereas the $x=10$ and $x=100$ columns probe
the residual high-rank tail and the large-$x$ saturation of the fitted
trajectory.}
\label{tab:s66_basis_target_rank}
\begin{tabular}{c l c rrrr}
\toprule
Case & Complex & Basis &
$E_{\mathrm{CCSD(T)}}^{\mathrm{int}}$ & $\Delta_4$ & $\Delta_{10}$ & $\Delta_{100}$ \\
\midrule
21 & AcNH$_2\cdots$AcNH$_2$ & DZ  & -12.3176 & -0.0237 & -0.0376 & -0.0384 \\
   &                         & TZ  & -14.8560 & -0.0320 & -0.0507 & -0.0517 \\
   &                          & QZ     &  -15.9335 & -0.0325 & -0.0511& -0.0521 \\
   &                         & ADZ & -14.6130 & -0.0422 & -0.0664 & -0.0678 \\
   &                         & ATZ & -15.7937 & -0.0408 & -0.0637 & -0.0650 \\
\midrule
24 & benzene$\cdots$benzene ($\pi\cdots\pi$) & DZ  &  0.0969 & -0.0304 & -0.0635 & -0.0653 \\
   &                                         & TZ  & -1.5221 &  0.0078 & -0.0108 & -0.0118 \\
   &                                         & ADZ & -2.0630 & -0.0432 & -0.0887 & -0.0913 \\
   &                                         & ATZ & -2.4238 &  0.0077 & -0.0120 & -0.0131 \\
\midrule
25 & pyridine$\cdots$pyridine ($\pi\cdots\pi$) & DZ  & -0.7944 & -0.0217 & -0.0526 & -0.0543 \\
   &                                           & TZ  & -2.4944 &  0.0512 &  0.0549 &  0.0552 \\
   &                                            & QZ & -3.1890 & 0.1007 & 0.1316 &  0.1335 \\
   &                                           & ADZ & -3.0554 & -0.0221 & -0.0584 & -0.0604 \\
   &                                           & ATZ & -3.4702 &  0.0555 &  0.0607 &  0.0612 \\
\midrule
30 & benzene$\cdots$ethene & DZ  &  0.5394 & -0.0128 & -0.0224 & -0.0229 \\
   &                       & TZ  & -0.5435 & -0.0801 & -0.1310 & -0.1340 \\
   &                       & ADZ & -0.9492 & -0.0369 & -0.0627 & -0.0641 \\
   &                       & ATZ & -1.2037 & -0.0926 & -0.1514 & -0.1547 \\
\midrule
54 & benzene$\cdots$water (OH$\cdots\pi$) & DZ  & -1.6935 & -0.0037 & -0.0072 & -0.0074 \\
   &                                      & TZ  & -2.5953 & -0.0072 & -0.0150 & -0.0154 \\
   &                                      & ADZ & -2.7123 & -0.0180 & -0.0308 & -0.0315 \\
   &                                      & ATZ & -3.1100 & -0.0087 & -0.0172 & -0.0177 \\
\midrule
57 & benzene$\cdots$peptide (NH$\cdots\pi$) & DZ  & -2.7271 & -0.0251 & -0.0462 & -0.0474 \\
   &                                        & TZ  & -4.3203 & -0.0365 & -0.0678 & -0.0695 \\
   &                                        & QZ & -4.8954 & -0.0239 & -0.0491&  -0.0505 \\
   &                                        & ADZ & -4.4596 & -0.0485 & -0.0844 & -0.0864 \\
   &                                        & ATZ & -5.0212 & -0.0409 & -0.0748 & -0.0767 \\
\midrule
58 & pyridine$\cdots$pyridine (CH$\cdots$N) & DZ  & -2.7381 & -0.0133 & -0.0239 & -0.0245 \\
   &                                        & TZ  & -3.5754 & -0.0050 & -0.0112 & -0.0116 \\
   &                                        & ADZ & -3.6048 & -0.0199 & -0.0344 & -0.0352 \\
   &                                        & ATZ & -4.0250 & -0.0077 & -0.0153 & -0.0157 \\
\bottomrule
\end{tabular}
\end{table}

Table~\ref{tab:s66_basis_target_rank} shows that the CP-corrected CCSD(T)
interaction energies have a pronounced basis set dependence, particularly for
the dispersion-dominated complexes. At the cc-pVDZ level, the
benzene$\cdots$benzene and benzene$\cdots$ethene complexes even have positive
interaction energies of $0.0969$ and $0.5394\,\mathrm{kcal\,mol^{-1}}$,
respectively, whereas the $\pi$-stacked pyridine dimer is only weakly bound,
with an interaction energy of $-0.7944\,\mathrm{kcal\,mol^{-1}}$. The \MERCE{} corrections should therefore be
interpreted as finite-basis corrections to the corresponding CCSD(T) result in each basis, rather than as direct estimates of the S66/CBS interaction energy.

The nominal $x=4$ corrections are small on the scale of the total interaction
energies. Across the reported basis sets and complexes, they remain within
about $0.1\,\mathrm{kcal\,mol^{-1}}$. Among the systems in Table~\ref{tab:s66_basis_target_rank}, sign changes in the
nominal $\Delta_4$ correction occur only for the two $\pi$-stacked dimers.
For benzene$\cdots$benzene, $\Delta_4$ is
$-0.0304$ and $-0.0432\,\mathrm{kcal\,mol^{-1}}$ in the DZ and ADZ bases,
respectively, but $+0.0078$ and $+0.0077\,\mathrm{kcal\,mol^{-1}}$ in the
corresponding TZ and ATZ bases. Similarly, for
pyridine$\cdots$pyridine, $\Delta_4$ changes from
$-0.0217$ and $-0.0221\,\mathrm{kcal\,mol^{-1}}$ in DZ and ADZ to
$+0.0512$ and $+0.0555\,\mathrm{kcal\,mol^{-1}}$ in TZ and ATZ.
Thus, for these two systems, the sign change is associated primarily with
the transition from double- to triple-$\zeta$ basis quality, reflecting the
basis dependence of the underlying MP2--CCSD--CCSD(T) trajectory. This
behavior is a property of the fixed-basis correction and should not be
interpreted as an error relative to the S66/CBS reference energy.

To assess the treatment of basis-set superposition error (BSSE), we also
evaluated the selected-S66 application using the no-CP convention. Across all
31 complex--basis entries in Table~\ref{tab:s66_basis_target_rank}, the RMS
CP/no-CP difference in the nominal $\Delta_4$ correction is
$0.0116\,\mathrm{kcal\,mol^{-1}}$, compared with an RMS CP shift of
$2.805\,\mathrm{kcal\,mol^{-1}}$ in the underlying CCSD(T) interaction
energies. The former is only $0.413\%$ of the latter, consistent with the
reduction of differential BSSE in explicit post-CCSD(T) contributions reported
by Fishman et al.\cite{Fishman2025BSSE} However, because $\Delta_4$ is itself
small, the sum of the absolute CP/no-CP changes is 29.3\% of the sum of
$|\Delta_4^{\mathrm{CP}}|$. The shift has the same sign in 30 of 31 entries.
Accordingly, CP sensitivity is small on the scale of the underlying CCSD(T)
interaction energy but not negligible relative to every predicted correction;
we retain CP as the primary convention. Because \MERCE{} predicts rather than
explicitly computes the post-CCSD(T) correction, the agreement with
Ref.~\cite{Fishman2025BSSE} is qualitative corroboration rather than a direct
high-rank benchmark. Complete CP/no-CP values, the three available QZ results,
basis-resolved statistics, and the formal nonadditivity audit are provided in
Tables~S41--S43.

The higher-$x$ columns support the trajectory-stability interpretation
introduced above. For several complexes, the $x=4\rightarrow x=10$ change
remains visible, indicating that the fitted curve has not fully saturated at
the nominal quadruple-excitation-level target. By contrast, the
$x=10\rightarrow x=100$ changes are at most a few thousandths of a
$\mathrm{kcal\,mol^{-1}}$ for the reported cases, showing that the fitted
trajectory is effectively saturated by $x=10$. These results characterize the
internal large-$x$ behavior of the \MERCE{} model and should not be interpreted
as benchmarked physical post-CCSDTQ contributions.

\section{Experimental Section}

\threesubsection{Computational Data}
The benchmark suite combines finite basis absolute correlation energy tests,
property-level energy-difference tests, size-consistency diagnostics, and
fixed-basis applications to noncovalent complexes. Two FCI-based datasets,
denoted FCI-full-core and FCI-frozen-core, monitor behavior near the exact
finite-basis limit. Cross-dataset robustness during model selection is assessed
using canonical CCSDTQ references for a W4-derived molecular
subset\cite{Karton2017} and a dataset of five- and six-membered conjugated
rings,\cite{Damour2021RingsFCI} together with approximate DLPNO-CCSDTQ
references for the CEPB subset and the original one-bond stretching
scans.\cite{CEPB,Jiang2026DLPNOCCSDTQ} The numbers of retained entries are
reported in Tables~\ref{tab:calibration} and \ref{tab:external}, and complete
dataset compositions are provided in the Supporting Information.

The counterpoise-corrected correlation contributions to the A24 noncovalent
interaction energies provide a late-stage property-level selection test using
published CCSDT(Q), CCSDT, CCSD(T), and DC-CCSDT
values.\cite{Lambie2025} Vertical ionization potentials provide
a separate energy-difference assessment constructed from finite-basis neutral
and cationic correlation energies. The S66 calculations use selected complexes
from the S66x8 geometry set at $r/r_{\mathrm e}=1.00$.\cite{Santra2022S66x8}
They are treated as fixed-basis application predictions rather than as
error benchmarks against the original S66 complete basis set reference
interaction energies.\cite{Rezac2011S66}

Electronic-structure calculations generated in this work used the 31 December
2024 release of MRCC,\cite{Mester2025MRCC,MRCCprogram} PySCF
2.9.0,\cite{Sun2018PySCF} and experimental full-DLPNO-CCSDTQ-enabled Psi4
builds.\cite{Smith2020Psi4,Jiang2026DLPNOCCSDTQ} MRCC was used for the
FCI-based developmental sets and the corresponding
vertical-ionization-potential calculations, including FCI, CCSDT, CCSDTQ,
CCSD(T), CCSD, and MP2 correlation energies. Consequently, all levels entering
a given neutral or cationic energy difference were obtained using the same
program.
The orbital reference was selected according to the electronic state and the
source-dataset protocol. Closed-shell records used an RHF reference, whereas
the open-shell developmental and ionization-potential records used restricted
open-shell Hartree--Fock (ROHF) or unrestricted Hartree--Fock (UHF), as
specified in the deposited inputs. Within each individual comparison, all
correlation levels used the same orbital reference, basis set, and
correlated-electron convention.

The A24 component calculations used frozen-core RHF/aug-cc-pVDZ in PySCF
2.9.0, whereas the selected-S66 calculations used all-electron RHF in PySCF
2.9.0. The CEPB DLPNO and original STRETCH calculations used an earlier
experimental full-DLPNO-CCSDTQ-enabled Psi4 build. The later LiF, CH$_3$F,
CH$_3$OH, and NH$_3$ scans used a repaired build incorporating a targeted
integral-record bookkeeping correction. This correction does not modify the
coupled cluster equations, screening thresholds, or convergence settings. The
deposited production inputs and complete outputs document all reported
calculations. For the repaired build, the repository additionally provides the
corrected source snapshot, source patch, complete revision identifiers, build
and smoke-test logs, source and executable checksums, and software-validation
records. The exact DLPNO thresholds and convergence settings are tabulated in
the Supporting Information. Literature-derived W4, Rings, and A24 high-rank
comparison data were used as reported in the corresponding source
publications.

The Supporting Information provides the candidate-function library, fitting
and rejection details, complete benchmark tables, A24 and
ionization-potential component data, size-consistency analyses, and the
hierarchical model-selection audit. The frozen implementation,
electronic-structure records, scripts, and numerical data required to
reproduce the analyses are available in the versioned Zenodo repository at
\url{https://doi.org/10.5281/zenodo.22975564}.

\threesubsection{Numerical Implementation and Model Selection}
The numerical search, fitting, and refinement workflows were implemented in
Python using NumPy and SciPy.\cite{Harris2020NumPy,Virtanen2020SciPy}
For each candidate model, the three system-dependent parameters were obtained
from the MP2, CCSD, and CCSD(T) correlation energies by multistart nonlinear
least-squares fitting. Candidate fits were rejected when they produced nonfinite
values, failed to reproduce the three input points within numerical tolerance, or,
for rational forms, introduced a real pole between the fitted effective-rank
interval and the target coordinate. The MP2--CCSD--CCSD(T)
correlation energy sequence was also required to satisfy the algebraic
ordering in Eq.~\ref{eq:input_ordering}. Entries violating this ordering were
treated as outside the algorithmic domain of the current \MERCE{} model rather
than being forced through the extrapolation. Passing this gate does not
certify accuracy or single-reference character.
This criterion excluded seven W4 entries from the model-selection statistics
and subsequently prevented \MERCE{} evaluation for 16 of the 17 LiF scan
geometries; only the shortest LiF geometry, at
$R/R_{\mathrm e}=0.80$, satisfied the required ordering. All such cases are
reported explicitly as domain rejections rather than as extrapolation
failures with numerical predictions.

Model selection was performed with the hierarchical protocol described in the
main text and documented in detail in the Supporting Information. Briefly,
candidate models were first required to give complete numerical coverage, then
screened for transferability across the cross-dataset model-selection benchmarks,  between-dataset behavior,
robustness on the FCI-based developmental sets, size-consistency defects, and
late-stage A24 property-level performance. Size consistency was assessed from artificial
noninteracting pairs by comparing the extrapolated supersystem result with the
sum of separately extrapolated fragment results. The final structural-simplicity and parameter filter excluded the quintic
inverse-polynomial family and edge-case mapping values before the A24 tie-break
selected the final lower-order inverse-cubic-linear \MERCE{} model.

\paragraph{Recommended use and failure reporting.}
The present \MERCE{} model should be applied only when the MP2, CCSD, and
CCSD(T) correlation energies are computed in the same basis set and with the
same core convention, and when the input sequence satisfies the expected
monotonic ordering. If the input ordering fails, if the MP2--CCSD increment is
numerically too small, or if the fitted rational form develops a pole between
the fitted interval and the target rank, the system should be reported as
outside the current model domain rather than forced through the extrapolation.
For composite systems, the size-consistency diagnostic
$\Delta_{\mathrm{SC}}$ should be reported when fragment data are available.

\threesubsection{Statistical Analysis}

No inferential hypothesis tests, stochastic resampling, or retrospective
outlier removal were performed. Each eligible molecule basis set combination
or scan geometry was treated as one numerical record. Performance was
summarized using the mean absolute error (MAE), root-mean-square (RMS) error,
maximum absolute error, and dimensionless MAE ratios defined in the main text.
Dataset sizes are reported in the corresponding tables. The geometric mean
used for cross-dataset ranking was evaluated from the positive normalized MAE
ratios according to Eq.~\ref{eq:gmean_test}. Records failing the predefined
algebraic applicability gates were reported as domain rejections rather than
removed as statistical outliers. Numerical analyses were implemented in
Python using NumPy and SciPy; the complete scripts and software-environment
records are provided in the versioned repository.
\section{Conclusions}

We have introduced \MERCE{} as a method-space counterpart of conventional
basis set extrapolation. Instead of extrapolating with respect to the basis
cardinal number, \MERCE{} constructs an adaptive effective-rank coordinate from
the MP2--CCSD--CCSD(T) correlation energy sequence. In this coordinate, MP2 and
CCSD(T) are assigned system-dependent effective ranks, CCSD is retained as an
anchor point with $x=2$, and a compact three-point form is used to estimate a high-rank
finite-basis correlation energy. The final representative \MERCE{} model is therefore not a single-metric curve fit, but the outcome of a hierarchical
selection protocol balancing transferability,  between-dataset robustness,
developmental-set behavior, size-consistency defects, late-stage A24 property-level performance,
and functional-form simplicity.

Across the benchmark sets considered here, the final model consistently reduces the CCSD(T) error and often improves on the corresponding CCSDT or
DLPNO-CCSDT baseline while using only MP2, CCSD, and CCSD(T) inputs. On the 23-entry A24 selection subset, it reaches an MAE of
$0.0060\,\kcalmol$ relative to CCSDT(Q), slightly below the DC-CCSDT value,
whereas DC-CCSDT retains the lower RMS and maximum errors. Vertical ionization
potentials provide a complementary energy-difference test, where the model 
approaches the accuracy of explicit high-rank coupled cluster corrections for
the small FCI-based systems. The fixed-basis S66x8 application further illustrates how \MERCE{} can provide
modest basis specific post-CCSD(T) corrections
for chemically recognizable noncovalent complexes when direct high-rank
references are not practical.
The additional coordinate scans provide the most direct assessment of
transfer after the model had been frozen. Along the predefined primary domains
of the CH$_3$F and CH$_3$OH stretching scans and across the complete rigid
NH$_3$ planarization scan, \MERCE{} substantially reduces deviations from
DLPNO-CCSDTQ relative to CCSD(T) and DLPNO-CCSDT. These results establish
transfer with respect to the approximate DLPNO-CCSDTQ comparator rather than
canonical CCSDTQ or FCI accuracy. The LiF scan provides an equally important
negative result: the MP2--CCSD--CCSD(T) sequence fails the predefined
applicability gate at all but one primary-domain geometry, and no extrapolated
values are forced through the rejected points. Taken together, these calculations demonstrate both useful transfer within the
accepted domain and automatic rejection of an ionic system for which the
current input trajectory is unsuitable.

The present formulation is not formally size-consistent because applying the
nonlinear \MERCE{} mapping once to a formal supersystem is not algebraically
equivalent to applying it separately to the noninteracting fragments and summing
the resulting energies. This limitation is explicitly
quantified rather than hidden. The largest size-consistency defects occur for
heterogeneous artificial pairs, especially mixed high/low effective-rank combinations for pair components,
but the defects are usually small relative to the corresponding pair-energy
errors. The artificial-pair and actual component analyses therefore define a
measurable nonadditivity diagnostic, not a calibrated reliability indicator.

Future development should focus on diagnostic-aware mappings for systems near
the single-reference boundary, uncertainty indicators, and alternative
method-space trajectories for cases in which the MP2--CCSD--CCSD(T) sequence is
not monotonic or no longer physically meaningful. The present results establish
the main principle: method space contains exploitable regularity, and this
regularity can be transformed into chemically meaningful accuracy gains using
only MP2, CCSD, and CCSD(T) input calculations.
They do not establish a universal correspondence between the empirical
effective-rank coordinate and the physical coupled cluster excitation rank,
nor do they demonstrate canonical FCI-level accuracy for chemically disjoint
systems outside the model-development data.

\medskip
\textbf{Supplementary Information} \par
Supplementary Information follows the main text in the assembled preprint. It includes the
candidate-function library, MERCE algorithm, fixed-rank and
effective-rank scans, complete benchmark tables, size-consistency pair data,
system-by-system A24 component energies, vertical ionization-potential tables, and the hierarchical selection audit.

\medskip
\textbf{Acknowledgements} \par
P.O.D. acknowledges funding by the projects for International Senior Scientists (Project
No.: W2531013) and for Outstanding Youth Scholars (Overseas, 2021) of the National Natural
Science Foundation of China, via the Lab project of the State Key Laboratory of Physical
Chemistry of Solid Surfaces. The authors gratefully acknowledge the Polish high-performance computing
infrastructure PLGrid (HPC Center: ACK Cyfronet AGH) for providing computer
facilities and support within the computational grant no. PLG/2025/018384.
S.~\'S. acknowledges partial financial support from the National Science Centre,
Poland (grant no. 2025/59/B/ST4/00523).

This work was supported by the PIAST-AI Factory project - PIAST AIF, co-financed by the EuroHPC Joint Undertaking through the European Union's Digital Europe Programme, and by Poland as the Participating Country, under Grant Agreement No. 101250730. Views and opinions expressed are, however, those of the author(s) only and do not necessarily reflect those of the European Union, EuroHPC Joint Undertaking or Poland. Neither the European Union, the EuroHPC Joint Undertaking nor Poland can be held responsible for them.

During the preparation of this work, the authors used ChatGPT and Codex
(OpenAI), Claude Code (Anthropic), Aitomistic Lab@XMU with a locally deployed
Qwen3.8-27B model, and AI Dral Group, an agentic system for scientific
research.
Because these services were updated over the course of the work, multiple
underlying model versions were used. The tools assisted with drafting and
linguistic revision of portions of the manuscript, development and debugging
of analysis and visualization code, consistency checks against the underlying
numerical data, and preparation of reproducibility documentation. The
electronic-structure data themselves were generated using the quantum-chemical
programs and computational procedures described in the Experimental Section
and Supporting Information. All AI-assisted text and code incorporated into
the work were critically reviewed by the authors, and the reported numerical
results were checked against the underlying quantum-chemical outputs and
reproducibility tests. The authors take full responsibility for the accuracy,
interpretation, and conclusions of the work.

\medskip
\textbf{Funding Statement} \par
This work was supported by the National Natural Science Foundation of China (projects for
International Senior Scientists, Project No. W2531013, and for Outstanding Youth Scholars,
Overseas, 2021) and the Lab project of the State Key Laboratory of Physical Chemistry of Solid
Surfaces; the Polish high-performance computing infrastructure PLGrid under computational grant
no. PLG/2025/018384; and the National Science Centre, Poland, under grant no. 2025/59/B/ST4/00523; and the PIAST-AI Factory project--PIAST AIF, co-financed by the EuroHPC
Joint Undertaking through the European Union's Digital Europe Programme and
by Poland as the Participating Country under Grant Agreement No.~101250730.

\medskip
\textbf{Conflict of Interest} \par
The authors declare no conflict of interest.

\medskip
\textbf{Data Availability Statement} \par
The frozen \MERCE{} implementation, machine-readable data,
electronic-structure inputs and outputs, and reproduction scripts supporting
this study are available in the versioned Zenodo repository at
\url{https://doi.org/10.5281/zenodo.22975564}. Third-party source
material is redistributed under its original license as documented in the
repository.

\medskip
\bibliographystyle{unsrtnat}
\bibliography{MERCE_references}

\begin{thebibliography}{44}
\providecommand{\natexlab}[1]{#1}
\providecommand{\url}[1]{\texttt{#1}}
\expandafter\ifx\csname urlstyle\endcsname\relax
  \providecommand{\doi}[1]{doi: #1}\else
  \providecommand{\doi}{doi: \begingroup \urlstyle{rm}\Url}\fi

\bibitem[Knowles and Handy(1984)]{KnowlesHandy1984}
Peter~J. Knowles and Nicholas~C. Handy.
\newblock A new determinant-based full configuration interaction method.
\newblock \emph{Chemical Physics Letters}, 111:\penalty0 315--321, 1984.
\newblock \doi{10.1016/0009-2614(84)85513-X}.

\bibitem[Purvis and Bartlett(1982)]{PurvisBartlett1982}
George~D. Purvis and Rodney~J. Bartlett.
\newblock A full coupled-cluster singles and doubles model: The inclusion of
  disconnected triples.
\newblock \emph{The Journal of Chemical Physics}, 76:\penalty0 1910--1918,
  1982.
\newblock \doi{10.1063/1.443164}.

\bibitem[Bartlett and Musia{\l}(2007)]{BartlettMusial2007}
Rodney~J. Bartlett and Monika Musia{\l}.
\newblock Coupled-cluster theory in quantum chemistry.
\newblock \emph{Reviews of Modern Physics}, 79:\penalty0 291--352, 2007.
\newblock \doi{10.1103/RevModPhys.79.291}.

\bibitem[K{\'a}llay and Surj{\'a}n(2001)]{Kallay2004MRCC}
Mih{\'a}ly K{\'a}llay and P{\'e}ter~R. Surj{\'a}n.
\newblock Higher excitations in coupled-cluster theory.
\newblock \emph{The Journal of Chemical Physics}, 115:\penalty0 2945--2954,
  2001.
\newblock \doi{10.1063/1.1383290}.

\bibitem[M{\o}ller and Plesset(1934)]{Moller1934}
Chr. M{\o}ller and M.~S. Plesset.
\newblock Note on an approximation treatment for many-electron systems.
\newblock \emph{Physical Review}, 46:\penalty0 618--622, 1934.
\newblock \doi{10.1103/PhysRev.46.618}.

\bibitem[Raghavachari et~al.(1989)Raghavachari, Trucks, Pople, and
  Head-Gordon]{Raghavachari1989}
Krishnan Raghavachari, Gary~W. Trucks, John~A. Pople, and Martin Head-Gordon.
\newblock A fifth-order perturbation comparison of electron correlation
  theories.
\newblock \emph{Chemical Physics Letters}, 157:\penalty0 479--483, 1989.
\newblock \doi{10.1016/S0009-2614(89)87395-6}.

\bibitem[Zhang et~al.(2026)Zhang, Xie, and Long]{Zhang2026Iodide}
Jiaying Zhang, Chaolu Xie, and Bo~Long.
\newblock Quantitative kinetics of iodide reactions: Method development and
  insights into higher-order coupled-cluster excitations.
\newblock \emph{The Journal of Physical Chemistry A}, 130\penalty0
  (34):\penalty0 6783--6793, 2026.
\newblock \doi{10.1021/acs.jpca.6c00791}.

\bibitem[Lyakh et~al.(2012)Lyakh, Musia{\l}, Lotrich, and Bartlett]{Lyakh2012}
Dmitry~I. Lyakh, Monika Musia{\l}, Victor~F. Lotrich, and Rodney~J. Bartlett.
\newblock Multireference nature of chemistry: The coupled-cluster view.
\newblock \emph{Chemical Reviews}, 112:\penalty0 182--243, 2012.
\newblock \doi{10.1021/cr2001417}.

\bibitem[Karton et~al.(2017)Karton, Sylvetsky, and Martin]{Karton2017}
Amir Karton, Nitai Sylvetsky, and Jan M.~L. Martin.
\newblock W4-17: A diverse and high-confidence dataset of atomization energies
  for benchmarking high-level electronic structure methods.
\newblock \emph{Journal of Computational Chemistry}, 38:\penalty0 2063--2075,
  2017.
\newblock \doi{10.1002/jcc.24854}.

\bibitem[Semidalas et~al.(2025)Semidalas, Boese, and
  Martin]{Semidalas2025PostCCSDT}
Emmanouil Semidalas, A.~Daniel Boese, and Jan M.~L. Martin.
\newblock Post-ccsd(t) corrections in the s66 noncovalent interactions
  benchmark.
\newblock \emph{Chemical Physics Letters}, 863:\penalty0 141874, 2025.
\newblock \doi{10.1016/j.cplett.2025.141874}.

\bibitem[Goodson(2002)]{Goodson2002}
David~Z. Goodson.
\newblock Extrapolating the coupled-cluster sequence toward the full
  configuration-interaction limit.
\newblock \emph{The Journal of Chemical Physics}, 116:\penalty0 6948--6956,
  2002.
\newblock \doi{10.1063/1.1462620}.

\bibitem[Goodson(2012)]{Goodson2012}
David~Z. Goodson.
\newblock Resummation methods.
\newblock \emph{WIREs Computational Molecular Science}, 2\penalty0
  (5):\penalty0 743--761, 2012.
\newblock \doi{10.1002/wcms.92}.

\bibitem[Gordon and Truhlar(1986)]{GordonTruhlar1986SAC}
Mark~S. Gordon and Donald~G. Truhlar.
\newblock Scaling all correlation energy in perturbation theory calculations of
  bond energies and barrier heights.
\newblock \emph{Journal of the American Chemical Society}, 108:\penalty0
  5412--5419, 1986.
\newblock \doi{10.1021/ja00278a007}.

\bibitem[Fast et~al.(1999)Fast, Corchado, S\'anchez, and
  Truhlar]{Fast1999MCSAC}
Patton~L. Fast, Jos\'e~C. Corchado, Mar\'ia~L. S\'anchez, and Donald~G.
  Truhlar.
\newblock Multi-coefficient correlation method for quantum chemistry.
\newblock \emph{The Journal of Physical Chemistry A}, 103:\penalty0 5129--5136,
  1999.
\newblock \doi{10.1021/jp9903460}.

\bibitem[Santra et~al.(2022)Santra, Semidalas, Mehta, Karton, and
  Martin]{Santra2022S66x8}
Golokesh Santra, Emmanouil Semidalas, Nisha Mehta, Amir Karton, and Jan M.~L.
  Martin.
\newblock {S66x8} noncovalent interactions revisited: new benchmark and
  performance of composite localized coupled-cluster methods.
\newblock \emph{Physical Chemistry Chemical Physics}, 24:\penalty0
  25555--25570, 2022.
\newblock \doi{10.1039/D2CP03938A}.

\bibitem[R{\v{e}}z{\'a}{\v{c}} et~al.(2011{\natexlab{a}})R{\v{e}}z{\'a}{\v{c}},
  Riley, and Hobza]{Rezac2011S66}
Jan R{\v{e}}z{\'a}{\v{c}}, Kevin~E. Riley, and Pavel Hobza.
\newblock {S66}: A well-balanced database of benchmark interaction energies
  relevant to biomolecular structures.
\newblock \emph{Journal of Chemical Theory and Computation}, 7:\penalty0
  2427--2438, 2011{\natexlab{a}}.
\newblock \doi{10.1021/ct2002946}.

\bibitem[R{\v{e}}z{\'a}{\v{c}} et~al.(2011{\natexlab{b}})R{\v{e}}z{\'a}{\v{c}},
  Riley, and Hobza]{Rezac2011S66Extensions}
Jan R{\v{e}}z{\'a}{\v{c}}, Kevin~E. Riley, and Pavel Hobza.
\newblock Extensions of the {S66} data set: More accurate interaction energies
  and angular-displaced nonequilibrium geometries.
\newblock \emph{Journal of Chemical Theory and Computation}, 7\penalty0
  (11):\penalty0 3466--3470, 2011{\natexlab{b}}.
\newblock \doi{10.1021/ct200523a}.

\bibitem[Kesharwani et~al.(2018)Kesharwani, Karton, Sylvetsky, and
  Martin]{Kesharwani2018S66F12}
Manoj~K. Kesharwani, Amir Karton, Nitai Sylvetsky, and Jan M.~L. Martin.
\newblock The {S66} non-covalent interactions benchmark reconsidered using
  explicitly correlated methods near the basis set limit.
\newblock \emph{Australian Journal of Chemistry}, 71\penalty0 (4):\penalty0
  238--248, 2018.
\newblock \doi{10.1071/CH17588}.

\bibitem[Sun et~al.(2026)Sun, Wang, and Wang]{Sun2026SmallBasisCorrection}
Tao Sun, Weizhou Wang, and Yi-Bo Wang.
\newblock Evaluation of the coupled-cluster correction term,
  $\delta_{\mathrm{mp2}}^{\mathrm{ccsd(t)}}$, calculated using small valence
  basis sets.
\newblock \emph{The Journal of Chemical Physics}, 165\penalty0 (11):\penalty0
  114105, 2026.
\newblock \doi{10.1063/5.0347054}.

\bibitem[Dunning(1989)]{Dunning1989}
Thom~H. Dunning.
\newblock Gaussian basis sets for use in correlated molecular calculations. i.
  the atoms boron through neon and hydrogen.
\newblock \emph{The Journal of Chemical Physics}, 90:\penalty0 1007--1023,
  1989.
\newblock \doi{10.1063/1.456153}.

\bibitem[Helgaker et~al.(1997)Helgaker, Klopper, Koch, and Noga]{Helgaker1997}
Trygve Helgaker, Wim Klopper, Henrik Koch, and Jozef Noga.
\newblock Basis-set convergence of correlated calculations on water.
\newblock \emph{The Journal of Chemical Physics}, 106:\penalty0 9639--9646,
  1997.
\newblock \doi{10.1063/1.473863}.

\bibitem[Jiang et~al.(2026)Jiang, Matthews, Poole, Briggs, Turney, Sherrill,
  and Schaefer]{Jiang2026DLPNOCCSDTQ}
Andy Jiang, Devin~A. Matthews, David Poole, Connor~G. Briggs, Justin~M. Turney,
  C.~David Sherrill, and Henry~F. Schaefer.
\newblock Local pair natural orbital-based coupled-cluster theory through full
  quadruples ({DLPNO--CCSDTQ}).
\newblock \emph{Journal of Chemical Theory and Computation}, 22\penalty0
  (6):\penalty0 2825--2845, 2026.
\newblock \doi{10.1021/acs.jctc.5c01910}.

\bibitem[Damour et~al.(2021)Damour, V{\'e}ril, Kossoski, Caffarel, Jacquemin,
  Scemama, and Loos]{Damour2021RingsFCI}
Yann Damour, Micka{\"e}l V{\'e}ril, F{\'a}bris Kossoski, Michel Caffarel, Denis
  Jacquemin, Anthony Scemama, and Pierre-Fran{\c c}ois Loos.
\newblock Accurate full configuration interaction correlation energy estimates
  for five- and six-membered rings.
\newblock \emph{The Journal of Chemical Physics}, 155:\penalty0 134104, 2021.
\newblock \doi{10.1063/5.0065314}.

\bibitem[Witkowski et~al.(2025)Witkowski, {\'S}miga, Hirata, Dral, and
  Grabowski]{CEPB}
Mateusz Witkowski, Szymon {\'S}miga, So~Hirata, Pavlo~O. Dral, and Ireneusz
  Grabowski.
\newblock Ultrafast correlation energy estimator.
\newblock \emph{The Journal of Physical Chemistry A}, 129\penalty0
  (38):\penalty0 8877--8890, 2025.
\newblock \doi{10.1021/acs.jpca.5c04423}.

\bibitem[Lambie et~al.(2025)Lambie, Rickert, Usvyat, Alavi, and
  Kats]{Lambie2025}
S.~Lambie, C.~Rickert, D.~Usvyat, A.~Alavi, and D.~Kats.
\newblock Benchmarking distinguishable cluster methods to platinum standard
  ccsdt(q) non-covalent interaction energies in the a24 dataset.
\newblock \emph{The Journal of Chemical Physics}, 163:\penalty0 111101, 2025.
\newblock \doi{10.1063/5.0280601}.

\bibitem[Sobol'(1967)]{Sobol1967}
Ilya~M. Sobol'.
\newblock On the distribution of points in a cube and the approximate
  evaluation of integrals.
\newblock \emph{USSR Computational Mathematics and Mathematical Physics},
  7\penalty0 (4):\penalty0 86--112, 1967.
\newblock \doi{10.1016/0041-5553(67)90144-9}.

\bibitem[Hauskrecht et~al.(2026)Hauskrecht, Lambie, Christlmaier, Schraivogel,
  Kats, and Alavi]{Hauskrecht2026A24}
Johannes Hauskrecht, Stephanie Lambie, Evelin Martine~Corvid Christlmaier,
  Thomas Schraivogel, Daniel Kats, and Ali Alavi.
\newblock First application of the transcorrelated method to noncovalent
  interactions: The a24 dataset, 2026.

\bibitem[{National Institute of Standards and Technology}(2022)]{NISTCCCBDB}
{National Institute of Standards and Technology}.
\newblock {NIST Computational Chemistry Comparison and Benchmark Database}.
\newblock NIST Standard Reference Database 101, Release 22, 2022.
\newblock URL \url{https://cccbdb.nist.gov/}.
\newblock Accessed 20 August 2026.

\bibitem[Lee and Taylor(1989)]{LeeTaylor1989}
Timothy~J. Lee and Peter~R. Taylor.
\newblock A diagnostic for determining the quality of single-reference electron
  correlation methods.
\newblock \emph{International Journal of Quantum Chemistry}, 36\penalty0
  (S23):\penalty0 199--207, 1989.
\newblock \doi{10.1002/qua.560360824}.

\bibitem[Janssen and Nielsen(1998)]{JanssenNielsen1998}
Curtis~L. Janssen and Ida M.~B. Nielsen.
\newblock New diagnostics for coupled-cluster and m{\o}ller--plesset
  perturbation theory.
\newblock \emph{Chemical Physics Letters}, 290:\penalty0 423--430, 1998.
\newblock \doi{10.1016/S0009-2614(98)00521-7}.

\bibitem[Nielsen and Janssen(1999)]{NielsenJanssen1999}
Ida M.~B. Nielsen and Curtis~L. Janssen.
\newblock Double-substitution-based diagnostics for coupled-cluster and
  m{\o}ller--plesset perturbation theory.
\newblock \emph{Chemical Physics Letters}, 310:\penalty0 568--576, 1999.
\newblock \doi{10.1016/S0009-2614(99)00737-8}.

\bibitem[Marquardt et~al.(2013)Marquardt, Sagui, Zheng, Thiel, Luckhaus,
  Yurchenko, Mariotti, and Quack]{Marquardt2013NH3}
Roberto Marquardt, Kenneth Sagui, Jingjing Zheng, Walter Thiel, David Luckhaus,
  Sergey~N. Yurchenko, Fabio Mariotti, and Martin Quack.
\newblock Global analytical potential energy surface for the electronic ground
  state of {NH3} from high level ab initio calculations.
\newblock \emph{The Journal of Physical Chemistry A}, 117\penalty0
  (32):\penalty0 7502--7522, 2013.
\newblock \doi{10.1021/jp4016728}.

\bibitem[Pople et~al.(1976)Pople, Binkley, and
  Seeger]{Pople1976SizeConsistency}
John~A. Pople, J.~Stephen Binkley, and Rolf Seeger.
\newblock Theoretical models incorporating electron correlation.
\newblock \emph{International Journal of Quantum Chemistry}, 10\penalty0
  (S10):\penalty0 1--19, 1976.
\newblock \doi{10.1002/qua.560100802}.

\bibitem[\'Smiga and Fabiano(2017)]{sscS}
Szymon \'Smiga and Eduardo Fabiano.
\newblock Approximate solution of coupled cluster equations: application to the
  coupled cluster doubles method and non-covalent interacting systems.
\newblock \emph{Physical Chemistry Chemical Physics}, 19\penalty0
  (44):\penalty0 30249--30260, 11 2017.
\newblock ISSN 1463-9076.
\newblock \doi{10.1039/c7cp06417a}.
\newblock URL \url{https://doi.org/10.1039/c7cp06417a}.

\bibitem[Vuckovic et~al.(2018)Vuckovic, Gori-Giorgi, Della~Sala, and
  Fabiano]{sscV}
Stefan Vuckovic, Paola Gori-Giorgi, Fabio Della~Sala, and Eduardo Fabiano.
\newblock Restoring size consistency of approximate functionals constructed
  from the adiabatic connection.
\newblock \emph{The Journal of Physical Chemistry Letters}, 9\penalty0
  (11):\penalty0 3137--3142, 06 2018.
\newblock ISSN 1948-7185.
\newblock \doi{10.1021/acs.jpclett.8b01054}.
\newblock URL \url{https://doi.org/10.1021/acs.jpclett.8b01054}.

\bibitem[Zamani et~al.(2026)Zamani, Zulueta, Ricciuti, Keith, and
  Carter-Fenk]{Zamani2026KSDensity}
Abdulrahman~Y. Zamani, Barbaro Zulueta, Andrew~M. Ricciuti, John~A. Keith, and
  Kay Carter-Fenk.
\newblock Improving coupled cluster theory for strongly correlated molecules
  with kohn--sham density encoding.
\newblock \emph{Nature Communications}, 2026.
\newblock \doi{10.1038/s41467-026-77540-x}.

\bibitem[Boys and Bernardi(1970)]{BoysBernardi1970}
S.~F. Boys and F.~Bernardi.
\newblock The calculation of small molecular interactions by the differences of
  separate total energies: Some procedures with reduced errors.
\newblock \emph{Molecular Physics}, 19\penalty0 (4):\penalty0 553--566, 1970.
\newblock \doi{10.1080/00268977000101561}.

\bibitem[Fishman et~al.(2025)Fishman, Semidalas, Shepelenko, and
  Martin]{Fishman2025BSSE}
Vladimir Fishman, Emmanouil Semidalas, Margarita Shepelenko, and Jan M.~L.
  Martin.
\newblock Does basis set superposition error significantly affect
  post-{CCSD(T)} corrections?
\newblock \emph{Journal of Computational Chemistry}, 46\penalty0 (1):\penalty0
  e70006, 2025.
\newblock \doi{10.1002/jcc.70007}.

\bibitem[Mester et~al.(2025)Mester, Nagy, Cs{\'o}ka, Gyevi-Nagy, Szab{\'o},
  Horv{\'a}th, Petrov, H{\'e}gely, Lad{\'o}czki, Samu, L{\H{o}}rincz, and
  K{\'a}llay]{Mester2025MRCC}
D{\'a}vid Mester, P{\'e}ter~R. Nagy, J{\'o}zsef Cs{\'o}ka, L{\'a}szl{\'o}
  Gyevi-Nagy, P.~Bern{\'a}t Szab{\'o}, R{\'e}ka~A. Horv{\'a}th, Kl{\'a}ra
  Petrov, Bal{\'a}zs H{\'e}gely, B{\'a}lint Lad{\'o}czki, Gergely Samu,
  B.~D{\'a}vid L{\H{o}}rincz, and Mih{\'a}ly K{\'a}llay.
\newblock Overview of developments in the mrcc program system.
\newblock \emph{The Journal of Physical Chemistry A}, 129:\penalty0 2086--2107,
  2025.
\newblock \doi{10.1021/acs.jpca.4c07807}.

\bibitem[K{\'a}llay et~al.(2026)K{\'a}llay, Nagy, Mester, Gyevi-Nagy,
  Cs{\'o}ka, Szab{\'o}, Rolik, Samu, H{\'e}gely, Lad{\'o}czki, Petrov, Csontos,
  Ganyecz, Ladj{\'a}nszki, Szegedy, Farkas, Mezei, Horv{\'a}th, and
  L{\H{o}}rincz]{MRCCprogram}
Mih{\'a}ly K{\'a}llay, P{\'e}ter~R. Nagy, D{\'a}vid Mester, L{\'a}szl{\'o}
  Gyevi-Nagy, J{\'o}zsef Cs{\'o}ka, P.~Bern{\'a}t Szab{\'o}, Zolt{\'a}n Rolik,
  Gergely Samu, Bal{\'a}zs H{\'e}gely, B{\'a}lint Lad{\'o}czki, Kl{\'a}ra
  Petrov, J{\'o}zsef Csontos, {\'A}d{\'a}m Ganyecz, Istv{\'a}n Ladj{\'a}nszki,
  L{\'o}r{\'a}nt Szegedy, M{\'a}t{\'e} Farkas, P{\'a}l~D. Mezei, R{\'e}ka~A.
  Horv{\'a}th, and B.~D{\'a}vid L{\H{o}}rincz.
\newblock {MRCC}, a quantum chemical program suite, 2026.
\newblock See \url{https://www.mrcc.hu}.

\bibitem[Sun et~al.(2018)Sun, Berkelbach, Blunt, Booth, Guo, Li, Liu, McClain,
  Sayfutyarova, Sharma, Wouters, and Chan]{Sun2018PySCF}
Qiming Sun, Timothy~C. Berkelbach, Nick~S. Blunt, George~H. Booth, Sheng Guo,
  Zhendong Li, Junzi Liu, James~D. McClain, Elvira~R. Sayfutyarova, Sandeep
  Sharma, Sebastian Wouters, and Garnet Kin-Lic Chan.
\newblock Pyscf: the python-based simulations of chemistry framework.
\newblock \emph{WIREs Computational Molecular Science}, 8:\penalty0 e1340,
  2018.
\newblock \doi{10.1002/wcms.1340}.

\bibitem[Smith et~al.(2020)Smith, Burns, Simmonett, Parrish, Schieber,
  Galvelis, Kraus, Kruse, Di~Remigio, Alenaizan, James, Lehtola, Misiewicz,
  Scheurer, Shaw, Schriber, Xie, Glick, Sirianni, O'Brien, Waldrop, Kumar,
  Hohenstein, Pritchard, Brooks, Schaefer, Sokolov, Patkowski, DePrince,
  Bozkaya, King, Evangelista, Turney, Crawford, and Sherrill]{Smith2020Psi4}
Daniel G.~A. Smith, Lori~A. Burns, Andrew~C. Simmonett, Robert~M. Parrish,
  Marc~C. Schieber, Raimondas Galvelis, Peter Kraus, Holger Kruse, Roberto
  Di~Remigio, Abdullah Alenaizan, Alan~M. James, Susi Lehtola, Jonathon~P.
  Misiewicz, Maximilian Scheurer, Robert~A. Shaw, Jeffrey~B. Schriber, Yu~Xie,
  Zachary~L. Glick, Dominic~A. Sirianni, Jonathan~S. O'Brien, Jonathan~M.
  Waldrop, Ashutosh Kumar, Edward~G. Hohenstein, Benjamin~P. Pritchard,
  Benjamin~R. Brooks, Henry~F. Schaefer, Alexei~Yu. Sokolov, Konrad Patkowski,
  A.~Eugene DePrince, U{\u g}ur Bozkaya, Rollin~A. King, Francesco~A.
  Evangelista, Justin~M. Turney, T.~Daniel Crawford, and C.~David Sherrill.
\newblock Psi4 1.4: Open-source software for high-throughput quantum chemistry.
\newblock \emph{The Journal of Chemical Physics}, 152:\penalty0 184108, 2020.
\newblock \doi{10.1063/5.0006002}.

\bibitem[Harris et~al.(2020)Harris, Millman, van~der Walt,
  et~al.]{Harris2020NumPy}
Charles~R. Harris, K.~Jarrod Millman, St{\'e}fan~J. van~der Walt, et~al.
\newblock Array programming with {NumPy}.
\newblock \emph{Nature}, 585:\penalty0 357--362, 2020.
\newblock \doi{10.1038/s41586-020-2649-2}.

\bibitem[Virtanen et~al.(2020)Virtanen, Gommers, Oliphant,
  et~al.]{Virtanen2020SciPy}
Pauli Virtanen, Ralf Gommers, Travis~E. Oliphant, et~al.
\newblock {SciPy} 1.0: Fundamental algorithms for scientific computing in
  python.
\newblock \emph{Nature Methods}, 17:\penalty0 261--272, 2020.
\newblock \doi{10.1038/s41592-019-0686-2}.

\end{thebibliography}


\begin{thebibliography}{18}
\providecommand{\natexlab}[1]{#1}
\providecommand{\url}[1]{\texttt{#1}}
\expandafter\ifx\csname urlstyle\endcsname\relax
  \providecommand{\doi}[1]{doi: #1}\else
  \providecommand{\doi}{doi: \begingroup \urlstyle{rm}\Url}\fi

\bibitem[Karton et~al.(2017)Karton, Sylvetsky, and Martin]{Karton2017}
Amir Karton, Nitai Sylvetsky, and Jan M.~L. Martin.
\newblock W4-17: A diverse and high-confidence dataset of atomization energies
  for benchmarking high-level electronic structure methods.
\newblock \emph{Journal of Computational Chemistry}, 38:\penalty0 2063--2075,
  2017.
\newblock \doi{10.1002/jcc.24854}.

\bibitem[Damour et~al.(2021)Damour, V{\'e}ril, Kossoski, Caffarel, Jacquemin,
  Scemama, and Loos]{Damour2021RingsFCI}
Yann Damour, Micka{\"e}l V{\'e}ril, F{\'a}bris Kossoski, Michel Caffarel, Denis
  Jacquemin, Anthony Scemama, and Pierre-Fran{\c c}ois Loos.
\newblock Accurate full configuration interaction correlation energy estimates
  for five- and six-membered rings.
\newblock \emph{The Journal of Chemical Physics}, 155:\penalty0 134104, 2021.
\newblock \doi{10.1063/5.0065314}.

\bibitem[Witkowski et~al.(2025)Witkowski, {\'S}miga, Hirata, Dral, and
  Grabowski]{CEPB}
Mateusz Witkowski, Szymon {\'S}miga, So~Hirata, Pavlo~O. Dral, and Ireneusz
  Grabowski.
\newblock Ultrafast correlation energy estimator.
\newblock \emph{The Journal of Physical Chemistry A}, 129\penalty0
  (38):\penalty0 8877--8890, 2025.
\newblock \doi{10.1021/acs.jpca.5c04423}.

\bibitem[Jiang et~al.(2026)Jiang, Matthews, Poole, Briggs, Turney, Sherrill,
  and Schaefer]{Jiang2026DLPNOCCSDTQ}
Andy Jiang, Devin~A. Matthews, David Poole, Connor~G. Briggs, Justin~M. Turney,
  C.~David Sherrill, and Henry~F. Schaefer.
\newblock Local pair natural orbital-based coupled-cluster theory through full
  quadruples ({DLPNO--CCSDTQ}).
\newblock \emph{Journal of Chemical Theory and Computation}, 22\penalty0
  (6):\penalty0 2825--2845, 2026.
\newblock \doi{10.1021/acs.jctc.5c01910}.

\bibitem[Lambie et~al.(2025)Lambie, Rickert, Usvyat, Alavi, and
  Kats]{Lambie2025}
S.~Lambie, C.~Rickert, D.~Usvyat, A.~Alavi, and D.~Kats.
\newblock Benchmarking distinguishable cluster methods to platinum standard
  ccsdt(q) non-covalent interaction energies in the a24 dataset.
\newblock \emph{The Journal of Chemical Physics}, 163:\penalty0 111101, 2025.
\newblock \doi{10.1063/5.0280601}.

\bibitem[Mester et~al.(2025)Mester, Nagy, Cs{\'o}ka, Gyevi-Nagy, Szab{\'o},
  Horv{\'a}th, Petrov, H{\'e}gely, Lad{\'o}czki, Samu, L{\H{o}}rincz, and
  K{\'a}llay]{Mester2025MRCC}
D{\'a}vid Mester, P{\'e}ter~R. Nagy, J{\'o}zsef Cs{\'o}ka, L{\'a}szl{\'o}
  Gyevi-Nagy, P.~Bern{\'a}t Szab{\'o}, R{\'e}ka~A. Horv{\'a}th, Kl{\'a}ra
  Petrov, Bal{\'a}zs H{\'e}gely, B{\'a}lint Lad{\'o}czki, Gergely Samu,
  B.~D{\'a}vid L{\H{o}}rincz, and Mih{\'a}ly K{\'a}llay.
\newblock Overview of developments in the mrcc program system.
\newblock \emph{The Journal of Physical Chemistry A}, 129:\penalty0 2086--2107,
  2025.
\newblock \doi{10.1021/acs.jpca.4c07807}.

\bibitem[K{\'a}llay et~al.(2026)K{\'a}llay, Nagy, Mester, Gyevi-Nagy,
  Cs{\'o}ka, Szab{\'o}, Rolik, Samu, H{\'e}gely, Lad{\'o}czki, Petrov, Csontos,
  Ganyecz, Ladj{\'a}nszki, Szegedy, Farkas, Mezei, Horv{\'a}th, and
  L{\H{o}}rincz]{MRCCprogram}
Mih{\'a}ly K{\'a}llay, P{\'e}ter~R. Nagy, D{\'a}vid Mester, L{\'a}szl{\'o}
  Gyevi-Nagy, J{\'o}zsef Cs{\'o}ka, P.~Bern{\'a}t Szab{\'o}, Zolt{\'a}n Rolik,
  Gergely Samu, Bal{\'a}zs H{\'e}gely, B{\'a}lint Lad{\'o}czki, Kl{\'a}ra
  Petrov, J{\'o}zsef Csontos, {\'A}d{\'a}m Ganyecz, Istv{\'a}n Ladj{\'a}nszki,
  L{\'o}r{\'a}nt Szegedy, M{\'a}t{\'e} Farkas, P{\'a}l~D. Mezei, R{\'e}ka~A.
  Horv{\'a}th, and B.~D{\'a}vid L{\H{o}}rincz.
\newblock {MRCC}, a quantum chemical program suite, 2026.
\newblock See \url{https://www.mrcc.hu}.

\bibitem[Sun et~al.(2018)Sun, Berkelbach, Blunt, Booth, Guo, Li, Liu, McClain,
  Sayfutyarova, Sharma, Wouters, and Chan]{Sun2018PySCF}
Qiming Sun, Timothy~C. Berkelbach, Nick~S. Blunt, George~H. Booth, Sheng Guo,
  Zhendong Li, Junzi Liu, James~D. McClain, Elvira~R. Sayfutyarova, Sandeep
  Sharma, Sebastian Wouters, and Garnet Kin-Lic Chan.
\newblock Pyscf: the python-based simulations of chemistry framework.
\newblock \emph{WIREs Computational Molecular Science}, 8:\penalty0 e1340,
  2018.
\newblock \doi{10.1002/wcms.1340}.

\bibitem[Helgaker et~al.(1997)Helgaker, Klopper, Koch, and Noga]{Helgaker1997}
Trygve Helgaker, Wim Klopper, Henrik Koch, and Jozef Noga.
\newblock Basis-set convergence of correlated calculations on water.
\newblock \emph{The Journal of Chemical Physics}, 106:\penalty0 9639--9646,
  1997.
\newblock \doi{10.1063/1.473863}.

\bibitem[Goodson(2002)]{Goodson2002}
David~Z. Goodson.
\newblock Extrapolating the coupled-cluster sequence toward the full
  configuration-interaction limit.
\newblock \emph{The Journal of Chemical Physics}, 116:\penalty0 6948--6956,
  2002.
\newblock \doi{10.1063/1.1462620}.

\bibitem[Aitken(1927)]{Aitken1927}
Alexander~C. Aitken.
\newblock Xxv.---on bernoulli's numerical solution of algebraic equations.
\newblock \emph{Proceedings of the Royal Society of Edinburgh}, 46:\penalty0
  289--305, 1927.
\newblock \doi{10.1017/S0370164600022070}.

\bibitem[Gordon and Truhlar(1986)]{GordonTruhlar1986SAC}
Mark~S. Gordon and Donald~G. Truhlar.
\newblock Scaling all correlation energy in perturbation theory calculations of
  bond energies and barrier heights.
\newblock \emph{Journal of the American Chemical Society}, 108:\penalty0
  5412--5419, 1986.
\newblock \doi{10.1021/ja00278a007}.

\bibitem[Fast et~al.(1999)Fast, Corchado, S\'anchez, and
  Truhlar]{Fast1999MCSAC}
Patton~L. Fast, Jos\'e~C. Corchado, Mar\'ia~L. S\'anchez, and Donald~G.
  Truhlar.
\newblock Multi-coefficient correlation method for quantum chemistry.
\newblock \emph{The Journal of Physical Chemistry A}, 103:\penalty0 5129--5136,
  1999.
\newblock \doi{10.1021/jp9903460}.

\bibitem[Lee and Taylor(1989)]{LeeTaylor1989}
Timothy~J. Lee and Peter~R. Taylor.
\newblock A diagnostic for determining the quality of single-reference electron
  correlation methods.
\newblock \emph{International Journal of Quantum Chemistry}, 36\penalty0
  (S23):\penalty0 199--207, 1989.
\newblock \doi{10.1002/qua.560360824}.

\bibitem[Janssen and Nielsen(1998)]{JanssenNielsen1998}
Curtis~L. Janssen and Ida M.~B. Nielsen.
\newblock New diagnostics for coupled-cluster and m{\o}ller--plesset
  perturbation theory.
\newblock \emph{Chemical Physics Letters}, 290:\penalty0 423--430, 1998.
\newblock \doi{10.1016/S0009-2614(98)00521-7}.

\bibitem[Nielsen and Janssen(1999)]{NielsenJanssen1999}
Ida M.~B. Nielsen and Curtis~L. Janssen.
\newblock Double-substitution-based diagnostics for coupled-cluster and
  m{\o}ller--plesset perturbation theory.
\newblock \emph{Chemical Physics Letters}, 310:\penalty0 568--576, 1999.
\newblock \doi{10.1016/S0009-2614(99)00737-8}.

\bibitem[Marquardt et~al.(2013)Marquardt, Sagui, Zheng, Thiel, Luckhaus,
  Yurchenko, Mariotti, and Quack]{Marquardt2013NH3}
Roberto Marquardt, Kenneth Sagui, Jingjing Zheng, Walter Thiel, David Luckhaus,
  Sergey~N. Yurchenko, Fabio Mariotti, and Martin Quack.
\newblock Global analytical potential energy surface for the electronic ground
  state of {NH3} from high level ab initio calculations.
\newblock \emph{The Journal of Physical Chemistry A}, 117\penalty0
  (32):\penalty0 7502--7522, 2013.
\newblock \doi{10.1021/jp4016728}.

\bibitem[Fishman et~al.(2025)Fishman, Semidalas, Shepelenko, and
  Martin]{Fishman2025BSSE}
Vladimir Fishman, Emmanouil Semidalas, Margarita Shepelenko, and Jan M.~L.
  Martin.
\newblock Does basis set superposition error significantly affect
  post-{CCSD(T)} corrections?
\newblock \emph{Journal of Computational Chemistry}, 46\penalty0 (1):\penalty0
  e70006, 2025.
\newblock \doi{10.1002/jcc.70007}.

\end{thebibliography}

\end{document}


\maketitle
\begin{center}
\small
\textsuperscript{1}Institute of Physics, Faculty of Physics, Astronomy, and
Informatics, Nicolaus Copernicus University in Toru\'n,
ul. Grudzi\k{a}dzka 5, 87-100 Toru\'n, Poland\\
\textsuperscript{2}Institute of Advanced Studies, Nicolaus Copernicus
University in Toru\'n, ul. Wile\'nska 4, 87-100 Toru\'n, Poland\\
\textsuperscript{3}State Key Laboratory of Physical Chemistry of Solid
Surfaces, Department of Chemistry, College of Chemistry and Chemical
Engineering, and Fujian Provincial Key Laboratory of Theoretical and
Computational Chemistry, Xiamen University, Xiamen, Fujian 361005, China\\
\textsuperscript{4}Aitomistic, Shenzhen 518000, China\\[0.35em]
\textsuperscript{*}Corresponding author: \href{mailto:ig@fizyka.umk.pl}{ig@fizyka.umk.pl}
\end{center}

\section*{Contents}
This Supporting Information is a numerical companion to the main text of
\textit{MERCE: Method-Space Correlation Energy Extrapolation toward High-Rank Coupled Cluster Accuracy}. The first sections explain how the
method was constructed and why several apparently attractive alternatives were
rejected. The later sections provide the numerical audit trail for the final
representative model, the expanded post-selection challenge, and the
additional molecular-coordinate assessments. Machine-readable source tables
and figure-generation scripts accompany this document.

\begin{enumerate}
\item Dataset roles, notation, and benchmark hierarchy.
\item Electronic-structure protocols and software provenance.
\item High-rank CCSD--CCSDT--CCSDTQ proof of principle tests.
\item Candidate extrapolation forms and the MERCE fitting/branch-selection algorithm.
\item Raw-rank scans without adaptive effective-rank mapping.
\item Reduced effective-rank scans and mapping comparison.
\item Full $(k_1,k_3)$ scan used to identify transferable candidates.
\item Size-consistency and conservative combined-error analysis for artificial noninteracting pairs.
\item A24 property-level model-selection statistics and system-by-system A24 component energies.
\item Vertical ionization potentials relative to full configuration
interaction (FCI) references.
\item Hierarchical selection audit and final MERCE model metrics.
\item Expanded post-selection functional-form and parameter-space challenge.
\item Comparison with alternative sequence transformations and refitted linear analogues.
\item Bond-stretching benchmarks and coupled cluster amplitude diagnostics.
\item Rigid ammonia umbrella-planarization assessment.
\item Counterpoise sensitivity of the selected S66 fixed-basis application.
\item Reproducibility notes and accompanying machine-readable files.
\end{enumerate}

\section*{Overview of the Supporting Information}

The main text presents \MERCE{} as a compact method-space extrapolation protocol.  The purpose of this Supporting Information is to make the selection of the final representative model transparent.  We therefore organize the material as a guided audit trail: first the reader sees the candidate functional forms and the branch-selection algorithm; then the raw-rank scans show why fixed MP2 and CCSD(T) ranks are insufficient; next the effective-rank scans and $(k_1,k_3)$ scan show how transferability improves; finally, size consistency, the A24 property-level assessment, ionization potentials, and the hierarchical selection audit document why the inverse-cubic-linear model was retained. A separate post-selection challenge then tests whether substantially broader functional-form, mapping, and parameter spaces provide a compelling reason to replace the established representative protocol.

Several large tables are included because they make the model-selection path reproducible.  They should be read together with the surrounding text: their role is not to rank every possible formula in isolation, but to show the successive narrowing of a broad candidate space toward a model that is accurate, transferable, and not dominated by a single favorable benchmark.

The final numerical section additionally compares counterpoise-corrected (CP)
and uncorrected (no-CP) predictions for the selected S66 application and
documents the corresponding basis-set superposition error (BSSE) sensitivity.

\section{Dataset roles and notation}
\label{sec:si_dataset_roles}

The benchmark sets play different roles in the \MERCE{} workflow.  The two
FCI-based developmental sets, FCI-full-core and FCI-frozen-core, probe whether
the extrapolation remains reasonable near the exact finite-basis limit for
small systems. The W4-derived subset,\cite{Karton2017} the Rings
dataset,\cite{Damour2021RingsFCI} the CEPB
subset,\cite{CEPB,Jiang2026DLPNOCCSDTQ} and the four original one-bond
stretching scans are cross-dataset model-selection benchmarks. A24\cite{Lambie2025}
and the vertical ionization potentials assess property-level energy
differences, while artificial noninteracting pairs are used only for
size-consistency diagnostics. Because these datasets contributed to screening,
ranking, or interpretation of candidate models, they are not described as
fully external validation sets. Additional molecular-coordinate scans generated only after the functional form
and global parameters had been frozen are discussed separately in the main text
and documented numerically below. LiF, CH$_3$F, and CH$_3$OH were absent from
the model-selection datasets. Although NH$_3$ molecular entries occurred in the
W4 and CEPB benchmarks, the umbrella-planarization geometries and energies were
generated only after model selection.

Some tables and released scripts preserve historical internal labels such as
\texttt{train\_frozen}, \texttt{train\_fullcore},
\texttt{Calibration\_FCI\_FROZEN}, and \texttt{calib\_worst}. These labels
refer to the FCI-frozen-core and FCI-full-core developmental sets or to metrics
computed from them. We retain the literal labels only where needed to match
the released machine-readable outputs; in the scientific prose and
interpretive captions we use ``FCI-based developmental sets.'' For the same
reason, the historical subscript ``calib'' is retained in quantities such as
$M_{\max,\mathrm{calib}}$.

All correlation energy errors in the monomer and artificial-pair benchmarks
are reported in $\microEh$ unless stated otherwise. A24 correlation
contributions to the interaction energies are reported in $\kcalmol$, and
vertical ionization potentials are reported in eV.

\section{Electronic-structure protocols and software provenance}
\label{sec:si_electronic_structure_provenance}

The FCI-developmental and vertical-ionization-potential calculations used the
31 December 2024 release of MRCC.\cite{Mester2025MRCC,MRCCprogram} Exact
production inputs and normally terminated outputs are deposited for every
entry. Each input records the geometry, charge, multiplicity, orbital
reference, basis, core treatment, and convergence settings. All levels entering
a given neutral--cation energy difference were evaluated in the same program
and orbital/core convention.

The 23 reference-bearing A24 component calculations used PySCF 2.9.0 with a
restricted Hartree--Fock (RHF) reference, aug-cc-pVDZ, and the frozen-core
approximation.\cite{Sun2018PySCF}
The dimer, isolated monomers, and both ghost-basis monomers were evaluated
separately. The self-consistent-field (SCF) energy threshold was $10^{-10}\Eh$ with at
most 200 cycles;
the CCSD energy and amplitude-norm thresholds were $10^{-8}\Eh$ and $10^{-6}$,
respectively. Case-specific frozen-orbital counts are retained in the deposited
scripts. The selected S66 calculations likewise used PySCF 2.9.0 and RHF, but
with all electrons correlated, in cc-pVDZ, cc-pVTZ, aug-cc-pVDZ, and
aug-cc-pVTZ; cc-pVQZ was additionally completed for cases 21, 25, and 57.

The CEPB DLPNO and original STRETCH calculations used an experimental
full-DLPNO-CCSDTQ Psi4 development build derived from upstream source revision
\texttt{4e782e1a69eb73578ee755fab7c1754553967357}. The archived program banner
indicates that this build also contained uncommitted local changes, whose
source diff is no longer available. The deposited production inputs, complete
outputs, and printed effective thresholds therefore provide the numerical
provenance for these calculations.

The additional LiF, CH$_3$F, CH$_3$OH, and NH$_3$ scans used a repaired build
derived from the same upstream source revision together with local repair
revision \texttt{11a82751f399289d7b699c41ff57f91612a8bcc8}. The deposited
source snapshot and patch reproduce a missing-QIA-record bookkeeping
correction; they do not modify the coupled cluster equations, amplitude
equations, screening thresholds, or convergence settings. The repository also
contains source-file and production-executable SHA-256 checksums, the complete
build log, a successful smoke test, and a completed CF$_4$/aug-cc-pVDZ
software-regression calculation. In this context, ``software regression''
denotes repetition of a calculation that had previously exposed the
implementation failure, followed by confirmation that the repaired build
passed the former failure point and terminated normally. The CF$_4$
calculation was used exclusively to validate the repaired implementation; it
did not enter any \MERCE{} dataset, fit, model-selection step, figure, or
performance statistic.

\begin{table}[htbp]
\centering
\caption{Resolved NORMAL full-DLPNO-CCSDTQ thresholds for the frozen CEPB
calculations. The complete outputs remain authoritative because they record
temporary internal prescreening changes and threshold resets.}
\label{tab:si_dlpno_thresholds}
\small
\begin{tabular}{lr}
\toprule
Quantity & Value \\
\midrule
\texttt{T\_CUT\_PNO} & $1.0\times10^{-8}$ \\
\texttt{T\_CUT\_TRACE} & $9.99\times10^{-1}$ \\
\texttt{T\_CUT\_ENERGY} & $9.97\times10^{-1}$ \\
\texttt{T\_CUT\_PAIRS} & $1.0\times10^{-8}$ \\
\texttt{MIN\_PNOS\_PAIR} & 5 \\
\texttt{T\_CUT\_PRE} & $1.0\times10^{-10}$ \\
\texttt{T\_CUT\_DO\_PRE} & $3.0\times10^{-2}$ \\
\texttt{T\_CUT\_MKN} & $1.0\times10^{-4}$ \\
\texttt{T\_CUT\_PNO\_MP2} & $1.0\times10^{-10}$ \\
\texttt{T\_CUT\_TRACE\_MP2} & $9.999\times10^{-1}$ \\
\texttt{T\_CUT\_ENERGY\_MP2} & $9.99\times10^{-1}$ \\
\texttt{T\_CUT\_DO} & $5.0\times10^{-3}$ \\
\texttt{T\_CUT\_DO\_ij}, \texttt{T\_CUT\_DO\_uv} & $1.0\times10^{-5}$ \\
\texttt{T\_CUT\_CLMO}, \texttt{T\_CUT\_CPAO} & $1.0\times10^{-3}$ \\
Strong/weak triples \texttt{T\_CUT\_TNO} & $1.0\times10^{-7}$ \\
Triples \texttt{T\_CUT\_DO}, \texttt{T\_CUT\_MKN} & $1.0\times10^{-2}$ \\
Full-quadruples \texttt{T\_CUT\_QNO\_FULL} & $3.33\times10^{-6}$ \\
\bottomrule
\end{tabular}
\end{table}

The CEPB calculations used RHF, PK integrals, no frozen core, a SAD guess, one
DLPNO coupled cluster thread, one density-fitting integral thread,
\texttt{maxiter}=300, energy convergence $10^{-9}\Eh$, density convergence
$10^{-7}$, eight DIIS vectors, zero damping and level shift, and the low-memory
overlap algorithm. The later coordinate scans used the same RHF/PK/all-electron
convention with energy convergence $10^{-8}\Eh$ and density convergence
$10^{-7}$ unless an individual deposited input states otherwise; the converged
NH$_3$ planar endpoint retry is labeled explicitly.

\section{High-rank proof of principle extrapolation}
\label{sec:si_high_rank_probe}

Before constructing the practical MP2--CCSD--CCSD(T) protocol, we tested whether a genuine coupled cluster excitation-rank sequence contains extrapolatable regularity.  
This diagnostic is related in spirit to conventional extrapolation strategies
for systematic quantum-chemical sequences, especially complete-basis-set
(CBS) extrapolation and correlation energy convergence
analyses.\cite{Helgaker1997,Goodson2002}
In this diagnostic exercise, CCSD, CCSDT, and CCSDTQ correlation energies were fitted with three-point functions and evaluated toward the FCI limit for the five-correlated-electron developmental systems.  This is not a practical production route, because CCSDTQ is already far too expensive for routine use.  Its purpose is conceptual: it shows that the correlation energy along the coupled cluster hierarchy can be smooth enough for short-sequence extrapolation.

\begin{table}[htbp]
\centering
\caption{Proof of principle performance of selected three-point extrapolation models for the CCSD/CCSDT/CCSDTQ $\rightarrow$ FCI scheme on the FCI-based developmental sets.  Only models that improve over the raw CCSDTQ baseline are shown.  Errors are summarized by the mean absolute error (MAE) and root mean square
(RMS) error and are given in $\microEh$.}
\label{tab:si_high_rank_probe}
\small
\begin{tabular}{lrr}
\toprule
Method / model & MAE [$\microEh$] & RMS [$\microEh$] \\
\midrule
\multicolumn{3}{l}{\textit{FCI-frozen-core developmental set}} \\
CCSDTQ baseline & 0.768 & 1.058 \\
Flexible exponential & 0.407 & 0.544 \\
Shifted sigmoid & 0.666 & 0.848 \\
Gompertz & 0.682 & 0.871 \\
Logistic offset & 0.697 & 0.892 \\
Feller-like & 0.698 & 0.894 \\
\addlinespace
\multicolumn{3}{l}{\textit{FCI-full-core developmental set}} \\
CCSDTQ baseline & 0.028 & 0.039 \\
Shifted sigmoid & 0.013 & 0.017 \\
Flexible exponential & 0.027 & 0.032 \\
\bottomrule
\end{tabular}
\end{table}

The very small baseline errors in \tabref{tab:si_high_rank_probe} also illustrate why the practical \MERCE{} model should not be judged by whether it surpasses explicit CCSDTQ for tiny systems.  The goal of the final protocol is instead to recover high-rank finite-basis information from the routinely accessible MP2, CCSD, and CCSD(T) sequence.
\section{Candidate extrapolation forms and MERCE algorithm}
\label{sec:si_candidate_forms_algorithm}

The main article focuses on the final representative MERCE model, but the model was not imposed a priori.  A broad library of three-point functions was screened to test whether the selected inverse-cubic-linear form is competitive across a chemically and numerically diverse design space.  This section records the functional forms and the fitting/branch-selection algorithm used in the scans.

\subsection{Candidate extrapolation forms}
\label{sec:si_candidate_forms}
Several entries in the library were motivated by standard correlation energy
extrapolation forms used in basis-set convergence studies, especially inverse
power-law and generalized Helgaker-type expressions.\cite{Helgaker1997,Goodson2002}
The remaining forms are used here as empirical smooth three-point continuation
functions rather than as claims of unique physical convergence laws. The complete candidate-function library used in the model-search stage is
listed in Table~\ref{tab:si_all_extrapolation_models}.

\begingroup
\small
\setlength{\tabcolsep}{4pt}
\renewcommand{\arraystretch}{1.15}
\begin{longtable}{p{0.31\textwidth}p{0.61\textwidth}}
\caption{Grouped list of three-point extrapolation forms tested in the
model-search stage. The names in the first column are descriptive of working labels used to organize the scan and should not be interpreted as formal
literature classifications. For example, ``Feller-like'' denotes an exponential
decay form inspired by common extrapolation practice, not a unique attribution
or a direct reproduction of a particular published protocol. The variable $x$
denotes the method-space rank or effective rank.}\label{tab:si_all_extrapolation_models}\\
\toprule
\textbf{Model / family} & \textbf{Equation for $\Ecorr(x)$} \\
\midrule
\endfirsthead
\caption[]{Grouped list of three-point extrapolation forms tested in the model-search stage (continued).}\\
\toprule
\textbf{Model / family} & \textbf{Equation for $\Ecorr(x)$} \\
\midrule
\endhead
\midrule
\multicolumn{2}{r}{Continued on next page}\\
\endfoot
\bottomrule
\endlastfoot

\multicolumn{2}{l}{\textit{Exponential and sigmoid-type models}} \\
Flexible exponential & $A+\exp(-Bx^C)$ \\
Feller-like & $A+B\exp(-Cx)$ \\
Gompertz & $A\exp[-B\exp(-Cx)]$ \\
Shifted sigmoid & $A/[1+\exp(-B(x-C))]$ \\
Logistic offset & $A+B/[1+\exp(Cx)]$ \\
Stretched Gaussian & $A+B\exp[-(x/C)^2]$ \\
Exponential over power & $A+B\exp(-x)/x^C$ \\
Double exponential I & $A+B\exp(-5x)+C\exp(-5x^2)$ \\
Double exponential II & $A+B\exp(-4.7x)+C\exp(-4.7x^2)$ \\
Double exponential III & $A+B\exp(-4.6x)+C\exp(-4.6x^2)$ \\
\addlinespace
\multicolumn{2}{l}{\textit{Inverse-power and rational models}} \\
Modified inverse-squared & $A+B/(x^2+C)$ \\
Inverse quadratic & $A+B/(x^2+Cx)$ \\
Generalized Helgaker & $A+B/x^C$ \\
Inverse-power saturated & $A+B/(x^C+1)$ \\
Shifted hyperbolic & $A+B/(x-C)$ \\
Pad\'e(0,1) & $A+B/(1+Cx)$ \\
Lorentzian & $A+B/[1+(Cx)^2]$ \\
Higher-order inverse-polynomial family & $A+B/(x^m+Cx^n+d)$ with $(m,n,d)$ equal to
$(3,0,0)$, $(3,1,0)$, $(3,1,1)$, $(3,2,0)$, $(3,2,1)$, $(4,0,0)$, $(4,1,0)$, $(4,1,1)$, $(4,2,0)$, $(4,2,1)$, $(4,3,0)$, $(4,3,1)$, $(5,0,0)$, $(5,1,0)$, $(5,1,1)$, $(5,2,0)$, $(5,2,1)$, $(5,3,0)$, $(5,3,1)$, $(5,4,0)$, and $(5,4,1)$ \\
\addlinespace
\multicolumn{2}{l}{\textit{Other smooth saturating models}} \\
Arctan decay & $A+(2B/\pi)\arctan[1/(Cx)]$ \\
Tanh decay & $A+B\tanh[1/(Cx)]$ \\
\end{longtable}
\endgroup

\subsection{Effective-rank mapping and branch selection}
\label{sec:si_algorithm}

For each system, the three input correlation energies are denoted
$\Ecorr_{\mathrm{MP2}}$, $\Ecorr_{\mathrm{CCSD}}$, and
$\Ecorr_{\mathrm{CCSD(T)}}$.  The present MERCE model first checks the regularity of the input trajectory and then assigns system-dependent effective ranks to MP2 and CCSD(T).  The CCSD rank is fixed at $\beta=2$.  The mapping is based on the adjacent increments
\begin{equation}
d_1=\Ecorr_{\mathrm{CCSD}}-\Ecorr_{\mathrm{MP2}},\qquad
d_2=\Ecorr_{\mathrm{CCSD(T)}}-\Ecorr_{\mathrm{CCSD}},
\end{equation}
and the ratio
\begin{equation}
 R=\frac{|d_2|}{|d_1|+\varepsilon_{\mathrm{num}}}.
\end{equation}
For the final calculator, $\varepsilon_{\mathrm{num}}=10^{-16}\Eh$. Inputs
with $|d_1|<10^{-14}\Eh$ are rejected before $R$ is formed. The regularizer
and rejection threshold serve different numerical purposes: the former avoids
an exactly zero denominator, whereas the latter prevents an ill-conditioned
ratio from entering the effective-rank mapping.
The effective ranks used in the scan are
\begin{align}
 t &= \frac{2}{\pi}\arctan(k_1R), &
 s &= \frac{2}{\pi}\arctan(k_3R),\\
 x_{\mathrm{MP2}}^{\mathrm{eff}} &= \alpha+(\beta-\alpha)t, &
 x_{\mathrm{CCSD}} &= \beta, &
 x_{\mathrm{CCSD(T)}}^{\mathrm{eff}} &= \gamma+(\gamma_{\max}-\gamma)s.
\end{align}
For the final representative MERCE model, $\alpha=1.85$, $\beta=2.00$, $\gamma=2.40$, $\gamma_{\max}=3.00$, $k_1=0.700$, and $k_3=4.250$.

\paragraph{Algorithm S1. General MERCE three-point extrapolation with effective ranks and branch selection.}
\begin{enumerate}
\item \textbf{Input:} $\Ecorr_{\mathrm{MP2}}$, $\Ecorr_{\mathrm{CCSD}}$, $\Ecorr_{\mathrm{CCSD(T)}}$, a candidate function $f(x;A,B,C)$, \\ mapping parameters $(\alpha,\beta,\gamma,\gamma_{\max},k_1,k_3)$, and target rank $x_{\mathrm{tar}}$.
\item Check the expected regular ordering $\Ecorr_{\mathrm{MP2}}>\Ecorr_{\mathrm{CCSD}}>
\Ecorr_{\mathrm{CCSD(T)}}$.  If this input hierarchy is not satisfied, return failure for the present model.
\item If $|d_1|<10^{-14}\Eh$, return failure; otherwise compute $d_2$ and $R$
with $\varepsilon_{\mathrm{num}}=10^{-16}\Eh$.
\item Compute $t$, $s$, and the three effective-rank positions $x_1=x_{\mathrm{MP2}}^{\mathrm{eff}}$, $x_2=\beta$, and $x_3=x_{\mathrm{CCSD(T)}}^{\mathrm{eff}}$.  Set $y_1=\Ecorr_{\mathrm{MP2}}$, $y_2=\Ecorr_{\mathrm{CCSD}}$, and $y_3=\Ecorr_{\mathrm{CCSD(T)}}$.
\item Fit the parameters $(A,B,C)$ by minimizing
\begin{equation}
\sum_{i=1}^{3}\left[f(x_i;A,B,C)-y_i\right]^2
\end{equation}
from the multistart guesses
\begin{equation}
\mathcal{G}=\{(1,1,1),(0.5,1,1),(1,0.5,1),(1,1,0.5),(-1,1,1),(1,-1,1),(1,1,-1)\}.
\end{equation}
\item Reject any branch for which the optimizer fails, any fitted value is non-finite, the target prediction $f(x_{\mathrm{tar}};A,B,C)$ is non-finite, or a rational-form pole lies between the fitted region and the target rank.
\item  Among the accepted branches, select the solution with the smallest residual
sum of squares. If several branches are numerically tied, choose the one with
the smallest $|f(x_{\mathrm{tar}};A,B,C)-\Ecorr_{\mathrm{CCSD(T)}}|$. Denote the
selected fitted parameters by $\widehat{A}$, $\widehat{B}$, and $\widehat{C}$.
\item \textbf{Output:} the extrapolated correlation energy \[
\Ecorr_{\mathrm{extr}}
=
f(x_{\mathrm{tar}};\widehat{A},\widehat{B},\widehat{C}),
\]
or return failure if no acceptable branch remains.
\end{enumerate}

For the final representative model discussed in the main text, the candidate function in Algorithm~S1 is the inverse-cubic-linear form
\begin{equation}
\Ecorr(x)=A+\frac{B}{x^3+Cx}.
\end{equation}
The target rank is $x_{\mathrm{tar}}=4$ when the finite-basis CCSDTQ-like limit is used as the target.

\section{Raw-rank scans without effective-rank mapping}

The raw-rank scans were used as an initial diagnostic of whether MP2, CCSD,
and CCSD(T) could be represented on a continuous, fixed method coordinate.
They showed that fixed, system-independent positions can give excellent
results on an individual cross-dataset model-selection benchmark, particularly
for some double-exponential forms on W4 or Rings, but that the same candidates
do not remain robust on the FCI-based developmental sets. This lack of
cross-dataset transferability motivated the subsequent introduction of the
adaptive effective-rank mapping.

For the raw-rank scan, MP2 and CCSD(T) were assigned fixed, system-independent
positions on the method-space coordinate. The scan used
$x_{\mathrm{MP2}}\in\{1.20,1.25,\ldots,1.90\}$ and
$x_{\mathrm{CCSD(T)}}\in\{2.30,2.35,\ldots,2.90\}$, with CCSD fixed at
$x=2.00$ and the target set to $x=4.0$. Tables~\ref{tab:w4_raw_index_top50} and
\ref{tab:ringsdamour_raw_index_top50} report the best 50 combinations for W4 and Rings,
respectively. These tables are not intended to identify the final model; rather,
they show that excellent single-set performance can be obtained by forms that
are not sufficiently transferable.

\begingroup
\footnotesize
\setlength{\tabcolsep}{4pt}
\begin{longtable}{r l l r r}

\caption{Top 50 fixed-rank model/rank combinations for the W4 subset.
Errors are relative to CCSDTQ and are reported in $\microEh$. The baseline
CCSDT error on the same valid W4 subset is $766.312\,\microEh$ (MAE) and
$962.215\,\microEh$ (RMS). The table is sorted by increasing MAE.}
\label{tab:w4_raw_index_top50} \\
\toprule
Rank & Variant & Model & MAE [$\microEh$] & RMS [$\microEh$] \\
\midrule
\endfirsthead

\caption[]{Top 50 model/rank combinations for the W4 raw-rank scan (continued).} \\
\toprule
Rank & Variant & Model & MAE [$\microEh$] & RMS [$\microEh$] \\
\midrule
\endhead

\midrule
\multicolumn{5}{r}{Continued on next page} \\
\endfoot

\bottomrule
\endlastfoot
1 & MP2\_1.20\_CCSDpT\_2.55 & Double exponential III & 412.299492 & 537.460773 \\
2 & MP2\_1.20\_CCSDpT\_2.55 & Double exponential II & 449.581089 & 598.596893 \\
3 & MP2\_1.20\_CCSDpT\_2.50 & Double exponential III & 451.738602 & 538.104647 \\
4 & MP2\_1.30\_CCSDpT\_2.55 & Double exponential III & 461.064984 & 604.556245 \\
5 & MP2\_1.25\_CCSDpT\_2.55 & Double exponential III & 462.920226 & 585.812331 \\
6 & MP2\_1.25\_CCSDpT\_2.50 & Double exponential II & 478.177106 & 586.662188 \\
7 & MP2\_1.20\_CCSDpT\_2.50 & Double exponential II & 479.317577 & 571.065768 \\
8 & MP2\_1.20\_CCSDpT\_2.60 & Double exponential III & 492.798843 & 662.283760 \\
9 & MP2\_1.25\_CCSDpT\_2.50 & Double exponential III & 505.644759 & 579.736341 \\
10 & MP2\_1.25\_CCSDpT\_2.55 & Double exponential II & 515.017965 & 706.364395 \\
11 & MP2\_1.30\_CCSDpT\_2.50 & Double exponential III & 515.429993 & 614.583539 \\
12 & MP2\_1.20\_CCSDpT\_2.50 & Double exponential I & 522.454156 & 684.975284 \\
13 & MP2\_1.25\_CCSDpT\_2.60 & Double exponential III & 538.670265 & 702.983885 \\
14 & MP2\_1.30\_CCSDpT\_2.60 & Double exponential III & 547.491897 & 724.729577 \\
15 & MP2\_1.20\_CCSDpT\_2.60 & Double exponential II & 556.711313 & 725.981864 \\
16 & MP2\_1.30\_CCSDpT\_2.55 & Double exponential II & 564.514244 & 755.905076 \\
17 & MP2\_1.35\_CCSDpT\_2.55 & Double exponential III & 570.490328 & 751.333245 \\
18 & MP2\_1.30\_CCSDpT\_2.50 & Double exponential II & 587.070798 & 723.418877 \\
19 & MP2\_1.35\_CCSDpT\_2.50 & Double exponential III & 596.500155 & 717.639939 \\
20 & MP2\_1.20\_CCSDpT\_2.55 & Double exponential I & 608.678339 & 829.068182 \\
21 & MP2\_1.20\_CCSDpT\_2.45 & Double exponential I & 614.447809 & 730.803709 \\
22 & MP2\_1.20\_CCSDpT\_2.45 & Double exponential II & 632.211227 & 757.566632 \\
23 & MP2\_1.25\_CCSDpT\_2.60 & Double exponential II & 650.740204 & 829.689700 \\
24 & MP2\_1.20\_CCSDpT\_2.65 & Double exponential III & 653.500280 & 818.621686 \\
25 & MP2\_1.35\_CCSDpT\_2.60 & Double exponential III & 656.423861 & 855.836781 \\
26 & MP2\_1.35\_CCSDpT\_2.50 & Double exponential II & 671.564712 & 830.712249 \\
27 & MP2\_1.30\_CCSDpT\_2.60 & Double exponential II & 681.668875 & 870.216854 \\
28 & MP2\_1.25\_CCSDpT\_2.45 & Double exponential II & 682.190786 & 794.410989 \\
29 & MP2\_1.25\_CCSDpT\_2.65 & Double exponential III & 696.114503 & 848.899632 \\
30 & MP2\_1.20\_CCSDpT\_2.45 & Double exponential III & 699.535793 & 828.232542 \\
31 & MP2\_1.30\_CCSDpT\_2.65 & Double exponential III & 704.251747 & 869.820183 \\
32 & MP2\_1.80\_CCSDpT\_2.90 & Shifted inverse quintic quartic & 705.012888 & 867.084159 \\
33 & MP2\_1.35\_CCSDpT\_2.55 & Double exponential II & 706.159465 & 892.508889 \\
34 & MP2\_1.80\_CCSDpT\_2.90 & Inverse quintic quartic & 706.473129 & 861.506999 \\
35 & MP2\_1.20\_CCSDpT\_2.65 & Double exponential II & 715.242136 & 864.787267 \\
36 & MP2\_1.25\_CCSDpT\_2.45 & Double exponential III & 734.097739 & 851.930190 \\
37 & MP2\_1.35\_CCSDpT\_2.45 & Double exponential II & 742.555485 & 880.614993 \\
38 & MP2\_1.30\_CCSDpT\_2.45 & Double exponential II & 746.245231 & 850.228260 \\
39 & MP2\_1.25\_CCSDpT\_2.45 & Double exponential I & 751.531633 & 919.145360 \\
40 & MP2\_1.30\_CCSDpT\_2.45 & Double exponential III & 752.387784 & 872.791712 \\
41 & MP2\_1.85\_CCSDpT\_2.90 & Inverse quintic quartic & 777.582487 & 999.444951 \\
42 & MP2\_1.80\_CCSDpT\_2.85 & Shifted inverse quintic quartic & 784.605393 & 970.712458 \\
43 & MP2\_1.25\_CCSDpT\_2.50 & Double exponential I & 785.203191 & 976.133952 \\
44 & MP2\_1.35\_CCSDpT\_2.45 & Double exponential III & 787.506423 & 915.983700 \\
45 & MP2\_1.80\_CCSDpT\_2.85 & Inverse quintic quartic & 789.226983 & 976.130916 \\
46 & MP2\_1.20\_CCSDpT\_2.70 & Double exponential III & 790.984938 & 944.961653 \\
47 & MP2\_1.75\_CCSDpT\_2.90 & Inverse quintic quartic & 793.657102 & 972.141229 \\
48 & MP2\_1.25\_CCSDpT\_2.65 & Double exponential II & 793.691014 & 945.832204 \\
49 & MP2\_1.35\_CCSDpT\_2.65 & Double exponential III & 800.854755 & 967.786810 \\
50 & MP2\_1.85\_CCSDpT\_2.90 & Shifted inverse quintic quartic & 801.039198 & 1019.782722 \\
\end{longtable}
\endgroup

\begingroup
\footnotesize
\setlength{\tabcolsep}{4pt}
\begin{longtable}{r l l r r}
\caption{Top 50 fixed-rank model/rank combinations for the Rings dataset.
Errors are relative to CCSDTQ and are reported in $\microEh$. The baseline
CCSDT error on the same valid Rings subset is $2704.627\,\microEh$ (MAE) and
$2751.515\,\microEh$ (RMS). The table is sorted by increasing MAE.}
\label{tab:ringsdamour_raw_index_top50} \\
\toprule
Rank & Variant & Model & MAE [$\microEh$] & RMS [$\microEh$] \\
\midrule
\endfirsthead

\caption[]{Top 50 model/rank combinations for the RingsDamour raw-rank scan (continued).} \\
\toprule
Rank & Variant & Model & MAE [$\microEh$] & RMS [$\microEh$] \\
\midrule
\endhead

\midrule
\multicolumn{5}{r}{Continued on next page} \\
\endfoot

\bottomrule
\endlastfoot
1 & MP2\_1.40\_CCSDpT\_2.55 & Double exponential II & 219.919792 & 268.524510 \\
2 & MP2\_1.20\_CCSDpT\_2.55 & Double exponential II & 220.432625 & 268.966779 \\
3 & MP2\_1.25\_CCSDpT\_2.55 & Double exponential II & 220.512438 & 269.087332 \\
4 & MP2\_1.30\_CCSDpT\_2.55 & Double exponential II & 220.633357 & 269.269958 \\
5 & MP2\_1.35\_CCSDpT\_2.55 & Double exponential II & 220.816617 & 269.546922 \\
6 & MP2\_1.45\_CCSDpT\_2.55 & Double exponential III & 264.826257 & 331.246905 \\
7 & MP2\_1.45\_CCSDpT\_2.55 & Double exponential II & 277.333595 & 340.405719 \\
8 & MP2\_1.30\_CCSDpT\_2.55 & Double exponential III & 296.749612 & 374.644114 \\
9 & MP2\_1.20\_CCSDpT\_2.55 & Double exponential III & 302.843745 & 375.141361 \\
10 & MP2\_1.25\_CCSDpT\_2.55 & Double exponential III & 303.057040 & 375.419711 \\
11 & MP2\_1.35\_CCSDpT\_2.55 & Double exponential III & 303.846818 & 376.453888 \\
12 & MP2\_1.35\_CCSDpT\_2.50 & Double exponential I & 363.323774 & 452.963839 \\
13 & MP2\_1.30\_CCSDpT\_2.50 & Double exponential I & 373.882419 & 457.151382 \\
14 & MP2\_1.20\_CCSDpT\_2.50 & Double exponential I & 374.052154 & 457.518006 \\
15 & MP2\_1.25\_CCSDpT\_2.50 & Double exponential I & 374.205111 & 457.697780 \\
16 & MP2\_1.30\_CCSDpT\_2.55 & Double exponential I & 392.276880 & 435.383817 \\
17 & MP2\_1.25\_CCSDpT\_2.55 & Double exponential I & 392.470824 & 435.581784 \\
18 & MP2\_1.20\_CCSDpT\_2.55 & Double exponential I & 392.597606 & 435.711635 \\
19 & MP2\_1.35\_CCSDpT\_2.55 & Double exponential I & 399.873228 & 442.345398 \\
20 & MP2\_1.35\_CCSDpT\_2.60 & Double exponential III & 413.473772 & 457.339194 \\
21 & MP2\_1.25\_CCSDpT\_2.60 & Double exponential III & 414.302278 & 458.202669 \\
22 & MP2\_1.20\_CCSDpT\_2.60 & Double exponential III & 414.526295 & 458.436778 \\
23 & MP2\_1.40\_CCSDpT\_2.55 & Double exponential III & 448.476123 & 536.146127 \\
24 & MP2\_1.30\_CCSDpT\_2.60 & Double exponential III & 495.374504 & 571.605390 \\
25 & MP2\_1.45\_CCSDpT\_2.60 & Double exponential III & 520.737433 & 564.686714 \\
26 & MP2\_1.35\_CCSDpT\_2.60 & Double exponential II & 554.298514 & 586.463172 \\
27 & MP2\_1.30\_CCSDpT\_2.60 & Double exponential II & 554.804380 & 586.924209 \\
28 & MP2\_1.25\_CCSDpT\_2.60 & Double exponential II & 555.138525 & 587.228936 \\
29 & MP2\_1.20\_CCSDpT\_2.60 & Double exponential II & 555.362894 & 587.433858 \\
30 & MP2\_1.40\_CCSDpT\_2.60 & Double exponential II & 575.171854 & 606.661875 \\
31 & MP2\_1.40\_CCSDpT\_2.60 & Double exponential III & 594.899685 & 723.137621 \\
32 & MP2\_1.40\_CCSDpT\_2.50 & Double exponential I & 710.945795 & 1327.133922 \\
33 & MP2\_1.45\_CCSDpT\_2.50 & Double exponential III & 728.067505 & 840.440157 \\
34 & MP2\_1.45\_CCSDpT\_2.60 & Double exponential II & 791.875393 & 1114.713115 \\
35 & MP2\_1.40\_CCSDpT\_2.50 & Double exponential II & 812.875064 & 900.183806 \\
36 & MP2\_1.90\_CCSDpT\_2.90 & Inverse quintic constant & 824.454716 & 1283.066556 \\
37 & MP2\_1.45\_CCSDpT\_2.50 & Double exponential II & 841.792830 & 953.020521 \\
38 & MP2\_1.90\_CCSDpT\_2.90 & Shifted inverse quintic linear & 858.503393 & 1324.839232 \\
39 & MP2\_1.90\_CCSDpT\_2.90 & Inverse quintic linear & 861.543168 & 1334.851047 \\
40 & MP2\_1.40\_CCSDpT\_2.55 & Double exponential I & 898.437671 & 1541.012286 \\
41 & MP2\_1.90\_CCSDpT\_2.90 & Shifted inverse quintic quadratic & 903.758735 & 1409.001712 \\
42 & MP2\_1.90\_CCSDpT\_2.90 & Inverse quintic quadratic & 908.731959 & 1427.266577 \\
43 & MP2\_1.20\_CCSDpT\_2.50 & Double exponential II & 910.721512 & 979.055853 \\
44 & MP2\_1.25\_CCSDpT\_2.50 & Double exponential II & 911.109982 & 979.436733 \\
45 & MP2\_1.30\_CCSDpT\_2.50 & Double exponential II & 911.689764 & 980.006445 \\
46 & MP2\_1.35\_CCSDpT\_2.50 & Double exponential II & 912.567555 & 980.869063 \\
47 & MP2\_1.35\_CCSDpT\_2.65 & Double exponential III & 918.808600 & 933.256186 \\
48 & MP2\_1.25\_CCSDpT\_2.65 & Double exponential III & 919.598876 & 934.024242 \\
49 & MP2\_1.30\_CCSDpT\_2.65 & Double exponential III & 934.167396 & 951.481090 \\
50 & MP2\_1.30\_CCSDpT\_2.60 & Double exponential I & 937.086829 & 951.065316 \\
\end{longtable}
\endgroup

Inspection of Tables~\ref{tab:w4_raw_index_top50} and
\ref{tab:ringsdamour_raw_index_top50} shows where the fixed-rank description
loses transferability. Double-exponential forms occupy many leading positions
in the Rings-only ranking, whereas the preferred families change when W4 and
the FCI-based developmental sets are considered simultaneously. After the
adaptive mapping is introduced, inverse-polynomial and shifted
inverse-polynomial forms become competitive across the broader selection
hierarchy. CEPB and the original STRETCH scans were subsequently included as
additional cross-dataset model-selection benchmarks.
\section{Reduced effective-rank scans and mapping comparison}

After the raw-rank scans, we tested whether the local increment ratio $R$ can improve transferability by moving MP2 and CCSD(T) in method space.  The first reduced scan used the same arctangent response for both methods, $t(R)=s(R)=2\pi^{-1}\arctan(R)$.  A second comparison kept the MP2 response
unchanged but allowed the CCSD(T) coordinate to respond more strongly,
$t(R)=2\pi^{-1}\arctan(R)$ and
$s(R)=2\pi^{-1}\arctan(2R)$. This comparison is empirical: it tests whether an
asymmetric mapping improves transferability, without assuming in advance that
one response form is physically unique.

Table~\ref{tab:top50_effective_arctan_models} shows the first effective-rank scan, in
which MP2 and CCSD(T) are moved with the same arctangent response,
$t(R)=s(R)=2\pi^{-1}\arctan(R)$. The ranking changes qualitatively relative to
the raw-rank scans: inverse-polynomial, shifted inverse-polynomial, and related
rational forms now dominate the top of the list. This indicates that the
adaptive coordinate removes much of the single-set advantage of the
double-exponential forms and favors extrapolation functions with more stable
cross-set behavior.

The effect of changing the arctangent response is summarized in
Tables~\ref{tab:arctan_same_vs_different}--\ref{tab:arctan_mapping_comparison_compact}.
The complete rankings for the equal-response and asymmetric-response scans are
reported in Tables~\ref{tab:top50_same_arctan_effective} and
\ref{tab:top50_diff_arctan_effective}, respectively.

\begingroup
\scriptsize
\begin{longtable}{rllrrrr}
\caption{Detailed list of the 50 best-performing models selected from the effective-rank scan with $t(R)=s(R)=\frac{2}{\pi}\arctan(R)$. Errors are reported in $\microEh$.}\\
\label{tab:top50_effective_arctan_models}\\
\toprule
Rank & Variant & Model & Train MAE & Train RMS & W4 MAE & W4 RMS \\
\midrule
\endfirsthead

\toprule
Rank & Variant & Model & Train MAE & Train RMS & W4 MAE & W4 RMS \\
\midrule
\endhead

\bottomrule
\endfoot
\small
1  & MP2\_1.85\_CCSDpT\_2.45 & Inverse cubic linear                & 137.625 & 161.810 & 355.461 & 436.868 \\
2  & MP2\_1.80\_CCSDpT\_2.50 & Inverse quartic linear              & 138.220 & 162.671 & 362.318 & 442.169 \\
3  & MP2\_1.85\_CCSDpT\_2.40 & Shifted inverse quartic linear      & 134.248 & 157.541 & 364.988 & 462.392 \\
4  & MP2\_1.90\_CCSDpT\_2.35 & Modified inverse-squared            & 125.626 & 145.292 & 365.528 & 468.087 \\
5  & MP2\_1.85\_CCSDpT\_2.55 & Shifted hyperbolic                  & 132.510 & 153.445 & 365.831 & 451.728 \\
6  & MP2\_1.80\_CCSDpT\_2.45 & Shifted inverse quartic quadratic & 136.902 & 160.941 & 366.265 & 444.950 \\
7  & MP2\_1.85\_CCSDpT\_2.40 & Inverse quartic linear              & 132.392 & 155.115 & 367.166 & 461.208 \\
8  & MP2\_1.85\_CCSDpT\_2.40 & Inverse cubic quadratic                 & 129.525 & 150.921 & 370.265 & 458.108 \\
9  & MP2\_1.90\_CCSDpT\_2.35 & Shifted hyperbolic                  & 120.059 & 136.982 & 374.219 & 460.337 \\
10 & MP2\_1.80\_CCSDpT\_2.50 & Inverse cubic quadratic                 & 133.703 & 156.067 & 375.834 & 458.250 \\
11 & MP2\_1.80\_CCSDpT\_2.45 & Shifted inverse cubic quadratic         & 130.944 & 152.118 & 376.803 & 454.678 \\
12 & MP2\_1.80\_CCSDpT\_2.45 & Inverse quartic quadratic           & 131.461 & 153.188 & 377.356 & 454.332 \\
13 & MP2\_1.80\_CCSDpT\_2.55 & Inverse cubic linear                & 135.844 & 159.144 & 378.452 & 466.619 \\
14 & MP2\_1.85\_CCSDpT\_2.50 & Modified inverse-squared            & 127.161 & 146.269 & 379.613 & 463.389 \\
15 & MP2\_1.80\_CCSDpT\_2.40 & Shifted inverse quintic quadratic  & 128.648 & 149.654 & 381.988 & 466.898 \\
16 & MP2\_1.85\_CCSDpT\_2.35 & Shifted inverse quartic quadratic & 124.368 & 143.954 & 383.204 & 492.865 \\
17 & MP2\_1.85\_CCSDpT\_2.45 & Inverse cubic constant                  & 124.128 & 142.294 & 383.382 & 462.663 \\
18 & MP2\_1.85\_CCSDpT\_2.35 & Shifted inverse quintic linear     & 125.152 & 145.319 & 383.738 & 500.529 \\
19 & MP2\_1.85\_CCSDpT\_2.35 & Inverse quintic linear              & 124.498 & 144.369 & 384.567 & 499.497 \\
20 & MP2\_1.85\_CCSDpT\_2.45 & Inverse quadratic                   & 123.472 & 141.504 & 385.580 & 466.139 \\
21 & MP2\_1.80\_CCSDpT\_2.40 & Inverse quintic quadratic           & 126.999 & 146.916 & 385.703 & 468.441 \\
22 & MP2\_1.90\_CCSDpT\_2.30 & Inverse cubic linear                & 123.057 & 142.760 & 386.368 & 525.139 \\
23 & MP2\_1.80\_CCSDpT\_2.45 & Inverse quintic constant                & 126.333 & 145.422 & 386.403 & 463.523 \\
24 & MP2\_1.75\_CCSDpT\_2.50 & Shifted inverse quintic linear     & 134.332 & 157.181 & 387.247 & 471.796 \\
25 & MP2\_1.90\_CCSDpT\_2.30 & Shifted inverse cubic linear        & 126.022 & 146.803 & 388.154 & 533.652 \\
26 & MP2\_1.85\_CCSDpT\_2.35 & Inverse quartic quadratic           & 121.111 & 138.732 & 388.859 & 488.709 \\
27 & MP2\_1.90\_CCSDpT\_2.30 & Inverse cubic constant                  & 114.939 & 130.345 & 389.827 & 496.482 \\
28 & MP2\_1.85\_CCSDpT\_2.40 & Inverse quartic constant                & 120.954 & 138.148 & 389.837 & 466.995 \\
29 & MP2\_1.85\_CCSDpT\_2.40 & Shifted inverse cubic linear        & 120.844 & 138.090 & 390.177 & 466.721 \\
30 & MP2\_1.90\_CCSDpT\_2.30 & Inverse quartic constant                & 125.110 & 145.569 & 390.285 & 535.253 \\
31 & MP2\_1.75\_CCSDpT\_2.50 & Inverse quintic linear              & 133.528 & 155.836 & 390.772 & 475.968 \\
32 & MP2\_1.80\_CCSDpT\_2.35 & Inverse quintic quartic             & 126.087 & 146.507 & 392.791 & 504.020 \\
33 & MP2\_1.80\_CCSDpT\_2.60 & Inverse cubic constant                  & 136.608 & 159.963 & 396.482 & 484.868 \\
34 & MP2\_1.85\_CCSDpT\_2.40 & Inverse cubic linear                & 134.889 & 205.593 & 398.441 & 475.229 \\
35 & MP2\_1.85\_CCSDpT\_2.35 & Inverse quintic constant                & 116.173 & 132.013 & 400.506 & 490.112 \\
36 & MP2\_1.90\_CCSDpT\_2.30 & Modified inverse-squared            & 116.373 & 130.649 & 403.474 & 489.511 \\
37 & MP2\_1.75\_CCSDpT\_2.40 & Inverse quintic quartic             & 127.649 & 147.342 & 404.086 & 482.640 \\
38 & MP2\_1.75\_CCSDpT\_2.45 & Shifted inverse quintic quadratic  & 127.194 & 146.385 & 405.508 & 490.986 \\
39 & MP2\_1.80\_CCSDpT\_2.35 & Shifted inverse quintic cubic      & 118.474 & 134.949 & 405.534 & 496.162 \\
40 & MP2\_1.85\_CCSDpT\_2.35 & Shifted inverse quartic linear      & 115.600 & 131.198 & 406.682 & 490.747 \\
41 & MP2\_1.85\_CCSDpT\_2.30 & Inverse quintic cubic               & 118.956 & 137.016 & 408.618 & 556.373 \\
42 & MP2\_1.85\_CCSDpT\_2.30 & Shifted inverse quintic quadratic  & 112.523 & 127.192 & 409.572 & 530.806 \\
43 & MP2\_1.85\_CCSDpT\_2.30 & Shifted inverse quintic cubic      & 122.320 & 141.954 & 409.591 & 565.596 \\
44 & MP2\_1.80\_CCSDpT\_2.50 & Inverse quartic constant                & 125.185 & 143.735 & 409.718 & 500.945 \\
45 & MP2\_1.85\_CCSDpT\_2.35 & Inverse quartic linear              & 115.732 & 131.224 & 410.232 & 493.291 \\
46 & MP2\_1.80\_CCSDpT\_2.60 & Inverse quadratic                   & 134.604 & 156.337 & 410.276 & 500.966 \\
47 & MP2\_1.80\_CCSDpT\_2.40 & Shifted inverse quintic linear     & 118.451 & 135.304 & 411.385 & 489.951 \\
48 & MP2\_1.80\_CCSDpT\_2.45 & Shifted inverse quartic linear      & 121.395 & 139.309 & 411.811 & 499.829 \\
49 & MP2\_1.85\_CCSDpT\_2.30 & Inverse quintic quadratic           & 112.012 & 126.611 & 411.824 & 529.055 \\
50 & MP2\_1.80\_CCSDpT\_2.35 & Inverse quintic cubic               & 116.347 & 132.336 & 412.755 & 500.102 \\
\end{longtable}
\endgroup

\newgeometry{hmargin=0.5cm}   

\begin{table}[ht]
\centering
\scriptsize
\caption{Comparison of two effective-rank mappings applied to the same set of
50 raw-rank-selected candidate models. Lower values of
$\mathrm{gmean}_{\mathrm{test}}$, \texttt{worst\_test}, and
\texttt{train\_mean} are better. Here, \texttt{train\_mean} is the mean of the
two FCI-based developmental-set MAE ratios. Historical column names are
retained to match the released analysis files. The median values were computed
over the listed top-50 models in each report.}
\label{tab:arctan_same_vs_different}
\begin{tabular}{llcccccc}
\toprule
Mapping & Top-ranked model & $\mathrm{gmean}_{\mathrm{test}}$ & worst$_{\mathrm{test}}$ & train$_{\mathrm{mean}}$ & med.\ $\mathrm{gmean}_{\mathrm{test}}$ & med.\ worst$_{\mathrm{test}}$ & med.\ train$_{\mathrm{mean}}$ \\
\midrule
$t(R)=s(R)=\frac{2}{\pi}\arctan(R)$
& modified inverse squared
& 0.2867 & 0.5861 & 0.2499 & 0.3508 & 0.6823 & 0.2482 \\
&MP2\_1.90\_CCSDpT\_2.35&&&&&& \\
\multirow{4}{*}{\makecell{$t(R)=\frac{2}{\pi}\arctan(R)$ \\$s(R)=\frac{2}{\pi}\arctan(2R)$}} 
&&&&&&& \\
& shifted inverse cubic quadratic
& 0.1932 & 0.3322 & 0.3017 & 0.2395 & 0.4082 & 0.2963 \\
& MP2\_1.80\_CCSDpT\_2.45&&&&&& \\
\bottomrule
\end{tabular}
\end{table}

\restoregeometry

The dataset-resolved error ratios for the two representative mapping choices
are reported in Table~\ref{tab:arctan_datasetwise_representatives}.

\begin{table}[ht]
\centering
\footnotesize
\caption{Dataset-specific MAE ratios, defined with respect to the corresponding CCSD(T) baseline, for the top-ranked representative model obtained with each mapping choice.}
\label{tab:arctan_datasetwise_representatives}
\begin{tabular}{lcc}
\toprule
Dataset & $\arctan(R),\,\arctan(R)$ & $\arctan(R),\,\arctan(2R)$ \\
\midrule
Calibration\_FCI\_FROZEN    & 0.231 & 0.286 \\
Calibration\_FCI\_FULLCORE  & 0.269 & 0.318 \\
C2H4\_dissociation       & 0.314 & 0.183 \\
CEPB                     & 0.245 & 0.166 \\
CH4\_dissociation        & 0.586 & 0.299 \\
H2O\_dissociation        & 0.234 & 0.187 \\
HF\_dissociation         & 0.414 & 0.332 \\
RingsDamour              & 0.126 & 0.069 \\
W4                       & 0.288 & 0.258 \\
\bottomrule
\end{tabular}
\end{table}

\begin{table}[ht]
\centering
\footnotesize
\caption{Aggregate comparison of two effective-rank mappings applied to the same set of 50 raw-rank-selected candidate models. Median values were computed over the listed top-50 models in each report. Lower values are better.}
\label{tab:arctan_mapping_comparison_compact}
\begin{tabular}{lcc}
\toprule
Metric & $\arctan(R),\,\arctan(R)$ & $\arctan(R),\,\arctan(2R)$ \\
\midrule
Top-ranked model
& modified inverse squared
& shifted inverse cubic quadratic \\

Variant
& MP2\_1.90\_CCSDpT\_2.35
& MP2\_1.80\_CCSDpT\_2.45 \\

Best $\mathrm{gmean}_{\mathrm{test}}$
& 0.2867
& 0.1932 \\

Best worst$_{\mathrm{test}}$
& 0.5861
& 0.3322 \\

Best train$_{\mathrm{mean}}$
& 0.2499
& 0.3017 \\

Best train$_{\mathrm{worst}}$
& 0.2685
& 0.3180 \\

Median $\mathrm{gmean}_{\mathrm{test}}$ in top 50
& 0.3508
& 0.2395 \\

Median worst$_{\mathrm{test}}$ in top 50
& 0.6823
& 0.4082 \\

Median train$_{\mathrm{mean}}$ in top 50
& 0.2482
& 0.2963 \\

\bottomrule
\end{tabular}
\end{table}

\newgeometry{hmargin=0.5cm}   
\begingroup
\scriptsize
\setlength{\tabcolsep}{4pt}
\begin{longtable}{r l l r r r r r r}
\caption{Top 50 models from the reduced-space effective-rank analysis with $t(R)=s(R)=\frac{2}{\pi}\arctan(R)$, ranked by full test-set coverage, $\mathrm{gmean}_{\mathrm{test}}$, worst-test error, average test rank, and mean developmental-set error ratio (historical column \texttt{train\_mean}). The candidate pool consists of the same 50 raw-rank-selected models discussed in the main text.}\label{tab:top50_same_arctan_effective} \\
\toprule
Rank & Variant & Model & $\mathrm{gmean}_{\mathrm{test}}$ & worst$_{\mathrm{test}}$ & avg\_rank & train$_{\mathrm{mean}}$ & train$_{\mathrm{worst}}$ & min\_cov \\
\midrule
\endfirsthead
\caption[]{Top 50 models from the reduced-space effective-rank analysis with $t(R)=s(R)=\frac{2}{\pi}\arctan(R)$, ranked by full test-set coverage, $\mathrm{gmean}_{\mathrm{test}}$, worst-test error, average test rank, and mean developmental-set error ratio (historical column \texttt{train\_mean}) (continued).} \\
\toprule
Rank & Variant & Model & $\mathrm{gmean}_{\mathrm{test}}$ & worst$_{\mathrm{test}}$ & avg\_rank & train$_{\mathrm{mean}}$ & train$_{\mathrm{worst}}$ & min\_cov \\
\midrule
\endhead
\midrule
\multicolumn{9}{r}{Continued on next page} \\
\endfoot
\bottomrule
\endlastfoot
1 & MP2\_1.90\_CCSDpT\_2.35 & Modified inverse-squared & 0.2867 & 0.5861 & 10.00 & 0.2499 & 0.2685 & 1.0000 \\
2 & MP2\_1.90\_CCSDpT\_2.30 & Shifted inverse cubic linear & 0.2888 & 0.5250 & 14.14 & 0.2496 & 0.2672 & 1.0000 \\
3 & MP2\_1.90\_CCSDpT\_2.30 & Inverse cubic linear & 0.2895 & 0.5523 & 12.71 & 0.2437 & 0.2608 & 1.0000 \\
4 & MP2\_1.90\_CCSDpT\_2.30 & Inverse quartic constant & 0.2918 & 0.5361 & 15.71 & 0.2476 & 0.2649 & 1.0000 \\
5 & MP2\_1.85\_CCSDpT\_2.35 & Shifted inverse quintic linear & 0.2929 & 0.6138 & 12.43 & 0.2487 & 0.2669 & 1.0000 \\
6 & MP2\_1.85\_CCSDpT\_2.40 & Shifted inverse quartic linear & 0.2941 & 0.5834 & 11.00 & 0.2710 & 0.2947 & 1.0000 \\
7 & MP2\_1.85\_CCSDpT\_2.35 & Inverse quintic linear & 0.2944 & 0.6209 & 13.57 & 0.2472 & 0.2652 & 1.0000 \\
8 & MP2\_1.85\_CCSDpT\_2.35 & Shifted inverse quartic quadratic & 0.2950 & 0.6269 & 13.64 & 0.2469 & 0.2648 & 1.0000 \\
9 & MP2\_1.85\_CCSDpT\_2.40 & Inverse quartic linear & 0.3004 & 0.5987 & 13.57 & 0.2670 & 0.2903 & 1.0000 \\
10 & MP2\_1.90\_CCSDpT\_2.30 & Inverse cubic constant & 0.3026 & 0.6769 & 18.57 & 0.2273 & 0.2430 & 1.0000 \\
11 & MP2\_1.80\_CCSDpT\_2.35 & Inverse quintic quartic & 0.3039 & 0.6437 & 18.86 & 0.2512 & 0.2702 & 1.0000 \\
12 & MP2\_1.85\_CCSDpT\_2.35 & Inverse quartic quadratic & 0.3070 & 0.6718 & 19.00 & 0.2390 & 0.2550 & 1.0000 \\
13 & MP2\_1.85\_CCSDpT\_2.30 & Inverse quintic cubic & 0.3149 & 0.5980 & 21.57 & 0.2347 & 0.2503 & 1.0000 \\
14 & MP2\_1.85\_CCSDpT\_2.30 & Shifted inverse quintic cubic & 0.3151 & 0.5613 & 21.71 & 0.2417 & 0.2581 & 1.0000 \\
15 & MP2\_1.90\_CCSDpT\_2.35 & Shifted hyperbolic & 0.3192 & 0.6585 & 17.43 & 0.2369 & 0.2528 & 1.0000 \\
16 & MP2\_1.85\_CCSDpT\_2.30 & Shifted inverse quintic quadratic & 0.3204 & 0.7163 & 23.57 & 0.2225 & 0.2378 & 1.0000 \\
17 & MP2\_1.85\_CCSDpT\_2.45 & Inverse cubic linear & 0.3210 & 0.5810 & 14.57 & 0.2796 & 0.3058 & 1.0000 \\
18 & MP2\_1.85\_CCSDpT\_2.30 & Inverse quintic quadratic & 0.3225 & 0.7348 & 25.93 & 0.2211 & 0.2360 & 1.0000 \\
19 & MP2\_1.80\_CCSDpT\_2.40 & Shifted inverse quintic quadratic & 0.3315 & 0.6717 & 21.43 & 0.2583 & 0.2797 & 1.0000 \\
20 & MP2\_1.85\_CCSDpT\_2.35 & Inverse quintic constant & 0.3322 & 0.7504 & 26.14 & 0.2297 & 0.2455 & 1.0000 \\
21 & MP2\_1.80\_CCSDpT\_2.35 & Shifted inverse quintic cubic & 0.3368 & 0.7503 & 28.00 & 0.2341 & 0.2501 & 1.0000 \\
22 & MP2\_1.80\_CCSDpT\_2.40 & Inverse quintic quadratic & 0.3407 & 0.6928 & 23.86 & 0.2537 & 0.2736 & 1.0000 \\
23 & MP2\_1.80\_CCSDpT\_2.45 & Shifted inverse quartic quadratic & 0.3454 & 0.6322 & 20.00 & 0.2783 & 0.3045 & 1.0000 \\
24 & MP2\_1.90\_CCSDpT\_2.30 & Modified inverse-squared & 0.3459 & 0.7926 & 29.86 & 0.2258 & 0.2372 & 1.0000 \\
25 & MP2\_1.80\_CCSDpT\_2.35 & Inverse quintic cubic & 0.3556 & 0.7965 & 33.43 & 0.2299 & 0.2456 & 1.0000 \\
26 & MP2\_1.85\_CCSDpT\_2.35 & Shifted inverse quartic linear & 0.3624 & 0.8052 & 32.86 & 0.2269 & 0.2410 & 1.0000 \\
27 & MP2\_1.85\_CCSDpT\_2.40 & Inverse quartic constant & 0.3650 & 0.7412 & 28.57 & 0.2386 & 0.2545 & 1.0000 \\
28 & MP2\_1.85\_CCSDpT\_2.35 & Inverse quartic linear & 0.3698 & 0.8220 & 35.00 & 0.2263 & 0.2395 & 1.0000 \\
29 & MP2\_1.85\_CCSDpT\_2.40 & Shifted inverse cubic linear & 0.3700 & 0.7426 & 29.71 & 0.2385 & 0.2544 & 1.0000 \\
30 & MP2\_1.80\_CCSDpT\_2.45 & Inverse quartic quadratic & 0.3718 & 0.6843 & 25.43 & 0.2658 & 0.2896 & 1.0000 \\
31 & MP2\_1.80\_CCSDpT\_2.50 & Inverse quartic linear & 0.3777 & 0.6372 & 21.71 & 0.2821 & 0.3097 & 1.0000 \\
32 & MP2\_1.75\_CCSDpT\_2.40 & Inverse quintic quartic & 0.3849 & 0.7539 & 33.71 & 0.2554 & 0.2757 & 1.0000 \\
33 & MP2\_1.80\_CCSDpT\_2.45 & Inverse quintic constant & 0.3890 & 0.7365 & 30.29 & 0.2517 & 0.2707 & 1.0000 \\
34 & MP2\_1.85\_CCSDpT\_2.40 & Inverse cubic linear & 0.3891 & 0.7800 & 33.71 & 0.2350 & 0.2508 & 1.0000 \\
35 & MP2\_1.85\_CCSDpT\_2.45 & Inverse cubic constant & 0.3986 & 0.7297 & 29.86 & 0.2454 & 0.2622 & 1.0000 \\
36 & MP2\_1.80\_CCSDpT\_2.40 & Shifted inverse quintic linear & 0.4010 & 0.8252 & 38.29 & 0.2339 & 0.2497 & 1.0000 \\
37 & MP2\_1.85\_CCSDpT\_2.45 & Inverse quadratic & 0.4056 & 0.7370 & 32.00 & 0.2436 & 0.2599 & 1.0000 \\
38 & MP2\_1.80\_CCSDpT\_2.50 & Inverse cubic quadratic & 0.4064 & 0.6802 & 27.64 & 0.2718 & 0.2975 & 1.0000 \\
39 & MP2\_1.85\_CCSDpT\_2.55 & Shifted hyperbolic & 0.4092 & 0.6282 & 25.86 & 0.2683 & 0.2926 & 1.0000 \\
40 & MP2\_1.85\_CCSDpT\_2.50 & Modified inverse-squared & 0.4171 & 0.6974 & 30.00 & 0.2535 & 0.2729 & 1.0000 \\
41 & MP2\_1.75\_CCSDpT\_2.50 & Shifted inverse quintic linear & 0.4259 & 0.7146 & 33.21 & 0.2735 & 0.2996 & 1.0000 \\
42 & MP2\_1.80\_CCSDpT\_2.55 & Inverse cubic linear & 0.4287 & 0.6614 & 30.57 & 0.2776 & 0.3050 & 1.0000 \\
43 & MP2\_1.75\_CCSDpT\_2.50 & Inverse quintic linear & 0.4314 & 0.7245 & 35.57 & 0.2714 & 0.2969 & 1.0000 \\
44 & MP2\_1.75\_CCSDpT\_2.45 & Shifted inverse quintic quadratic & 0.4360 & 0.7882 & 38.71 & 0.2539 & 0.2736 & 1.0000 \\
45 & MP2\_1.80\_CCSDpT\_2.45 & Shifted inverse quartic linear & 0.4482 & 0.8251 & 41.43 & 0.2396 & 0.2556 & 1.0000 \\
46 & MP2\_1.80\_CCSDpT\_2.60 & Inverse cubic constant & 0.4601 & 0.6497 & 37.14 & 0.2796 & 0.3076 & 1.0000 \\
47 & MP2\_1.80\_CCSDpT\_2.50 & Inverse quartic constant & 0.4621 & 0.7880 & 41.29 & 0.2477 & 0.2649 & 1.0000 \\
48 & MP2\_1.80\_CCSDpT\_2.60 & Inverse quadratic & 0.4798 & 0.6705 & 41.00 & 0.2739 & 0.3000 & 1.0000 \\
49 & MP2\_1.85\_CCSDpT\_2.40 & Inverse cubic quadratic & 0.3109 & 0.6254 & 14.71 & 0.2605 & 0.2824 & 0.8889 \\
50 & MP2\_1.80\_CCSDpT\_2.45 & Shifted inverse cubic quadratic & 0.3794 & 0.6898 & 26.00 & 0.2644 & 0.2876 & 0.9444 \\
\end{longtable}
\endgroup
\restoregeometry

\newgeometry{hmargin=0.5cm}  
\begingroup
\scriptsize
\setlength{\tabcolsep}{4pt}
\begin{longtable}{r l l r r r r r r}
\caption{Top 50 models from the reduced-space effective-rank analysis with $t(R)=\frac{2}{\pi}\arctan(R)$ and $s(R)=\frac{2}{\pi}\arctan(2R)$, ranked by full test-set coverage, $\mathrm{gmean}_{\mathrm{test}}$, worst-test error, average test rank, and mean developmental-set error ratio (historical column \texttt{train\_mean}). The candidate pool consists of the same 50 raw-rank-selected models discussed in the main text.}\label{tab:top50_diff_arctan_effective} \\
\toprule
Rank & Variant & Model & $\mathrm{gmean}_{\mathrm{test}}$ & worst$_{\mathrm{test}}$ & avg\_rank & train$_{\mathrm{mean}}$ & train$_{\mathrm{worst}}$ & min\_cov \\
\midrule
\endfirsthead

\caption[]{Top 50 models from the reduced-space effective-rank analysis with $t(R)=\frac{2}{\pi}\arctan(R)$ and $s(R)=\frac{2}{\pi}\arctan(2R)$, ranked by full test-set coverage, $\mathrm{gmean}_{\mathrm{test}}$, worst-test error, average test rank, and mean developmental-set error ratio (continued).} \\
\toprule
Rank & Variant & Model & $\mathrm{gmean}_{\mathrm{test}}$ & worst$_{\mathrm{test}}$ & avg\_rank & train$_{\mathrm{mean}}$ & train$_{\mathrm{worst}}$ & min\_cov \\
\midrule
\endhead

\midrule
\multicolumn{9}{r}{Continued on next page} \\
\endfoot

\bottomrule
\endlastfoot
1 & MP2\_1.80\_CCSDpT\_2.45 & Shifted inverse cubic quadratic & 0.1932 & 0.3322 & 11.86 & 0.3017 & 0.3180 & 1.0000 \\
2 & MP2\_1.80\_CCSDpT\_2.50 & Inverse quartic linear & 0.1946 & 0.3126 & 13.00 & 0.3159 & 0.3341 & 1.0000 \\
3 & MP2\_1.80\_CCSDpT\_2.45 & Inverse quartic quadratic & 0.1954 & 0.3439 & 12.29 & 0.3050 & 0.3204 & 1.0000 \\
4 & MP2\_1.80\_CCSDpT\_2.45 & Shifted inverse quartic quadratic & 0.1981 & 0.3534 & 13.14 & 0.3207 & 0.3346 & 1.0000 \\
5 & MP2\_1.80\_CCSDpT\_2.45 & Inverse quintic constant & 0.1985 & 0.3366 & 14.57 & 0.2873 & 0.3029 & 1.0000 \\
6 & MP2\_1.85\_CCSDpT\_2.45 & Inverse cubic linear & 0.1992 & 0.3566 & 14.00 & 0.3176 & 0.3329 & 1.0000 \\
7 & MP2\_1.75\_CCSDpT\_2.45 & Shifted inverse quintic quadratic & 0.2001 & 0.3484 & 15.57 & 0.2911 & 0.3074 & 1.0000 \\
8 & MP2\_1.85\_CCSDpT\_2.45 & Inverse cubic constant & 0.2052 & 0.3637 & 18.00 & 0.2709 & 0.2894 & 1.0000 \\
9 & MP2\_1.85\_CCSDpT\_2.40 & Inverse cubic linear & 0.2086 & 0.3571 & 18.43 & 0.2581 & 0.2725 & 1.0000 \\
10 & MP2\_1.85\_CCSDpT\_2.45 & Inverse quadratic & 0.2097 & 0.3721 & 18.71 & 0.2672 & 0.2859 & 1.0000 \\
11 & MP2\_1.75\_CCSDpT\_2.50 & Shifted inverse quintic linear & 0.2106 & 0.3457 & 16.86 & 0.3080 & 0.3263 & 1.0000 \\
12 & MP2\_1.85\_CCSDpT\_2.40 & Shifted inverse cubic linear & 0.2111 & 0.3644 & 18.00 & 0.2687 & 0.2831 & 1.0000 \\
13 & MP2\_1.75\_CCSDpT\_2.50 & Inverse quintic linear & 0.2136 & 0.3539 & 18.43 & 0.3051 & 0.3237 & 1.0000 \\
14 & MP2\_1.85\_CCSDpT\_2.40 & Inverse quartic constant & 0.2153 & 0.3706 & 20.00 & 0.2707 & 0.2844 & 1.0000 \\
15 & MP2\_1.80\_CCSDpT\_2.40 & Shifted inverse quintic linear & 0.2163 & 0.3671 & 20.14 & 0.2623 & 0.2747 & 1.0000 \\
16 & MP2\_1.80\_CCSDpT\_2.45 & Shifted inverse quartic linear & 0.2170 & 0.4074 & 20.14 & 0.2627 & 0.2799 & 1.0000 \\
17 & MP2\_1.75\_CCSDpT\_2.40 & Inverse quintic quartic & 0.2181 & 0.3874 & 18.86 & 0.3064 & 0.3171 & 1.0000 \\
18 & MP2\_1.90\_CCSDpT\_2.35 & Shifted hyperbolic & 0.2273 & 0.3975 & 20.86 & 0.2691 & 0.2821 & 1.0000 \\
19 & MP2\_1.80\_CCSDpT\_2.40 & Inverse quintic quadratic & 0.2324 & 0.4034 & 22.29 & 0.3025 & 0.3127 & 1.0000 \\
20 & MP2\_1.85\_CCSDpT\_2.50 & Modified inverse-squared & 0.2345 & 0.3891 & 23.14 & 0.2736 & 0.2954 & 1.0000 \\
21 & MP2\_1.80\_CCSDpT\_2.40 & Shifted inverse quintic quadratic & 0.2352 & 0.4089 & 23.57 & 0.3087 & 0.3184 & 1.0000 \\
22 & MP2\_1.85\_CCSDpT\_2.40 & Inverse quartic linear & 0.2359 & 0.4136 & 24.57 & 0.3144 & 0.3249 & 1.0000 \\
23 & MP2\_1.85\_CCSDpT\_2.40 & Shifted inverse quartic linear & 0.2379 & 0.4182 & 25.14 & 0.3192 & 0.3291 & 1.0000 \\
24 & MP2\_1.85\_CCSDpT\_2.35 & Inverse quartic linear & 0.2412 & 0.4109 & 24.14 & 0.2467 & 0.2571 & 1.0000 \\
25 & MP2\_1.90\_CCSDpT\_2.35 & Modified inverse-squared & 0.2428 & 0.4271 & 26.64 & 0.2949 & 0.3056 & 1.0000 \\
26 & MP2\_1.85\_CCSDpT\_2.35 & Shifted inverse quartic linear & 0.2438 & 0.4155 & 25.00 & 0.2507 & 0.2606 & 1.0000 \\
27 & MP2\_1.80\_CCSDpT\_2.55 & Inverse cubic linear & 0.2445 & 0.3748 & 24.00 & 0.3015 & 0.3247 & 1.0000 \\
28 & MP2\_1.80\_CCSDpT\_2.50 & Inverse quartic constant & 0.2456 & 0.4320 & 25.00 & 0.2701 & 0.2905 & 1.0000 \\
29 & MP2\_1.85\_CCSDpT\_2.55 & Shifted hyperbolic & 0.2468 & 0.3732 & 25.43 & 0.2856 & 0.3103 & 1.0000 \\
30 & MP2\_1.90\_CCSDpT\_2.30 & Modified inverse-squared & 0.2485 & 0.4247 & 26.57 & 0.2332 & 0.2441 & 1.0000 \\
31 & MP2\_1.80\_CCSDpT\_2.35 & Inverse quintic cubic & 0.2614 & 0.4485 & 29.86 & 0.2697 & 0.2763 & 1.0000 \\
32 & MP2\_1.85\_CCSDpT\_2.35 & Inverse quintic constant & 0.2619 & 0.4459 & 30.07 & 0.2657 & 0.2731 & 1.0000 \\
33 & MP2\_1.85\_CCSDpT\_2.35 & Inverse quartic quadratic & 0.2664 & 0.4640 & 31.71 & 0.2898 & 0.2964 & 1.0000 \\
34 & MP2\_1.80\_CCSDpT\_2.35 & Shifted inverse quintic cubic & 0.2667 & 0.4612 & 31.14 & 0.2833 & 0.2890 & 1.0000 \\
35 & MP2\_1.85\_CCSDpT\_2.35 & Shifted inverse quartic quadratic & 0.2708 & 0.4784 & 33.43 & 0.3033 & 0.3088 & 1.0000 \\
36 & MP2\_1.90\_CCSDpT\_2.30 & Inverse cubic constant & 0.2739 & 0.4760 & 33.14 & 0.2656 & 0.2711 & 1.0000 \\
37 & MP2\_1.85\_CCSDpT\_2.35 & Inverse quintic linear & 0.2778 & 0.4911 & 34.29 & 0.3054 & 0.3098 & 1.0000 \\
38 & MP2\_1.85\_CCSDpT\_2.35 & Shifted inverse quintic linear & 0.2782 & 0.4932 & 35.14 & 0.3077 & 0.3118 & 1.0000 \\
39 & MP2\_1.80\_CCSDpT\_2.35 & Inverse quintic quartic & 0.2788 & 0.5025 & 36.00 & 0.3170 & 0.3190 & 1.0000 \\
40 & MP2\_1.90\_CCSDpT\_2.30 & Shifted inverse cubic linear & 0.2824 & 0.5347 & 39.71 & 0.3149 & 0.3150 & 1.0000 \\
41 & MP2\_1.80\_CCSDpT\_2.60 & Inverse cubic constant & 0.2844 & 0.4070 & 30.00 & 0.2976 & 0.3235 & 1.0000 \\
42 & MP2\_1.90\_CCSDpT\_2.30 & Inverse cubic linear & 0.2846 & 0.5250 & 38.00 & 0.3057 & 0.3073 & 1.0000 \\
43 & MP2\_1.90\_CCSDpT\_2.30 & Inverse quartic constant & 0.2869 & 0.5382 & 41.00 & 0.3130 & 0.3131 & 1.0000 \\
44 & MP2\_1.85\_CCSDpT\_2.30 & Shifted inverse quintic cubic & 0.2941 & 0.5852 & 42.86 & 0.3191 & 0.3257 & 1.0000 \\
45 & MP2\_1.80\_CCSDpT\_2.60 & Inverse quadratic & 0.3003 & 0.4273 & 32.29 & 0.2897 & 0.3159 & 1.0000 \\
46 & MP2\_1.85\_CCSDpT\_2.30 & Inverse quintic cubic & 0.3023 & 0.5758 & 42.00 & 0.3070 & 0.3117 & 1.0000 \\
47 & MP2\_1.85\_CCSDpT\_2.30 & Inverse quintic quadratic & 0.3058 & 0.5318 & 38.43 & 0.2640 & 0.2644 & 1.0000 \\
48 & MP2\_1.85\_CCSDpT\_2.30 & Shifted inverse quintic quadratic & 0.3068 & 0.5367 & 39.57 & 0.2691 & 0.2693 & 1.0000 \\
49 & MP2\_1.80\_CCSDpT\_2.50 & Inverse cubic quadratic & 0.2097 & 0.3428 & 17.71 & 0.3021 & 0.3220 & 0.9167 \\
50 & MP2\_1.85\_CCSDpT\_2.40 & Inverse cubic quadratic & 0.2271 & 0.4002 & 20.29 & 0.3054 & 0.3171 & 0.9821 \\
\end{longtable}
\endgroup
\restoregeometry

The asymmetric mapping reduces the best $\mathrm{gmean}_{\mathrm{test}}$ from
0.2867 to 0.1932 and the best $\mathrm{worst}_{\mathrm{test}}$ from 0.5861 to
0.3322. The median values over the top 50 models improve in the same direction,
showing that the effect is not limited to a single favorable candidate. The
dataset-resolved ratios show that the largest gains occur for the cross-dataset model-selection benchmarks, including CEPB, Rings, and the one-bond stretching
scans. This provides the numerical motivation for scanning independent $k_1$
and $k_3$ response rates in the next stage.

\section{Full $(k_1,k_3)$ scan}

The full $(k_1,k_3)$ scan refines the effective-rank mapping for the 50 most
promising model families identified in the reduced analysis. The important
result is not only the identity of the first-ranked candidate, but also the
structure of the high-performing parameter region. In the mapping,
$\alpha=x_{\mathrm{MP2}}^{\mathrm{eff}}(R=0)$ and
$\gamma=x_{\mathrm{CCSD(T)}}^{\mathrm{eff}}(R=0)$ define the base effective
positions of MP2 and CCSD(T), respectively. Among the top-ranked models in
Table~\ref{tab:si_top50_k1k3_scan_monomer}, the successful MP2 base positions
are concentrated in the interval $\alpha=1.75$--$1.90$, close to the upper end
of the scanned range $1.20$--$1.90$. By contrast, the successful CCSD(T) base
positions are concentrated in the interval $\gamma=2.35$--$2.50$, close to the
lower part of the scanned range $2.30$--$2.90$. In every listed top-ranked
model, the MP2 response rate is smaller than the CCSD(T) response rate,
$k_1<k_3$. Thus, the full scan supports an asymmetric effective-rank mapping as
an empirical transferability result, not as a formal physical constraint.
The corresponding size-consistency statistics for the same 50 top-ranked
models are reported in Table~\ref{tab:si_top50_k1k3_scan_sc}.

\begingroup
\scriptsize
\setlength{\tabcolsep}{4pt}
\begin{longtable}{r c c l c c c c c}
\caption{Top 50 models from the full $(k_1,k_3)$ scan, sorted by
$\mathrm{gmean}_{\mathrm{all}}$. The column
$(x_{\mathrm{MP2}},x_{\mathrm{CCSD(T)}})$ lists the base effective-rank
parameters $(\alpha,\gamma)$, where
$\alpha=x_{\mathrm{MP2}}^{\mathrm{eff}}(R=0)$ and
$\gamma=x_{\mathrm{CCSD(T)}}^{\mathrm{eff}}(R=0)$. The column $(k_1,k_3)$ gives
the arctangent response rates. The successful models concentrate at
$\alpha=1.75$--$1.90$ and $\gamma=2.35$--$2.50$, and all listed top-ranked
models satisfy $k_1<k_3$.} 
\label{tab:si_top50_k1k3_scan_monomer}\\
\toprule
Rank & $(x_{\mathrm{MP2}};x_{\mathrm{CCSD(T)}})$ & $(k_1,k_3)$ & Model & $\mathrm{gmean}_{\mathrm{all}}$ & $\mathrm{gmean}_{\mathrm{test}}$ & $\mathrm{worst}_{\mathrm{test}}$ & train$_{\mathrm{frozen}}$ & train$_{\mathrm{fullcore}}$ \\
\midrule
\endfirsthead
\caption[]{All 50 top-ranked models from the full $(k_1,k_3)$ scan (continued).} \\
\toprule
Rank & $(x_{\mathrm{MP2}};x_{\mathrm{CCSD(T)}})$ & $(k_1,k_3)$ & Model & $\mathrm{gmean}_{\mathrm{all}}$ & $\mathrm{gmean}_{\mathrm{test}}$ & $\mathrm{worst}_{\mathrm{test}}$ & train$_{\mathrm{frozen}}$ & train$_{\mathrm{fullcore}}$ \\
\midrule
\endhead
\midrule
\multicolumn{9}{r}{Continued on next page} \\
\endfoot
\bottomrule
\endlastfoot
1 & (1.90; 2.35) & (0.600, 4.000) & Shifted hyperbolic & 0.1487 & 0.1171 & 0.2603 & 0.3512 & 0.3355 \\
2 & (1.85; 2.40) & (0.700, 4.250) & Inverse cubic linear & 0.1496 & 0.1180 & 0.2592 & 0.3533 & 0.3339 \\
3 & (1.80; 2.45) & (0.700, 4.250) & Inverse quintic constant & 0.1503 & 0.1158 & 0.2586 & 0.3838 & 0.3642 \\
4 & (1.90; 2.35) & (0.600, 3.750) & Shifted hyperbolic & 0.1508 & 0.1202 & 0.2602 & 0.3381 & 0.3288 \\
5 & (1.85; 2.40) & (0.700, 4.500) & Inverse cubic linear & 0.1513 & 0.1187 & 0.2649 & 0.3653 & 0.3413 \\
6 & (1.85; 2.40) & (0.600, 4.250) & Inverse quartic constant & 0.1513 & 0.1183 & 0.2691 & 0.3697 & 0.3466 \\
7 & (1.75; 2.45) & (0.800, 4.000) & Shifted inverse quintic quadratic & 0.1517 & 0.1171 & 0.2621 & 0.3849 & 0.3670 \\
8 & (1.85; 2.40) & (0.700, 4.000) & Shifted inverse cubic linear & 0.1520 & 0.1201 & 0.2630 & 0.3551 & 0.3373 \\
9 & (1.80; 2.45) & (0.700, 3.750) & Shifted inverse quartic quadratic & 0.1523 & 0.1164 & 0.2602 & 0.3956 & 0.3856 \\
10 & (1.80; 2.40) & (0.600, 3.500) & Inverse quintic quadratic & 0.1528 & 0.1185 & 0.2609 & 0.3836 & 0.3615 \\
11 & (1.85; 2.45) & (0.700, 3.750) & Inverse cubic linear & 0.1532 & 0.1183 & 0.2605 & 0.3795 & 0.3765 \\
12 & (1.80; 2.45) & (0.700, 4.000) & Inverse quartic quadratic & 0.1533 & 0.1181 & 0.2671 & 0.3884 & 0.3755 \\
13 & (1.80; 2.45) & (0.700, 4.000) & Inverse quintic constant & 0.1537 & 0.1201 & 0.2560 & 0.3719 & 0.3562 \\
14 & (1.80; 2.45) & (0.800, 4.500) & Shifted inverse quartic linear & 0.1538 & 0.1214 & 0.2576 & 0.3616 & 0.3442 \\
15 & (1.85; 2.45) & (0.700, 4.000) & Inverse cubic linear & 0.1540 & 0.1184 & 0.2663 & 0.3903 & 0.3835 \\
16 & (1.80; 2.50) & (0.800, 4.500) & Inverse quartic linear & 0.1541 & 0.1174 & 0.2637 & 0.4017 & 0.3966 \\
17 & (1.85; 2.45) & (0.800, 4.500) & Inverse cubic constant & 0.1542 & 0.1219 & 0.2616 & 0.3548 & 0.3461 \\
18 & (1.80; 2.50) & (0.800, 4.250) & Inverse quartic linear & 0.1543 & 0.1183 & 0.2596 & 0.3922 & 0.3888 \\
19 & (1.75; 2.40) & (0.700, 3.250) & Inverse quintic quartic & 0.1543 & 0.1201 & 0.2643 & 0.3809 & 0.3619 \\
20 & (1.85; 2.40) & (0.700, 4.250) & Shifted inverse cubic linear & 0.1545 & 0.1217 & 0.2691 & 0.3676 & 0.3456 \\
21 & (1.80; 2.40) & (0.600, 4.000) & Shifted inverse quintic linear & 0.1549 & 0.1227 & 0.2807 & 0.3644 & 0.3359 \\
22 & (1.75; 2.50) & (0.800, 4.500) & Shifted inverse quintic linear & 0.1551 & 0.1185 & 0.2665 & 0.4028 & 0.3926 \\
23 & (1.85; 2.40) & (0.700, 4.000) & Inverse cubic linear & 0.1552 & 0.1246 & 0.2576 & 0.3406 & 0.3274 \\
24 & (1.85; 2.40) & (0.700, 4.000) & Inverse quartic constant & 0.1553 & 0.1231 & 0.2706 & 0.3612 & 0.3399 \\
25 & (1.80; 2.45) & (0.800, 4.000) & Shifted inverse cubic quadratic & 0.1556 & 0.1206 & 0.2716 & 0.3855 & 0.3733 \\
26 & (1.85; 2.40) & (0.600, 4.000) & Inverse quartic constant & 0.1558 & 0.1239 & 0.2721 & 0.3569 & 0.3380 \\
27 & (1.75; 2.50) & (0.900, 4.250) & Inverse quintic linear & 0.1558 & 0.1201 & 0.2685 & 0.3927 & 0.3826 \\
28 & (1.80; 2.40) & (0.700, 3.750) & Shifted inverse quintic linear & 0.1560 & 0.1247 & 0.2628 & 0.3535 & 0.3298 \\
29 & (1.75; 2.50) & (0.900, 4.250) & Shifted inverse quintic linear & 0.1561 & 0.1201 & 0.2708 & 0.3962 & 0.3863 \\
30 & (1.75; 2.50) & (0.800, 4.500) & Inverse quintic linear & 0.1563 & 0.1201 & 0.2687 & 0.3992 & 0.3888 \\
31 & (1.80; 2.40) & (0.600, 3.250) & Inverse quintic quadratic & 0.1565 & 0.1234 & 0.2657 & 0.3676 & 0.3513 \\
32 & (1.80; 2.40) & (0.700, 4.000) & Shifted inverse quintic linear & 0.1569 & 0.1245 & 0.2712 & 0.3683 & 0.3377 \\
33 & (1.85; 2.40) & (0.700, 3.750) & Shifted inverse cubic linear & 0.1573 & 0.1266 & 0.2583 & 0.3418 & 0.3307 \\
34 & (1.85; 2.45) & (0.800, 4.500) & Inverse quadratic & 0.1574 & 0.1259 & 0.2628 & 0.3476 & 0.3412 \\
35 & (1.80; 2.45) & (0.700, 3.500) & Shifted inverse quartic quadratic & 0.1575 & 0.1224 & 0.2609 & 0.3831 & 0.3772 \\
36 & (1.80; 2.45) & (0.800, 3.750) & Shifted inverse cubic quadratic & 0.1575 & 0.1234 & 0.2652 & 0.3736 & 0.3656 \\
37 & (1.75; 2.45) & (0.800, 3.750) & Shifted inverse quintic quadratic & 0.1577 & 0.1241 & 0.2583 & 0.3719 & 0.3586 \\
38 & (1.85; 2.40) & (0.600, 4.250) & Shifted inverse cubic linear & 0.1580 & 0.1256 & 0.2881 & 0.3633 & 0.3435 \\
39 & (1.85; 2.40) & (0.700, 3.750) & Inverse quartic constant & 0.1589 & 0.1279 & 0.2632 & 0.3477 & 0.3327 \\
40 & (1.90; 2.35) & (0.600, 3.500) & Shifted hyperbolic & 0.1590 & 0.1298 & 0.2637 & 0.3246 & 0.3219 \\
41 & (1.75; 2.50) & (0.900, 4.000) & Shifted inverse quintic linear & 0.1594 & 0.1241 & 0.2651 & 0.3856 & 0.3789 \\
42 & (1.75; 2.45) & (0.800, 4.250) & Shifted inverse quintic quadratic & 0.1598 & 0.1242 & 0.2684 & 0.3974 & 0.3752 \\
43 & (1.75; 2.50) & (0.900, 4.000) & Inverse quintic linear & 0.1599 & 0.1249 & 0.2629 & 0.3821 & 0.3757 \\
44 & (1.85; 2.45) & (0.800, 4.250) & Inverse cubic constant & 0.1599 & 0.1287 & 0.2628 & 0.3445 & 0.3396 \\
45 & (1.80; 2.45) & (0.800, 3.750) & Inverse quartic quadratic & 0.1600 & 0.1254 & 0.2735 & 0.3804 & 0.3696 \\
46 & (1.75; 2.50) & (0.800, 4.250) & Shifted inverse quintic linear & 0.1603 & 0.1245 & 0.2688 & 0.3925 & 0.3842 \\
47 & (1.75; 2.50) & (0.900, 4.500) & Inverse quintic linear & 0.1603 & 0.1237 & 0.2735 & 0.4030 & 0.3909 \\
48 & (1.80; 2.45) & (0.700, 3.750) & Inverse quartic quadratic & 0.1604 & 0.1262 & 0.2703 & 0.3763 & 0.3677 \\
49 & (1.85; 2.45) & (0.700, 3.500) & Inverse cubic linear & 0.1604 & 0.1265 & 0.2593 & 0.3679 & 0.3693 \\
50 & (1.80; 2.45) & (0.800, 4.250) & Shifted inverse quartic linear & 0.1606 & 0.1292 & 0.2593 & 0.3502 & 0.3371 \\
\end{longtable}
\endgroup

\begingroup
\scriptsize
\setlength{\tabcolsep}{4pt}
\begin{longtable}{r c c l c c c c c}
\caption{Size-consistency statistics for all 50 top-ranked models from the full $(k_1,k_3)$ scan. Here $\mathrm{SC}_{\mathrm{all}}$ denotes the overall size-consistency MAE, $Q_{\mathrm{all}}=\mathrm{SC}_{\mathrm{all}}/\mathrm{pair\_MAE}_{\mathrm{all}}$, and the last three columns give the corresponding W4 values. All size-consistency and pair-MAE quantities are reported in $\microEh$.}\label{tab:si_top50_k1k3_scan_sc} \\
\toprule
Rank & $(x_{\mathrm{MP2}};x_{\mathrm{CCSD(T)}})$ & $(k_1,k_3)$ & Model & $\mathrm{SC}_{\mathrm{all}}$ & $Q_{\mathrm{all}}$ & $\mathrm{SC}_{\mathrm{W4}}$ & $\mathrm{pair\_MAE}_{\mathrm{W4}}$ & $Q_{\mathrm{W4}}$ \\
\midrule
\endfirsthead
\caption[]{Size-consistency statistics for all 50 top-ranked models from the full $(k_1,k_3)$ scan (continued).} \\
\toprule
Rank & $(x_{\mathrm{MP2}};x_{\mathrm{CCSD(T)}})$ & $(k_1,k_3)$ & Model & $\mathrm{SC}_{\mathrm{all}}$ & $Q_{\mathrm{all}}$ & $\mathrm{SC}_{\mathrm{W4}}$ & $\mathrm{pair\_MAE}_{\mathrm{W4}}$ & $Q_{\mathrm{W4}}$ \\
\midrule
\endhead
\midrule
\multicolumn{9}{r}{Continued on next page} \\
\endfoot
\bottomrule
\endlastfoot
1 & (1.90; 2.35) & (0.600, 4.000) & Shifted hyperbolic & 32.989 & 0.074 & 47.0 & 532.6 & 0.088 \\
2 & (1.85; 2.40) & (0.700, 4.250) & Inverse cubic linear & 38.040 & 0.087 & 54.5 & 521.3 & 0.105 \\
3 & (1.80; 2.45) & (0.700, 4.250) & Inverse quintic constant & 31.765 & 0.071 & 45.3 & 528.1 & 0.086 \\
4 & (1.90; 2.35) & (0.600, 3.750) & Shifted hyperbolic & 34.785 & 0.082 & 49.7 & 509.9 & 0.098 \\
5 & (1.85; 2.40) & (0.700, 4.500) & Inverse cubic linear & 36.431 & 0.081 & 52.1 & 540.8 & 0.096 \\
6 & (1.85; 2.40) & (0.600, 4.250) & Inverse quartic constant & 28.593 & 0.065 & 40.3 & 515.3 & 0.078 \\
7 & (1.75; 2.45) & (0.800, 4.000) & Shifted inverse quintic quadratic & 39.953 & 0.090 & 57.4 & 526.6 & 0.109 \\
8 & (1.85; 2.40) & (0.700, 4.000) & Shifted inverse cubic linear & 38.575 & 0.087 & 55.4 & 533.6 & 0.104 \\
9 & (1.80; 2.45) & (0.700, 3.750) & Shifted inverse quartic quadratic & 35.947 & 0.080 & 51.5 & 525.8 & 0.098 \\
10 & (1.80; 2.40) & (0.600, 3.500) & Inverse quintic quadratic & 28.529 & 0.062 & 40.2 & 543.6 & 0.074 \\
11 & (1.85; 2.45) & (0.700, 3.750) & Inverse cubic linear & 42.223 & 0.097 & 61.0 & 515.6 & 0.118 \\
12 & (1.80; 2.45) & (0.700, 4.000) & Inverse quartic quadratic & 35.869 & 0.083 & 51.2 & 506.4 & 0.101 \\
13 & (1.80; 2.45) & (0.700, 4.000) & Inverse quintic constant & 33.549 & 0.078 & 48.1 & 507.3 & 0.095 \\
14 & (1.80; 2.45) & (0.800, 4.500) & Shifted inverse quartic linear & 43.343 & 0.103 & 62.4 & 500.7 & 0.125 \\
15 & (1.85; 2.45) & (0.700, 4.000) & Inverse cubic linear & 39.728 & 0.088 & 57.2 & 536.8 & 0.107 \\
16 & (1.80; 2.50) & (0.800, 4.500) & Inverse quartic linear & 43.924 & 0.099 & 63.3 & 519.1 & 0.122 \\
17 & (1.85; 2.45) & (0.800, 4.500) & Inverse cubic constant & 49.535 & 0.116 & 71.7 & 507.4 & 0.141 \\
18 & (1.80; 2.50) & (0.800, 4.250) & Inverse quartic linear & 45.921 & 0.107 & 66.4 & 503.3 & 0.132 \\
19 & (1.75; 2.40) & (0.700, 3.250) & Inverse quintic quartic & 37.958 & 0.086 & 53.9 & 521.6 & 0.103 \\
20 & (1.85; 2.40) & (0.700, 4.250) & Shifted inverse cubic linear & 36.639 & 0.079 & 52.5 & 556.0 & 0.094 \\
21 & (1.80; 2.40) & (0.600, 4.000) & Shifted inverse quintic linear & 27.846 & 0.064 & 39.1 & 507.6 & 0.077 \\
22 & (1.75; 2.50) & (0.800, 4.500) & Shifted inverse quintic linear & 38.394 & 0.089 & 55.0 & 499.0 & 0.110 \\
23 & (1.85; 2.40) & (0.700, 4.000) & Inverse cubic linear & 40.047 & 0.095 & 57.6 & 500.8 & 0.115 \\
24 & (1.85; 2.40) & (0.700, 4.000) & Inverse quartic constant & 37.114 & 0.080 & 53.4 & 563.1 & 0.095 \\
25 & (1.80; 2.45) & (0.800, 4.000) & Shifted inverse cubic quadratic & 43.579 & 0.094 & 62.9 & 552.7 & 0.114 \\
26 & (1.85; 2.40) & (0.600, 4.000) & Inverse quartic constant & 29.585 & 0.070 & 41.9 & 494.9 & 0.085 \\
27 & (1.75; 2.50) & (0.900, 4.250) & Inverse quintic linear & 48.082 & 0.106 & 69.7 & 536.2 & 0.130 \\
28 & (1.80; 2.40) & (0.700, 3.750) & Shifted inverse quintic linear & 33.548 & 0.074 & 48.0 & 547.1 & 0.088 \\
29 & (1.75; 2.50) & (0.900, 4.250) & Shifted inverse quintic linear & 47.623 & 0.104 & 69.0 & 544.3 & 0.127 \\
30 & (1.75; 2.50) & (0.800, 4.500) & Inverse quintic linear & 38.794 & 0.090 & 55.6 & 492.5 & 0.113 \\
31 & (1.80; 2.40) & (0.600, 3.250) & Inverse quintic quadratic & 29.750 & 0.069 & 42.1 & 509.9 & 0.083 \\
32 & (1.80; 2.40) & (0.700, 4.000) & Shifted inverse quintic linear & 31.482 & 0.066 & 44.9 & 575.6 & 0.078 \\
33 & (1.85; 2.40) & (0.700, 3.750) & Shifted inverse cubic linear & 41.079 & 0.097 & 59.2 & 510.1 & 0.116 \\
34 & (1.85; 2.45) & (0.800, 4.500) & Inverse quadratic & 50.943 & 0.123 & 73.7 & 490.9 & 0.150 \\
35 & (1.80; 2.45) & (0.700, 3.500) & Shifted inverse quartic quadratic & 38.408 & 0.090 & 55.2 & 500.8 & 0.110 \\
36 & (1.80; 2.45) & (0.800, 3.750) & Shifted inverse cubic quadratic & 46.383 & 0.104 & 67.2 & 529.2 & 0.127 \\
37 & (1.75; 2.45) & (0.800, 3.750) & Shifted inverse quintic quadratic & 42.492 & 0.100 & 61.2 & 502.8 & 0.122 \\
38 & (1.85; 2.40) & (0.600, 4.250) & Shifted inverse cubic linear & 30.395 & 0.071 & 42.9 & 493.1 & 0.087 \\
39 & (1.85; 2.40) & (0.700, 3.750) & Inverse quartic constant & 39.683 & 0.089 & 57.3 & 537.4 & 0.107 \\
40 & (1.90; 2.35) & (0.600, 3.500) & Shifted hyperbolic & 37.114 & 0.091 & 53.3 & 486.1 & 0.110 \\
41 & (1.75; 2.50) & (0.900, 4.000) & Shifted inverse quintic linear & 50.085 & 0.113 & 72.7 & 524.6 & 0.139 \\
42 & (1.75; 2.45) & (0.800, 4.250) & Shifted inverse quintic quadratic & 37.954 & 0.082 & 54.3 & 549.3 & 0.099 \\
43 & (1.75; 2.50) & (0.900, 4.000) & Inverse quintic linear & 50.557 & 0.116 & 73.4 & 516.9 & 0.142 \\
44 & (1.85; 2.45) & (0.800, 4.250) & Inverse cubic constant & 51.590 & 0.124 & 74.8 & 491.9 & 0.152 \\
45 & (1.80; 2.45) & (0.800, 3.750) & Inverse quartic quadratic & 46.321 & 0.099 & 67.1 & 556.7 & 0.121 \\
46 & (1.75; 2.50) & (0.800, 4.250) & Shifted inverse quintic linear & 40.129 & 0.095 & 57.7 & 483.4 & 0.119 \\
47 & (1.75; 2.50) & (0.900, 4.500) & Inverse quintic linear & 45.973 & 0.098 & 66.5 & 554.6 & 0.120 \\
48 & (1.80; 2.45) & (0.700, 3.750) & Inverse quartic quadratic & 37.863 & 0.090 & 54.3 & 485.5 & 0.112 \\
49 & (1.85; 2.45) & (0.700, 3.500) & Inverse cubic linear & 45.403 & 0.108 & 65.7 & 493.5 & 0.133 \\
50 & (1.80; 2.45) & (0.800, 4.250) & Shifted inverse quartic linear & 45.424 & 0.111 & 65.6 & 484.1 & 0.135 \\
\end{longtable}
\endgroup

\section{Size-consistency and conservative combined-error analysis}

Because the \MERCE{} mapping is nonlinear and system dependent, size
consistency cannot be assumed. This section therefore separates three related
quantities: the formal size-consistency defect, the ordinary pair-energy error
relative to the additive high-level reference, and a conservative combined
error that penalizes both effects.

For a pair of fragments $A$ and $B$, the formal noninteracting supersystem
$A+B$ is constructed by adding the corresponding MP2, CCSD, CCSD(T), and
reference correlation energies of the fragments. The \MERCE{} prediction is
then evaluated in two ways. In the first route, the extrapolation is applied
once to the combined $A+B$ input sequence. In the second route, the same
extrapolation is applied separately to $A$ and $B$, and the two extrapolated
energies are summed. The difference between these two routes defines the
size-consistency defect, $\Delta_{\mathrm{SC}}$. The ordinary pair-energy error
relative to the additive high-level reference is denoted
$\Delta_{\mathrm{pair}}$.

For the conservative stress-test analysis, we also use
\[
\Delta_{\mathrm{comb}}=
|\Delta_{\mathrm{pair}}|+|\Delta_{\mathrm{SC}}|,
\]
which penalizes both the prediction error for the artificial pair and the
nonadditivity introduced by the nonlinear extrapolation.
The artificial-pair construction is used here only as an internal diagnostic of
additivity for the nonlinear extrapolation functional. The underlying MP2,
CCSD, and CCSD(T) input energies are size-consistent for noninteracting
fragments when evaluated with a consistent orbital space and computational
protocol; the defect analyzed below is introduced by the system-dependent
nonlinear mapping, not by the input methods themselves.

\subsection{Effective-rank mismatch analysis for the final \MERCE{} model}

The largest size-consistency defects occur for heterogeneous artificial pairs,
most prominently in the W4-derived subset. Here, ``heterogeneous'' does not
refer primarily to chemical composition, but to the effective-rank response of
the two fragments. The most problematic cases are mixed high-$R$+low-$R$
combinations, where one fragment samples the saturated part of the arctangent
mapping while the other remains in the low-response region.

If the fragments are extrapolated separately, each fragment is mapped according
to its own local increment ratio, $R_A$ or $R_B$. In the formal supersystem
calculation, however, the mapping is controlled by the ratio formed from the
summed increments,
\[
R_{A+B} =
\frac{|d_2(A)+d_2(B)|}
{|d_1(A)+d_1(B)|+\varepsilon_{\mathrm{num}}},
\]
which generally does not reproduce the two fragment-specific responses.
Nonadditivity is therefore expected to be most visible when the fragments
occupy different nonlinear regions of the effective-rank mapping.

For the category-resolved W4 analysis, each monomer was classified according to
the tertiles of $\ln R$ over the retained W4 monomer set. The cutoffs are
\[
\ln R_{\mathrm{low}}=-0.8674,\qquad
\ln R_{\mathrm{high}}=0.0840,
\]
corresponding to
\[
R_{\mathrm{low}}=0.4200,\qquad
R_{\mathrm{high}}=1.0876.
\]
Monomers with $\ln R\leq \ln R_{\mathrm{low}}$ were assigned to the low-$R$
category, monomers with
$\ln R_{\mathrm{low}}<\ln R<\ln R_{\mathrm{high}}$ to the mid-$R$ category,
and monomers with $\ln R\geq \ln R_{\mathrm{high}}$ to the high-$R$ category.
Pair categories such as mixed high-$R$+low-$R$ were then assigned from the two
fragment labels.

For the final representative \MERCE{} model, the absolute size-consistency
defect in the W4 artificial-pair set correlates strongly with the fragment
mismatch measure $|\ln R_A-\ln R_B|$,
\[
r\!\left(|\ln R_A-\ln R_B|,|\Delta_{\mathrm{SC}}|\right)=0.812.
\]
By contrast, the ordinary pair-energy error has no analogous positive
correlation with this mismatch measure,
\[
r\!\left(|\ln R_A-\ln R_B|,|\Delta_{\mathrm{pair}}|\right)=-0.178.
\]
Thus, the size-consistency defect is specifically associated with a mismatch in
fragment effective-rank response, not simply with large pair-energy errors.

\begin{table}[t]
\centering
\caption{Category-resolved size-consistency analysis for W4 artificial pairs
using the final representative \MERCE{} model only. Low-, mid-, and high-$R$
categories are defined by tertiles of $\ln R$ over the W4 monomer set, with
cutoffs $\ln R=-0.8674$ and $\ln R=0.0840$, corresponding to
$R=0.4200$ and $R=1.0876$. Size-consistency and pair-error quantities are
reported in $\microEh$.}
\label{tab:w4_sc_categories_final_model}
\begin{tabular}{lrrrrr}
\toprule
Category & $N$ & SC MAE & SC RMS & pair MAE & mean $|\Delta \ln R|$ \\
\midrule
low+low        & 171 & 18.856  & 29.577  & 375.566 & 0.716 \\
high+high      & 171 & 36.810  & 57.569  & 474.587 & 0.925 \\
mixed high+low & 361 & 109.808 & 152.344 & 351.747 & 2.750 \\
mid+low        & 342 & 17.958  & 29.132  & 579.521 & 1.327 \\
mid+high       & 342 & 83.365  & 120.666 & 607.641 & 1.422 \\
mid+mid        & 153 & 0.904   & 1.432   & 813.207 & 0.335 \\
\bottomrule
\end{tabular}
\end{table}

Table~\ref{tab:w4_sc_categories_final_model} shows that the strongest
nonadditivity is concentrated in mixed high-$R$+low-$R$ pairs. For the final
representative \MERCE{} model, the SC MAE in this category is
$109.8\,\microEh$, compared with $36.8\,\microEh$ for high+high pairs,
$18.9\,\microEh$ for low+low pairs, and only $0.9\,\microEh$ for mid+mid
pairs. The same table also shows that the largest size-consistency defects are
not identical to the largest pair-energy errors: mid+mid pairs have the largest
pair MAE in this category analysis but almost no size-consistency defect.

\subsection{Aggregate W4 stress test over ten high-performing models}

The final-model analysis above gives the numbers discussed in the main text.
To test whether the same pattern is specific to the selected inverse-cubic-linear
model or more general across the high-performing region, we repeated the
category analysis for ten high-performing models from the full $(k_1,k_3)$
scan. In this aggregate analysis, the reported $N$ values count model--pair
entries rather than unique W4 pairs.

\begin{table}[t]
\centering
\caption{Category-resolved conservative combined-error summary for W4
artificial pairs, aggregated over ten high-performing models from the full
$(k_1,k_3)$ scan. The combined model error is defined as
$\Delta_{\mathrm{comb}}=
|\Delta_{\mathrm{pair}}|+|\Delta_{\mathrm{SC}}|$. Low-, mid-, and high-$R$
categories are defined by tertiles of $\ln R$ over the W4 monomer set, with
cutoffs $\ln R=-0.8674$ and $\ln R=0.0840$. Because the table aggregates over
ten models, the reported $N$ values count model--pair entries rather than
unique W4 pairs. All error quantities are reported in $\microEh$.}
\label{tab:w4_sc_categories_top10}
\begin{tabular}{lrrrrrr}
\toprule
Category & $N$ & model pair MAE & SC MAE & combined MAE
& CCSD(T) MAE & CCSDT MAE \\
\midrule
low+low        & 1710 & 398.1 & 18.5  & 416.6 & 1550.1 & 387.1 \\
high+high      & 1710 & 446.0 & 37.4  & 483.4 & 2883.7 & 2487.3 \\
mixed high+low & 3610 & 347.6 & 106.8 & 454.4 & 2216.9 & 1437.2 \\
mid+low        & 3420 & 586.4 & 19.4  & 605.8 & 2381.6 & 1060.6 \\
mid+high       & 3420 & 599.2 & 81.3  & 680.6 & 3048.4 & 2110.7 \\
mid+mid        & 1530 & 814.2 & 1.9   & 816.0 & 3213.0 & 1734.0 \\
\bottomrule
\end{tabular}
\end{table}

The aggregate top-10 analysis gives the same qualitative conclusion as the
final-model analysis. Mixed high-$R$+low-$R$ pairs again have the largest
size-consistency component, with an aggregate SC MAE of
$106.8\,\microEh$, whereas matched categories such as low+low, high+high, and
mid+mid remain much less affected. The numerical values differ slightly from
the final-model values in Table~\ref{tab:w4_sc_categories_final_model} because
Table~\ref{tab:w4_sc_categories_top10} averages over ten models rather than
reporting the selected \MERCE{} model alone.

The per-model contrast confirms that the mixed-category enhancement of
nonadditivity is systematic across the high-performing region. For all ten
models, the mixed-category SC MAE is several times larger than the matched
category SC MAE. For the final representative inverse-cubic-linear model, the overall W4
size-consistency MAE is $54.5\,\microEh$. The mixed high-$R$+low-$R$ category
has an SC MAE of $109.8\,\microEh$, whereas the pooled same-regime categories
(low+low, mid+mid, and high+high) have an SC MAE of $19.5\,\microEh$. Thus, the
high-$R$+low-$R$ category has a size-consistency defect about 5.63 times larger
than the matched same-regime categories.

The detailed entries with the largest absolute size-consistency defects, the
per-model conservative combined-error summary, and the largest combined-error
cases are reported in Tables~\ref{tab:w4_worst_sc_combined},
\ref{tab:w4_combined_per_model_summary}, and
\ref{tab:w4_worst_combined}, respectively.
They show that the largest size-consistency defects are dominated by pairs
involving fragments with very different $R$ values, whereas the largest
combined errors can also reflect intrinsically difficult monomer or pair
energetics. Therefore, $\Delta_{\mathrm{SC}}$ and
$\Delta_{\mathrm{pair}}$ diagnose different aspects of the extrapolation error.

\begin{landscape}
\scriptsize
\setlength{\LTleft}{0pt}
\setlength{\LTright}{0pt}
\begin{longtable}{l l l l r r r r r c}
\caption{Detailed conservative combined-error analysis for the W4 pairs with the largest size-consistency defects.}\label{tab:w4_worst_sc_combined}\\
\toprule
Variant & Model & Pair & Category & $|SC|$ & $|\Delta_{\mathrm{pair}}|$ & $\Delta_{\mathrm{comb}}$ & $|\Delta_{\mathrm{CCSD(T)}}|$ & $|\Delta_{\mathrm{CCSDT}}|$ & Better than  \\
& & & & [$\mu E_h$] & [$\mu E_h$] & [$\mu E_h$] & [$\mu E_h$] & [$\mu E_h$] &CCSDT \\
\midrule
\endfirsthead
\toprule
Variant & Model & Pair & Category & $|SC|$ & $|\Delta_{\mathrm{pair}}|$ & $\Delta_{\mathrm{comb}}$ & $|\Delta_{\mathrm{CCSD(T)}}|$ & $|\Delta_{\mathrm{CCSDT}}|$ & Better than CCSDT \\
& & & & [$\mu E_h$] & [$\mu E_h$] & [$\mu E_h$] & [$\mu E_h$] & [$\mu E_h$] & \\
\midrule
\endhead
\midrule
\multicolumn{10}{r}{Continued on next page}\\
\endfoot
\bottomrule
\endlastfoot
$(1.80; 2.50),\,(0.800, 4.250)$ & Inverse quartic linear & w417\_alh3 + w417\_f2o & mixed high+low & 543.4 & 120.4 & 663.7 & 2677.0 & 2199.4 & yes \\
$(1.80; 2.50),\,(0.800, 4.500)$ & Inverse quartic linear & w417\_alh3 + w417\_f2o & mixed high+low & 526.1 & 142.3 & 668.4 & 2677.0 & 2199.4 & yes \\
$(1.80; 2.50),\,(0.800, 4.250)$ & Inverse quartic linear & w417\_bh + w417\_f2o & mixed high+low & 524.0 & 137.6 & 661.6 & 2737.5 & 2240.1 & yes \\
$(1.85; 2.45),\,(0.700, 3.750)$ & Inverse cubic linear & w417\_alh3 + w417\_f2o & mixed high+low & 513.4 & 166.0 & 679.5 & 2677.0 & 2199.4 & yes \\
$(1.80; 2.50),\,(0.800, 4.250)$ & Inverse quartic linear & w417\_f2o + w417\_sih4 & mixed high+low & 510.5 & 261.0 & 771.5 & 3043.8 & 2258.9 & yes \\
$(1.80; 2.50),\,(0.800, 4.500)$ & Inverse quartic linear & w417\_bh + w417\_f2o & mixed high+low & 507.8 & 160.1 & 667.9 & 2737.5 & 2240.1 & yes \\
$(1.80; 2.50),\,(0.800, 4.250)$ & Inverse quartic linear & w417\_alh + w417\_f2o & mixed high+low & 502.8 & 156.5 & 659.3 & 2650.1 & 2212.0 & yes \\
$(1.80; 2.50),\,(0.800, 4.500)$ & Inverse quartic linear & w417\_f2o + w417\_sih4 & mixed high+low & 494.9 & 288.0 & 783.0 & 3043.8 & 2258.9 & yes \\
$(1.75; 2.45),\,(0.800, 4.000)$ & Shifted inverse quintic quadratic & w417\_alh3 + w417\_f2o & mixed high+low & 492.3 & 186.1 & 678.4 & 2677.0 & 2199.4 & yes \\
$(1.85; 2.45),\,(0.700, 3.750)$ & Inverse cubic linear & w417\_bh + w417\_f2o & mixed high+low & 491.8 & 184.2 & 676.0 & 2737.5 & 2240.1 & yes \\
$(1.80; 2.50),\,(0.800, 4.500)$ & Inverse quartic linear & w417\_alh + w417\_f2o & mixed high+low & 488.8 & 175.1 & 663.9 & 2650.1 & 2212.0 & yes \\
$(1.80; 2.50),\,(0.800, 4.250)$ & Inverse quartic linear & w417\_f2o + w417\_sih & mixed high+low & 479.3 & 439.5 & 918.8 & 3092.0 & 2243.0 & yes \\
$(1.85; 2.45),\,(0.700, 3.750)$ & Inverse cubic linear & w417\_alh + w417\_f2o & mixed high+low & 477.5 & 197.9 & 675.4 & 2650.1 & 2212.0 & yes \\
$(1.80; 2.50),\,(0.800, 4.250)$ & Inverse quartic linear & w417\_bh3 + w417\_f2o & mixed high+low & 476.7 & 88.6 & 565.4 & 2679.3 & 2198.2 & yes \\
$(1.80; 2.50),\,(0.800, 4.250)$ & Inverse quartic linear & w417\_ch + w417\_f2o & mixed high+low & 475.3 & 268.4 & 743.7 & 2946.6 & 2242.7 & yes \\
$(1.85; 2.40),\,(0.700, 4.250)$ & Inverse cubic linear & w417\_alh3 + w417\_f2o & mixed high+low & 474.9 & 181.9 & 656.8 & 2677.0 & 2199.4 & yes \\
$(1.85; 2.45),\,(0.700, 3.750)$ & Inverse cubic linear & w417\_f2o + w417\_sih4 & mixed high+low & 469.8 & 312.4 & 782.2 & 3043.8 & 2258.9 & yes \\
$(1.75; 2.45),\,(0.800, 4.000)$ & Shifted inverse quintic quadratic & w417\_bh + w417\_f2o & mixed high+low & 469.6 & 204.7 & 674.2 & 2737.5 & 2240.1 & yes \\
$(1.80; 2.50),\,(0.800, 4.500)$ & Inverse quartic linear & w417\_f2o + w417\_sih & mixed high+low & 467.4 & 461.0 & 928.4 & 3092.0 & 2243.0 & yes \\
$(1.80; 2.50),\,(0.800, 4.500)$ & Inverse quartic linear & w417\_bh3 + w417\_f2o & mixed high+low & 464.8 & 108.6 & 573.4 & 2679.3 & 2198.2 & yes \\
$(1.80; 2.50),\,(0.800, 4.500)$ & Inverse quartic linear & w417\_ch + w417\_f2o & mixed high+low & 463.9 & 290.2 & 754.2 & 2946.6 & 2242.7 & yes \\
$(1.75; 2.45),\,(0.800, 4.000)$ & Shifted inverse quintic quadratic & w417\_alh + w417\_f2o & mixed high+low & 459.4 & 217.0 & 676.4 & 2650.1 & 2212.0 & yes \\
$(1.80; 2.50),\,(0.800, 4.250)$ & Inverse quartic linear & w417\_f2o + w417\_ph3 & mixed high+low & 458.0 & 361.6 & 819.6 & 3318.7 & 2419.2 & yes \\
$(1.85; 2.40),\,(0.700, 4.250)$ & Inverse cubic linear & w417\_bh + w417\_f2o & mixed high+low & 452.0 & 200.4 & 652.4 & 2737.5 & 2240.1 & yes \\
$(1.80; 2.50),\,(0.800, 4.500)$ & Inverse quartic linear & w417\_f2o + w417\_ph3 & mixed high+low & 451.0 & 386.9 & 837.9 & 3318.7 & 2419.2 & yes \\
$(1.80; 2.45),\,(0.700, 3.750)$ & Shifted inverse quartic quadratic & w417\_alh3 + w417\_f2o & mixed high+low & 449.5 & 166.6 & 616.1 & 2677.0 & 2199.4 & yes \\
$(1.80; 2.45),\,(0.700, 4.000)$ & Inverse quartic quadratic & w417\_alh3 + w417\_f2o & mixed high+low & 448.3 & 108.7 & 557.0 & 2677.0 & 2199.4 & yes \\
$(1.80; 2.50),\,(0.800, 4.250)$ & Inverse quartic linear & w417\_ch2ch + w417\_f2o & mixed high+low & 447.3 & 588.5 & 1035.8 & 4124.9 & 2829.7 & yes \\
$(1.85; 2.45),\,(0.700, 3.750)$ & Inverse cubic linear & w417\_bh3 + w417\_f2o & mixed high+low & 446.5 & 132.5 & 579.0 & 2679.3 & 2198.2 & yes \\
$(1.85; 2.45),\,(0.700, 3.750)$ & Inverse cubic linear & w417\_f2o + w417\_sih & mixed high+low & 445.6 & 485.7 & 931.4 & 3092.0 & 2243.0 & yes \\
$(1.85; 2.40),\,(0.700, 4.250)$ & Inverse cubic linear & w417\_alh + w417\_f2o & mixed high+low & 444.7 & 211.6 & 656.3 & 2650.1 & 2212.0 & yes \\
$(1.80; 2.50),\,(0.800, 4.500)$ & Inverse quartic linear & w417\_ch2ch + w417\_f2o & mixed high+low & 444.7 & 618.5 & 1063.1 & 4124.9 & 2829.7 & yes \\
$(1.75; 2.45),\,(0.800, 4.000)$ & Shifted inverse quintic quadratic & w417\_f2o + w417\_sih4 & mixed high+low & 444.2 & 335.7 & 779.9 & 3043.8 & 2258.9 & yes \\
$(1.80; 2.50),\,(0.800, 4.250)$ & Inverse quartic linear & w417\_cch + w417\_f2o & mid+high & 441.7 & 1231.0 & 1672.7 & 4855.6 & 3299.2 & yes \\
$(1.85; 2.45),\,(0.700, 3.750)$ & Inverse cubic linear & w417\_ch + w417\_f2o & mixed high+low & 440.6 & 315.2 & 755.8 & 2946.6 & 2242.7 & yes \\
$(1.80; 2.50),\,(0.800, 4.500)$ & Inverse quartic linear & w417\_cch + w417\_f2o & mid+high & 439.6 & 1261.0 & 1700.6 & 4855.6 & 3299.2 & yes \\
$(1.80; 2.50),\,(0.800, 4.250)$ & Inverse quartic linear & w417\_f2o + w417\_hs & mixed high+low & 436.5 & 358.8 & 795.3 & 3229.2 & 2371.4 & yes \\
$(1.80; 2.50),\,(0.800, 4.250)$ & Inverse quartic linear & w417\_f2o + w417\_ssh & mid+high & 432.6 & 758.1 & 1190.7 & 4592.0 & 3223.5 & yes \\
$(1.80; 2.50),\,(0.800, 4.500)$ & Inverse quartic linear & w417\_f2o + w417\_hs & mixed high+low & 431.4 & 380.6 & 812.0 & 3229.2 & 2371.4 & yes \\
$(1.80; 2.50),\,(0.800, 4.500)$ & Inverse quartic linear & w417\_f2o + w417\_ssh & mid+high & 431.3 & 788.8 & 1220.1 & 4592.0 & 3223.5 & yes \\
\end{longtable}
\end{landscape}

\begin{table}[htbp]
\centering
\caption{Per-model conservative combined-error summary for the top 10 W4 models.}
\label{tab:w4_combined_per_model_summary}
\scriptsize
\begin{tabular}{l l r r r r r}
\toprule
Variant & Model & model pair MAE & $SC$ MAE & combined MAE & CCSDT MAE & better than  \\
& & [$\mu E_h$] & [$\mu E_h$] & [$\mu E_h$] & [$\mu E_h$] &CCSDT fraction \\
\midrule
$(1.75; 2.45),\,(0.800, 4.000)$ & Shifted inverse quintic quadratic & 526.6 & 57.4 & 584.0 & 1532.6 & 0.906 \\
$(1.80; 2.45),\,(0.700, 3.750)$ & Shifted inverse quartic quadratic & 525.8 & 51.5 & 577.3 & 1532.6 & 0.903 \\
$(1.80; 2.45),\,(0.700, 4.000)$ & Inverse quartic quadratic & 506.4 & 51.2 & 557.6 & 1532.6 & 0.905 \\
$(1.80; 2.45),\,(0.700, 4.250)$ & Inverse quintic constant & 528.1 & 45.3 & 573.5 & 1532.6 & 0.903 \\
$(1.80; 2.50),\,(0.800, 4.250)$ & Inverse quartic linear & 503.3 & 66.4 & 569.6 & 1532.6 & 0.908 \\
$(1.80; 2.50),\,(0.800, 4.500)$ & Inverse quartic linear & 519.1 & 63.3 & 582.4 & 1532.6 & 0.899 \\
$(1.85; 2.40),\,(0.600, 4.250)$ & Inverse quartic constant & 515.3 & 40.3 & 555.7 & 1532.6 & 0.905 \\
$(1.85; 2.40),\,(0.700, 4.250)$ & Inverse cubic linear & 521.3 & 54.5 & 575.8 & 1532.6 & 0.910 \\
$(1.85; 2.45),\,(0.700, 3.750)$ & Inverse cubic linear & 515.6 & 61.0 & 576.6 & 1532.6 & 0.906 \\
$(1.90; 2.35),\,(0.600, 4.000)$ & Shifted hyperbolic & 532.6 & 47.0 & 579.6 & 1532.6 & 0.902 \\
\bottomrule
\end{tabular}
\end{table}

\begin{landscape}
\scriptsize
\setlength{\LTleft}{0pt}
\setlength{\LTright}{0pt}
\begin{longtable}{l l l l r r r r r c}
\caption{Detailed conservative combined-error analysis for the W4 pairs with the largest combined error.}\label{tab:w4_worst_combined}\\
\toprule
Variant & Model & Pair & Category & $|SC|$ & $|\Delta_{\mathrm{pair}}|$ & $\Delta_{\mathrm{comb}}$ & $|\Delta_{\mathrm{CCSD(T)}}|$ & $|\Delta_{\mathrm{CCSDT}}|$ & Better than CCSDT \\
& & & & [$\mu E_h$] & [$\mu E_h$] & [$\mu E_h$] & [$\mu E_h$] & [$\mu E_h$] & \\
\midrule
\endfirsthead
\toprule
Variant & Model & Pair & Category & $|SC|$ & $|\Delta_{\mathrm{pair}}|$ & $\Delta_{\mathrm{comb}}$ & $|\Delta_{\mathrm{CCSD(T)}}|$ & $|\Delta_{\mathrm{CCSDT}}|$ & Better than CCSDT \\
& & & & [$\mu E_h$] & [$\mu E_h$] & [$\mu E_h$] & [$\mu E_h$] & [$\mu E_h$] & \\
\midrule
\endhead
\midrule
\multicolumn{10}{r}{Continued on next page}\\
\endfoot
\bottomrule
\endlastfoot
$(1.90; 2.35),\,(0.600, 4.000)$ & Shifted hyperbolic & w417\_b2 + w417\_bn3pi & mid+mid & 0.0 & 3171.1 & 3171.1 & 5726.9 & 2757.3 & no \\
$(1.80; 2.45),\,(0.700, 4.250)$ & Inverse quintic constant & w417\_b2 + w417\_bn3pi & mid+mid & 0.0 & 3158.6 & 3158.6 & 5726.9 & 2757.3 & no \\
$(1.80; 2.45),\,(0.700, 3.750)$ & Shifted inverse quartic quadratic & w417\_b2 + w417\_bn3pi & mid+mid & 0.0 & 3149.9 & 3149.9 & 5726.9 & 2757.3 & no \\
$(1.85; 2.40),\,(0.600, 4.250)$ & Inverse quartic constant & w417\_b2 + w417\_bn3pi & mid+mid & 0.0 & 3143.9 & 3143.9 & 5726.9 & 2757.3 & no \\
$(1.90; 2.35),\,(0.600, 4.000)$ & Shifted hyperbolic & w417\_b2 + w417\_cch & mid+mid & 5.2 & 3138.4 & 3143.6 & 5956.2 & 3048.7 & no \\
$(1.75; 2.45),\,(0.800, 4.000)$ & Shifted inverse quintic quadratic & w417\_b2 + w417\_bn3pi & mid+mid & 0.0 & 3140.4 & 3140.4 & 5726.9 & 2757.3 & no \\
$(1.85; 2.40),\,(0.700, 4.250)$ & Inverse cubic linear & w417\_b2 + w417\_bn3pi & mid+mid & 0.0 & 3133.2 & 3133.2 & 5726.9 & 2757.3 & no \\
$(1.80; 2.45),\,(0.700, 4.250)$ & Inverse quintic constant & w417\_b2 + w417\_cch & mid+mid & 4.6 & 3124.1 & 3128.7 & 5956.2 & 3048.7 & no \\
$(1.85; 2.40),\,(0.600, 4.250)$ & Inverse quartic constant & w417\_b2 + w417\_cch & mid+mid & 7.4 & 3118.7 & 3126.1 & 5956.2 & 3048.7 & no \\
$(1.80; 2.45),\,(0.700, 3.750)$ & Shifted inverse quartic quadratic & w417\_b2 + w417\_cch & mid+mid & 4.9 & 3117.2 & 3122.2 & 5956.2 & 3048.7 & no \\
$(1.85; 2.45),\,(0.700, 3.750)$ & Inverse cubic linear & w417\_b2 + w417\_bn3pi & mid+mid & 0.0 & 3111.0 & 3111.0 & 5726.9 & 2757.3 & no \\
$(1.80; 2.45),\,(0.700, 4.000)$ & Inverse quartic quadratic & w417\_b2 + w417\_bn3pi & mid+mid & 0.0 & 3106.1 & 3106.1 & 5726.9 & 2757.3 & no \\
$(1.80; 2.50),\,(0.800, 4.500)$ & Inverse quartic linear & w417\_b2 + w417\_bn3pi & mid+mid & 0.0 & 3105.0 & 3105.0 & 5726.9 & 2757.3 & no \\
$(1.75; 2.45),\,(0.800, 4.000)$ & Shifted inverse quintic quadratic & w417\_b2 + w417\_cch & mid+mid & 3.7 & 3099.6 & 3103.2 & 5956.2 & 3048.7 & no \\
$(1.85; 2.40),\,(0.700, 4.250)$ & Inverse cubic linear & w417\_b2 + w417\_cch & mid+mid & 3.3 & 3088.2 & 3091.5 & 5956.2 & 3048.7 & no \\
$(1.80; 2.45),\,(0.700, 4.000)$ & Inverse quartic quadratic & w417\_b2 + w417\_cch & mid+mid & 5.6 & 3078.4 & 3084.0 & 5956.2 & 3048.7 & no \\
$(1.80; 2.50),\,(0.800, 4.500)$ & Inverse quartic linear & w417\_b2 + w417\_cch & mid+mid & 0.9 & 3078.2 & 3079.1 & 5956.2 & 3048.7 & no \\
$(1.80; 2.50),\,(0.800, 4.250)$ & Inverse quartic linear & w417\_b2 + w417\_bn3pi & mid+mid & 0.0 & 3068.7 & 3068.7 & 5726.9 & 2757.3 & no \\
$(1.85; 2.45),\,(0.700, 3.750)$ & Inverse cubic linear & w417\_b2 + w417\_cch & mid+mid & 1.6 & 3063.6 & 3065.2 & 5956.2 & 3048.7 & no \\
$(1.80; 2.50),\,(0.800, 4.250)$ & Inverse quartic linear & w417\_b2 + w417\_cch & mid+mid & 0.4 & 3030.6 & 3031.0 & 5956.2 & 3048.7 & yes \\
$(1.90; 2.35),\,(0.600, 4.000)$ & Shifted hyperbolic & w417\_b2 + w417\_p2 & mid+high & 4.7 & 2835.0 & 2839.7 & 6013.2 & 4220.1 & yes \\
$(1.75; 2.45),\,(0.800, 4.000)$ & Shifted inverse quintic quadratic & w417\_b2 + w417\_p2 & mid+high & 17.2 & 2815.4 & 2832.6 & 6013.2 & 4220.1 & yes \\
$(1.80; 2.45),\,(0.700, 4.250)$ & Inverse quintic constant & w417\_b2 + w417\_p2 & mid+high & 7.7 & 2824.0 & 2831.7 & 6013.2 & 4220.1 & yes \\
$(1.85; 2.40),\,(0.700, 4.250)$ & Inverse cubic linear & w417\_b2 + w417\_p2 & mid+high & 14.3 & 2810.3 & 2824.6 & 6013.2 & 4220.1 & yes \\
$(1.80; 2.45),\,(0.700, 3.750)$ & Shifted inverse quartic quadratic & w417\_b2 + w417\_p2 & mid+high & 8.2 & 2807.9 & 2816.1 & 6013.2 & 4220.1 & yes \\
$(1.85; 2.45),\,(0.700, 3.750)$ & Inverse cubic linear & w417\_b2 + w417\_p2 & mid+high & 15.6 & 2786.8 & 2802.4 & 6013.2 & 4220.1 & yes \\
$(1.80; 2.50),\,(0.800, 4.500)$ & Inverse quartic linear & w417\_b2 + w417\_p2 & mid+high & 23.3 & 2759.8 & 2783.2 & 6013.2 & 4220.1 & yes \\
$(1.85; 2.40),\,(0.600, 4.250)$ & Inverse quartic constant & w417\_b2 + w417\_p2 & mid+high & 2.1 & 2778.6 & 2780.7 & 6013.2 & 4220.1 & yes \\
$(1.80; 2.50),\,(0.800, 4.250)$ & Inverse quartic linear & w417\_b2 + w417\_p2 & mid+high & 23.7 & 2725.8 & 2749.6 & 6013.2 & 4220.1 & yes \\
$(1.80; 2.45),\,(0.700, 4.000)$ & Inverse quartic quadratic & w417\_b2 + w417\_p2 & mid+high & 8.7 & 2740.4 & 2749.1 & 6013.2 & 4220.1 & yes \\
$(1.90; 2.35),\,(0.600, 4.000)$ & Shifted hyperbolic & w417\_b2 + w417\_s2 & mid+mid & 0.1 & 2705.6 & 2705.7 & 5770.7 & 3337.4 & yes \\
$(1.80; 2.45),\,(0.700, 4.250)$ & Inverse quintic constant & w417\_b2 + w417\_s2 & mid+mid & 0.1 & 2690.9 & 2690.9 & 5770.7 & 3337.4 & yes \\
$(1.80; 2.45),\,(0.700, 3.750)$ & Shifted inverse quartic quadratic & w417\_b2 + w417\_s2 & mid+mid & 0.1 & 2679.9 & 2680.0 & 5770.7 & 3337.4 & yes \\
$(1.85; 2.40),\,(0.600, 4.250)$ & Inverse quartic constant & w417\_b2 + w417\_s2 & mid+mid & 0.2 & 2671.0 & 2671.2 & 5770.7 & 3337.4 & yes \\
$(1.75; 2.45),\,(0.800, 4.000)$ & Shifted inverse quintic quadratic & w417\_b2 + w417\_s2 & mid+mid & 0.0 & 2670.0 & 2670.0 & 5770.7 & 3337.4 & yes \\
$(1.90; 2.35),\,(0.600, 4.000)$ & Shifted hyperbolic & w417\_b2 + w417\_ssh & mid+mid & 2.6 & 2666.9 & 2669.5 & 5692.6 & 2973.0 & yes \\
$(1.85; 2.40),\,(0.700, 4.250)$ & Inverse cubic linear & w417\_b2 + w417\_s2 & mid+mid & 0.0 & 2661.9 & 2661.9 & 5770.7 & 3337.4 & yes \\
$(1.80; 2.45),\,(0.700, 4.250)$ & Inverse quintic constant & w417\_b2 + w417\_ssh & mid+mid & 2.2 & 2651.6 & 2653.8 & 5692.6 & 2973.0 & yes \\
$(1.85; 2.40),\,(0.600, 4.250)$ & Inverse quartic constant & w417\_b2 + w417\_ssh & mid+mid & 3.7 & 2643.0 & 2646.7 & 5692.6 & 2973.0 & yes \\
$(1.80; 2.45),\,(0.700, 3.750)$ & Shifted inverse quartic quadratic & w417\_b2 + w417\_ssh & mid+mid & 2.4 & 2643.4 & 2645.9 & 5692.6 & 2973.0 & yes \\
\end{longtable}
\end{landscape}

\section{A24 property-level model-selection statistics}

A24 is used as a late-stage property-level model-selection benchmark rather
than as the primary fitting target. This is a deliberately stringent test
because interaction energies are small differences of total or correlation
energies and are sensitive to error cancellation. The complete A24 scan below
shows that several candidates reach similar MAEs, so A24 alone does not uniquely
define the final model. Instead, it acts as one component of the selection
hierarchy: a candidate must be transferable, robust, size-consistency
controlled, and then competitive on A24.
The A24 assessment uses the CCSDT(Q), CCSDT, CCSD(T), and DC-CCSDT
correlation contributions to the interaction energies reported by Lambie
et al.\cite{Lambie2025} It contains the 23 entries for which that study reports
the complete high-rank comparison ladder. Case~17 is omitted because its
CCSDT(Q), CCSDT, and DC-CCSDT values are not reported; case~18 is retained.
This omission is inherited from the source data and was not selected according
to the performance of \MERCE{}.

\begin{landscape}
\scriptsize
\setlength{\LTleft}{0pt}
\setlength{\LTright}{0pt}
\begin{longtable}{r l l r r r r r}
\caption{Performance of the 50 best effective-rank extrapolation models on the
23 reference-bearing A24 entries. Error statistics are given relative to the
published CCSDT(Q) correlation contributions to the interaction energies from
Ref.~\cite{Lambie2025}. Case~17 lacks the required high-rank reference values
and is omitted.}\\
\label{tab:a24_all50_models}\\
\toprule
Rank & Variant $(x_{\mathrm{MP2}};x_{\mathrm{CCSD(T)}}),(k_1,k_3)$ & Model & MAE & RMSE & MAXE & wins vs CCSD(T) & wins vs DC-CCSDT \\
& & & [$\kcalmol$] & [$\kcalmol$] & [$\kcalmol$] & /23 & /23 \\
\midrule
\endfirsthead
\toprule
Rank & Variant $(x_{\mathrm{MP2}};x_{\mathrm{CCSD(T)}}),(k_1,k_3)$ & Model & MAE & RMSE & MAXE & wins vs CCSD(T) & wins vs DC-CCSDT \\
& & & (kcal/mol) & (kcal/mol) & (kcal/mol) & /23 & /23 \\
\midrule
\endhead
\midrule
\multicolumn{8}{r}{Continued on next page}\\
\endfoot
\bottomrule
\endlastfoot
1 & $(1.75; 2.50),\,(0.800, 4.250)$ & Shifted inverse quintic linear & 0.0056 & 0.0136 & 0.0611 & 19 & 14 \\
2 & $(1.80; 2.50),\,(0.800, 4.500)$ & Inverse quartic linear & 0.0056 & 0.0133 & 0.0596 & 19 & 11 \\
3 & $(1.75; 2.50),\,(0.800, 4.500)$ & Inverse quintic linear & 0.0057 & 0.0137 & 0.0616 & 19 & 12 \\
4 & $(1.75; 2.50),\,(0.800, 4.500)$ & Shifted inverse quintic linear & 0.0057 & 0.0137 & 0.0615 & 19 & 12 \\
5 & $(1.80; 2.45),\,(0.800, 4.500)$ & Shifted inverse quartic linear & 0.0057 & 0.0132 & 0.0587 & 20 & 13 \\
6 & $(1.85; 2.45),\,(0.700, 4.000)$ & Inverse cubic linear & 0.0058 & 0.0133 & 0.0591 & 19 & 11 \\
7 & $(1.75; 2.50),\,(0.900, 4.500)$ & Inverse quintic linear & 0.0059 & 0.0129 & 0.0569 & 20 & 11 \\
8 & $(1.75; 2.45),\,(0.800, 4.000)$ & Shifted inverse quintic quadratic & 0.0059 & 0.0134 & 0.0589 & 19 & 13 \\
9 & $(1.80; 2.50),\,(0.800, 4.250)$ & Inverse quartic linear & 0.0060 & 0.0133 & 0.0593 & 20 & 10 \\
10 & $(1.85; 2.40),\,(0.700, 4.250)$ & Inverse cubic linear & 0.0060 & 0.0134 & 0.0588 & 19 & 11 \\
11 & $(1.80; 2.45),\,(0.800, 4.000)$ & Shifted inverse cubic quadratic & 0.0060 & 0.0130 & 0.0572 & 20 & 11 \\
12 & $(1.80; 2.45),\,(0.800, 4.250)$ & Shifted inverse quartic linear & 0.0061 & 0.0133 & 0.0583 & 20 & 10 \\
13 & $(1.85; 2.40),\,(0.700, 4.000)$ & Inverse cubic linear & 0.0061 & 0.0134 & 0.0584 & 19 & 13 \\
14 & $(1.85; 2.40),\,(0.700, 4.000)$ & Shifted inverse cubic linear & 0.0061 & 0.0133 & 0.0578 & 19 & 13 \\
15 & $(1.85; 2.45),\,(0.800, 4.500)$ & Inverse cubic constant & 0.0062 & 0.0131 & 0.0574 & 20 & 10 \\
16 & $(1.85; 2.45),\,(0.800, 4.500)$ & Inverse quadratic & 0.0062 & 0.0132 & 0.0580 & 20 & 10 \\
17 & $(1.85; 2.40),\,(0.700, 4.250)$ & Shifted inverse cubic linear & 0.0062 & 0.0133 & 0.0582 & 19 & 12 \\
18 & $(1.80; 2.45),\,(0.700, 4.000)$ & Inverse quartic quadratic & 0.0062 & 0.0139 & 0.0614 & 19 & 12 \\
19 & $(1.80; 2.45),\,(0.700, 3.750)$ & Inverse quartic quadratic & 0.0062 & 0.0138 & 0.0609 & 19 & 11 \\
20 & $(1.80; 2.45),\,(0.700, 3.750)$ & Shifted inverse quartic quadratic & 0.0062 & 0.0137 & 0.0602 & 19 & 13 \\
21 & $(1.75; 2.50),\,(0.900, 4.250)$ & Shifted inverse quintic linear & 0.0062 & 0.0130 & 0.0565 & 20 & 12 \\
22 & $(1.75; 2.45),\,(0.800, 3.750)$ & Shifted inverse quintic quadratic & 0.0062 & 0.0134 & 0.0584 & 20 & 10 \\
23 & $(1.80; 2.45),\,(0.700, 4.000)$ & Inverse quintic constant & 0.0063 & 0.0136 & 0.0594 & 19 & 13 \\
24 & $(1.75; 2.45),\,(0.800, 4.250)$ & Shifted inverse quintic quadratic & 0.0063 & 0.0135 & 0.0593 & 19 & 11 \\
25 & $(1.75; 2.50),\,(0.900, 4.250)$ & Inverse quintic linear & 0.0063 & 0.0130 & 0.0566 & 20 & 11 \\
26 & $(1.85; 2.40),\,(0.700, 4.500)$ & Inverse cubic linear & 0.0063 & 0.0134 & 0.0591 & 19 & 10 \\
27 & $(1.80; 2.45),\,(0.700, 4.250)$ & Inverse quintic constant & 0.0063 & 0.0136 & 0.0598 & 19 & 12 \\
28 & $(1.85; 2.40),\,(0.700, 4.000)$ & Inverse quartic constant & 0.0063 & 0.0132 & 0.0566 & 19 & 13 \\
29 & $(1.85; 2.45),\,(0.700, 3.750)$ & Inverse cubic linear & 0.0063 & 0.0134 & 0.0587 & 19 & 10 \\
30 & $(1.90; 2.35),\,(0.600, 3.750)$ & Shifted hyperbolic & 0.0064 & 0.0137 & 0.0596 & 19 & 13 \\
31 & $(1.90; 2.35),\,(0.600, 4.000)$ & Shifted hyperbolic & 0.0065 & 0.0137 & 0.0600 & 19 & 12 \\
32 & $(1.80; 2.45),\,(0.700, 3.500)$ & Shifted inverse quartic quadratic & 0.0065 & 0.0137 & 0.0597 & 19 & 11 \\
33 & $(1.85; 2.40),\,(0.700, 3.750)$ & Shifted inverse cubic linear & 0.0066 & 0.0133 & 0.0573 & 20 & 12 \\
34 & $(1.85; 2.45),\,(0.800, 4.250)$ & Inverse cubic constant & 0.0066 & 0.0132 & 0.0570 & 19 & 10 \\
35 & $(1.80; 2.45),\,(0.800, 3.750)$ & Shifted inverse cubic quadratic & 0.0066 & 0.0132 & 0.0568 & 20 & 10 \\
36 & $(1.80; 2.45),\,(0.800, 3.750)$ & Inverse quartic quadratic & 0.0067 & 0.0131 & 0.0558 & 19 & 12 \\
37 & $(1.90; 2.35),\,(0.600, 3.500)$ & Shifted hyperbolic & 0.0067 & 0.0137 & 0.0591 & 19 & 12 \\
38 & $(1.75; 2.50),\,(0.900, 4.000)$ & Shifted inverse quintic linear & 0.0067 & 0.0131 & 0.0561 & 19 & 10 \\
39 & $(1.75; 2.50),\,(0.900, 4.000)$ & Inverse quintic linear & 0.0068 & 0.0132 & 0.0562 & 19 & 10 \\
40 & $(1.85; 2.40),\,(0.700, 3.750)$ & Inverse quartic constant & 0.0068 & 0.0133 & 0.0561 & 20 & 12 \\
41 & $(1.85; 2.40),\,(0.600, 4.000)$ & Inverse quartic constant & 0.0069 & 0.0142 & 0.0614 & 19 & 11 \\
42 & $(1.85; 2.45),\,(0.700, 3.500)$ & Inverse cubic linear & 0.0069 & 0.0135 & 0.0582 & 19 & 9 \\
43 & $(1.80; 2.40),\,(0.700, 3.750)$ & Shifted inverse quintic linear & 0.0071 & 0.0136 & 0.0570 & 19 & 11 \\
44 & $(1.85; 2.40),\,(0.600, 4.250)$ & Inverse quartic constant & 0.0073 & 0.0144 & 0.0618 & 19 & 11 \\
45 & $(1.80; 2.40),\,(0.700, 4.000)$ & Shifted inverse quintic linear & 0.0073 & 0.0136 & 0.0575 & 19 & 10 \\
46 & $(1.85; 2.40),\,(0.600, 4.250)$ & Shifted inverse cubic linear & 0.0073 & 0.0147 & 0.0634 & 18 & 10 \\
47 & $(1.75; 2.40),\,(0.700, 3.250)$ & Inverse quintic quartic & 0.0077 & 0.0146 & 0.0607 & 19 & 10 \\
48 & $(1.80; 2.40),\,(0.600, 3.250)$ & Inverse quintic quadratic & 0.0080 & 0.0146 & 0.0603 & 19 & 10 \\
49 & $(1.80; 2.40),\,(0.600, 3.500)$ & Inverse quintic quadratic & 0.0081 & 0.0148 & 0.0609 & 18 & 10 \\
50 & $(1.80; 2.40),\,(0.600, 4.000)$ & Shifted inverse quintic linear & 0.0084 & 0.0151 & 0.0627 & 17 & 9 \\
\midrule
\multicolumn{3}{l}{CCSD(T) baseline from Ref.~\cite{Lambie2025}} & 0.0133 & 0.0160 & 0.0357 & -- & -- \\
\multicolumn{3}{l}{CCSDT baseline from Ref.~\cite{Lambie2025}} & 0.0171 & 0.0233 & 0.0575 & -- & -- \\
\multicolumn{3}{l}{DC-CCSDT baseline from Ref.~\cite{Lambie2025}} & 0.0070 & 0.0127 & 0.0375 & -- & -- \\
\end{longtable}
\end{landscape}

The final inverse-cubic-linear model is not the absolute A24 minimum in
Table~\ref{tab:a24_all50_models}. This is intentional: A24 is used as a
late-stage property-level assessment and tie-break among models that have
already passed transferability, developmental-set, and size-consistency screens.
The final model therefore sacrifices a negligible amount of A24 MAE relative to
the best A24 entry in order to retain a simpler and more transferable functional
form. For the selected model, the MAE, RMS error, and maximum absolute error are
$0.0060$, $0.0134$, and $0.0588\,\kcalmol$, respectively, compared with
$0.0070$, $0.0127$, and $0.0375\,\kcalmol$ for DC-CCSDT. \MERCE{} has the
smaller absolute error for 11 of the 23 entries. Thus, the MAE is modestly
smaller, whereas the RMS error, maximum error, and entrywise win count do not
support an unqualified claim of superiority over DC-CCSDT

\subsection{System-by-system A24 results for the final MERCE model}
\label{sec:si_a24_system_by_system}

Table~\ref{tab:si_a24_detailed_final_model} gives the system-by-system A24 interaction-energy results for the final MERCE model.  This table complements Table~\ref{tab:a24_all50_models}, which reports aggregate statistics for the 50 shortlisted candidates.

\begin{table*}[htbp]
\centering
\caption{Detailed A24 benchmark results for the final \MERCE{} model. The selected protocol uses
the inverse-cubic-linear form with $x_{\mathrm{MP2}}=1.85$,
$x_{\mathrm{CCSD(T)}}=2.40$, $k_1=0.700$, and $k_3=4.250$.
The CCSDT(Q), CCSDT, CCSD(T), and DC-CCSDT values are taken from the
A24 benchmark reference study. Error columns are defined relative to
CCSDT(Q). All interaction energies and errors are given in $\kcalmol$.}
\label{tab:si_a24_detailed_final_model}
\scriptsize
\setlength{\tabcolsep}{4pt}
\resizebox{\textwidth}{!}{
\begin{tabular}{c l >{\bfseries}r r r r >{\bfseries}r r r r >{\bfseries}r}
\toprule
& & \multicolumn{5}{c}{$E^{\mathrm{corr,int}}$ [$\kcalmol$]} &
\multicolumn{4}{c}{$E_{\mathrm{method}}^{\mathrm{corr,int}}
-E_{\mathrm{CCSDT(Q)}}^{\mathrm{corr,int}}$ [$\kcalmol$]} \\
\cmidrule(lr){3-7}\cmidrule(lr){8-11}
ID & Label & CCSDT(Q) & CCSDT & CCSD(T) & DC-CCSDT & MERCE model
& $\delta$CCSDT & $\delta$CCSD(T) & $\delta$DC-CCSDT
& $\delta{\mathrm{MERCE}}$ \\
\midrule
1 & Water-ammonia & -1.2012 & -1.1865 & -1.1892 & -1.2006 & -1.2018 & 0.0147 & 0.0120 & 0.0006 & -0.0006 \\
2 & Water dimer & -0.7361 & -0.7270 & -0.7336 & -0.7358 & -0.7442 & 0.0091 & 0.0025 & 0.0003 & -0.0081 \\
3 & HCN dimer & -0.2469 & -0.2410 & -0.2491 & -0.2478 & -0.2497 & 0.0059 & -0.0022 & -0.0009 & -0.0028 \\
4 & HF dimer & -0.2145 & -0.2082 & -0.2213 & -0.2122 & -0.2279 & 0.0063 & -0.0068 & 0.0023 & -0.0134 \\
5 & Ammonia dimer & -1.2245 & -1.2139 & -1.2127 & -1.2245 & -1.2238 & 0.0106 & 0.0118 & 0.0000 & 0.0007 \\
6 & HF-methane & -0.9453 & -0.9398 & -0.9342 & -0.9470 & -0.9431 & 0.0055 & 0.0111 & -0.0017 & 0.0022 \\
7 & Ammonia-methane & -0.6442 & -0.6392 & -0.6340 & -0.6444 & -0.6434 & 0.0050 & 0.0102 & -0.0002 & 0.0008 \\
8 & Water-methane & -0.5365 & -0.5342 & -0.5305 & -0.5381 & -0.5364 & 0.0023 & 0.0060 & -0.0016 & 0.0001 \\
9 & Formaldehyde dimer & -1.1048 & -1.0833 & -1.1405 & -1.1146 & -1.1636 & 0.0215 & -0.0357 & -0.0098 & -0.0588 \\
10 & Water-ethene & -1.2133 & -1.1987 & -1.2011 & -1.2134 & -1.2113 & 0.0146 & 0.0122 & -0.0001 & 0.0020 \\
11 & Formaldehyde-ethene & -1.2353 & -1.2149 & -1.2255 & -1.2291 & -1.2346 & 0.0204 & 0.0098 & 0.0062 & 0.0007 \\
12 & Ethyne dimer (C2v) & -0.7458 & -0.7228 & -0.7363 & -0.7329 & -0.7448 & 0.0230 & 0.0095 & 0.0129 & 0.0010 \\
13 & Ammonia-ethene & -1.1053 & -1.0893 & -1.0934 & -1.1004 & -1.1028 & 0.0160 & 0.0119 & 0.0049 & 0.0025 \\
14 & Ethene dimer (C2v) & -1.7653 & -1.7348 & -1.7468 & -1.7521 & -1.7610 & 0.0305 & 0.0185 & 0.0132 & 0.0043 \\
15 & Methane-ethene & -0.7657 & -0.7554 & -0.7561 & -0.7615 & -0.7651 & 0.0103 & 0.0096 & 0.0042 & 0.0006 \\
16 & Borane-methane & -2.1842 & -2.1714 & -2.1512 & -2.1834 & -2.1932 & 0.0128 & 0.0330 & 0.0008 & -0.0090 \\
18 & Methane-ethane (Cs) & -1.3848 & -1.3755 & -1.3718 & -1.3836 & -1.3843 & 0.0093 & 0.0130 & 0.0012 & 0.0005 \\
19 & Methane dimer & -0.8933 & -0.8880 & -0.8845 & -0.8930 & -0.8920 & 0.0053 & 0.0088 & 0.0003 & 0.0013 \\
20 & Ar-methane & -0.5902 & -0.5874 & -0.5859 & -0.5903 & -0.5900 & 0.0028 & 0.0043 & -0.0001 & 0.0002 \\
21 & Ar-ethene & -0.6140 & -0.6073 & -0.6095 & -0.6112 & -0.6137 & 0.0067 & 0.0045 & 0.0028 & 0.0003 \\
22 & Ethene-ethyne & -2.3032 & -2.2498 & -2.2800 & -2.2712 & -2.2922 & 0.0534 & 0.0232 & 0.0320 & 0.0110 \\
23 & Ethene dimer (D2h) & -2.6930 & -2.6426 & -2.6685 & -2.6661 & -2.6868 & 0.0504 & 0.0245 & 0.0269 & 0.0062 \\
24 & Ethyne dimer (D2h) & -2.0675 & -2.0100 & -2.0433 & -2.0300 & -2.0560 & 0.0575 & 0.0242 & 0.0375 & 0.0115 \\
\midrule
\multicolumn{7}{r}{MAE} & 0.0171 & 0.0133 & 0.0070 & 0.0060 \\
\multicolumn{7}{r}{RMSE} & 0.0233 & 0.0160 & 0.0127 & 0.0134 \\
\multicolumn{7}{r}{MAXE} & 0.0575 & 0.0357 & 0.0375 & 0.0588 \\
\bottomrule
\end{tabular}
}
\end{table*}

\subsection{Formal nonadditivity in the A24 component calculations}
\label{sec:si_a24_property_nonadditivity}

Because the final mapping is nonlinear in the three input correlation
energies, applying it once to the componentwise sum of two fragment input
sequences need not give the same result as extrapolating the fragments
separately and summing the two predictions. For an interaction convention
$Y\in\{\mathrm{CP},\mathrm{no\text{-}CP}\}$, we quantify this formal offset as
\begin{equation}
\Delta_{\mathrm{SC}}^{Y}(A,B;x)
=
E_{\mathrm{MERCE}}^{Y}(A+B;x)
-
E_{\mathrm{MERCE}}^{Y}(A;x)
-
E_{\mathrm{MERCE}}^{Y}(B;x).
\label{eq:si_property_nonadditivity}
\end{equation}
Here, $E_{\mathrm{MERCE}}^{Y}(A+B;x)$ is obtained by applying the
\MERCE{} mapping at target coordinate $x$ to the componentwise sums of the
MP2, CCSD, and CCSD(T) correlation energies of fragments $A$ and $B$ under
convention $Y$. The remaining two terms are obtained by extrapolating the
fragment input sequences separately. A dissociation-offset-corrected
interaction prediction is obtained by subtracting this formal offset from the
raw \MERCE{} interaction contribution.

For the 23 CP-corrected A24 entries, the mean absolute, RMS, and maximum
absolute values of $\Delta_{\mathrm{SC}}$ are 2.667, 7.044, and
$31.999\,\microEh$, respectively. The corresponding no-CP values are 2.736,
7.156, and $32.475\,\microEh$. Removing the CP formal offset changes the A24
MAE from 0.006029 to $0.006442\,\kcalmol$, the RMS error from 0.013381 to
$0.013406\,\kcalmol$, and the maximum absolute error from 0.058767 to
$0.058767\,\kcalmol$ at the shown precision. The number of entries with a
smaller absolute error than DC-CCSDT changes from 11 to 10 of 23. The aggregate
conclusion is therefore stable, although the entrywise offsets are not always
negligible relative to the small post-CCSD(T) correction.

\begin{longtable}{r l r r r}
\caption{Entrywise CP formal nonadditivity and its effect on the A24 error.
$\Delta_{\mathrm{SC}}$ is given in $\microEh$; raw and offset-corrected errors
relative to CCSDT(Q) are in $\kcalmol$.}\\
\label{tab:si_a24_property_nonadditivity}\\
\toprule
ID & Complex & $\Delta_{\mathrm{SC}}^{\mathrm{CP}}$ & Raw error & Corrected error \\
\midrule
\endfirsthead
\multicolumn{5}{c}{\tablename\ \thetable\ --- continued}\\
\toprule
ID & Complex & $\Delta_{\mathrm{SC}}^{\mathrm{CP}}$ & Raw error & Corrected error \\
\midrule
\endhead
1 & Water--ammonia & 2.145 & -0.0006 & -0.0019 \\
2 & Water dimer & 0.000 & -0.0081 & -0.0081 \\
3 & HCN dimer & -0.014 & -0.0028 & -0.0027 \\
4 & HF dimer & -0.001 & -0.0134 & -0.0134 \\
5 & Ammonia dimer & 0.000 & 0.0007 & 0.0007 \\
6 & HF--methane & -0.060 & 0.0022 & 0.0022 \\
7 & Ammonia--methane & -4.243 & 0.0008 & 0.0035 \\
8 & Water--methane & -1.236 & 0.0001 & 0.0009 \\
9 & Formaldehyde dimer & -0.000 & -0.0588 & -0.0588 \\
10 & Water--ethene & 2.325 & 0.0020 & 0.0006 \\
11 & Formaldehyde--ethene & 5.580 & 0.0007 & -0.0028 \\
12 & Ethyne dimer ($C_{2v}$) & -0.000 & 0.0010 & 0.0010 \\
13 & Ammonia--ethene & 0.016 & 0.0025 & 0.0025 \\
14 & Ethene dimer ($C_{2v}$) & 0.001 & 0.0043 & 0.0043 \\
15 & Methane--ethene & -4.715 & 0.0006 & 0.0036 \\
16 & Borane--methane & -31.999 & -0.0090 & 0.0111 \\
18 & Methane--ethane ($C_s$) & -2.354 & 0.0005 & 0.0020 \\
19 & Methane dimer & -0.000 & 0.0013 & 0.0013 \\
20 & Ar--methane & -0.210 & 0.0002 & 0.0003 \\
21 & Ar--ethene & -1.204 & 0.0003 & 0.0011 \\
22 & Ethene--ethyne & 5.230 & 0.0110 & 0.0078 \\
23 & Ethene dimer ($D_{2h}$) & -0.000 & 0.0062 & 0.0062 \\
24 & Ethyne dimer ($D_{2h}$) & -0.000 & 0.0115 & 0.0115 \\
\bottomrule
\end{longtable}

\section{Vertical ionization potentials}
\label{sec:si_vertical_ip}

Vertical ionization potentials provide a different type of energy difference,
involving neutral and cationic states. Table~\ref{tab:si_vip_errors_fci_ev}
reports the FCI reference IPs and method errors in eV, using
$\Delta IP=IP(\mathrm{method})-IP(\mathrm{FCI})$. The MERCE model gives an MAE of
0.0041 eV and an RMS error of 0.0088 eV, substantially smaller than the
corresponding CCSD(T), CCSD, MP2, and HF errors and comparable to the high-rank
coupled cluster corrections for this small FCI-based set.

\begin{table}[htbp]\centering\begin{threeparttable}
\caption{Vertical ionization potentials (IP) in eV. Errors relative to FCI are defined as
$\Delta IP = IP(\mathrm{method}) - IP(\mathrm{FCI})$.}
\label{tab:si_vip_errors_fci_ev}

\setlength{\tabcolsep}{4pt}
\small

\sisetup{
  round-mode = places,
  round-precision = 4
}

\begin{tabular}{l l
  >{\bfseries}S[table-format=2.4]
  S[table-format=+1.4]
  S[table-format=+1.4]
  S[table-format=+1.4]
  S[table-format=+1.4]
  S[table-format=+1.4]
  S[table-format=+1.4]
  >{\bfseries}S[table-format=+1.4]
}
\toprule
\multirow{2}{*}{System} &
\multirow{2}{*}{Basis} &
\multicolumn{1}{c}{REF} &
\multicolumn{7}{c}{$\Delta IP$ [eV]} \\
\cmidrule(lr){3-3}\cmidrule(lr){4-10}
& & {FCI} & {CCSDTQ} & {CCSDT} & {CCSD(T)} & {CCSD} & {MP2} & {HF} & {MERCE model} \\
\midrule

$\mathrm{BH/BH^+}$                  & aVTZ &  9.7525 &  0.0000 & -0.0026 & -0.0066 & -0.0458 & -0.3490 & -1.3123 &  0.0002 \\
$\mathrm{C/C^+}$                    & aVTZ & 11.2089 &  0.0000 & -0.0013 & -0.0062 & -0.0326 &  0.0559 & -0.4238 & -0.0029 \\
$\mathrm{BeH_2/BeH_2^+}$              & aVTZ &  9.1640 &  0.0000 & -0.0006 & -0.0025 & -0.0149 & -0.1656 & -0.7896 &  0.0000 \\
$\mathrm{Li_2/Li_2^+}$                & aVTZ &  5.0389 &  0.0000 & -0.0002 & -0.0013 & -0.0091 & -0.2621 & -0.8385 &  0.0004 \\
$\mathrm{O/O^+}$                    & aVTZ & 13.4259 &  0.0000 & -0.0011 & -0.0038 & -0.0527 & -0.0770 & -1.4459 &  0.0018 \\
$\mathrm{O/O^+}$                    & aVQZ & 13.5202 &  0.0000 & -0.0015 & -0.0040 & -0.0572 & -0.0734 & -1.5522 &  0.0017 \\
$\mathrm{B_2/B_2^+}$                  & aVTZ &  8.3465 & -0.0027 & -0.0426 & -0.0432 & -0.2129 &  0.0034 & -0.9431 & -0.0273 \\
$\mathrm{CH_2/CH_2^+}$                & aVTZ & 10.3260 &  0.0000 & -0.0017 & -0.0040 & -0.0457 & -0.0296 & -0.7645 &  0.0007 \\
$\mathrm{PH/PH^+}$                  & aVTZ & 10.1404 & -0.0001 & -0.0023 & -0.0062 & -0.0506 &  0.0283 & -0.5216 & -0.0017 \\
$\mathrm{SiH_2/SiH_2^+}$              & aVTZ &  8.5868 &  0.0000 & -0.0046 & -0.0121 & -0.0642 & -0.2577 & -1.1487 & -0.0045 \\
\midrule

& & \multicolumn{1}{c}{\textbf{MAE}}
  & 0.0003 & 0.0058 & 0.0090 & 0.0586 & 0.1302 & 0.9740 & \textbf{0.0041} \\
& & \multicolumn{1}{c}{\textbf{RMS}}
  & 0.0009 & 0.0136 & 0.0148 & 0.0797 & 0.1733 & 1.0389 & \textbf{0.0088} \\

\bottomrule
\end{tabular}
\end{threeparttable}\end{table}

Because the target coordinate is empirical, we shifted the neutral and cation
targets together by a common $\Delta x$ without refitting any global parameter.
Table~\ref{tab:si_ip_target_sensitivity} shows that the aggregate result is
insensitive to shifts of up to one half-coordinate unit.

\begin{table}[htbp]
\centering
\caption{Sensitivity of the ten-record vertical-ionization-potential statistics
to a common shift of the neutral and cation target coordinates.}
\label{tab:si_ip_target_sensitivity}
\small
\begin{tabular}{rrrr}
\toprule
$\Delta x$ & MAE [eV] & RMS [eV] & Maximum absolute error [eV] \\
\midrule
$-0.50$ & 0.004067 & 0.008731 & 0.026936 \\
$-0.25$ & 0.004097 & 0.008802 & 0.027153 \\
$0.00$ & 0.004121 & 0.008849 & 0.027298 \\
$+0.25$ & 0.004138 & 0.008881 & 0.027394 \\
$+0.50$ & 0.004151 & 0.008902 & 0.027457 \\
\bottomrule
\end{tabular}
\end{table}

The largest \MERCE{} ionization-potential error occurs for
B$_2$/B$_2^+$, where the signed error is $-0.0273$ eV. This single system
dominates the \MERCE{} RMS error and contributes substantially to the MAE. This
behavior is chemically meaningful rather than merely numerical: B$_2$ is a more
demanding small molecule with stronger nondynamical-correlation character than
most other entries in this set. The outlier therefore marks a boundary case for
the present single-reference MP2--CCSD--CCSD(T) trajectory.

\begin{table}[t]
\centering
\caption{Exploratory vertical ionization-potential error statistics after
excluding the B$_2$/B$_2^+$ boundary case. The primary statistics reported in
the main text retain all systems. Errors are reported in eV relative to
finite-basis FCI; this secondary analysis is provided only to show the
influence of one large residual on the aggregate RMS error.}
\label{tab:ip_no_b2}
\begin{tabular}{lrrrrrrr}
\toprule
Statistic & CCSDTQ & CCSDT & CCSD(T) & CCSD & MP2 & HF & \MERCE{} \\
\midrule
MAE & 0.0000 & 0.0018 & 0.0052 & 0.0414 & 0.1443 & 0.9775 & 0.0015 \\
RMS & 0.0000 & 0.0021 & 0.0060 & 0.0451 & 0.1827 & 1.0491 & 0.0021 \\
\bottomrule
\end{tabular}
\end{table}

After excluding B$_2$/B$_2^+$, the \MERCE{} MAE and RMS decrease to
0.0015 eV and 0.0021 eV, respectively. The method then approaches the CCSDT
error level for this subset and remains clearly more accurate than CCSD(T),
CCSD, MP2, and HF. The corresponding ionization-potential error statistics after excluding this
outlier are given in Table~\ref{tab:ip_no_b2}.

\section{Hierarchical selection audit}

\subsection{Stepwise selection of the final representative model}
\label{sec:si_stepwise_selection_final_model}

Table~\ref{tab:si_selection_summary} summarizes the hierarchical screening procedure used to identify the final representative MERCE model from the 50 candidates retained after the full $(k_1,k_3)$ scan. For compactness, the full list of the initial 50 candidates is not repeated here, because it is already given in Table~\ref{tab:si_top50_k1k3_scan_monomer}. Table~\ref{tab:si_selection_audit} provides the exact survivor ranks after each nontrivial step of the decision tree, where the rank numbers refer to Table~\ref{tab:si_top50_k1k3_scan_monomer}. The numerical characteristics of the finally selected model are collected in Table~\ref{tab:si_final_selected_model_metrics}.

\begin{table}[htbp]
\centering
\caption{Summary of the hierarchical selection protocol used to identify the
final representative model. The quantities are defined as in the main text:
$\mathrm{gmean}_{\mathrm{test}}$ is the geometric mean of normalized
cross-dataset MAE ratios, $M_{\max,\mathrm{test}}$ is their maximum,
$M_{\max,\mathrm{calib}}$ is the maximum normalized MAE over the two
FCI-based developmental sets, and $Q_{\mathrm{all}}$ is the relative
size-consistency measure. Historical script labels are given in parentheses
where useful for matching the released audit files.}
\label{tab:si_selection_summary}
\scriptsize
\begin{tabular}{c p{6.8cm} p{3.7cm} c}
\toprule
Step & Criterion & Numerical condition / cutoff & Survivors \\
\midrule
1 & Full coverage screen & Require $\mathrm{min\_cov}=1$ & 50 / 50 \\
2 & Best-half transferability screen & Keep the half of models with the smallest $\mathrm{gmean}_{\mathrm{test}}$; cutoff $=0.1219$ & 25 / 50 \\
3a & Best-half maximum cross-dataset screen & Keep the half with the smallest $M_{\max,\mathrm{test}}$ (script: \texttt{worst\_test}); cutoff $=0.2630$ & 13 / 25 \\
3b & Best-half maximum developmental-set screen & Keep the half with the smallest $M_{\max,\mathrm{calib}}$ (script: \texttt{calib\_worst}); cutoff $=0.3719$ & 7 / 13 \\
4 & Best-half size-consistency screen & Keep the half of models with the smallest $Q_{\mathrm{all}}$; cutoff $=0.0870$ & 4 / 7 \\
5 & A24 property-level screen & Require $\mathrm{A24\_MAE}<0.0070\ \mathrm{kcal\ mol^{-1}}$ & 4 / 4 \\
6 & Structural and parameter filter & Reject quintic-family models; retain shifted-hyperbolic candidates; reject edge cases with $k_1\le 0.4$ or $k_3\ge 4.5$ when applicable & 3 / 4 \\
7 & A24 tie-break & Retain models with minimal $\mathrm{A24\_MAE}=0.0060\ \mathrm{kcal\ mol^{-1}}$ & 1 / 3 \\
\bottomrule
\end{tabular}
\end{table}

\begin{table}[htbp]
\centering
\caption{Audit trail of the hierarchical selection. The listed numbers are the \texttt{general\_rank} values from Table~\ref{tab:si_top50_k1k3_scan_monomer}. Step~1 is not repeated because all 50 candidates survive and therefore coincide with the complete list given in Table~\ref{tab:si_top50_k1k3_scan_monomer}.}
\label{tab:si_selection_audit}
\scriptsize
\begin{tabular}{c p{13.2cm}}
\toprule
Step & Surviving \texttt{general\_rank} values \\
\midrule
2 & 3, 9, 1, 7, 16, 2, 12, 6, 11, 18, 15, 10, 22, 5, 8, 13, 19, 27, 29, 30, 4, 25, 14, 20, 17 \\
3a & 13, 14, 3, 2, 18, 4, 9, 1, 11, 10, 17, 7, 8 \\
3b & 4, 1, 2, 17, 8, 14, 13 \\
4 & 1, 13, 4, 2 \\
5 & 1, 13, 4, 2 \\
6 & 1, 4, 2 \\
7 & 2 \\
\bottomrule
\end{tabular}
\end{table}

\begin{table}[htbp]
\centering
\caption{Numerical characteristics of the final representative model selected by the hierarchical protocol.}
\label{tab:si_final_selected_model_metrics}
\scriptsize
\begin{tabular}{l l}
\toprule
Quantity & Value \\
\midrule
\texttt{general\_rank} & 2 \\
Variant & \texttt{MP2\_1.85\_CCSDpT\_2.40\_k1\_0.700\_k3\_4.250} \\
Model & \texttt{inverse\_cubic\_linear} \\
$(\alpha,\gamma)$ & $(1.85,\ 2.40)$ \\
$(k_1,k_3)$ & $(0.700,\ 4.250)$ \\
$\mathrm{gmean}_{\mathrm{test}}$ & 0.1180 \\
$M_{\max,\mathrm{test}}$ & 0.2592 \\
FCI-frozen-core developmental-set ratio & 0.3533 \\
FCI-full-core developmental-set ratio & 0.3339 \\
$M_{\max,\mathrm{calib}}$ & 0.3533 \\
A24 MAE [$\kcalmol$] & 0.0060 \\
A24 RMS error [$\kcalmol$] & 0.0134 \\
$Q_{\mathrm{all}}$ & 0.087 \\
$\mathrm{SC}_{\mathrm{all}}$ [$\microEh$] & 38.040 \\
\bottomrule
\end{tabular}
\end{table}

\paragraph{Comment on the final representative model.}
The final model retained by the hierarchical protocol is intentionally not
chosen by a single metric alone. It is the inverse-cubic-linear model with
$(\alpha,\gamma)=(1.85,2.40)$ and
$(k_1,k_3)=(0.700,4.250)$. Its role in the main text is that of a representative
compromise among transferability, worst-case robustness, developmental-set
behavior, size consistency, and A24 property-level performance.

\subsection{Sensitivity to benchmark weighting and selection metrics}
\label{sec:si_selection_sensitivity}

The primary cross-dataset score used in the main text gives equal weight to the seven individual
datasets, so the four original STRETCH scans jointly contribute four logarithmic
terms. To test whether this convention determined the selected model, we
reranked all 960 candidates with complete dataset-level summary coverage under two
alternatives: (i) the four STRETCH scans were collapsed into one family by
geometrically averaging their four normalized MAE ratios before combining that
family with Rings, W4, and CEPB; and (ii) each dataset was weighted by its
number of accepted entries. Table~\ref{tab:si_weighting_sensitivity} reports the
rank of the selected model under all three definitions. These are sensitivity
analyses of the existing candidate pool, not independent validation tests.

\begin{table}[htbp]
\centering
\caption{Rank of the selected inverse-cubic-linear model among the 960 candidates with complete dataset-level summary coverage under three cross-dataset weighting conventions.}
\label{tab:si_weighting_sensitivity}
\small
\begin{tabular}{lrr}
\toprule
Definition & Score & Rank / 960 \\
\midrule
Seven datasets, equal weight (primary) & 0.118029 & 6 \\
Four families, STRETCH collapsed & 0.129400 & 4 \\
Entry-count weighted & 0.158899 & 7 \\
\bottomrule
\end{tabular}
\end{table}

The archive contains 960 candidate identities with dataset-level summaries for
all nine benchmark groups. Three candidates lose one individual benchmark
point, leaving 957 candidates with complete pointwise coverage. We performed a
leave-one-family-out reranking of these 957 candidates. This operation keeps
the candidate library and all
earlier design decisions fixed and is therefore a selection-stability audit,
not nested validation. The results are summarized in
Table~\ref{tab:si_leave_one_family_out}.

\begin{table}[htbp]
\centering
\caption{Leave-one-family-out reranking of the 957 candidates with complete
pointwise coverage. For each row, candidates were ranked using only the
retained benchmark families. The final-model rank is its position in this
reduced ranking. The retained-family winner is the candidate ranked first
without using the omitted family; its performance on that family was evaluated
only after ranking. It is therefore not necessarily the best-performing
candidate on the omitted family. Reported values are MAE ratios relative to
CCSD(T), with values below unity indicating improvement over CCSD(T).}
\label{tab:si_leave_one_family_out}
\small
\begin{tabular}{lrrr}
\toprule
Omitted family &
\makecell{Final-model rank\\(of 957)} &
\makecell{Final-model\\omitted-family ratio} &
\makecell{Retained-family winner:\\omitted-family ratio}\\
\midrule
Rings & 11 & 0.0539 & 0.0548 \\
W4 & 7 & 0.2593 & 0.2586 \\
CEPB & 10 & 0.1921 & 0.1978 \\
STRETCH (four scans) & 101 & 0.1044 & 0.1922 \\
\bottomrule
\end{tabular}
\end{table}

All 25 leading retained-family candidates improve on CCSD(T) for each omitted
family. Removing STRETCH changes the rank most strongly, but the final model
has a substantially smaller omitted-family STRETCH ratio than the candidate selected
without those scans. This behavior is consistent with the final model being a
balanced representative rather than a reduced-objective optimum.

To avoid treating multiple basis-set entries for one species as independent,
we also performed a system-cluster bootstrap with 20,000 fixed-seed
replicates. All basis entries belonging to a resampled species were retained
together. Table~\ref{tab:si_cluster_bootstrap} gives percentile intervals for
the \MERCE{}/CCSD(T) MAE ratio.

\begin{table}[htbp]
\centering
\caption{System-cluster bootstrap of the \MERCE{}/CCSD(T) MAE ratio. The
intervals describe variation within the present benchmark collection after
model selection and are not nested-validation confidence intervals.}
\label{tab:si_cluster_bootstrap}
\small
\begin{tabular}{lrrrr}
\toprule
Dataset & Entries & System clusters & MAE ratio & 95\% interval \\
\midrule
FCI frozen core & 66 & 11 & 0.3533 & [0.2968, 0.4027] \\
FCI full core & 41 & 6 & 0.3339 & [0.2140, 0.4468] \\
Rings & 12 & 12 & 0.0539 & [0.0335, 0.0765] \\
W4 & 56 & 56 & 0.2593 & [0.2112, 0.3064] \\
CEPB & 18 & 15 & 0.1921 & [0.1396, 0.2509] \\
\bottomrule
\end{tabular}
\end{table}

We do not assign an analogous inferential interval to an individual coordinate
scan because its geometry points are correlated observations along one path,
not independent chemical replicates.

The relative size-consistency screen is similarly insensitive to replacing
$Q_{\mathrm{all}}$ by an absolute defect measure. Among the seven candidates
entering Step~4, the best-half survivor set is \texttt{general\_rank}
$\{1,13,4,2\}$ under either $Q_{\mathrm{all}}$ or
$\mathrm{SC}_{\mathrm{all}}$. For the final model, the artificial-pair mean
absolute, RMS, and maximum absolute size-consistency defects are 38.040,
79.298, and 474.888~$\microEh$, respectively.

\subsection{Pointwise coverage and error tails}
\label{sec:si_pointwise_tails}

Dataset-level MAEs can conceal incomplete coverage or isolated large errors.
Table~\ref{tab:si_pointwise_tails} therefore reports accepted-point coverage
and pointwise error statistics for the final model. ``Improved'' is the
fraction of accepted entries whose absolute \MERCE{} error is smaller than
the corresponding CCSD(T) error. W4 contains 65 attempted entries: two lack
the required reference value and seven fail the monotonic-input gate, leaving
56 accepted entries. The other listed datasets have complete coverage of their
deposited analysis domains.

\begin{landscape}
\begin{table}[htbp]
\centering
\caption{Coverage and pointwise signed-error statistics for the final model.
All errors are in $\microEh$. The P95 column is the 95th percentile of the
absolute error.}
\label{tab:si_pointwise_tails}
\scriptsize
\begin{tabular}{lrrrrrrrrrrr}
\toprule
Dataset & Attempted & No ref. & Gate reject & Accepted & Mean signed & MAE & Median AE & RMS & P95 AE & Max. AE & Improved \\
\midrule
Rings & 12 & 0 & 0 & 12 & 14.08 & 148.96 & 128.21 & 181.78 & 313.15 & 367.59 & 1.000 \\
W4 & 65 & 2 & 7 & 56 & 248.61 & 328.87 & 205.32 & 481.08 & 1054.46 & 1835.17 & 1.000 \\
CEPB & 18 & 0 & 0 & 18 & 138.39 & 237.81 & 173.10 & 294.05 & 550.28 & 629.00 & 1.000 \\
HF stretch & 15 & 0 & 0 & 15 & -44.34 & 124.62 & 91.31 & 220.22 & 344.64 & 781.44 & 0.933 \\
H$_2$O stretch & 13 & 0 & 0 & 13 & -70.58 & 121.28 & 79.18 & 216.34 & 383.66 & 725.16 & 1.000 \\
CH$_4$ stretch & 12 & 0 & 0 & 12 & -2.98 & 18.16 & 14.57 & 22.70 & 45.16 & 50.43 & 1.000 \\
C$_2$H$_4$ stretch & 10 & 0 & 0 & 10 & 77.85 & 169.13 & 110.90 & 226.70 & 448.68 & 507.18 & 1.000 \\
\bottomrule
\end{tabular}
\end{table}
\end{landscape}

\subsection{Target-coordinate ladder and algebraic conditioning}
\label{sec:si_rank_ladder_conditioning}

The target coordinate is an empirical continuation coordinate rather than an
exact excitation-rank label. Without refitting, we evaluated the final model at
$x=3$, 4, and 5 against CCSDT, CCSDTQ, and FCI, respectively, for the two
five-correlated-electron developmental sets. The results in
Table~\ref{tab:si_rank_ladder} show progressive improvement along this ladder,
while not implying a rigorous identity between $x$ and the coupled cluster
truncation rank.

\begin{table}[htbp]
\centering
\caption{No-refit target-coordinate ladder for the FCI developmental sets.
Errors are in $\microEh$.}
\label{tab:si_rank_ladder}
\small
\begin{tabular}{llrrr}
\toprule
Dataset & Comparison & $x$ & MAE & Maximum absolute error \\
\midrule
FCI frozen core & CCSDT  & 3 & 293.398 & 553.252 \\
FCI frozen core & CCSDTQ & 4 & 222.039 & 465.811 \\
FCI frozen core & FCI    & 5 & 191.886 & 414.119 \\
FCI full core   & CCSDT  & 3 & 142.954 & 275.488 \\
FCI full core   & CCSDTQ & 4 & 95.045 & 223.886 \\
FCI full core   & FCI    & 5 & 84.001 & 209.321 \\
\bottomrule
\end{tabular}
\end{table}

Finally, the released algebraic audit covers all 243 accepted records used in
the two FCI developmental sets and seven cross-dataset benchmarks. The
unreduced inverse-cubic--linear equation has two real algebraic roots for every
record. After the finite-denominator and no-pole continuation rules are
applied, 113 records retain one admissible branch and 130 retain two. In every
case, the released multistart solver agrees with an independent polynomial-root
enumeration of the admissible minimum-target-correction branch; the maximum
prediction difference is $2.30\times10^{-5}\,\microEh$. When two admissible
branches remain, the median separation between their target predictions is
$1.14\times10^{6}\,\microEh$, confirming that the deterministic
minimum-target-correction rule in Algorithm~S1 is essential rather than a
cosmetic tie-break.

For the selected branches, no fitted trajectory has a stationary point between
the last input coordinate and the target, and no positive denominator pole
occurs beyond the target. The maximum interpolation residual is
$8.68\times10^{-11}\Eh$. Perturbing each input energy by $10^{-8}\Eh$ changes
the prediction by at most $1.236\times10^{-8}\Eh$ (maximum local amplification
1.236). These checks establish algebraic well-conditioning on the accepted
records; they do not establish chemical single-reference validity outside the
gated domain.

\section{Expanded Post-Selection Functional-Form and Parameter-Space Challenge}
\label{sec:si_expanded_challenge}

The representative \MERCE{} protocol was first established using the
effective-rank construction and hierarchical model-selection procedure
described in the main text and in the preceding sections of this Supporting
Information. We subsequently performed an expanded post-selection challenge to
determine whether the reported result depended narrowly on the original
discrete functional-form library or parameter grid. This analysis was designed
as a robustness and model-identifiability test. It did not use the additional
post-selection coordinate scans and did not alter the frozen \MERCE{} protocol.

The expanded challenge is deliberately distinguished from a claim of exhaustive
optimization. No finite library can cover every possible three-point analytic
continuation. The objective was instead to test a substantially broader set of
smooth formula families and effective-rank mappings, identify tradeoffs among
the established development criteria, and determine whether a clearly superior
replacement for the selected model emerged.

\subsection{Expanded functional-form library}
\label{sec:si_expanded_shapes}

The search contained 180 three-point extrapolation shapes. The principal family,
comprising 156 shapes, was
\begin{equation}
 \Ecorr(x)=A+\frac{B}{(x+\delta)^p+C(x+\delta)^q},\qquad q<p,
 \label{eq:si_expanded_denominator}
\end{equation}
with
\begin{align}
p&\in\{2,2.5,2.75,3,3.25,3.5,4,4.5\},\\
q&\in\{0,0.5,1,1.5,2\},\\
\delta&\in\{-0.25,0,0.5,1\},
\end{align}
subject to the domain conditions $q<p$ and $x+\delta>0$ throughout the fitted
and evaluated interval. The selected inverse-cubic-linear \MERCE{} expression
is the member $(p,q,\delta)=(3,1,0)$ of this generalized family.

The remaining 24 shapes were structurally different pole-free linear
three-point families: 11 two-power forms,
\begin{equation}
 \Ecorr(x)=A+B(x+\delta)^{-p}+C(x+\delta)^{-q},
\end{equation}
nine exponential/power hybrids,
\begin{equation}
 \Ecorr(x)=A+B\exp(-\rho x)+Cx^{-p},
\end{equation}
and four biexponential forms,
\begin{equation}
 \Ecorr(x)=A+B\exp(-\rho_1x)+C\exp(-\rho_2x).
\end{equation}
For every system and candidate, the coefficients $A$, $B$, and $C$ were fitted
from the MP2, CCSD, and CCSD(T) correlation energies. Only the analytic shape
and the global effective-rank mapping parameters were shared among systems.

\subsection{Effective-rank mappings and Sobol parameter sampling}
\label{sec:si_expanded_mappings}

For each system, the increment ratio was
\begin{equation}
 R=\frac{|\Ecorr_{\mathrm{CCSD(T)}}-\Ecorr_{\mathrm{CCSD}}|}
 {|\Ecorr_{\mathrm{CCSD}}-\Ecorr_{\mathrm{MP2}}|+10^{-15}\Eh},
\end{equation}
and the three coordinates were
\begin{align}
x_{\mathrm{MP2}}^{\mathrm{eff}}&=
\alpha+(2-\alpha)s(k_1R),\\
x_{\mathrm{CCSD}}&=2,\\
x_{\mathrm{CCSD(T)}}^{\mathrm{eff}}&=
\gamma+(3-\gamma)s(k_3R).
\end{align}

The $10^{-15}\Eh$ regularizer records the historical value used by the
expanded-search code. The released calculator instead uses
$\varepsilon_{\mathrm{num}}=10^{-16}\Eh$ together with the explicit
$|d_1|<10^{-14}\Eh$ rejection in Section~\ref{sec:si_algorithm}. We recomputed
all 243 accepted developmental/model-selection entries and 50 complete,
algebraically valid later-scan entries with both regularizers while leaving the
fit and branch policy unchanged. The smallest observed $|d_1|$ was
$2.58\times10^{-5}\Eh$, and the maximum prediction change was only
$1.85\times10^{-7}\,\microEh$ ($1.85\times10^{-13}\Eh$). Thus, the historical
regularizer difference is numerically immaterial for the released predictions;
it does not change the searched functional families or parameter bounds.

Six bounded saturation mappings were tested:
\begin{align}
s_{\mathrm{atan}}(z)&=\frac{2}{\pi}\arctan z,&
s_{\mathrm{rational}}(z)&=\frac{z}{1+z},\\
s_{\mathrm{exponential}}(z)&=1-\exp(-z),&
s_{\mathrm{Hill},n}(z)&=\frac{z^n}{1+z^n},
\end{align}
where $n\in\{0.75,1.25,1.50\}$ for the three Hill variants.

For every shape--mapping pair, 1024 parameter combinations were generated with
a four-dimensional scrambled Sobol low-discrepancy sequence. The fixed seed was
20260813, and the bounds were
\begin{equation}
\begin{split}
1.60&\leq\alpha\leq1.95,\qquad
2.20\leq\gamma\leq2.70,\\
0.30&\leq k_1\leq1.20,\qquad
1.50\leq k_3\leq6.00.
\end{split}
\label{eq:si_expanded_bounds}
\end{equation}
Sobol sampling provides more uniform coverage of a bounded multidimensional
domain than independent pseudorandom sampling at the same cost. The
power-of-two sample size, $1024=2^{10}$, preserves the balance properties of
the sequence. The same normalized Sobol points were used for every
shape--mapping pair and were scaled linearly to Eq.~\ref{eq:si_expanded_bounds}.
The parameter combination of the selected MERCE protocol was added explicitly as an internal
control. The total attempted candidate count was therefore
\begin{equation}
180\times6\times1024+1=1{,}105{,}921.
\end{equation}

\subsection{Benchmark datasets, metrics, and numerical safeguards}
\label{sec:si_expanded_metrics}

The search used the same two FCI-based developmental sets and seven
cross-dataset model-selection benchmarks used in the established workflow. It did not load A24 or any additional post-selection stretching data. Records
were included only when
$\Ecorr_{\mathrm{MP2}}>\Ecorr_{\mathrm{CCSD}}>\Ecorr_{\mathrm{CCSD(T)}}$.
The resulting sizes and CCSD(T) baseline MAEs are given in
Table~\ref{tab:si_expanded_datasets}.

\begin{table}[htbp]
\centering
\caption{Benchmark data used in the expanded post-selection challenge. The
target is the \MERCE{} evaluation rank. Baseline MAEs are absolute
correlation energy errors of CCSD(T).}
\label{tab:si_expanded_datasets}
\small
\begin{tabular}{lrrr}
\toprule
Dataset & $N$ & Target $x$ & CCSD(T) MAE [$\microEh$]\\
\midrule
FCI-frozen-core & 66 & 5 & 543.083\\
FCI-full-core   & 41 & 5 & 251.600\\
Rings           & 12 & 4 & 2762.884\\
W4              & 56 & 4 & 1268.553\\
CEPB            & 18 & 4 & 1237.746\\
HF stretch      & 15 & 4 & 487.378\\
H$_2$O stretch  & 13 & 4 & 703.737\\
CH$_4$ stretch  & 12 & 4 & 721.051\\
C$_2$H$_4$ stretch & 10 & 4 & 1579.394\\
\bottomrule
\end{tabular}
\end{table}

For each cross-dataset benchmark $d$, the normalized error ratio was
\begin{equation}
M_d=
\frac{\operatorname{MAE}_d(\mathrm{candidate})}
{\operatorname{MAE}_d(\mathrm{CCSD(T)})}.
\end{equation}
For each FCI-based developmental set $c$, the corresponding ratio was
\begin{equation}
M_c^{\mathrm{calib}}=
\frac{\operatorname{MAE}_c(\mathrm{candidate})}
{\operatorname{MAE}_c(\mathrm{CCSD(T)})}.
\end{equation}
For the seven cross-dataset benchmarks
$\mathcal D_{\mathrm{test}}$, the three selection coordinates were
\begin{align}
\mathrm{gmean}_{\mathrm{test}}
&=
\exp\left[
\frac{1}{7}
\sum_{d\in\mathcal D_{\mathrm{test}}}\ln M_d
\right],\\
M_{\max,\mathrm{test}}
&=
\max_{d\in\mathcal D_{\mathrm{test}}}M_d,\\
M_{\max,\mathrm{calib}}
&=
\max_{c\in\mathcal D_{\mathrm{calib}}}M_c^{\mathrm{calib}}.
\end{align}
The existing FCI safeguard was applied without redefinition: candidates were
eligible only when $M_{\max,\mathrm{calib}}<0.3719$.

A candidate was retained only if it produced a valid prediction for every
record in all nine datasets. Linear three-point families were solved directly;
design matrices with condition number greater than $10^{12}$ were rejected.
For Eq.~\ref{eq:si_expanded_denominator}, all real algebraic branches for $C$
were examined. A branch had to reproduce the three input energies within
$10^{-9}\Eh$, remain finite, and contain no real denominator pole between the
smallest fitted effective rank and the target. When multiple branches were tied
within the fitting tolerance, the branch giving the smallest absolute target
correction relative to CCSD(T) was selected, consistent with the released
calculator.

Of the 1,105,921 attempted candidates, 1,063,197 (96.14\%) were algebraically
valid on every datum; 42,724 were rejected by the numerical safeguards. The FCI
criterion retained 389,535 eligible candidates.

\subsection{Pareto analysis, holdout rankings, and continuous refinement}
\label{sec:si_expanded_postprocessing}

An exact nondominated set was constructed by simultaneously minimizing
$\mathrm{gmean}_{\mathrm{test}}$, $M_{\max,\mathrm{test}}$, and $M_{\max,\mathrm{calib}}$. Candidate
$a$ dominated candidate $b$ only if it was no worse in all three coordinates
and strictly better in at least one. The Pareto set contained 168 candidates.

Ranking sensitivity was examined by seven leave-one-dataset-out analyses. For
each held-out dataset, candidates were ranked by the geometric mean over the
remaining six cross-dataset benchmarks, and the leading 25 were retained. This
is a stability analysis of the established benchmark suite; it does not make
the held-out datasets independent external tests.

To determine whether the best Sobol candidates merely reflected finite sampling
resolution, eligible candidates were sorted by $\mathrm{gmean}_{\mathrm{test}}$. The leading
candidate from each distinct combination of functional shape, mapping family,
and mapping exponent was retained until 25 starting regions had been collected.
For each fixed shape--mapping region, the four global parameters were refined
with bounded, derivative-free Powell optimization. The bounds remained those
in Eq.~\ref{eq:si_expanded_bounds}; the minimized objective was
\begin{equation}
F=\mathrm{gmean}_{\mathrm{test}}
+100\max(0,M_{\max,\mathrm{calib}}-0.3719)^2
+10\max(0,M_{\max,\mathrm{test}}-0.30)^2.
\label{eq:si_refinement_objective}
\end{equation}
The calculation used at most 120 Powell iterations, a parameter tolerance of
$10^{-4}$, and a function tolerance of $10^{-6}$. This was a targeted local
refinement of 25 leading regions, not proof of a global continuous minimum.

\subsection{Artificial-pair size-consistency audit}
\label{sec:si_expanded_size}

All 168 Pareto candidates were tested on deterministic adjacent noninteracting
pairs within every benchmark dataset, including wraparound. If $M$ denotes a
MERCE prediction, each defect was
\begin{equation}
\Delta_{AB}=|M(A+B)-M(A)-M(B)|.
\end{equation}
The audit comprised 243 artificial pairs. These are algebraic size-consistency
diagnostics obtained from existing component energies; no additional
electronic-structure calculations were performed.

\subsection{Expanded-search results and tradeoffs}
\label{sec:si_expanded_results}

Table~\ref{tab:si_expanded_key_models} compares the selected \MERCE{} protocol
with the best coarse and continuously refined expanded candidates according to
$\mathrm{gmean}_{\mathrm{test}}$. The alternatives modestly improve the aggregate cross-dataset score but move closer to the FCI safeguard boundary.

\begin{table}[htbp]
\centering
\caption{Key results from the expanded challenge. The coarse and refined rows
are selected by the cross-dataset geometric-mean score and are not proposed
as replacement protocols.}
\label{tab:si_expanded_key_models}
\small
\begin{tabular}{lrrr}
\toprule
Candidate & $\mathrm{gmean}_{\mathrm{test}}$ & $M_{\max,\mathrm{test}}$ & $M_{\max,\mathrm{calib}}$\\
\midrule
Selected \MERCE{} & 0.118029 & 0.259250 & 0.353327\\
Best coarse expanded & 0.113956 & 0.251951 & 0.367074\\
Best refined expanded & 0.113270 & 0.250993 & 0.370580\\
\bottomrule
\end{tabular}
\end{table}

The best coarse candidate uses the generalized denominator
$(p,q,\delta)=(2,1,-0.25)$, the rational mapping, and
$(\alpha,\gamma,k_1,k_3)=(1.83018,2.43840,1.08266,5.21489)$. Its continuously
refined counterpart gives
$(\alpha,\gamma,k_1,k_3)=(1.83074,2.43881,1.05686,5.31395)$. The decrease from
$\mathrm{gmean}_{\mathrm{test}}=0.113956$ to 0.113270 is only 0.60\%, showing that the Sobol
point had already located the high-performing region closely.

The high-performance region is not localized to a single formula. Among
eligible candidates, 11 lie within 1\% of the lowest coarse
$\mathrm{gmean}_{\mathrm{test}}$, 185 within 5\%, 653 within 10\%, and 2410 within 20\%.
The leading 100 candidates contain many generalized denominator shapes and are
divided between rational (53) and arctangent (47) mappings. These observations
support a plateau interpretation rather than a uniquely identifiable optimum.

\begin{figure}[htbp]
\centering
\includegraphics[width=0.78\linewidth]{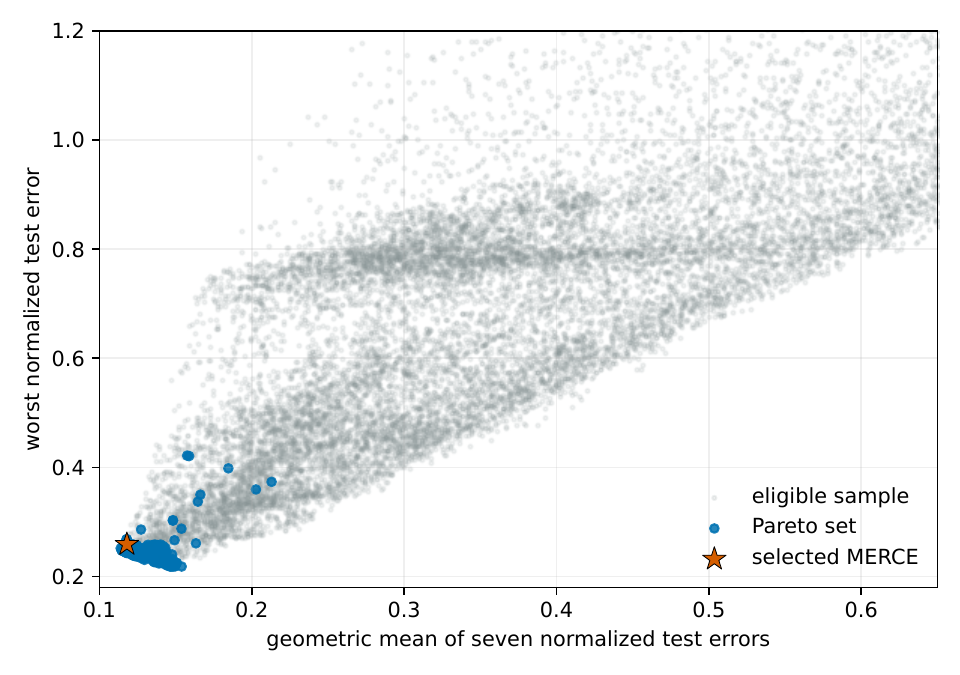}
\caption{Cross-dataset error tradeoff among candidates passing the FCI
developmental-set safeguard in the expanded search. The gray cloud is a
deterministic subsample of the 389,535 eligible candidates; all 168 Pareto
candidates and the selected \MERCE{} model are shown.}
\label{fig:si_expanded_tradeoff}
\end{figure}

The artificial-pair audit also reveals a tradeoff. The selected model has a
mean defect of 16.770 $\microEh$ and maximum defect of 282.374 $\microEh$ under
the expanded audit definition. The best coarse expanded candidate has a lower
mean defect of 6.445 $\microEh$ and maximum defect of 102.105 $\microEh$.
Accordingly, the expanded candidate improves this particular diagnostic;
size-consistency performance is not a reason to claim that the established
model is the numerical optimum.

\subsection{Separate A24 assessment of expanded representatives}
\label{sec:si_expanded_a24}

A24 was not used in the million-candidate development search. Afterward, five
representative parameterizations were evaluated from saved counterpoise-corrected
component calculations for the 23 systems with usable CCSDT(Q) references. In
the notation below, $D(p,q,\delta)$ denotes the generalized-denominator form
in Eq.~\ref{eq:si_expanded_denominator} with the specified values of
$(p,q,\delta)$. ``Sobol'' indicates an evaluated point from the discrete
low-discrepancy sample, whereas ``refined'' indicates a parameterization
obtained by the continuous optimization described in
Section~\ref{sec:si_expanded_postprocessing}. Table~\ref{tab:si_expanded_a24_models}
defines every parameterization included in this follow-up assessment. The computed
and literature CCSD(T) interaction energies agree to an MAE of 0.000026 and
maximum difference of 0.000053 $\kcalmol$. The computed CCSD(T) baseline MAE
relative to CCSDT(Q) is 0.013272 $\kcalmol$.

\begin{table}[htbp]
\centering
\caption{Complete definitions of the five parameterizations assessed on A24.
The mapping exponent is $n=1$ for both the rational and arctangent mappings.}
\label{tab:si_expanded_a24_models}
\small
\begin{tabular}{llllrrrr}
\toprule
ID & Denominator & Mapping & Stage & $\alpha$ & $\gamma$ & $k_1$ & $k_3$\\
\midrule
M0 & $D(3,1,0)$ & arctangent & selected MERCE & 1.85000 & 2.40000 & 0.70000 & 4.25000\\
M1 & $D(2,1.5,0)$ & rational & Sobol & 1.84690 & 2.34342 & 0.91891 & 5.23336\\
M2 & $D(3,1,0.5)$ & rational & Sobol & 1.84690 & 2.34342 & 0.91891 & 5.23336\\
M3 & $D(3,1,0.5)$ & rational & refined & 1.82371 & 2.41721 & 1.07677 & 6.00000\\
M4 & $D(3,0,-0.25)$ & arctangent & refined & 1.85485 & 2.33185 & 0.70210 & 5.44782\\
\bottomrule
\end{tabular}
\end{table}

\begin{table}[htbp]
\centering
\caption{A24 performance of the five expanded-search representatives. Errors
are relative to the published CCSDT(Q) interaction energies and are given in
$\kcalmol$.}
\label{tab:si_expanded_a24}
\small
\begin{tabular}{lrrr}
\toprule
ID & MAE & RMSE & MAXE\\
\midrule
M0 & 0.006029 & 0.013381 & 0.058767\\
M1 & 0.006303 & 0.013118 & 0.057195\\
M2 & 0.006359 & 0.013130 & 0.057177\\
M3 & 0.006408 & 0.013228 & 0.057776\\
M4 & 0.007142 & 0.013853 & 0.059893\\
\bottomrule
\end{tabular}
\end{table}

\begin{figure}[htbp]
\centering
\includegraphics[width=0.82\linewidth]{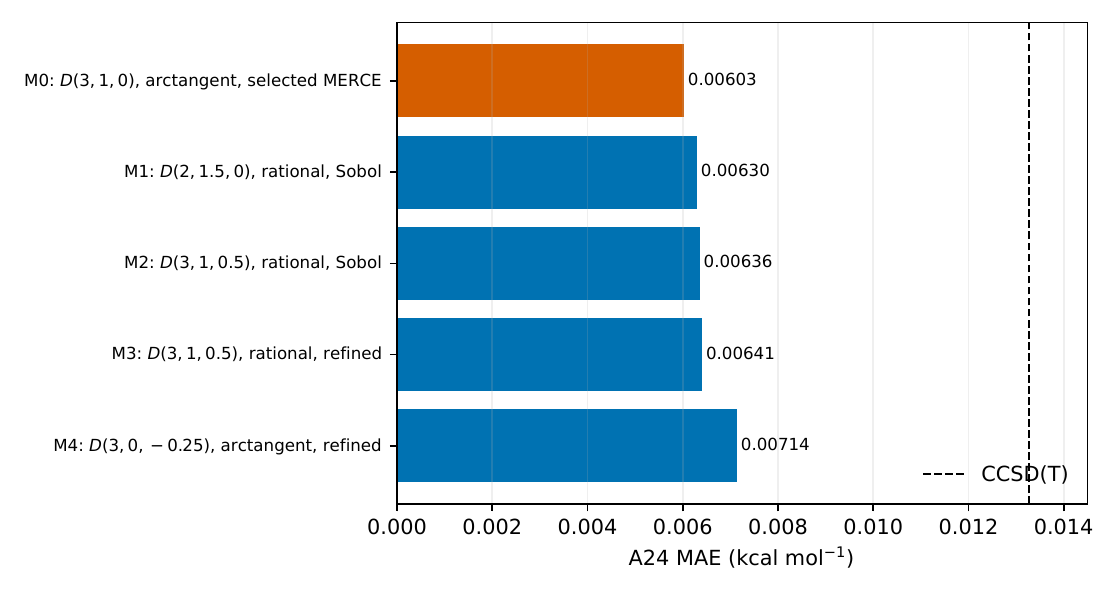}
\caption{A24 MAEs for the five expanded-search representatives. All five
improve upon CCSD(T), while M0, the selected \MERCE{} model, has the lowest
A24 MAE within this follow-up set. The parameterizations M0--M4 are defined in
Table~\ref{tab:si_expanded_a24_models}.}
\label{fig:si_expanded_a24}
\end{figure}

The A24 conclusion is limited to these five representatives. It does not establish that every Pareto-optimal candidate, or every candidate
passing the developmental-set safeguard, would have a higher A24 MAE. It establishes that the expanded representatives carried into
this property-level assessment did not provide a uniform improvement over the
selected protocol.

\subsection{Interpretation and reproduction}
\label{sec:si_expanded_interpretation}

The expanded search did not identify the selected parameters as a unique
numerical optimum. Some alternatives modestly improve the aggregate
development score or the artificial-pair size-consistency diagnostic. Those
gains are accompanied by weaker FCI-developmental safeguards, and the five
expanded representatives assessed on A24 do not improve the selected model's
A24 MAE. The inverse-cubic-linear, arctangent-mapping model is therefore
retained as a structurally simple and balanced representative of a broad
high-performance region. This result does not imply that no alternative can
improve an individual metric or that a better formula cannot exist outside the
explored families and bounds.

The search used Python 3, NumPy 2.4.2, and SciPy 1.17.1. The production
calculation was divided into 32 deterministic single-core chunks and required
12.33 aggregate CPU hours, excluding queueing time. The workflow comprised
input validation, a small-scale verification calculation, the 32-part coarse
search, aggregation and Pareto analysis, continuous refinement of 25 selected
regions, and the artificial-pair size-consistency audit.
The authoritative outputs contain 389,535 eligible candidates, 168 Pareto
candidates, 175 leave-one-dataset-out top-25 records, 25 refined candidates,
and 168 size-consistency records. The source code, nine input data tables,
machine-readable outputs, and SHA-256 manifest are deposited in the versioned
repository so that all aggregate values can be regenerated without new
quantum-chemical calculations.

\section{Comparison with Alternative Models}
\label{sec:si_literature_baselines}

After the \MERCE{} functional form, effective-rank mapping, global parameters,
and selection procedure had been frozen, we performed a separate comparison
with structurally different estimators. This comparison was designed to test
whether the favorable results were restricted to the family of nonlinear
three-point forms explored during \MERCE{} development. None of the models in
this section participated in the original selection of the \MERCE{} protocol,
and the results did not alter that protocol.

\subsection{Parameter-free sequence transformations}

Goodson introduced continued-fraction and Pad\'e-type transformations for the
SCF--CCSD--CCSD(T) sequence.\cite{Goodson2002} Define
\begin{equation}
\delta_1=E_{\mathrm{HF}},\qquad
\delta_2=\Ecorr_{\mathrm{CCSD}},\qquad
\delta_3=\Ecorr_{\mathrm{CCSD(T)}}-\Ecorr_{\mathrm{CCSD}}.
\label{eq:si_goodson_deltas}
\end{equation}
where $E_{\mathrm{HF}}$ is the total Hartree--Fock energy and $\Ecorr$ denotes a correlation energy. The three Goodson estimates evaluated here are
\begin{align}
E_{\mathrm{CCcf}}
&=
\frac{\delta_1}
{1-\dfrac{\delta_2/\delta_1}{1-\delta_3/\delta_2}},
\label{eq:si_goodson_cf}\\
E_{\mathrm{CCr}}
&=
\delta_1
\frac{1+\delta_2/\delta_1-\delta_3/\delta_2}
{1-\delta_3/\delta_2}
=
\delta_1+\frac{\delta_2}{1-\delta_3/\delta_2},
\label{eq:si_goodson_r}\\
E_{\mathrm{CCq}}
&=
\delta_1+
\frac{1}{2}\frac{\delta_2^2}{\delta_3}
\left[1-\sqrt{1-4\delta_3/\delta_2}\right].
\label{eq:si_goodson_q}
\end{align}
Equation~\ref{eq:si_goodson_r} shows both Goodson's original rational
expression and the algebraically equivalent form used in the implementation.
The reported correlation energy estimate associated with each transformation
is
$\widehat{E}^{\mathrm{corr}}_{\mathcal X}
=E_{\mathcal X}-E_{\mathrm{HF}}$,
where $\mathcal X\in\{\mathrm{CCcf},\mathrm{CCr},\mathrm{CCq}\}$. The
continued-fraction, rational, and quadratic expressions are denoted
Goodson-CF, Goodson-R, and Goodson-Q, respectively. A prediction is rejected
at a numerical pole, and Goodson-Q is additionally rejected when its
discriminant is negative. These transformations do not use MP2 and do not
contain a target-rank argument: they estimate a sequence limit rather than a
rank-matched $x=4$ or $x=5$ energy.

As a generic sequence-acceleration control, we also evaluated the signed
Aitken $\Delta^2$ transformation.\cite{Aitken1927} For $S_0=\Ecorr_{\mathrm{MP2}}$, $S_1=\Ecorr_{\mathrm{CCSD}}$, and
$S_2=\Ecorr_{\mathrm{CCSD(T)}}$, with $d_0=S_1-S_0$ and $d_1=S_2-S_1$, it is
\begin{equation}
\widehat{E}^{\mathrm{corr}}_{\mathrm{Aitken}}
=
S_2+\frac{d_1^2}{d_0-d_1}.
\label{eq:si_aitken}
\end{equation}
Aitken acceleration was not introduced as a coupled cluster correlation energy
model and is included here only as a general three-sequence numerical control.
No absolute values or denominator clipping were used; a result was rejected
only when $|d_0-d_1|<10^{-14}\,E_{\mathrm h}$.

\clearpage
\subsection{Refitted linear analogues}

Scaling all correlation energy and multi-coefficient correlation approaches
were originally developed for specific energetic targets and basis-set
protocols.\cite{GordonTruhlar1986SAC,Fast1999MCSAC} Their published
coefficients were therefore not transferred to the present fixed basis
correlation energy problem. Instead, we defined and refitted the following
zero-intercept analogues:
\begin{align}
\widehat{E}^{\mathrm{corr}}_{\mathrm{SAC}}
&=c\,\Ecorr_{\mathrm{CCSD(T)}},
\label{eq:si_sac}\\
\widehat{E}^{\mathrm{corr}}_{\mathrm{scaled(T)}}
&=\Ecorr_{\mathrm{CCSD(T)}}
+a\left(
\Ecorr_{\mathrm{CCSD(T)}}-\Ecorr_{\mathrm{CCSD}}
\right),
\label{eq:si_scaledt}\\
\widehat{E}^{\mathrm{corr}}_{\mathrm{MCSAC}}
&=c_1\Ecorr_{\mathrm{MP2}}
+c_2\left(
\Ecorr_{\mathrm{CCSD}}-\Ecorr_{\mathrm{MP2}}
\right)
+c_3\left(
\Ecorr_{\mathrm{CCSD(T)}}-\Ecorr_{\mathrm{CCSD}}
\right).
\label{eq:si_mcsac}
\end{align}
The zero-intercept restriction makes all three expressions homogeneous in the
component correlation energies. Consequently, they are exactly additive for
noninteracting fragments when the component methods are additive. The labels
SAC, scaled-(T), and MCSAC below refer only to these explicitly defined
analogues and must not be interpreted as applications of the historical
parameterizations.

Two fitting objectives were used. The suffix ``FCI'' denotes ordinary least
squares over the 107 common-domain entries in the two FCI-based developmental
sets. The suffix ``BAL'' denotes a dataset-balanced fit over all 243 common-domain
entries from the two FCI-based developmental sets and seven cross-dataset
model-selection benchmarks,
\begin{equation}
\mathcal L_{\mathrm{BAL}}
=\frac{1}{9}\sum_{D=1}^{9}\frac{1}{N_D}
\sum_{i\in D}
\left(
\widehat{E}^{\mathrm{corr}}_i
-
E_i^{\mathrm{corr,ref}}
\right)^2,
\label{eq:si_balanced_objective}
\end{equation}
so that each of the nine datasets has equal total weight. Because the BAL models are fitted to the same combined benchmark population
used in their aggregate assessment, the corresponding common-domain
statistics are fit-aligned rather than independent validation results.

\begin{table}[htbp]
\centering
\caption{Coefficients of the zero-intercept linear analogues. FCI denotes a
fit to the 107 common-domain FCI entries; BAL denotes the dataset-balanced fit
in Equation~\ref{eq:si_balanced_objective}.}
\label{tab:si_literature_baseline_coefficients}
\small
\begin{tabular}{llrrrr}
\toprule
Fit & Model & $c$ or $a$ & $c_1$ & $c_2$ & $c_3$\\
\midrule
FCI & SAC        & 1.00449945 & --- & --- & ---\\
FCI & scaled-(T) & 0.23347328 & --- & --- & ---\\
FCI & MCSAC      & --- & 0.99731462 & 1.01672109 & 1.19434567\\
BAL & SAC        & 1.00321713 & --- & --- & ---\\
BAL & scaled-(T) & 0.08257320 & --- & --- & ---\\
BAL & MCSAC      & --- & 0.99899372 & 1.01874695 & 1.08720905\\
\bottomrule
\end{tabular}
\end{table}

\subsection{Common-domain and post-selection results}

The common-domain comparison uses the same entries as the \MERCE{} analysis. For the Rings, W4, CEPB, HF, H$_2$O, CH$_4$, and C$_2$H$_4$
benchmarks, the aggregate quantity is the geometric mean of the seven
dataset-level MAE ratios relative to CCSD(T), identical to
$\mathrm{gmean}_{\mathrm{test}}$ in the main text. A24 contains the 23 entries
with available CCSDT(Q) references, and the ionization-potential assessment
contains ten neutral--cation pairs. The CH$_3$F and CH$_3$OH statistics use
the predefined 13-point primary domains,
$0.80\leq R/R_{\mathrm e}\leq1.60$, exactly as in the main text. Thus, the
strong-stretch points are not mixed into the primary-domain MAEs.

\begin{table}[htbp]
\centering
\caption{Comparison with literature-inspired and refitted baselines. Lower
values are better. A24 MAEs are in $\kcalmol$, ionization-potential (IP) MAEs
are in eV, and the CH$_3$F and CH$_3$OH primary-domain MAEs are in
$\microEh$. The FCI and BAL suffixes identify the fitting objectives defined
above; they do not denote historical published coefficients.}
\label{tab:si_literature_baseline_summary}
\scriptsize
\resizebox{\textwidth}{!}{%
\begin{tabular}{lrrrrr}
\toprule
Model & $\mathrm{gmean}_{\mathrm{test}}$ & A24 MAE & IP MAE &
CH$_3$F MAE & CH$_3$OH MAE\\
\midrule
CCSD(T)        & 1.0000  & 0.013272 & 0.008986 & 1097.1   & 1316.8\\
\MERCE{}       & 0.1180  & 0.006029 & 0.004121 &  212.4   &  242.7\\
Goodson-CF     & 0.9252  & 0.070344 & 0.005201 & 1259.0   & 1541.8\\
Goodson-R      & 0.5457  & 0.010571 & 0.005510 &  548.9   &  645.0\\
Goodson-Q      & 0.2952  & 0.018858 & 0.003777 &  216.0   &  126.3\\
Aitken         & 37.9850 & 6.888190 & 0.033610 & 156848.5 & 141160.9\\
SAC-FCI        & 0.5461  & 0.008848 & 0.006463 & 1149.4   &  931.2\\
scaled-(T)-FCI & 1.7596  & 0.033676 & 0.003317 & 2566.0   & 2731.4\\
MCSAC-FCI      & 0.6921  & 0.018587 & 0.003567 &  970.4   & 1122.2\\
SAC-BAL        & 0.2489  & 0.010058 & 0.006517 &  509.1   &  487.2\\
scaled-(T)-BAL & 0.1895  & 0.007240 & 0.005008 &  211.3   &  211.9\\
MCSAC-BAL      & 0.1407  & 0.007415 & 0.005688 &  165.3   &  193.7\\
\bottomrule
\end{tabular}%
}
\end{table}

Among all models in Table~\ref{tab:si_literature_baseline_summary}, \MERCE{}
retains the lowest cross-dataset aggregate,
$\mathrm{gmean}_{\mathrm{test}}=0.1180$. The closest tested alternative is
MCSAC-BAL at 0.1407, although this value is obtained from a
selection-aligned fit. The strongest parameter-free Goodson transformation is
Goodson-Q at 0.2952. \MERCE{} also gives the lowest A24 MAE among the tested
models, $0.006029\,\kcalmol$.

The ordering is not universal. Goodson-Q and the FCI-fitted scaled-(T) and
MCSAC analogues give smaller ionization-potential MAEs than \MERCE{}. On the
additional stretching scans, MCSAC-BAL gives smaller primary-domain MAEs for
both CH$_3$F and CH$_3$OH, while Goodson-Q is similar to \MERCE{} for CH$_3$F
and has a smaller CH$_3$OH MAE. The latter comparison also illustrates why a
single average should not be interpreted in isolation: for CH$_3$OH,
Goodson-Q has an RMS error of $270.3\,\microEh$ and a maximum absolute error of
$927.0\,\microEh$, compared with $264.3$ and $382.6\,\microEh$, respectively,
for \MERCE{}. For CH$_3$F, the corresponding Goodson-Q maximum error is
$1487.4\,\microEh$, compared with $678.4\,\microEh$ for \MERCE{}.

These results do not establish that \MERCE{} is optimal for every observable
or molecular coordinate. They show instead that its advantage is not confined
to comparisons among closely related \MERCE{} functional forms: it remains the
best-balanced model under the cross-dataset aggregation used for selection and
under the A24 assessment, while several simpler estimators provide competitive
or superior results for individual targets. The complete pointwise
predictions, native- and common-domain summaries, available strong-stretch
statistics, formal nonadditivity analyses, fitted coefficients, and scripts
used to regenerate this comparison are supplied in the deposited repository.

\section{Bond-stretching benchmarks and amplitude diagnostics}
\label{sec:si_stretch_diagnostics}

This section supplies the diagnostic profiles and complete coordinate records
referenced in the main text. Throughout, the ordinate is the signed
correlation energy deviation from DLPNO-CCSDTQ. For these closed-shell scans,
the plotted values are those printed after the final converged canonical CCSD
amplitude iteration of the corresponding Psi4 CCSD(T) calculation. With
converged single amplitudes $t_i^a$ and $N_{\mathrm e}$ correlated electrons,
the implementation uses\cite{LeeTaylor1989}
\begin{equation}
T_1=
\left(
\frac{1}{N_{\mathrm e}}\sum_{ia}|t_i^a|^2
\right)^{1/2}.
\end{equation}
For $D_2$, it forms the occupied- and virtual-space doubles-amplitude
contractions\cite{JanssenNielsen1998,NielsenJanssen1999}
\begin{align}
T^{(\mathrm o)}_{ij}&=\sum_{kab}t_{ik}^{ab}t_{jk}^{ab},&
T^{(\mathrm v)}_{ab}&=\sum_{ijc}t_{ij}^{ac}t_{ij}^{bc},
\end{align}
and reports the square root of the largest eigenvalue found in either matrix.
The exact program convention is stated because numerical diagnostic values
can depend on normalization and implementation details. Both quantities are
shown only as descriptive indicators of changing electronic structure;
neither was used retrospectively to choose the displayed primary domains,
accept or reject individual points, or refit \MERCE{}.

\subsection{Original STRETCH model-selection benchmarks}

\begin{figure}[htbp]
\centering
\includegraphics[width=0.96\linewidth]{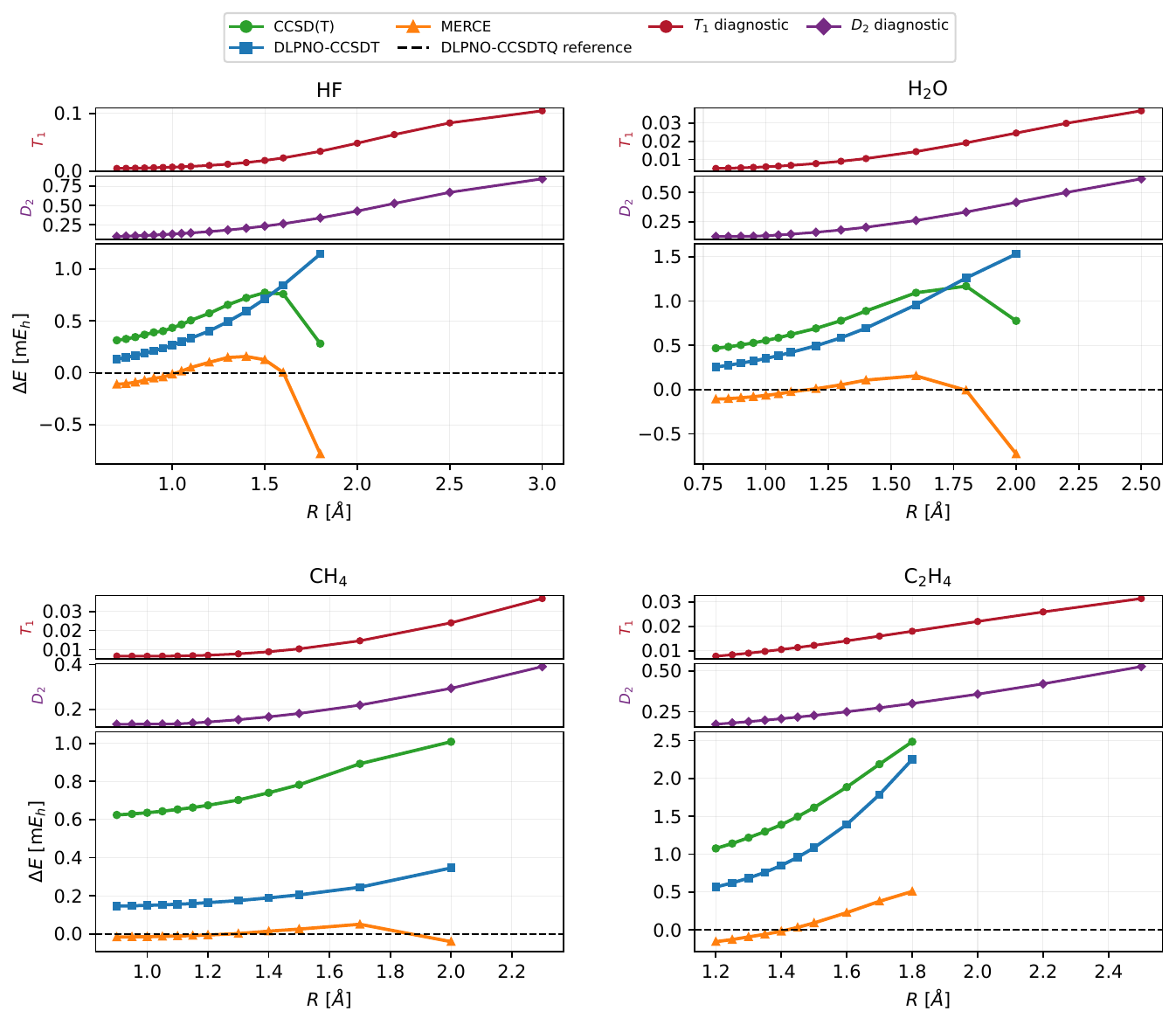}
\caption{Original STRETCH benchmarks with $T_1$ and $D_2$ diagnostic profiles.
Signed deviations are relative to DLPNO-CCSDTQ. The plotted assessment domains
match those used in the model-selection workflow; diagnostic values outside
those domains are retained in the accompanying TSV file.}
\label{fig:si_original_stretch_diagnostics}
\end{figure}

Figure~\ref{fig:si_original_stretch_diagnostics} reports the original HF,
H$_2$O, CH$_4$, and C$_2$H$_4$ scans. The energy-error panels cover the domains
used in model selection, while the accompanying machine-readable diagnostic
table also preserves the available strongly stretched points. The simultaneous
growth of the diagnostics and the loss of accuracy at long bond lengths are
consistent with the expected breakdown of an MP2--CCSD--CCSD(T)-based
single-reference trajectory. This association is interpretive rather than a
diagnostic threshold for accepting or rejecting individual calculations.

\subsection{Additional rigid bond-stretching scans}

The additional scans were generated after the functional form, mapping,
parameters, target rank, and domain checks had been frozen. Each scan used the
17 predefined ratios $R/R_{\mathrm e}=0.80$, 0.85, 0.90, 0.95, 1.00, 1.05,
1.10, 1.15, 1.20, 1.30, 1.40, 1.50, 1.60, 1.75, 2.00, 2.25, and 2.50. The
primary assessment domain ended at $R/R_{\mathrm e}=1.60$; the final four
points were retained as a separate strong-stretch stress test.

\begin{figure}[htbp]
\centering
\includegraphics[width=0.94\linewidth]{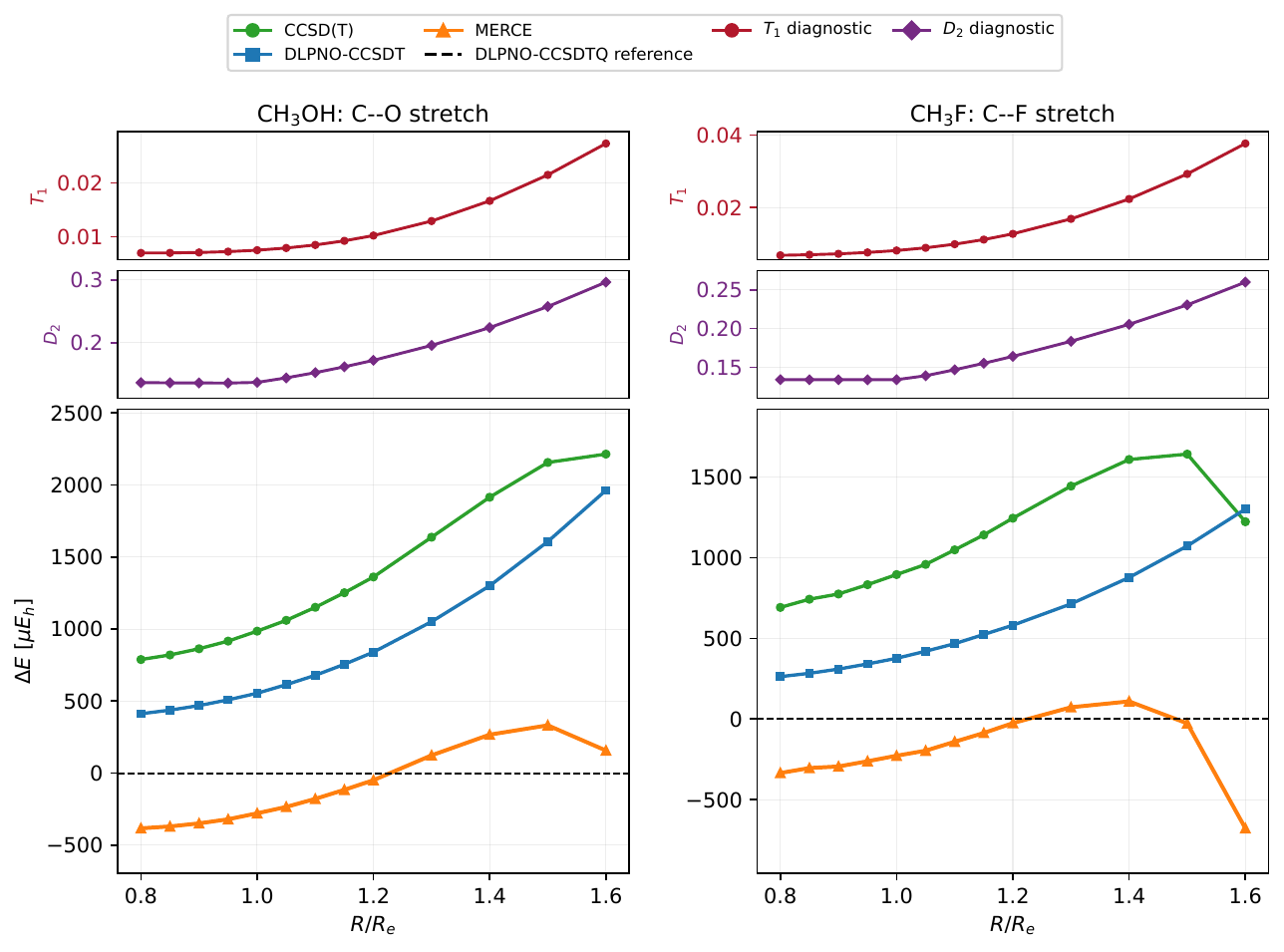}
\caption{Additional CH$_3$OH C--O and CH$_3$F C--F rigid stretches over the
predefined primary domain, with $T_1$ and $D_2$ diagnostic profiles. Deviations
are relative to DLPNO-CCSDTQ.}
\label{fig:si_additional_stretch_diagnostics}
\end{figure}

Within the 13-point primary domains, MERCE gives MAEs of 212.4 and
242.7~$\microEh$ for CH$_3$F and CH$_3$OH, respectively, compared with 1097.1
and 1316.8~$\microEh$ for CCSD(T), and 579.8 and 860.9~$\microEh$ for
DLPNO-CCSDT. Figure~\ref{fig:si_additional_stretch_diagnostics} adds the
corresponding amplitude diagnostics. At the longest separations, the MERCE and
CCSD(T) deviations increase sharply, providing a deliberately retained failure
test rather than evidence for extending the primary domain.

LiF provides an explicit domain-of-applicability failure. Sixteen of its 17
points violate the required ordering
$\Ecorr_{\mathrm{MP2}}>\Ecorr_{\mathrm{CCSD}}>\Ecorr_{\mathrm{CCSD(T)}}$ and are therefore
rejected before extrapolation. Only $R/R_{\mathrm e}=0.80$ is accepted, with a
MERCE deviation of 13.9~$\microEh$; consequently, no LiF aggregate MERCE error
is reported. Table~\ref{tab:si_external_stretch_full} retains every accepted, rejected, and
failed-reference record. The unavailable DLPNO-CCSDTQ comparisons for
CH$_3$F at $R/R_{\mathrm e}=2.50$ and CH$_3$OH at
$R/R_{\mathrm e}=1.75$ are retained rather than being treated as zero error or
silently removed.

\begin{landscape}
\begingroup\scriptsize\setlength{\tabcolsep}{3.2pt}
\begin{longtable}{llrrrrrr}
\caption{Complete rigid-stretch audit. Deviations are relative to DLPNO-CCSDTQ and are in $\microEh$. Blank entries denote unavailable reference deviations; status fields preserve failed references and MERCE domain rejections.}\label{tab:si_external_stretch_full}\\
\toprule System & $R/R_{\mathrm e}$ & CCSD(T) & DLPNO-CCSDT & MERCE & $T_1$ & $D_2$ & Status \\
\midrule\endfirsthead
\caption[]{Complete rigid-stretch audit. Deviations are relative to DLPNO-CCSDTQ and are in $\microEh$. Blank entries denote unavailable reference deviations; status fields preserve failed references and MERCE domain rejections. (continued).}\\
\toprule System & $R/R_{\mathrm e}$ & CCSD(T) & DLPNO-CCSDT & MERCE & $T_1$ & $D_2$ & Status \\
\midrule\endhead
\bottomrule\endfoot
CH3F & 0.80 & 692.6 & 263.0 & -335.1 & 0.00680 & 0.13394 & valid \\
CH3F & 0.85 & 743.3 & 283.5 & -304.3 & 0.00696 & 0.13384 & valid \\
CH3F & 0.90 & 775.7 & 309.8 & -294.5 & 0.00721 & 0.13381 & valid \\
CH3F & 0.95 & 833.6 & 341.5 & -262.1 & 0.00758 & 0.13383 & valid \\
CH3F & 1.00 & 896.4 & 376.7 & -227.8 & 0.00812 & 0.13389 & valid \\
CH3F & 1.05 & 959.2 & 420.6 & -196.6 & 0.00887 & 0.13900 & valid \\
CH3F & 1.10 & 1049.6 & 467.5 & -141.0 & 0.00987 & 0.14671 & valid \\
CH3F & 1.15 & 1142.4 & 524.6 & -86.7 & 0.01114 & 0.15506 & valid \\
CH3F & 1.20 & 1246.2 & 581.9 & -25.5 & 0.01272 & 0.16399 & valid \\
CH3F & 1.30 & 1445.3 & 714.3 & 72.8 & 0.01685 & 0.18347 & valid \\
CH3F & 1.40 & 1610.3 & 878.2 & 109.3 & 0.02235 & 0.20537 & valid \\
CH3F & 1.50 & 1643.8 & 1073.6 & -27.0 & 0.02926 & 0.23042 & valid \\
CH3F & 1.60 & 1224.0 & 1302.7 & -678.4 & 0.03768 & 0.25982 & valid \\
CH3F & 1.75 & -1021.8 & 1762.1 & -3462.8 & 0.05326 & 0.31575 & valid \\
CH3F & 2.00 & -17877.6 & 2756.1 & -22234.2 & 0.08687 & 0.45670 & valid \\
CH3F & 2.25 & -75469.1 & 3514.8 & -84155.0 & 0.12564 & 0.66819 & valid \\
CH3F & 2.50 & -- & -- & -- & 0.16053 & 0.91465 & reference failed \\
CH3OH & 0.80 & 788.4 & 412.4 & -382.6 & 0.00698 & 0.13664 & valid \\
CH3OH & 0.85 & 820.3 & 437.4 & -369.5 & 0.00699 & 0.13631 & valid \\
CH3OH & 0.90 & 863.3 & 469.0 & -348.2 & 0.00707 & 0.13609 & valid \\
CH3OH & 0.95 & 916.2 & 508.9 & -319.8 & 0.00723 & 0.13596 & valid \\
CH3OH & 1.00 & 985.3 & 554.6 & -278.4 & 0.00750 & 0.13701 & valid \\
CH3OH & 1.05 & 1060.4 & 614.1 & -234.1 & 0.00790 & 0.14423 & valid \\
CH3OH & 1.10 & 1150.7 & 679.0 & -178.1 & 0.00847 & 0.15252 & valid \\
CH3OH & 1.15 & 1251.8 & 755.4 & -115.4 & 0.00924 & 0.16183 & valid \\
CH3OH & 1.20 & 1361.4 & 839.9 & -48.7 & 0.01022 & 0.17213 & valid \\
CH3OH & 1.30 & 1637.0 & 1051.1 & 123.8 & 0.01291 & 0.19590 & valid \\
CH3OH & 1.40 & 1914.6 & 1300.1 & 267.4 & 0.01667 & 0.22427 & valid \\
CH3OH & 1.50 & 2155.7 & 1606.9 & 332.0 & 0.02150 & 0.25774 & valid \\
CH3OH & 1.60 & 2213.7 & 1962.3 & 157.1 & 0.02734 & 0.29664 & valid \\
CH3OH & 1.75 & -- & -- & -- & 0.03755 & 0.36495 & reference failed \\
CH3OH & 2.00 & -5684.9 & 3241.3 & -9418.0 & 0.05579 & 0.49706 & valid \\
CH3OH & 2.25 & -20871.0 & 3414.9 & -26055.3 & 0.07117 & 0.62786 & valid \\
CH3OH & 2.50 & -39275.3 & 3272.2 & -45654.2 & 0.08124 & 0.73337 & valid \\
LiF & 0.80 & 401.4 & 258.6 & -13.9 & 0.01082 & 0.11057 & valid \\
LiF & 0.85 & 443.4 & 288.5 & -- & 0.01124 & 0.11112 & MERCE rejected \\
LiF & 0.90 & 459.6 & 306.6 & -- & 0.01171 & 0.11155 & MERCE rejected \\
LiF & 0.95 & 483.7 & 330.9 & -- & 0.01219 & 0.11186 & MERCE rejected \\
LiF & 1.00 & 501.2 & 355.4 & -- & 0.01268 & 0.11210 & MERCE rejected \\
LiF & 1.05 & 507.3 & 378.5 & -- & 0.01317 & 0.11227 & MERCE rejected \\
LiF & 1.10 & 532.2 & 397.2 & -- & 0.01367 & 0.11238 & MERCE rejected \\
LiF & 1.15 & 560.5 & 424.7 & -- & 0.01417 & 0.11244 & MERCE rejected \\
LiF & 1.20 & 580.2 & 440.4 & -- & 0.01467 & 0.11335 & MERCE rejected \\
LiF & 1.30 & 585.8 & 448.1 & -- & 0.01573 & 0.11665 & MERCE rejected \\
LiF & 1.40 & 554.4 & 435.7 & -- & 0.01687 & 0.11956 & MERCE rejected \\
LiF & 1.50 & 561.5 & 439.5 & -- & 0.01812 & 0.12204 & MERCE rejected \\
LiF & 1.60 & 564.0 & 437.8 & -- & 0.01949 & 0.12412 & MERCE rejected \\
LiF & 1.75 & 533.1 & 436.9 & -- & 0.02177 & 0.12664 & MERCE rejected \\
LiF & 2.00 & 499.0 & 399.6 & -- & 0.02627 & 0.12990 & MERCE rejected \\
LiF & 2.25 & 400.7 & 369.3 & -- & 0.03217 & 0.13296 & MERCE rejected \\
LiF & 2.50 & 194.4 & 329.2 & -- & 0.04113 & 0.13737 & MERCE rejected \\
\end{longtable}
\endgroup
\end{landscape}

\section{Rigid NH$_3$ umbrella-planarization assessment}
\label{sec:si_nh3}

The NH$_3$ scan probes a deformation distinct from bond stretching. Its
dimensionless coordinate is $\lambda=h/h_{\mathrm e}$, where $h$ is the
perpendicular displacement of N from the plane of the three H atoms and
$h_{\mathrm e}=0.37635446$~\AA{} for the reference pyramidal geometry. The
N--H distances are held fixed at 1.0128~\AA{}; $\lambda=1$ is the pyramidal
($C_{3v}$) endpoint and $\lambda=0$ is the rigid planar ($D_{3h}$) endpoint.
The reference structural parameters are those of Marquardt et al.; the
intermediate rigid geometries were generated in this work.\cite{Marquardt2013NH3}

\begin{figure}[htbp]
\centering
\includegraphics[width=0.72\linewidth]{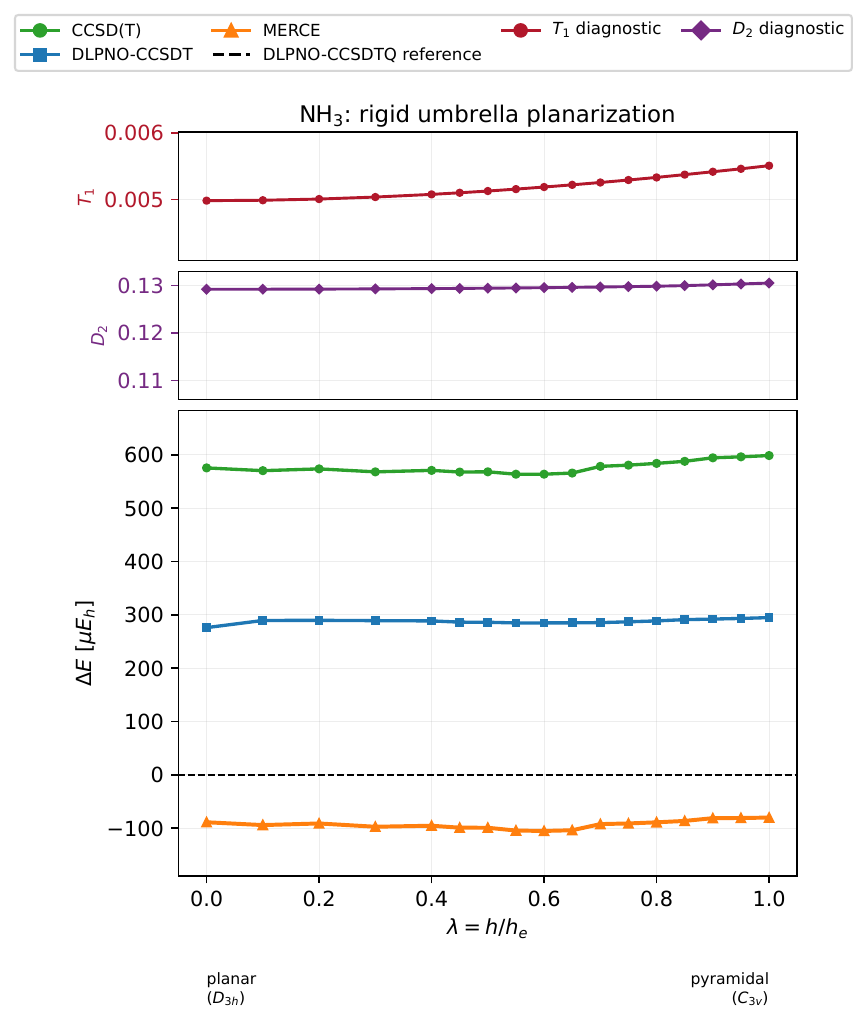}
\caption{Rigid NH$_3$ umbrella-planarization deviations relative to
DLPNO-CCSDTQ, together with $T_1$ and $D_2$ diagnostics. The scan runs from the
rigid planar structure at $\lambda=0$ to the reference pyramidal structure at
$\lambda=1$.}
\label{fig:si_nh3_diagnostics}
\end{figure}

All 17 predefined geometries have converged DLPNO-CCSDTQ comparison values.
The MERCE MAE, RMS deviation, and maximum absolute deviation are 93.0, 93.4,
and 105.3~$\microEh$, respectively. The corresponding values are 577.1, 577.2,
and 598.9~$\microEh$ for CCSD(T), and 287.8, 287.9, and 294.9~$\microEh$ for
DLPNO-CCSDT. Figure~\ref{fig:si_nh3_diagnostics} shows that the MERCE deviation
remains nearly uniform across this rigid planarization path.

\begin{landscape}
\begingroup\scriptsize\setlength{\tabcolsep}{3.2pt}
\begin{longtable}{llrrrrrr}
\caption{Complete NH$_3$ rigid umbrella-planarization audit. Deviations are relative to DLPNO-CCSDTQ and are in $\microEh$.}\label{tab:si_nh3_full}\\
\toprule System & $\lambda=h/h_{\mathrm e}$ & CCSD(T) & DLPNO-CCSDT & MERCE & $T_1$ & $D_2$ & Status \\
\midrule\endfirsthead
\caption[]{Complete NH$_3$ rigid umbrella-planarization audit. Deviations are relative to DLPNO-CCSDTQ and are in $\microEh$. (continued).}\\
\toprule System & $\lambda=h/h_{\mathrm e}$ & CCSD(T) & DLPNO-CCSDT & MERCE & $T_1$ & $D_2$ & Status \\
\midrule\endhead
\bottomrule\endfoot
NH3 & 1.00 & 598.9 & 294.9 & -80.2 & 0.00551 & 0.13049 & valid \\
NH3 & 0.95 & 596.5 & 293.3 & -81.0 & 0.00546 & 0.13030 & valid \\
NH3 & 0.90 & 594.6 & 292.1 & -81.3 & 0.00542 & 0.13012 & valid \\
NH3 & 0.85 & 588.1 & 291.1 & -86.5 & 0.00537 & 0.12995 & valid \\
NH3 & 0.80 & 584.2 & 288.7 & -89.1 & 0.00533 & 0.12982 & valid \\
NH3 & 0.75 & 580.9 & 287.0 & -91.2 & 0.00529 & 0.12974 & valid \\
NH3 & 0.70 & 578.6 & 285.5 & -92.4 & 0.00526 & 0.12966 & valid \\
NH3 & 0.65 & 566.1 & 285.1 & -104.0 & 0.00522 & 0.12958 & valid \\
NH3 & 0.60 & 563.9 & 284.9 & -105.3 & 0.00519 & 0.12952 & valid \\
NH3 & 0.55 & 563.8 & 284.9 & -104.6 & 0.00516 & 0.12946 & valid \\
NH3 & 0.50 & 568.4 & 286.2 & -99.3 & 0.00513 & 0.12941 & valid \\
NH3 & 0.45 & 567.9 & 286.4 & -99.2 & 0.00510 & 0.12936 & valid \\
NH3 & 0.40 & 571.1 & 288.7 & -95.5 & 0.00508 & 0.12932 & valid \\
NH3 & 0.30 & 568.3 & 289.2 & -97.4 & 0.00504 & 0.12926 & valid \\
NH3 & 0.20 & 573.9 & 289.8 & -91.3 & 0.00501 & 0.12922 & valid \\
NH3 & 0.10 & 570.5 & 289.5 & -94.3 & 0.00499 & 0.12919 & valid \\
NH3 & 0.00 & 575.7 & 276.1 & -89.0 & 0.00498 & 0.12918 & valid \\
\end{longtable}
\endgroup
\end{landscape}

\section{Counterpoise sensitivity of the selected S66 fixed-basis application}
\label{sec:si_s66_cp_sensitivity}

The selected S66 table in the main text reports Boys--Bernardi
counterpoise-corrected (CP) interaction energies and \MERCE{} corrections. To
determine whether the predicted post-CCSD(T) contribution depends materially
on the treatment of basis-set superposition error (BSSE), we additionally
evaluated the same quantities using the uncorrected (no-CP) interaction-energy
convention. For method $M$, the two interaction energies are defined as
\begin{align}
E_{M,\mathrm{no\text{-}CP}}^{\mathrm{int}}
&=E_M(AB)-E_M(A)-E_M(B),\\
E_{M,\mathrm{CP}}^{\mathrm{int}}
&=E_M(AB)-E_M(A^{AB})-E_M(B^{AB}),
\end{align}
where $A^{AB}$ and $B^{AB}$ denote monomers evaluated in the complete dimer
basis, including the ghost functions of the partner fragment. For
$Y\in\{\mathrm{no\text{-}CP},\mathrm{CP}\}$, the predicted post-CCSD(T) correction
and its differential CP shift are defined as
\begin{align}
\Delta_x^Y
&=E_{\mathrm{MERCE},Y}^{\mathrm{int}}(x)
-E_{\mathrm{CCSD(T)},Y}^{\mathrm{int}},\\
\delta_{\mathrm{CP}}\Delta_x
&=\Delta_x^{\mathrm{CP}}-\Delta_x^{\mathrm{no\text{-}CP}}.
\label{eq:si_s66_delta_cp}
\end{align}
The corresponding CP shift in the underlying CCSD(T) interaction energy is
\begin{equation}
\delta_{\mathrm{CP}}E_{\mathrm{CCSD(T)}}^{\mathrm{int}}
=
E_{\mathrm{CCSD(T)},\mathrm{CP}}^{\mathrm{int}}
-
E_{\mathrm{CCSD(T)},\mathrm{no\text{-}CP}}^{\mathrm{int}}.
\end{equation}

The analysis contains all 31 complex--basis entries reported in the main-text
table. These comprise cc-pVDZ, cc-pVTZ, aug-cc-pVDZ, and aug-cc-pVTZ results
for all seven selected complexes, together with cc-pVQZ results for cases 21,
25, and 57. The $x=4$ result is the nominal CCSDTQ-like prediction, whereas
the $x=10$ and $x=100$ values retain their role as internal large-$x$
trajectory diagnostics rather than physical coupled cluster truncation
levels.

\begin{table}[htbp]
\centering
\small
\setlength{\tabcolsep}{4.0pt}
\caption{Basis-resolved sensitivity of the nominal $\Delta_4$ correction to
the CP convention. MAD, RMS, and maximum refer to
$\delta_{\mathrm{CP}}\Delta_4$. RMS$_{\mathrm{CCSD(T)}}$ is the RMS value of
$\delta_{\mathrm{CP}}E_{\mathrm{CCSD(T)}}^{\mathrm{int}}$, and the ratio is
the former RMS divided by the latter. Energies are in $\kcalmol$ unless
$\microEh$ is specified. QZ statistics contain three entries; each other
basis contains all seven selected complexes.}
\label{tab:si_s66_cp_summary}
\begin{tabular}{lrrrrrr}
\toprule
Basis & $N$ & MAD [$\microEh$] & RMS & Maximum &
RMS$_{\mathrm{CCSD(T)}}$ & Ratio [\%] \\
\midrule
DZ  & 7 & 20.12 & 0.01349 & 0.02185 & 3.182 & 0.424 \\
TZ  & 7 & 10.15 & 0.00700 & 0.01214 & 2.096 & 0.334 \\
QZ  & 3 &  2.55 & 0.00175 & 0.00250 & 0.914 & 0.192 \\
ADZ & 7 & 27.54 & 0.01839 & 0.02582 & 3.381 & 0.544 \\
ATZ & 7 &  7.36 & 0.00485 & 0.00737 & 2.921 & 0.166 \\
\bottomrule
\end{tabular}
\end{table}

Across all 31 entries, the mean signed and mean absolute values of
$\delta_{\mathrm{CP}}\Delta_4$ are $+0.00929$ and
$0.00939\,\kcalmol$, respectively; the latter corresponds to
$14.96\,\microEh$. The pooled RMS and maximum absolute differences are
$0.01158$ and $0.02582\,\kcalmol$, respectively. CP does not change the sign
of $\Delta_4$ in any entry. The signed difference is positive in 30 entries
and negative in one, showing that the small effect is systematic rather than
produced by random cancellation.

The RMS CP shift in the underlying CCSD(T) interaction energies is
$2.8047\,\kcalmol$. Consequently, the RMS differential CP effect on the
predicted $\Delta_4$ is only $0.413\%$ of the underlying CCSD(T) shift, or
smaller by a factor of 242. This pronounced scale separation is consistent
with the explicit post-CCSD(T) analysis of Fishman et al., who found that CP
corrections to post-CCSD(T) contributions in the A24x8
noncovalent-interaction set are approximately two orders of magnitude smaller
than the corresponding CP corrections to CCSD(T) interaction
energies.\cite{Fishman2025BSSE} Their analysis included basis sets ranging
from cc-pVDZ(no d) to larger unaugmented and half-augmented basis sets. The
present selected S66 results show analogous behavior for \MERCE-predicted
total post-CCSD(T) corrections, including the three cc-pVQZ calculations.
Because the systems, basis-set families, and evaluated quantities differ, the
agreement constitutes qualitative corroboration rather than a direct
numerical comparison or an explicit high-rank benchmark.

The small absolute difference should not be confused with a negligible
relative change in every post-CCSD(T) correction. Because $\Delta_4$ itself is
small,
\begin{equation}
\frac{
\sum_i\left|\delta_{\mathrm{CP}}\Delta_{4,i}\right|
}{
\sum_i\left|\Delta_{4,i}^{\mathrm{CP}}\right|
}
=29.3\%.
\end{equation}
Thus, the CP dependence is more than two orders of magnitude smaller than the
CCSD(T) CP shift, but it can remain a noticeable fraction of a correction
whose absolute magnitude is only a few hundredths of a $\kcalmol$.

\begingroup
\scriptsize
\setlength{\tabcolsep}{3.2pt}
\begin{longtable}{@{}r p{3.80cm} l rr rr rr rr@{}}
\caption{CP and no-CP fixed-basis \MERCE{} post-CCSD(T) corrections for all
available selected S66 complex--basis entries. Energies are given in
kcal mol$^{-1}$. For each interaction-energy convention
$Y\in\{\mathrm{no\text{-}CP},\mathrm{CP}\}$,
$E_{\mathrm{CCSD(T)},Y}^{\mathrm{int}}$ is the total CCSD(T) interaction
energy and
$\Delta_x^Y=
E_{\mathrm{MERCE},Y}^{\mathrm{int}}(x)
-E_{\mathrm{CCSD(T)},Y}^{\mathrm{int}}$.
QZ results are available for cases 21, 25, and 57 only.}
\label{tab:table4_cp_nocp}\\
\toprule
& & & \multicolumn{2}{c}{$E_{\mathrm{CCSD(T)}}^{\mathrm{int}}$} & \multicolumn{2}{c}{$\Delta_4$} & \multicolumn{2}{c}{$\Delta_{10}$} & \multicolumn{2}{c}{$\Delta_{100}$} \\
\cmidrule(lr){4-5}\cmidrule(lr){6-7}\cmidrule(lr){8-9}\cmidrule(l){10-11}
Case & Complex & Basis & no-CP & CP & no-CP & CP & no-CP & CP & no-CP & CP \\
\midrule
\endfirsthead
\multicolumn{11}{c}{\tablename\ \thetable\ --- continued}\\
\toprule
& & & \multicolumn{2}{c}{$E_{\mathrm{CCSD(T)}}^{\mathrm{int}}$} & \multicolumn{2}{c}{$\Delta_4$} & \multicolumn{2}{c}{$\Delta_{10}$} & \multicolumn{2}{c}{$\Delta_{100}$} \\
\cmidrule(lr){4-5}\cmidrule(lr){6-7}\cmidrule(lr){8-9}\cmidrule(l){10-11}
Case & Complex & Basis & no-CP & CP & no-CP & CP & no-CP & CP & no-CP & CP \\
\midrule
\endhead
21 & AcNH$_2\cdots$AcNH$_2$ & DZ & -18.3794 & -12.3176 & -0.0456 & -0.0237 & -0.0727 & -0.0376 & -0.0742 & -0.0384 \\
 &  & TZ & -18.1025 & -14.8560 & -0.0441 & -0.0320 & -0.0703 & -0.0507 & -0.0718 & -0.0517 \\
 &  & QZ & -17.0747 & -15.9335 & -0.0350 & -0.0325 & -0.0560 & -0.0511 & -0.0571 & -0.0521 \\
 &  & ADZ & -17.2817 & -14.6130 & -0.0570 & -0.0422 & -0.0895 & -0.0664 & -0.0913 & -0.0678 \\
 &  & ATZ & -19.1451 & -15.7937 & -0.0465 & -0.0408 & -0.0723 & -0.0637 & -0.0737 & -0.0650 \\
\addlinespace[2pt]
24 & benzene$\cdots$benzene ($\pi\cdots\pi$) & DZ & -2.3414 & 0.0969 & -0.0430 & -0.0304 & -0.0847 & -0.0635 & -0.0870 & -0.0653 \\
 &  & TZ & -3.6570 & -1.5221 & 0.0005 & 0.0078 & -0.0227 & -0.0108 & -0.0239 & -0.0118 \\
 &  & ADZ & -6.4858 & -2.0630 & -0.0680 & -0.0432 & -0.1273 & -0.0887 & -0.1306 & -0.0913 \\
 &  & ATZ & -5.6410 & -2.4238 & 0.0022 & 0.0077 & -0.0207 & -0.0120 & -0.0219 & -0.0131 \\
\addlinespace[2pt]
25 & pyridine$\cdots$pyridine ($\pi\cdots\pi$) & DZ & -3.5256 & -0.7944 & -0.0350 & -0.0217 & -0.0747 & -0.0526 & -0.0769 & -0.0543 \\
 &  & TZ & -4.5997 & -2.4944 & 0.0458 & 0.0512 & 0.0459 & 0.0549 & 0.0461 & 0.0552 \\
 &  & QZ & -3.9861 & -3.1890 & 0.1023 & 0.1007 & 0.1333 & 0.1316 & 0.1352 & 0.1335 \\
 &  & ADZ & -7.4148 & -3.0554 & -0.0440 & -0.0221 & -0.0929 & -0.0584 & -0.0956 & -0.0604 \\
 &  & ATZ & -6.4616 & -3.4702 & 0.0519 & 0.0555 & 0.0549 & 0.0607 & 0.0553 & 0.0612 \\
\addlinespace[2pt]
30 & benzene$\cdots$ethene & DZ & -0.9585 & 0.5394 & -0.0193 & -0.0128 & -0.0334 & -0.0224 & -0.0342 & -0.0229 \\
 &  & TZ & -1.6780 & -0.5435 & -0.0846 & -0.0801 & -0.1384 & -0.1310 & -0.1415 & -0.1340 \\
 &  & ADZ & -2.9963 & -0.9492 & -0.0484 & -0.0369 & -0.0807 & -0.0627 & -0.0825 & -0.0641 \\
 &  & ATZ & -2.7338 & -1.2037 & -0.0955 & -0.0926 & -0.1560 & -0.1514 & -0.1595 & -0.1547 \\
\addlinespace[2pt]
54 & benzene$\cdots$water (OH$\cdots\pi$) & DZ & -3.3558 & -1.6935 & -0.0114 & -0.0037 & -0.0196 & -0.0072 & -0.0201 & -0.0074 \\
 &  & TZ & -4.2492 & -2.5953 & -0.0126 & -0.0072 & -0.0241 & -0.0150 & -0.0247 & -0.0154 \\
 &  & ADZ & -4.7233 & -2.7123 & -0.0313 & -0.0180 & -0.0512 & -0.0308 & -0.0524 & -0.0315 \\
 &  & ATZ & -4.9427 & -3.1100 & -0.0122 & -0.0087 & -0.0225 & -0.0172 & -0.0230 & -0.0177 \\
\addlinespace[2pt]
57 & benzene$\cdots$peptide (NH$\cdots\pi$) & DZ & -5.3523 & -2.7271 & -0.0405 & -0.0251 & -0.0713 & -0.0462 & -0.0730 & -0.0474 \\
 &  & TZ & -6.4080 & -4.3203 & -0.0441 & -0.0365 & -0.0800 & -0.0678 & -0.0820 & -0.0695 \\
 &  & QZ & -5.6484 & -4.8954 & -0.0247 & -0.0239 & -0.0510 & -0.0491 & -0.0525 & -0.0505 \\
 &  & ADZ & -9.0484 & -4.4596 & -0.0743 & -0.0485 & -0.1248 & -0.0844 & -0.1276 & -0.0864 \\
 &  & ATZ & -9.2088 & -5.0212 & -0.0482 & -0.0409 & -0.0858 & -0.0748 & -0.0879 & -0.0767 \\
\addlinespace[2pt]
58 & pyridine$\cdots$pyridine (CH$\cdots$N) & DZ & -5.7063 & -2.7381 & -0.0242 & -0.0133 & -0.0420 & -0.0239 & -0.0430 & -0.0245 \\
 &  & TZ & -5.2645 & -3.5754 & -0.0071 & -0.0050 & -0.0151 & -0.0112 & -0.0156 & -0.0116 \\
 &  & ADZ & -5.8456 & -3.6048 & -0.0287 & -0.0199 & -0.0487 & -0.0344 & -0.0498 & -0.0352 \\
 &  & ATZ & -6.4672 & -4.0250 & -0.0114 & -0.0077 & -0.0208 & -0.0153 & -0.0213 & -0.0157 \\
\bottomrule
\end{longtable}
\endgroup

\subsection{Formal nonadditivity in the selected S66 component calculations}
\label{sec:si_s66_property_nonadditivity}

Applying Eq.~\ref{eq:si_property_nonadditivity} to the actual selected S66
component records gives a CP mean absolute defect of $10.551\,\microEh$, an
RMS defect of $27.537\,\microEh$, and a maximum absolute defect of
$107.832\,\microEh$ over the 31 complex--basis entries. The corresponding
no-CP values are 10.474, 27.383, and $107.531\,\microEh$. Identical-fragment
dimers have a zero defect to numerical precision because the same nonlinear
mapping is applied to two equal component vectors; the larger offsets arise
for chemically dissimilar fragments. The three largest CP offsets are the
benzene--ethene complex in aug-cc-pVTZ ($-107.832\,\microEh$), the same
complex in cc-pVTZ ($-95.822\,\microEh$), and benzene--peptide in cc-pVQZ
($-32.122\,\microEh$). These values are properties of the nonlinear estimator,
not errors against canonical S66/CBS reference energies.

\begin{landscape}
\begin{longtable}{r p{5.0cm} l rr}
\caption{Entrywise CP formal nonadditivity for the selected S66 application.
The predicted post-CCSD(T) correction $\Delta_4^{\mathrm{CP}}$ and
$\Delta_{\mathrm{SC}}^{\mathrm{CP}}$ are given in $\microEh$.}\\
\label{tab:si_s66_property_nonadditivity}\\
\toprule
Case & Complex & Basis & $\Delta_4^{\mathrm{CP}}$ & $\Delta_{\mathrm{SC}}^{\mathrm{CP}}$ \\
\midrule
\endfirsthead
\multicolumn{5}{c}{\tablename\ \thetable\ --- continued}\\
\toprule
Case & Complex & Basis & $\Delta_4^{\mathrm{CP}}$ & $\Delta_{\mathrm{SC}}^{\mathrm{CP}}$ \\
\midrule
\endhead
21 & AcNH$_2\cdots$AcNH$_2$ & aug-cc-pVDZ & -67.263 & 0.000 \\
21 & AcNH$_2\cdots$AcNH$_2$ & aug-cc-pVTZ & -64.975 & 0.000 \\
21 & AcNH$_2\cdots$AcNH$_2$ & cc-pVDZ & -37.827 & 0.000 \\
21 & AcNH$_2\cdots$AcNH$_2$ & cc-pVQZ & -51.867 & -0.000 \\
21 & AcNH$_2\cdots$AcNH$_2$ & cc-pVTZ & -51.014 & -0.000 \\
24 & benzene$\cdots$benzene ($\pi\cdots\pi$) & aug-cc-pVDZ & -68.838 & 0.000 \\
24 & benzene$\cdots$benzene ($\pi\cdots\pi$) & aug-cc-pVTZ & 12.261 & 0.000 \\
24 & benzene$\cdots$benzene ($\pi\cdots\pi$) & cc-pVDZ & -48.450 & 0.000 \\
24 & benzene$\cdots$benzene ($\pi\cdots\pi$) & cc-pVTZ & 12.382 & 0.000 \\
25 & pyridine$\cdots$pyridine ($\pi\cdots\pi$) & aug-cc-pVDZ & -35.163 & 0.000 \\
25 & pyridine$\cdots$pyridine ($\pi\cdots\pi$) & aug-cc-pVTZ & 88.379 & 0.000 \\
25 & pyridine$\cdots$pyridine ($\pi\cdots\pi$) & cc-pVDZ & -34.571 & 0.000 \\
25 & pyridine$\cdots$pyridine ($\pi\cdots\pi$) & cc-pVQZ & 160.471 & -0.000 \\
25 & pyridine$\cdots$pyridine ($\pi\cdots\pi$) & cc-pVTZ & 81.546 & -0.000 \\
30 & benzene$\cdots$ethene & aug-cc-pVDZ & -58.758 & -6.789 \\
30 & benzene$\cdots$ethene & aug-cc-pVTZ & -147.521 & -107.832 \\
30 & benzene$\cdots$ethene & cc-pVDZ & -20.390 & 8.838 \\
30 & benzene$\cdots$ethene & cc-pVTZ & -127.625 & -95.822 \\
54 & benzene$\cdots$water (OH$\cdots\pi$) & aug-cc-pVDZ & -28.678 & -4.028 \\
54 & benzene$\cdots$water (OH$\cdots\pi$) & aug-cc-pVTZ & -13.919 & -0.346 \\
54 & benzene$\cdots$water (OH$\cdots\pi$) & cc-pVDZ & -5.952 & 1.214 \\
54 & benzene$\cdots$water (OH$\cdots\pi$) & cc-pVTZ & -11.463 & -4.993 \\
57 & benzene$\cdots$peptide (NH$\cdots\pi$) & aug-cc-pVDZ & -77.339 & -10.155 \\
57 & benzene$\cdots$peptide (NH$\cdots\pi$) & aug-cc-pVTZ & -65.146 & -25.783 \\
57 & benzene$\cdots$peptide (NH$\cdots\pi$) & cc-pVDZ & -39.944 & -2.209 \\
57 & benzene$\cdots$peptide (NH$\cdots\pi$) & cc-pVQZ & -38.143 & -32.122 \\
57 & benzene$\cdots$peptide (NH$\cdots\pi$) & cc-pVTZ & -58.089 & -26.942 \\
58 & pyridine$\cdots$pyridine (CH$\cdots$N) & aug-cc-pVDZ & -31.710 & 0.000 \\
58 & pyridine$\cdots$pyridine (CH$\cdots$N) & aug-cc-pVTZ & -12.340 & 0.000 \\
58 & pyridine$\cdots$pyridine (CH$\cdots$N) & cc-pVDZ & -21.213 & 0.000 \\
58 & pyridine$\cdots$pyridine (CH$\cdots$N) & cc-pVTZ & -8.001 & -0.000 \\
\bottomrule
\end{longtable}
\end{landscape}

\section{Reproducibility notes}
\label{sec:si_reproducibility}

For each entry in the monomer, energy-difference, and artificial-pair analyses, the required numerical inputs are the MP2, CCSD, and CCSD(T) correlation energies computed in the same one-particle basis and with the same frozen-core convention.  The final representative \MERCE{} model first checks the monotonic input ordering, computes $d_1$, $d_2$, and $R$, assigns effective ranks using the fixed global parameters, fits the inverse-cubic-linear form, and evaluates the chosen target rank.  Systems that fail the monotonicity, numerical-stability, or rational-branch checks should be reported as outside the current model domain rather than forced through the extrapolation.

The scripts used to generate the numerical tables preserve some internal labels from the development stage.  In particular, \texttt{train\_frozen} and \texttt{train\_fullcore} correspond to the FCI-frozen-core and FCI-full-core developmental sets discussed in the main text.  The final representative model is identified by the internal variant name
\begin{equation}
\texttt{MP2\_1.85\_CCSDpT\_2.40\_k1\_0.700\_k3\_4.250}
\end{equation}
and by the inverse-cubic-linear function
\begin{equation}
\Ecorr(x)=A+\frac{B}{x^3+Cx}.
\end{equation}
The fitted coefficients $A$, $B$, and $C$ are system dependent, while the mapping parameters are fixed globally.

The versioned repository contains (i) the standalone final calculator and unit
tests; (ii) all tabulated MP2, CCSD, CCSD(T), and reference energies used in
the analyses; (iii) the original and expanded model-search programs and
authoritative machine-readable outputs; (iv) figure-generation scripts and
source tables; (v) complete MRCC inputs and normally terminated outputs for
all 107 FCI developmental records and the ten neutral--cation pairs used for
ionization potentials; (vi) all 18 CEPB DLPNO-CCSDTQ input/output pairs,
together with the 21-row source spreadsheet and an explicit
included/excluded-row manifest; (vii) the A24 and selected S66 component
records; and (viii) representative MRCC, original-build Psi4, repaired-build
Psi4, and PySCF inputs. A validation script checks the raw-to-tabulated
correspondence, including every one of the five energy columns retained from
the CEPB spreadsheet. The repository also contains the repaired Psi4 source snapshot, patch,
applicable LGPL notices, checksums, build record, smoke test, and CF$_4$
software-regression output described in
Section~\ref{sec:si_electronic_structure_provenance}.

The implementation, scripts, and numerical records are archived in the
versioned Zenodo repository at
\url{https://doi.org/10.5281/zenodo.22975564}.

\bibliographystyle{unsrtnat}
\bibliography{MERCE_references}